\documentclass[11pt,a4paper]{article}
\pdfoutput=1
\usepackage{amsfonts}
\usepackage{dsfont}
\usepackage{amssymb}
\usepackage[centertags]{amsmath}
\usepackage{graphicx}
\usepackage{hyperref}
\usepackage{booktabs}
\usepackage{theorem}
\usepackage{array}
\usepackage{epsfig}
\usepackage{colordvi}
\usepackage{color}
\usepackage{ulem}
\usepackage[small,bf]{caption}
\usepackage{cite}
\usepackage{color}
\usepackage{subfigure}
\usepackage{pdflscape}
\usepackage{multirow}
\usepackage{enumitem}
\usepackage{bbm}
\def\1{\mathbf{1}}
\def\3{\mathbf{3}}
\def\2{\mathbf{2}}

\allowdisplaybreaks[1]

\usepackage{a4wide}
\newcommand{\bec}{\begin{cases}}
\newcommand{\eec}{\end{cases}}
\newcommand{\beq}{\begin{equation*}}
\newcommand{\eeq}{\end{equation*}}
\newcommand{\be}{\begin{equation}}
\newcommand{\ee}{\end{equation}}
\newcommand{\ba}{\begin{eqnarray}}
\newcommand{\ea}{\end{eqnarray}}

\DeclareMathOperator{\diag}{diag}

\makeatletter

\newcommand{\Rmnum}[1]{\expandafter\@slowromancap\romannumeral #1@}
\makeatother

\begin{document}

\begin{titlepage}

\vspace*{-15mm}
\begin{flushright}
SISSA 39/2015/FISI\\
IPMU15-0161\\
arXiv:1509.02502
\end{flushright}
\vspace*{0.7cm}

\begin{center}
{\bf\Large {Leptonic Dirac CP Violation Predictions from }}\\ 
[4mm]
{\bf\Large {Residual Discrete Symmetries }}\\
[5.5mm]
\vspace{0.4cm} I.~Girardi$\mbox{}^{a)}$, S.~T.~Petcov$\mbox{}^{a,b)}$,\footnote{Also at: Institute of Nuclear Research and Nuclear Energy,
Bulgarian Academy of Sciences, 1784 Sofia, Bulgaria.}
Alexander J.~Stuart$\mbox{}^{a)}$   and A.~V.~Titov$\mbox{}^{a)}$ 
\\[1mm]
\end{center}
\vspace*{0.5cm}
\centerline{$^{a}$ \it SISSA/INFN, Via Bonomea 265, 34136 Trieste, Italy}
\vspace*{0.2cm}
\centerline{$^{b}$ \it Kavli IPMU (WPI), University of Tokyo,
5-1-5 Kashiwanoha, 277-8583 Kashiwa, Japan}
\vspace*{1.20cm}

\begin{abstract}
\noindent
Assuming that the observed pattern of 3-neutrino mixing is related 
to the existence of a (lepton) flavour symmetry, corresponding to 
a non-Abelian discrete symmetry group $G_f$, and that $G_f$ is broken 
to specific residual symmetries $G_e$ and $G_\nu$ of 
the charged lepton and neutrino mass terms, 
we derive sum rules for the cosine of the Dirac phase 
$\delta$ of the neutrino mixing matrix $U$. 
The residual symmetries considered are:
i) $G_e = Z_2$ and $G_{\nu} = Z_n$, $n > 2$ or $Z_n \times Z_m$, $n,m \geq 2$;
ii) $G_e = Z_n$, $n > 2$ or $Z_n \times Z_m$, $n,m \geq 2$ and 
$G_{\nu} = Z_2$;
iii) $G_e = Z_2$ and $G_{\nu} = Z_2$;
iv) $G_e$ is fully broken and $G_{\nu} = Z_n$, $n > 2$ or 
$Z_n \times Z_m$, $n,m \geq 2$; and 
v) $G_e = Z_n$, $n > 2$ or $Z_n \times Z_m$, 
$n,m \geq 2$ and $G_{\nu}$ is fully broken. 
For given $G_e$ and $G_\nu$, the sum rules for $\cos\delta$ 
thus derived are exact, within the approach employed, 
and are valid, in particular, for any $G_f$ containing 
$G_e$ and $G_\nu$ as subgroups. 
We identify the cases when the value of $\cos\delta$ 
cannot be determined, or cannot be uniquely determined,  
without making additional assumptions on unconstrained parameters. 
In a large class of cases considered the value of $\cos\delta$ 
can be unambiguously predicted once the flavour symmetry $G_f$ 
is fixed. We present predictions for $\cos\delta$ in these cases  
for the flavour symmetry groups $G_f = S_4$, $A_4$, $T^\prime$ and $A_5$, 
requiring that the measured values of the 3-neutrino mixing parameters 
$\sin^2\theta_{12}$, $\sin^2\theta_{13}$ and $\sin^2\theta_{23}$, 
taking into account their respective $3\sigma$ uncertainties, are 
successfully reproduced.
\end{abstract}

\vspace{0.5cm}

\end{titlepage}
\setcounter{footnote}{0}

%%%%%%%%%%%%%%%%%%%%%%%%%%%%%%%%%
\section{Introduction}
\label{sec:intr}
%%%%%%%%%%%%%%%%%%%%%%%%%%%%%%%%%

 The discrete symmetry approach to understanding the observed 
pattern of 3-neutrino mixing (see, e.g., \cite{PDG2014}), 
which is widely explored at present (see, e.g., 
\cite{Ishimori:2010au,Altarelli:2010gt,King:2013eh, King:2014nza}),
leads to specific correlations between the values of 
at least some of the mixing angles 
of the Pontecorvo, Maki, Nakagawa, Sakata (PMNS) 
neutrino mixing matrix $U$ and,  
either to specific fixed trivial or maximal values of 
the CP violation (CPV) phases present in  $U$ 
(see, e.g., 
\cite{King:2013vna,Ballett:2015wia,Hagedorn:2014wha,Li:2015jxa,DiIura:2015kfa}
and references quoted therein), or 
to a correlation between the values of the neutrino 
mixing angles and of the Dirac CPV phase of $U$  
\cite{Marzocca:2013cr,Petcov:2014laa,Girardi:2014faa,Girardi:2015vha,Turner:2015uta}.\footnote{In the case of massive neutrinos being Majorana particles
one can obtain under specific conditions also 
correlations between the values of the two Majorana 
CPV phases present in the neutrino mixing matrix 
\cite{Bilenky:1980cx} and of the three neutrino mixing angles 
and of the Dirac CPV phase \cite{Petcov:2014laa,Girardi:2013sza}.} 
As a consequence of this correlation the cosine of the 
Dirac CPV phase $\delta$ of the PMNS matrix $U$ 
can be expressed in terms of the three 
neutrino mixing angles of $U$ 
\cite{Marzocca:2013cr,Petcov:2014laa,Girardi:2014faa,Girardi:2015vha}, 
i.e., one obtains a sum rule for $\cos\delta$. 
This sum rule depends on 
the underlying discrete symmetry used to 
derive the observed pattern of neutrino mixing and  
on the type of breaking of the symmetry necessary to 
reproduce the measured values of the neutrino mixing angles. 
It depends also on the assumed status  
of the CP symmetry before the breaking of 
the underlying discrete symmetry.   

   The approach of interest is based on the assumption 
of the existence at some energy scale of a (lepton) flavour  
symmetry corresponding to a non-Abelian discrete group $G_f$. 
Groups that have been considered in the literature 
include $S_4$, $A_4$, $T'$, $A_5$, $D_{n}$ (with $n=10,12$) 
and $\Delta(6n^2)$, to name several. 
The choice of these groups is related to the fact that 
they lead to values of the neutrino mixing angles,  
which can differ from the measured values at most 
by subleading perturbative corrections.
For instance, the groups $A_4$, $S_4$ and $T'$ are commonly utilised to generate
tri-bimaximal (TBM) mixing \cite{TBM}; the group $S_4$ can also be used to 
generate  bimaximal (BM) mixing  \cite{BM};\footnote{Bimaximal mixing can also be 
a consequence of 
the conservation of the lepton charge
$L' = L_e - L_{\mu} - L_{\tau}$ (LC) \cite{SPPD82}, 
supplemented by a $\mu - \tau$ symmetry.} 
$A_5$ can be utilised to generate golden ratio type A (GRA) 
\cite{Datta:2003qg,Everett:2008et,GRAM} mixing;
and the groups $D_{10}$ and $D_{12}$ can lead to  golden ratio type B (GRB) \cite{GRBM} 
and hexagonal (HG) \cite{HGM} mixing.

   The flavour symmetry group $G_f$ can be broken, in general, 
to different symmetry subgroups $G_e$ and $G_{\nu}$ 
of the charged lepton and neutrino mass terms,
respectively. 
$G_e$ and $G_{\nu}$ are usually called ``residual symmetries'' 
of the charged lepton and neutrino mass matrices.  
Given $G_f$, which is usually assumed to be discrete, 
typically there are more than one 
(but still a finite number of) possible 
residual symmetries $G_e$ and $G_{\nu}$.
The subgroup $G_e$, in particular, can be trivial, i.e., 
$G_f$ can be completely broken in the process of 
generation of the charged lepton mass term.

 The residual symmetries can constrain the forms of the 
$3\times 3$ unitary matrices $U_{e}$ and $U_{\nu}$, 
which diagonalise the charged lepton and 
neutrino mass matrices, 
and the product of which represents the 
PMNS matrix:
%%%%%%%%%%%%%%%%%%%%%%%%%%%%%%%%%%%%%%
\begin{equation}
U = U_e^{\dagger}\, U_{\nu}\,.
\label{PMNS1}
\end{equation}
%%%%%%%%%%%%%%%%%%%%%%%%%%
%
Thus, by constraining the form of the matrices 
$U_{e}$ and $U_{\nu}$, the residual symmetries 
constrain also the form of the PMNS matrix $U$.

In general, there are two cases of residual 
symmetry $G_\nu$ for the neutrino Majorana mass term
when a portion of $G_f$ is left unbroken in the neutrino sector. 
They characterise two possible
approaches~---~direct and semi-direct \cite{King:2013eh}~--- 
in making predictions for the 
neutrino mixing observables using 
discrete flavour symmetries:   $G_\nu$ 
can either be a $Z_2\times Z_2$ symmetry    
(which sometimes is identified in the literature 
with the Klein four group),
or a $Z_2$ symmetry. In models 
based on the semi-direct approach, where 
$G_\nu = Z_2$, the matrix $U_{\nu}$ 
contains two free parameters, i.e., one angle and one phase,
as long as the neutrino Majorana mass term does not 
have additional ``accidental'' symmetries, 
e.g., the $\mu - \tau$ symmetry.
In such a case as well  as in the case of 
$G_\nu = Z_2\times Z_2$, the matrix 
$U_{\nu}$ is completely determined by symmetries  
up to  re-phasing on the right and permutations of columns.  
The latter can be fixed by considering a specific model.  
It is also important to note here that in this approach Majorana 
phases are undetermined. 

  In the general case of absence of constraints,  
the PMNS matrix can be parametrised 
in terms of the parameters of $U_{e}$ and $U_{\nu}$ 
as follows \cite{Frampton:2004ud}:
%%%%%%%%%%%%%%%%%%%%%%%%%%%%%%
\begin{equation}
U =  U_e^{\dagger}\, U_{\nu} = 
(\tilde{U}_{e})^\dagger\, \Psi \tilde{U}_{\nu} \, Q_0\,.
\label{PMNS2}
\end{equation}
%%%%%%%%%%%%%%%%%%%%%%%%%%%%
%
Here $\tilde{U}_e$ and $\tilde{U}_\nu$ 
are CKM-like $3\times 3$ unitary matrices 
and  $\Psi$ and $Q_0$ are given by:
%%%%%%%%%%%%%%%%%%%%%%%%
\begin{equation} 
\Psi =
{\rm diag} \left(1,\text{e}^{-i \psi}, \text{e}^{-i \omega} \right)\,,~~
Q_0 = {\rm diag} \left(1,\text{e}^{i \frac{\xi_{21}}{2}}, 
\text{e}^{i \frac{\xi_{31}}{2}} \right)\,,
\label{PsieQ0}
\end{equation}
%%%%%%%%%%%%%%%%%%%%%%%%%%%%%%
%
where $\psi$, $\omega$, $\xi_{21}$ and $\xi_{31}$  are phases which contribute to physical CPV phases.
Thus, in general, each of the two phase matrices 
$\Psi$ and $Q_0$ contain two physical CPV phases.  
The phases in $Q_0$ contribute to the Majorana phases \cite{Bilenky:1980cx} 
in the PMNS matrix (see further)
and can appear in 
eq.~(\ref{PMNS2}) as a result of the diagonalisation of 
 the neutrino Majorana mass term, while 
the phases in  $\Psi$ can result 
from the charged lepton sector 
($U_e^{\dagger} = (\tilde{U}_{e})^\dagger\, \Psi$),
from the neutrino sector  
($U_{\nu} = \Psi \tilde{U}_{\nu} Q_0$), 
or can receive contributions from both sectors.

As is well known, the $3\times 3$ unitary PMNS matrix $U$ can be parametrised
in terms of three neutrino mixing angles and, 
depending on whether the massive neutrinos are Dirac or 
Majorana particles, by one Dirac CPV phase, or by one Dirac 
and two Majorana \cite{Bilenky:1980cx} CPV phases:
%%%%%%%%%%%%%%%%%%%%%%%%%%%%%%%%%%%%%%
\begin{equation}
U = U_e^{\dagger}\, U_{\nu} = VQ\,,~~
Q = {\rm diag}\left(1, e^{i \frac{\alpha_{21}}{2}}, e^{i \frac{\alpha_{31}}{2}}\right)\,,
\label{eq:VQ}
\end{equation}
%%%%%%%%%%%%%%%%%%%%%%%%%%
%
where $\alpha_{21,31}$ are the two Majorana CPV phases 
and $V$ is a CKM-like matrix. In the standard 
parametrisation of the PMNS matrix \cite{PDG2014}, 
which we are going to use in what follows, 
$V$ has the form:  
%%%%%%%%%%%%%%%%%%%%%%%%%%%%%%%%%%%
\begin{equation}
\begin{array}{c}
\label{eq:Vpara}
V = \left(\begin{array}{ccc}
 c_{12} c_{13} & s_{12} c_{13} & s_{13} e^{-i \delta}  \\[0.2cm]
 -s_{12} c_{23} - c_{12} s_{23} s_{13} e^{i \delta}
 & c_{12} c_{23} - s_{12} s_{23} s_{13} e^{i \delta}
 & s_{23} c_{13} 
\\[0.2cm]
 s_{12} s_{23} - c_{12} c_{23} s_{13} e^{i \delta} &
 - c_{12} s_{23} - s_{12} c_{23} s_{13} e^{i \delta}
 & c_{23} c_{13} 
\\
  \end{array}
\right)\,,
\end{array}
\end{equation}
%%%%%%%%%%%%%%%%%%%%%%%%%%%%%%%%%
%
\noindent
where
$0 \leq \delta \leq 2\pi$ is the Dirac CPV phase and
we have used the standard notation
$c_{ij} = \cos\theta_{ij}$,
$s_{ij} = \sin\theta_{ij}$ with
$0 \leq  \theta_{ij} \leq \pi/2$.  Notice that if CP invariance holds, 
then we have
$\delta =0,\pi,2\pi$, with
the values 0 and $2\pi$ being physically
indistinguishable, and 
$\alpha_{21} = k\pi$, $\alpha_{31} = k'\pi$,
$k,k'=0,1,2$~\footnote{In the case of the type I seesaw 
mechanism of neutrino mass generation
the range in which  $\alpha_{21}$ 
and $\alpha_{31}$ vary is 
$[0,4\pi]$ \cite{Molinaro:2008rg}. Thus, 
in this case  $\alpha_{21}$ 
and $\alpha_{31}$ possess CP-conserving values for 
$k,k'=0,1,2,3,4$.}.
Therefore, the neutrino mixing observables 
are the three mixing angles, $\theta_{12}$, $\theta_{13}$, 
$\theta_{23}$, the Dirac phase $\delta$ and, 
if the massive neutrinos are Majorana particles, the 
Majorana phases $\alpha_{21}$ and $\alpha_{31}$.
 
 The neutrino mixing parameters
$\sin^2\theta_{12}$, $\sin^2\theta_{23}$ and $\sin^2\theta_{13}$,
which will be  relevant for our further discussion,
have been determined with a relatively good 
precision in the recent global analyses of the neutrino 
oscillation data \cite{Capozzi:2013csa,Gonzalez-Garcia:2014bfa}.
For the best fit values and the 
3$\sigma$ allowed ranges of 
$\sin^2\theta_{12}$, $\sin^2\theta_{23}$ and $\sin^2\theta_{13}$,
the authors of ref. \cite{Capozzi:2013csa} have obtained:
%%%%%%%%%%%%%%%%%%%%%%%
\begin{eqnarray}
\label{th12values}
(\sin^2 \theta_{12})_{\rm BF} = 0.308\,,~~~~
 0.259 \leq \sin^2 \theta_{12} \leq 0.359\,,\\ [0.30cm]
\label{th23values}
(\sin^2\theta_{23})_{\rm BF} = 0.437~(0.455)\,,~~~~
 0.374~(0.380) \leq \sin^2\theta_{23} \leq 0.626~(0.641)\,,\\[0.30cm]
\label{th13values}
(\sin^2\theta_{13})_{\rm BF} = 0.0234~(0.0240)\,,~~~~
0.0176~(0.0178) \leq \sin^2\theta_{13} \leq 0.0295~(0.0298)\,.
\end{eqnarray}
%%%%%%%%%%
%
Here the values (values in brackets)
correspond to neutrino mass spectrum with normal ordering
(inverted ordering) (see, e.g., \cite{PDG2014}),
denoted further as the NO (IO) spectrum.

In ref.~\cite{Petcov:2014laa} 
(see also \cite{Marzocca:2013cr,Girardi:2014faa,Girardi:2015vha}) we have considered 
the cases when, as a consequence of 
underlying and residual symmetries, 
the matrix $U_{\nu}$, and more 
specifically, the matrix $\tilde{U}_{\nu}$ 
in eq.~(\ref{PMNS2}), has the 
i) TBM, ii) BM, iii) GRA, iv) GRB and 
v) HG forms. For all these forms 
we have $\tilde{U}_{\nu} = R_{23}(\theta^\nu_{23})
R_{12}(\theta^\nu_{12})$ with 
$\theta^\nu_{23} = -\,\pi/4$, $R_{23}$ and $R_{12}$ 
being $3\times 3$ orthogonal matrices 
describing rotations in the 2-3 and 1-2 planes:
%%%%%%%%%%%%%%%%%%%%%%%%%%%%%%%%%%%%%%%%%
\begin{equation}
\tilde{U}_{\nu} =  R_{23} \left ( \theta^\nu_{23} \right)
R_{12}\left( \theta^\nu_{12}\right) = 
 \begin{pmatrix}
\cos \theta^{\nu}_{12} & \sin \theta^{\nu}_{12} & 0 \vspace{0.2cm} \\
- \dfrac{\sin \theta^{\nu}_{12}}{\sqrt{2}} &
\dfrac{\cos \theta^{\nu}_{12}}{\sqrt{2}} &
- \dfrac{1}{\sqrt{2}} \vspace{0.2cm} \\
- \dfrac{\sin \theta^{\nu}_{12}}{\sqrt{2}}  &
\dfrac{\cos \theta^{\nu}_{12}}{\sqrt{2}} &
\dfrac{1}{\sqrt{2}}
\end{pmatrix} \;.
\label{Unu1}
\end{equation}
%%%%%%%%%%%%%%%%%%%%%%%%%
%
The value of the angle  $\theta^{\nu}_{12}$, and thus of  
$\sin^2\theta^{\nu}_{12}$, depends on the form of $\tilde{U}_{\nu}$.
For the TBM, BM, GRA, GRB and HG forms we have: 
i)  $\sin^2\theta^{\nu}_{12} = 1/3$ (TBM),
ii)  $\sin^2\theta^{\nu}_{12} = 1/2$ (BM),
iii)  $\sin^2\theta^{\nu}_{12} =  (2 + r)^{-1} \cong 0.276$ (GRA),
$r$ being the golden ratio, $r = (1 +\sqrt{5})/2$,
iv) $\sin^2\theta^{\nu}_{12} = (3 - r)/4 \cong 0.345$ (GRB), and
v) $\sin^2\theta^{\nu}_{12} = 1/4$ (HG).

  The TBM form of $\tilde{U}_{\nu}$, for example,  
can be obtained from a $G_f = A_4$ symmetry,
when the residual symmetry is $G_\nu = Z_2$, i.e.~the $S$ generator of 
$A_4$ (see Appendix \ref{app:A4}) is unbroken. In this case there is an additional accidental 
$\mu-\tau$ symmetry, which together with the $Z_2$ symmetry 
leads to the TBM form of $\tilde{U}_{\nu}$ (see, e.g., \cite{Altarelli:2010gt}).
The TBM  form can also be derived from $G_f = T'$ with $G_\nu = Z_2$, provided that 
the left-handed (LH) charged leptons and neutrinos each transform as triplets  of 
$T^{\prime}$ and the  $TST^2$ element of $T^{\prime}$ is unbroken, 
see Appendix~\ref{app:A4} for further explanation.
Indeed when working with 3-dimensional and 1-dimensional
representations of $T^{\prime}$, there is no way to distinguish 
$T^{\prime}$ from $A_4$ \cite{Feruglio:2007uu}.
Finally, one can obtain BM mixing from, e.g., the $G_f = S_4$ symmetry, 
when the residual symmetry is $G_{\nu} = Z_2$. 
There is an accidental $\mu-\tau$ symmetry in this case as well
\cite{Altarelli:2009gn}.

 For all the forms of $\tilde{U}_{\nu}$ 
considered in \cite{Petcov:2014laa} and listed 
above we have i) $\theta^{\nu}_{13} = 0$, which should be corrected 
to the measured value of $\theta_{13} \cong 0.15$, 
and ii) $\sin^2\theta^{\nu}_{23} = 0.5$, which might also need
to be corrected if it is firmly established that 
$\sin^2\theta_{23}$ deviates significantly from 0.5.
In the case of the BM and HG forms, the values of 
$\sin^2\theta^{\nu}_{12}$ lie outside the current 
$3\sigma$ allowed ranges of $\sin^2\theta_{12}$ and 
have also to be corrected.

The requisite corrections 
are provided by the matrix $U_e$, or equivalently, by $\tilde{U}_e$. 
The approach followed in 
\cite{Petcov:2014laa,Marzocca:2013cr,Girardi:2014faa,Girardi:2015vha}
corresponds to the case of a trivial subgroup $G_e$, i.e., 
of $G_f$ completely broken by the charged lepton mass term.
In this case the matrix $\tilde{U}_e$ is unconstrained 
and was chosen in \cite{Petcov:2014laa} on phenomenological grounds 
to have the following two forms: 
%%%%%%%%%%%%%%%%%%%%%%%%%%%%%%%
\begin{eqnarray}
\label{Ue12}
\tilde{U}_{e} = R^{-1}_{12}(\theta^e_{12})\,,\\[0.30cm]
\tilde{U}_{e} = R^{-1}_{23}(\theta^e_{23})\,R^{-1}_{12}(\theta^e_{12})\,.
\label{Ue2312}
\end{eqnarray}
%%%%%%%%%%%%%%%%%%%%%%%%
%
These two forms appear in a large class of theoretical models 
of flavour and theoretical studies, in which the generation of charged lepton masses
is an integral part (see, e.g., \cite{Gehrlein:2014wda,Meroni:2012ty,
Marzocca:2011dh,Antusch:2012fb,Chen:2009gf,Girardi:2013sza,Marzocca:2014tga}).

In this setting with $\tilde{U}_{\nu}$ having one 
of the five symmetry forms, TBM, BM, GRA, GRB and HG, and 
$\tilde{U}_{e}$ given by eq.~(\ref{Ue2312}),
the Dirac phase $\delta$ of the PMNS matrix 
was shown in \cite{Petcov:2014laa} to satisfy 
the following sum rule:\footnote{The sum rule is given 
in the standard parametrisation 
of the PMNS matrix (see, e.g., \cite{PDG2014}).}
%%%%%%%%%%%%%%%%%%%%%%%%%%%%%%%%%%%%
\begin{equation}
\cos\delta =  \frac{\tan\theta_{23}}{\sin2\theta_{12}\sin\theta_{13}}\,
\left [\cos2\theta^{\nu}_{12} + 
\left (\sin^2\theta_{12} - \cos^2\theta^{\nu}_{12} \right )\,
 \left (1 - \cot^2\theta_{23}\,\sin^2\theta_{13}\right )\right ]\,.
\label{cosdthnu}
\end{equation}
%%%%%%%%%%%%%%%%%%%%%%%%%%%%%%%%%%%%
%
Within the approach employed this sum rule is exact.\footnote{For the TBM and BM forms of $\tilde{U}_{\nu}$,  
and for $\tilde{U}_{e}$ given in eq.~(\ref{Ue2312}), 
it was first derived in ref. \cite{Marzocca:2013cr}.}
It is valid, in particular, for any value of the angle $\theta^{\nu}_{23}$ \cite{Girardi:2015vha}.\footnote{The two forms of  $\tilde{U}_{e}$
in eqs.~(\ref{Ue12}) and 
(\ref{Ue2312}) lead, in particular, 
to different predictions  
for $\sin^2\theta_{23}$: for $\theta^{\nu}_{23} = -\pi/4$ in the case of 
eq.~(\ref{Ue12}) we have 
$\sin^2\theta_{23} \cong 0.5(1-\sin^2\theta_{13})$,
while if $\tilde{U}_{e}$ is given by eq.~(\ref{Ue2312}), 
$\sin^2\theta_{23}$ can deviate significantly from 0.5.}
In \cite{Petcov:2014laa}, by using the sum rule in eq.~(\ref{cosdthnu}), 
predictions for $\cos\delta$ and $\delta$ 
were obtained in the TBM, BM, GRA, GRB and HG cases 
for the best fit values of $\sin^2\theta_{12}$, 
 $\sin^2\theta_{23}$ and  $\sin^2\theta_{13}$.
The results thus obtained permitted to conclude that 
a sufficiently precise measurement of $\cos\delta$ 
would allow to discriminate between the different forms 
of $\tilde{U}_{\nu}$ considered. 

Statistical analyses of predictions of the sum rule 
given in eq.~(\ref{cosdthnu}) 
i) for $\delta$ and for the $J_{\rm CP}$ factor,  
which determines the magnitude of CP-violating effects 
in neutrino oscillations \cite{PKSP3nu88}, 
using the current uncertainties 
in the determination of $\sin^2\theta_{12}$, $\sin^2\theta_{13}$, 
$\sin^2\theta_{23}$ and $\delta$ from \cite{Capozzi:2013csa},  
and ii) for $\cos\delta$ 
using the prospective uncertainties  
on $\sin^2\theta_{12}$, $\sin^2\theta_{13}$ and $\sin^2\theta_{23}$, 
were performed in \cite{Girardi:2014faa} 
for the five symmetry forms~---~BM (LC), 
TBM, GRA, GRB and HG~---~of $\tilde{U}_{\nu}$. 

 In \cite{Girardi:2015vha} we extended the 
analyses performed in 
\cite{Petcov:2014laa,Girardi:2014faa} by obtaining  
sum rules for $\cos\delta$ for the following forms of 
the matrices $\tilde{U}_e$ and  $\tilde{U}_\nu$:\footnote{We performed in \cite{Girardi:2015vha}
a systematic analysis of the forms of 
$\tilde{U}_e$ and  $\tilde{U}_\nu$, 
for which sum rules for $\cos\delta$ of the 
type of eq.~(\ref{cosdthnu}) could be derived,
but did not exist in the literature.} 
\begin{itemize}
\item[A.] $\tilde{U}_\nu = R_{23}(\theta^{\nu}_{23})R_{12}(\theta^{\nu}_{12})$ 
with  $\theta^{\nu}_{23} = -\pi/4$ and $\theta^{\nu}_{12}$
as dictated by TBM, BM, GRA, GRB or HG mixing, and 
i)  $\tilde{U}_e =  R^{-1}_{13}(\theta^{e}_{13})$, 
ii) $\tilde{U}_e =  R^{-1}_{23}(\theta^{e}_{23})R^{-1}_{13}(\theta^{e}_{13})$,
and iii) $\tilde{U}_e = R^{-1}_{13}(\theta^{e}_{13})
R^{-1}_{12}(\theta^{e}_{12})$; 
\item[B.] $\tilde{U}_\nu = R_{23}(\theta^{\nu}_{23})R_{13}(\theta^{\nu}_{13}) 
 R_{12}(\theta^{\nu}_{12})$ with $\theta^{\nu}_{23}$, 
$\theta^{\nu}_{13}$ and $\theta^{\nu}_{12}$ 
fixed by arguments associated with symmetries, and 
iv) $\tilde{U}_e = R^{-1}_{12}(\theta^{e}_{12})$, 
and v) $\tilde{U}_e = R^{-1}_{13}(\theta^{e}_{13})$.
\end{itemize}
The sum rules for $\cos\delta$ were derived 
first for $\theta^{\nu}_{23} = -\,\pi/4$
for the cases listed in point A, 
and for the specific values of (some of) the angles 
in  $\tilde U_{\nu}$, characterising 
the cases listed in point B, as well as for 
arbitrary fixed values of all angles 
contained in $\tilde U_{\nu}$.
Predictions for $\cos\delta$ and $J_{\rm CP}$
($\cos\delta$) were also obtained 
performing statistical analyses 
utilising the  current (the prospective) uncertainties 
in the determination of 
$\sin^2\theta_{12}$, $\sin^2\theta_{13}$, $\sin^2\theta_{23}$ and $\delta$ 
($\sin^2\theta_{12}$, $\sin^2\theta_{13}$ and $\sin^2\theta_{23}$). 

In the present article we extend the analyses performed in 
\cite{Petcov:2014laa,Girardi:2014faa,Girardi:2015vha}
to the following cases:
\begin{enumerate}[label=\theenumi)]
\item $G_e = Z_2$ and $G_{\nu} = Z_n$, $n > 2$ or $Z_n \times Z_m$, $n,m \geq 2$;
\item $G_e = Z_n$, $n > 2$ or $Z_n \times Z_m$, $n,m \geq 2$ and $G_{\nu} = Z_2$;
\item $G_e = Z_2$ and $G_{\nu} = Z_2$;
\item $G_e$ is fully broken and $G_{\nu} = Z_n$, $n > 2$ or $Z_n \times Z_m$, $n,m \geq 2$;
\item $G_e = Z_n$, $n > 2$ or $Z_n \times Z_m$, $n,m \geq 2$ and $G_{\nu}$ is fully broken.
\end{enumerate}
In the case of $G_e = Z_2$ ($G_{\nu} = Z_2$) the matrix $U_e$ ($U_{\nu}$) is determined up to a
$U(2)$ transformation in the degenerate subspace, since the representation
matrix of the generator of the residual symmetry has degenerate eigenvalues.
On the contrary, when the residual symmetry is large enough, namely,
$G_e = Z_n$, $n > 2$ or $Z_n \times Z_m$, $n,m \geq 2$
and $G_{\nu} = Z_2 \times Z_2$
($G_{\nu} = Z_n$, $n > 2$ or $Z_n \times Z_m$, $n,m \geq 2$) for Majorana
(Dirac) neutrinos, the matrices $U_e$ and $U_{\nu}$ are fixed
(up to diagonal phase matrices on the right, which are either unphysical
for Dirac neutrinos, or contribute to the Majorana phases otherwise, and permutations of columns)
by the residual symmetries of the charged lepton and neutrino mass matrices.
In the case when the discrete symmetry $G_f$ is fully broken in one
of the two sectors, the corresponding mixing matrix $U_e$ or $U_{\nu}$ is unconstrained and 
contains in general three angles and six phases.

Our article is organised as follows.
In Section~\ref{sec:prelim} we describe the parametrisations 
of the PMNS matrix depending on the residual symmetries
$G_e$ and $G_{\nu}$ considered above.
In Sections~\ref{sec:GeZGnuZZ}, \ref{sec:GeZorZZGnuZ}
and \ref{sec:GeZ2GnuZ2} we consider the breaking patterns
1), 2), 3) and derive sum rules for $\cos \delta$.
At the end of each of these sections we present 
numerical predictions for $\cos \delta$ in the cases
of the flavour symmetry groups $G_f = A_4$, $T^{\prime}$, $S_4$ and $A_5$.
In Section~\ref{sec:sumcom} we provide a summary of the
sum rules derived in Sections~\ref{sec:GeZGnuZZ}~--~\ref{sec:GeZ2GnuZ2}.
Further, in Sections~\ref{sec:Gebroken} and \ref{sec:Gnubroken}
we derive the sum rules for the cases 4) and 5), respectively.
In these cases the value of $\cos \delta$ cannot be fixed
without additional assumptions on the unconstrained matrix
$U_e$ or $U_{\nu}$. The cases studied in \cite{Girardi:2015vha}
belong to the ones considered in Section~\ref{sec:Gebroken},
where the particular forms of the matrix $U_e$,
leading to sum rules of interest, have been considered.
In Section~\ref{sec:predictions} we present the summary of 
the numerical results.
Section~\ref{sec:summary} contains the conclusions.
Appendices~\ref{app:A4}, \ref{app:ParU}, \ref{app:Gebroken}, \ref{app:A5andGCP}
and \ref{sec:genstatement}
contain technical details related to the study.

%%%%%%%%%%%%%%%%%%%%%%%%%%%%%%%%%
\section{Preliminary Considerations}
\label{sec:prelim}
%%%%%%%%%%%%%%%%%%%%%%%%%%%%%%%%%

 As was already mentioned in the Introduction, the residual symmetries 
of the charged lepton and neutrino mass matrices constrain the forms
of the matrices $U_e$ and $U_{\nu}$ and, thus, the form 
of the PMNS matrix $U$.
To be more specific, if the charged lepton mass term is written 
in the left-right convention,
the matrix $U_e$ diagonalises the hermitian matrix $M_e M^\dagger_e$, 
$U^\dagger_e M_e M^\dagger_e U_e = {\rm diag}(m^2_e,m^2_{\mu},m^2_{\tau})$, 
$M_e$ being the charged lepton mass matrix.
If $G_e$ is the residual symmetry group  
of $M_e M^\dagger_e$ we have:
%%%%%%%%%%%%%%%%%%%%%%%%%
\be
\rho(g_e)^{\dagger} M_e M_e^{\dagger} \rho(g_e) = M_e M_e^{\dagger} \,,
\label{GeMe}
\ee
%%%%%%%%%%%%%%%%%%%%%%%%%%
%
where $g_e$ is an element of $G_e$, 
$\rho$ is a unitary representation of 
$G_f$ and $\rho(g_e)$ gives 
the action of $G_e$ on the  
LH components 
of the charged lepton fields 
having as mass matrix $M_e$.
As can be seen from eq.~(\ref{GeMe}),
the matrices $\rho(g_e)$ and $M_e M_e^{\dagger}$
commute, implying that they are diagonalised by the
same matrix $U_e$.

Similarly, if $G_\nu$ is the residual symmetry 
of the neutrino Majorana mass matrix $M_{\nu}$ one has:
%%%%%%%%%%%%%%%%%%%%%%%%%
\be
\rho(g_{\nu})^T M_{\nu} \rho(g_{\nu}) = M_{\nu} \,,
\label{GnuMnu}
\ee
%%%%%%%%%%%%%%%%%%%%%%%%%%
%
where $g_\nu$ is an element of $G_\nu$, 
$\rho$ is a unitary representation of 
$G_f$ under which the LH flavour neutrino fields 
$\nu_{lL}(x)$, $l=e,\mu,\tau$, transform,
and $\rho(g_{\nu})$ determines the action 
of $G_\nu$ on  $\nu_{lL}(x)$.
It is not difficult to show that also in this case the matrices 
$\rho(g_{\nu})$ and $M_{\nu}^{\dagger} M_{\nu}$~\footnote{The right-left convention
for the neutrino mass term is assumed.}
commute, and therefore they can be diagonalised
simultaneously by the same matrix $U_{\nu}$.
In the case of Dirac neutrinos eq.~(\ref{GnuMnu}) is modified 
as follows: 
\begin{equation}
\rho(g_{\nu})^{\dagger} M_{\nu}^{\dagger} M_{\nu} \rho(g_{\nu}) = M_{\nu}^{\dagger} M_{\nu}.
\end{equation}
The types of residual symmetries allowed in this case and discussed below are the same 
as those of the charged lepton mass term. 

In many cases studied in the literature 
(e.g., in the cases of $G_f = S_4$, $A_4$, $T'$, $A_5$) 
$\rho(g_f)$, $g_f$ being an element of $G_f$,
is assumed to be a 3-dimensional representation 
of $G_f$ because one aims at  unification
of the three flavours (e.g., three lepton families) 
at high energy scales, where the flavour symmetry 
group $G_f$ is unbroken.

  At low energies the flavour symmetry group $G_f$ has
necessarily to be broken to residual symmetries $G_e$ and $G_{\nu}$,
which act on the LH charged leptons and LH neutrinos as follows:
\begin{equation*}
l_L \rightarrow \rho(g_e) l_L \,,
\quad
\nu_{lL} \rightarrow \rho(g_{\nu}) \nu_{lL}\,,
\end{equation*}
where $g_e$ and $g_{\nu}$ are the elements of the
residual symmetry groups $G_e$ and $G_{\nu}$, respectively,
and $l_L = (e_L, \mu_L,\tau_L)^T$, $\nu_{lL} = (\nu_{eL},\nu_{\mu L},\nu_{\tau L})^T$. 
%
%%%%%%%%%%%%%%%%%%%%%%%%%%%%%%%%%%%%%%

The largest possible exact 
symmetry of the Majorana mass matrix $M_{\nu}$ 
having three non-zero and non-degenerate eigenvalues, 
is a $Z_2 \times Z_2 \times Z_2$ symmetry. The largest possible
exact symmetry of the Dirac mass matrix $M_e$ is $U(1)\times U(1)\times U(1)$.
Restricting ourselves to the case in which 
$G_f$ is a subgroup of $SU(3)$ instead of $U(3)$, 
the indicated largest possible exact symmetries reduce 
respectively to $Z_2 \times Z_2$ and  $U(1)\times U(1)$ 
because of the {\it special} determinant condition imposed from $SU(3)$.   
The residual symmetries $G_e$ and
$G_{\nu}$, being subgroups of $G_f$ (unless there are accidental symmetries), 
should also be contained in 
$U(1) \times U(1)$ and $Z_2 \times Z_2$ ($U(1) \times U(1)$) for massive 
Majorana (Dirac) neutrinos, respectively.

 If $G_e = Z_n$, $n > 2$ or $Z_n \times Z_m$, $n,m \geq 2$,
the matrix $U_e$ is fixed by the matrix $\rho(g_e)$ (up to multiplication by 
diagonal phase matrices on the right 
and permutations of columns), 
$U_e = U_e^{\circ}$. 
In the case of a smaller symmetry, i.e., $G_e = Z_2$, $U_e$
is defined up to a $U(2)$ transformation in the degenerate subspace,
because in this case $\rho(g_e)$ 
has two degenerate eigenvalues.
Therefore,
%%%%%%%%%%%%%%%%%%%%%%%%
\begin{equation*}
U_e = U_e^{\circ} U_{ij}(\theta^e_{ij},\delta^e_{ij}) \Psi_k \Psi_l\,,
\end{equation*}
%%%%%%%%%%%%%%%%%%%%%%%%
%
where $U_{ij}$ is a complex rotation in the $i$-$j$ plane and $\Psi_k$, $\Psi_l$ 
are diagonal phase matrices,
%%%%%%%%%%%%%%%%%%%%%%%%
\begin{equation} 
\Psi_1 = \diag \left(e^{i \psi_1},1,  1 \right)\,,
\quad
\Psi_2 = \diag \left(1,e^{i \psi_2}, 1 \right)\,,
\quad 
\Psi_3 = \diag \left(1, 1, e^{i \psi_3} \right)\,.
\label{eq:Psimatricesfree}
\end{equation}
%%%%%%%%%%%%%%%%%%%%%%%%%%%%%
%
The angle $\theta^e_{ij}$ and the phases 
$\delta^e_{ij}$, $\psi_1$, $\psi_2$ and $\psi_3$ are 
free parameters. As an example of the explicit form 
of $ U_{ij}(\theta^a_{ij},\delta^a_{ij})$,
we give the expression of the matrix $U_{12}  (\theta^a_{12}, \delta^a_{12})$:
%%%%%%%%%%%%%%%%%%%%%%%%%%%%%%
\be
U_{12}  (\theta^a_{12}, \delta^a_{12}) = \begin{pmatrix}
\cos \theta^a_{12} & \sin \theta^a_{12}  e^{-i \delta^a_{12}} & 0\\
- \sin \theta^a_{12}  e^{i \delta^a_{12}} & \cos \theta^a_{12} & 0\\
0 & 0 & 1 \end{pmatrix} \;,
\label{eq:U12comrot}
\ee
%%%%%%%%%%%%%%%%%%%%%%%%%%%%%
%
where $a = e,\nu, \circ$. The indices $e$, $\nu$ indicate the free parameters, 
while ``$\circ$" indicates the angles and phases which are fixed.
The complex rotation matrices  
$U_{23} (\theta^a_{23},\delta^a_{23})$ and $U_{13} (\theta^a_{13}, \delta^a_{13})$
are defined in an analogous way.
The real rotation matrices $R_{ij}(\theta^a_{ij})$ can be obtained
from $U_{ij}(\theta^a_{ij}, \delta^a_{ij})$ setting $\delta^a_{ij}$ to zero, i.e., 
$R_{ij}(\theta^a_{ij}) = U_{ij}(\theta^a_{ij}, 0)$.
In the absence of a residual symmetry no constraints are present
for the mixing matrix $U_e$, which can be in general expressed
in terms of three rotation angles and six phases. 

Similar considerations apply to the neutrino sector. 
If $G_{\nu} = Z_n$, $n > 2$ or $Z_n \times Z_m$, $n,m \geq 2$
for Dirac neutrinos, or $G_{\nu}= Z_2\times Z_2$ for Majorana neutrinos,
the matrix $U_{\nu}$ is fixed up to permutations of columns
and right multiplication by diagonal phase matrices by the residual 
symmetry, i.e., $U_{\nu} = U_{\nu}^{\circ}$.
If the symmetry is smaller, $G_{\nu} = Z_2$, then 
%%%%%%%%%%%%%%%%%%%%%%%%%%%%
\begin{equation*}
U_{\nu} = U_{\nu}^{\circ} U_{ij}(\theta^{\nu}_{ij},\delta^{\nu}_{ij}) \Psi_k \Psi_l\,.
\end{equation*}
%%%%%%%%%%%%%%%%%%%%%%%%%%%%
%
Obviously, in the absence of a residual symmetry, $U_{\nu}$ is unconstrained.  
In all the cases considered above where $G_e$
and $G_{\nu}$ are non-trivial, the matrices $\rho(g_e)$ and $\rho(g_{\nu})$
are diagonalised by $U_e^{\circ}$ and $U_{\nu}^{\circ}$:
%%%%%%%%%%%%%%%%%%%%%%%%%%%%%%%
\begin{equation*}
(U_e^{\circ})^{\dagger} \rho(g_e) U_e^{\circ} = \rho(g_e)^{\diag}
\quad \mbox{and} \quad
(U_{\nu}^{\circ})^{\dagger} \rho(g_{\nu}) U_{\nu}^{\circ} = \rho(g_{\nu})^{\diag}\,.
\end{equation*}
%%%%%%%%%%%%%%%%%%%%%%%%%%%%%%%%
%

In what follows we define $U^{\circ}$ as the matrix fixed by 
the residual symmetries,
which, in general, gets contributions from both the charged lepton and
neutrino sectors, $U^{\circ} = (U_e^{\circ})^{\dagger} U_{\nu}^{\circ}$.
Since $U^{\circ}$ is a unitary $3 \times 3$ matrix, we will parametrise  
it in terms of three angles and six phases. These, however,
as we are going to explain, reduce effectively to three angles
and one phase, since the other five phases 
contribute to the Majorana phases
of the PMNS mixing matrix, unphysical charged lepton phases 
and/or to a redefinition of
the free parameters contained in $U_e$ and $U_{\nu}$.
Furthermore, we will use the notation 
$\theta^e_{ij}$, $\theta^{\nu}_{ij}$, $\delta^e_{ij}$, $\delta^{\nu}_{ij}$ 
for the free angles and phases 
contained in $U$, while
the parameters marked with a circle contained in $U^{\circ}$, e.g., 
$\theta^{\circ}_{ij}$, $\delta^{\circ}_{ij}$,
are fixed by the residual symmetries.

In the case when $G_e = Z_2$ and $G_{\nu} = Z_n$, $n > 2$ or $Z_n \times Z_m$, $n,m \geq 2$
for massive Dirac neutrinos, or $G_{\nu} = Z_2\times Z_2$ for Majorana neutrinos, we have:
%%%%%%%%%%%%%%%%%%%%%%%%%%%%%%%%%%%
\begin{align}
U & = U_{ij}(\theta^e_{ij}, \delta^e_{ij}) \Psi_j^{\circ} U^\circ(\theta^{\circ}_{12}, \theta^{\circ}_{13}, \theta^{\circ}_{23}, \{ \delta^{\circ}_{kl} \}) Q_0 \nonumber \\
& = \Psi_j^{\circ} U_{ij}(\theta^e_{ij}, \delta^e_{ij} - \psi^{\circ}_j) U^\circ(\theta^{\circ}_{12}, \theta^{\circ}_{13}, \theta^{\circ}_{23}, \{ \delta^{\circ}_{kl} \}) Q_0 \,,
\label{eq:UZ2eKnu0}
\end{align}
%%%%%%%%%%%%%%%%%%%%%%%%%%%%
%
where $(ij) = (12), (13), (23)$ and 
$\{ \delta^{\circ}_{kl} \} = \{ \delta^{\circ}_{12}, \delta^{\circ}_{13}, \delta^{\circ}_{23} \}$. 
The unitary matrix $U^{\circ}$ contains three angles and three phases, since the 
additional three phases can be absorbed by redefining
the charged lepton fields and the free parameter $\delta^e_{ij}$ 
(see below).
Here $\Psi^{\circ}_j$ is a diagonal matrix containing a fixed phase 
in the $j$-th position. Namely,
%%%%%%%%%%%%%%%%%%%%%%%%
\begin{equation} 
\Psi_1^{\circ} = \diag \left(e^{i \psi_1^{\circ}},1,  1 \right)\,,
\quad
\Psi_2^{\circ} = \diag \left(1,e^{i \psi_2^{\circ}}, 1 \right)\,,
\quad 
\Psi_3^{\circ} = \diag \left(1, 1, e^{i \psi_3^{\circ}} \right)\,.
\label{eq:Psimatrices}
\end{equation}
%%%%%%%%%%%%%%%%%%%%%%%%%%%%%%
%
The matrix $Q_0$, defined in eq.~(\ref{PsieQ0}), is a diagonal matrix
containing two free parameters contributing to the Majorana phases.
Since the presence of the phase $\psi^{\circ}_j$ amounts to 
a redefinition of the free parameter $\delta^e_{ij}$, 
we denote $(\delta^e_{ij} - \psi^{\circ}_j)$ as $\delta^e_{ij}$. 
This allows us to employ the following parametrisation for $U$:
%%%%%%%%%%%%%%%%%%%%%%%%%%%%
\be
U = U_{ij}(\theta^e_{ij}, \delta^e_{ij}) U^\circ(\theta^{\circ}_{12}, \theta^{\circ}_{13}, \theta^{\circ}_{23}, \delta^{\circ}_{kl}) Q_0 \,,
\label{eq:UZ2eKnu}
\ee
%%%%%%%%%%%%%%%%%%%%%%%%%
%
where the unphysical phase matrix $\Psi^{\circ}_j$ on the left has been 
removed by charged lepton re-phasing 
and the set of three phases $\{ \delta^{\circ}_{kl}  \}$ reduces to only one phase, 
$\delta^{\circ}_{kl}$,  since the other two contribute to redefinitions of $Q_0$, $\delta^e_{ij}$
and to unphysical phases.
The possible forms of the matrix $U^{\circ}$, which we are going to
employ, are given in Appendix~\ref{app:ParU}.

For the breaking patterns $G_e = Z_n$, $n > 2$ or $Z_n \times Z_m$, $n,m \geq 2$
and $G_{\nu} = Z_2$, valid for both Majorana and Dirac neutrinos, we have:
%%%%%%%%%%%%%%%%%%%%%%%%%%%%%%%%%%%%%%%%%%
\begin{align}
 U & = U^\circ(\theta^{\circ}_{12}, \theta^{\circ}_{13}, \theta^{\circ}_{23}, \delta^{\circ}_{kl} ) \Psi^{\circ}_i \Psi^{\circ}_j  U_{ij}(\theta^{\nu}_{ij}, \delta^{\nu}_{ij}) Q_0 \nonumber \\
& = U^\circ(\theta^{\circ}_{12}, \theta^{\circ}_{13}, \theta^{\circ}_{23},  \delta^{\circ}_{kl} ) U_{ij}(\theta^{\nu}_{ij}, \delta^{\nu}_{ij} - \psi_i^{\circ} + \psi^{\circ}_j) \Psi^{\circ}_i \Psi^{\circ}_j Q_0 \,,
\end{align}
%%%%%%%%%%%%%%%%%%%%%%%%%%%
%
where $(ij) = (12), (13), (23)$, and the two free phases, 
which contribute to the Majorana phases of the PMNS matrix if the massive neutrinos are Majorana
particles, have been included in the diagonal phase matrix $Q_0$. 
Notice that if neutrinos are assumed to be Dirac instead of Majorana, then the matrix $Q_0$ can be removed through re-phasing
of the Dirac neutrino fields.
Without loss of generality we can redefine the combination 
$\delta^{\nu}_{ij} - \psi^{\circ}_i + \psi^{\circ}_j$ as
$\delta^{\nu}_{ij}$ and the combination 
$\Psi^{\circ}_i \Psi^{\circ}_j Q_0$ as $Q_0$, so that
the following parametrisation of $U$ is obtained:
%%%%%%%%%%%%%%%%%%%%%%%%%%%%%%%%%%%%%
\be
U = U^\circ(\theta^{\circ}_{12}, \theta^{\circ}_{13}, \theta^{\circ}_{23}, \delta^{\circ}_{kl} ) U_{ij}(\theta^{\nu}_{ij}, \delta^{\nu}_{ij}) Q_0 \,.
\label{eq:UZneZ2nu}
\ee
%%%%%%%%%%%%%%%%%%%%%%%%%%
%

In the case of $G_e = Z_2$ and $G_{\nu} = Z_2$ for both Dirac and Majorana neutrinos, we can write
%%%%%%%%%%%%%%%%%%%%%%%%%%%%%%%%%%%
\begin{align}
U & = U_{ij}(\theta^e_{ij}, \delta^e_{ij}) \Psi_j^{\circ} U^\circ(\theta^{\circ}_{12}, \theta^{\circ}_{13}, \theta^{\circ}_{23}, \delta^{\circ}_{kl} ) \Psi_r^{\circ} \Psi_s^{\circ} U_{rs}(\theta^{\nu}_{rs}, \delta^{\nu}_{rs}) Q_0 \nonumber \\
& =  \Psi_j^{\circ} U_{ij}(\theta^e_{ij}, \delta^e_{ij} - \psi^{\circ}_j)  U^\circ(\theta^{\circ}_{12}, \theta^{\circ}_{13}, \theta^{\circ}_{23}, \delta^{\circ}_{kl} ) U_{rs}(\theta^{\nu}_{rs}, \delta^{\nu}_{rs} - \psi^{\circ}_r + \psi^{\circ}_s) \Psi_r^{\circ} \Psi_s^{\circ} Q_0 \,,
\label{eq:Z2eZ2nu0}
\end{align}
%%%%%%%%%%%%%%%%%%%%%%%%%%%%%%%%%
%
with $(ij) = (12), (13), (23)$, $(rs) = (12), (13), (23)$. The phase 
matrices $\Psi_i^{\circ}$ are defined as in eq.~(\ref{eq:Psimatrices}).
Similarly to the previous cases, we can redefine the parameters in such a way 
that $U$ can be cast in the following form:
%%%%%%%%%%%%%%%%%%%%%%%%%%%%%%%
 \begin{align}
U  = U_{ij}(\theta^e_{ij}, \delta^e_{ij})  U^\circ(\theta^{\circ}_{12}, \theta^{\circ}_{13}, \theta^{\circ}_{23}, \delta^{\circ}_{kl} ) U_{rs}(\theta^{\nu}_{rs}, \delta^{\nu}_{rs}) Q_0 \,,
\label{eq:Z2eZ2nu}
\end{align}
where $Q_0$ can be phased away if neutrinos are assumed to be Dirac particles.\footnote{We will not repeat this statement further, but it should be always understood 
that if the massive neutrinos are Dirac fermions, then two phases in the matrix
$Q_0$ are unphysical and can be removed from $U$ by a re-phasing of the
Dirac neutrino fields.}
%%%%%%%%%%%%%%%%%%%%%%%%%%%%%%
%

If $G_e$ is fully broken and $G_{\nu} = Z_n$, $n > 2$ or $Z_n \times Z_m$, 
$n,m \geq 2$ for Dirac neutrinos or $G_{\nu}=Z_2\times Z_2$ for Majorana neutrinos, the form of $U$ reads
%%%%%%%%%%%%%%%%%%%%%%%%%%%%%%%
\be
U = U(\theta^e_{12},\theta^e_{13},\theta^e_{23}, \delta^e_{rs}) \Psi_2 \Psi_3 U^\circ(\theta^{\circ}_{12}, \theta^{\circ}_{13}, \theta^{\circ}_{23}, \{ \delta^{\circ}_{kl} \}) Q_0 \,,
\label{eq:U0eKnu}
\ee
%%%%%%%%%%%%%%%%%%%%%%%%%%%%%%%%%%%%
%
where the phase matrices $\Psi_2$ and $\Psi_3$ are defined as in eq.~(\ref{eq:Psimatricesfree}). Notice that in general we can effectively parametrise $U^{\circ}$ 
in terms of three angles and one phase 
since of the set of three phases 
$\{ \delta^{\circ}_{kl} \}$, two contribute to a 
redefinition of the matrices $Q_0$, $\Psi_2$ and $\Psi_3$. 
Furthermore, under the additional assumptions 
on the form of $U(\theta^e_{12},\theta^e_{13},\theta^e_{23}, \delta^e_{rs})$
and also taking $\{ \delta^{\circ}_{kl} \}=0$, 
the form of $U$ given in eq.~(\ref{eq:U0eKnu}) 
leads to the sum rules derived in \cite{Petcov:2014laa,Girardi:2015vha}.
In the numerical analyses performed in 
\cite{Petcov:2014laa,Girardi:2014faa,Girardi:2015vha},
the angles $\theta^{\circ}_{ij}$
have been set, in particular, to the values corresponding to the
TBM, BM (LC), GRA, GRB and HG symmetry forms.

Finally for the breaking patterns $G_e = Z_n$, $n > 2$ or $Z_n \times Z_m$, $n,m \geq 2$ and
$G_{\nu}$ fully broken when considering both Dirac and Majorana neutrino possibilities, 
the form of $U$  can be derived from eq.~(\ref{eq:U0eKnu}) by interchanging 
the fixed and the free parameters. Namely,
\be
U = U^\circ(\theta^{\circ}_{12}, \theta^{\circ}_{13}, \theta^{\circ}_{23}, \delta^{\circ}_{kl} ) \Psi_2 \Psi_3 U(\theta^{\nu}_{12}, \theta^{\nu}_{13}, \theta^{\nu}_{23}, \delta^{\nu}_{rs}) Q_0 \,.
\label{eq:UeZorZZUnu0}
\ee
The cases found in eqs.~(\ref{eq:UZ2eKnu}), (\ref{eq:UZneZ2nu}), (\ref{eq:Z2eZ2nu}), 
(\ref{eq:U0eKnu}) and (\ref{eq:UeZorZZUnu0}) are summarised in 
Table~\ref{tab:dof1}.
The reduction of the number of free parameters indicated with arrows corresponds
to a redefinition of the charged lepton fields.
%%%%%%%%%%%%%%%%%%%%%%%%%%
\begin{table}[h]
\centering
\renewcommand*{\arraystretch}{1.2}
\renewcommand{\tabcolsep}{4.6pt}
\begin{tabular}{ccccc}
\hline
& & & & \\ [-10pt]
$G_e \subset G_f$ &  $G_{\nu} \subset G_f$ & \phantom{1} $U_e$ d.o.f. & $U_{\nu}$ d.o.f. & $U$ d.o.f.  \\ [6pt]
\bottomrule
& & & & \\ [-10pt]
fully broken & fully broken & $9 \rightarrow 6$ & $9 \rightarrow 8$ & $12 \rightarrow 4$ $( + 2)$  \\ [6pt]
$Z_2$ & fully broken & $4 \rightarrow 2$ & $9 \rightarrow 8$ & $10 \rightarrow 4$ $( + 2)$  \\ [6pt]
$\left \{ \begin{tabular}{c} 
$Z_n$, $n > 2$ or \\ 
$Z_n \times Z_m$, $n,m \geq 2$ \\
\end{tabular} \right \}$ 
& fully broken & 0 & $9 \rightarrow 8$ & $\phantom{0}8 \rightarrow 4$ $( + 2)$  \\ [15pt]
fully broken & $Z_2$ & $9 \rightarrow 6$ & 4  & $10 \rightarrow 4$ $( + 2)$ \\ [6pt]
fully broken &
$\left \{ \begin{tabular}{c} 
$Z_n$, $n > 2$ or \\ 
$Z_n \times Z_m$, $n,m \geq 2$ \\
\end{tabular}  \right \}$
& $9 \rightarrow 6$  & 2 & $\phantom{0}8 \rightarrow 4$ $( + 2)$  \\ [15pt]
$Z_2$ & $Z_2$ & $4 \rightarrow 2$ & 4  & $4$ $( + 2)$  \\ [6pt]
$\left \{ \begin{tabular}{c} 
$Z_n$, $n > 2$ or \\ 
$Z_n \times Z_m$, $n,m \geq 2$ \\
\end{tabular} \right \}$
& $Z_2$ & 0 & 4 & $2$ $( + 2)$  \\ [15pt]
$Z_2$ &
$\left \{ \begin{tabular}{c} 
$Z_n$, $n > 2$ or \\ 
$Z_n \times Z_m$, $n,m \geq 2$ \\
\end{tabular} \right \}$
& $4 \rightarrow 2$  & 2  & $2$ $( + 2)$  \\ [15pt]
$\left \{ \begin{tabular}{c} 
$Z_n$, $n > 2$ or \\ 
$Z_n \times Z_m$, $n,m \geq 2$ \\
\end{tabular} \right \}$
& 
$\left \{ \begin{tabular}{c} 
$Z_n$, $n > 2$ or \\ 
$Z_n \times Z_m$, $n,m \geq 2$ \\
\end{tabular} \right \}$
& 0 & 2 & $0$ $( + 2)$  \\ [15pt]
\hline
\end{tabular}
\caption{Number of effective free parameters, degrees of freedom (d.o.f.), contained in $U$
relevant for the PMNS angles and the Dirac phase (and Majorana phases)
in the cases of the different breaking patterns of $G_f$ to $G_e$ and $G_{\nu}$.
Arrows indicate the reduction of the number of parameters, which can be
absorbed with a redefinition of the charged lepton fields. 
}
\label{tab:dof1}
\end{table}
%%%%%%%%%%%%%%%%%%%%%%%%%%

In the  breaking patterns considered, it may be also possible
to impose a generalised CP (GCP) symmetry. An example
of how imposing a GCP affects the sum rules is shown
in Appendix~\ref{app:A5andGCP}.
In the case in which a GCP symmetry 
is preserved in the neutrino sector
we have for the neutrino Majorana mass matrix \cite{Branco:1986gr}:
%%%%%%%%%%%%%%%%%%%%%%%%%%%%%%%%%%%%%
\be
X_i^{T} M_{\nu} X_i = M_{\nu}^* \,.
\label{eq:GCPnu}
\ee
%%%%%%%%%%%%%%%%%%%%%%%%%%%
%
Since the matrix $X_i$ is symmetric there exists 
a unitary matrix $\Omega_i$ such that 
$X_i = \Omega_i \, \Omega_i^T$ and $\Omega_i^T M_{\nu} \Omega_i$ is real.
Therefore when GCP is preserved in the neutrino sector, the phases
in the matrix $U_{\nu}$ can be fixed.
An alternative possibility is that GCP is preserved in the 
charged lepton sector, which leads to the condition \cite{Branco:1986gr}:
%%%%%%%%%%%%%%%%%%%%%%%%%%%%%%%%%%%
\be
(X^e_i)^{\dagger} M_e M_e^{\dagger} X_i^e = (M_e M_e^{\dagger})^* \,.
\label{eq:GCPe}
\ee
%%%%%%%%%%%%%%%%%%%%%%%%%%%%%%%
%
% is quite similar. 
Since $(X^e_i)^T = X^e_i$, the phases in the matrix $U_e$
are fixed, because $(\Omega^e_i)^{\dagger} M_e M_e^{\dagger} \Omega^e_i$ is real.
The fact that the matrices $X_i$, if GCP is preserved in the neutrino sector, or
$X^e_i$ if it is preserved in the charged lepton sector, 
are symmetric matrices can be proved applying the GCP 
transformation twice.
In the first case,
eq.~(\ref{eq:GCPnu}) allows one to derive
the general form of $X_i$ \cite{Everett:2015oka,Chen:2014wxa,Feruglio:2012cw}:
%%%%%%%%%%%%%%%%%%%%%%%%%%%%%%%
\be
X_i = U_{\nu} X_i^{\rm diag} U_{\nu}^T\,,
\label{eq:Xnu}
\ee
%%%%%%%%%%%%%%%%%%%%%%%%%%%%%%%%%%
%
while in the latter case
\be
X_i^e = U_e (X_i^e)^{\rm diag} U_e^T \,.
\label{eq:Xe}
\ee
%%%%%%%%%%%%%%%%%%%%%%
%
Equations~(\ref{eq:Xnu}) and (\ref{eq:Xe}) imply that $X_i$ and $X_i^e$ are
symmetric matrices.\footnote{This fact can be also derived from 
the requirement that the GCP transformations contain 
the physical CP transformation, i.e., the GCP transformations applied twice 
to a field should give the field itself \cite{Feruglio:2012cw,Holthausen:2012dk,Chen:2014tpa}:
\be
\phi(x) \rightarrow X_{\bf r} \phi^*(x_{\rm p}) \rightarrow X_{\bf r} X_{\bf r}^* \phi(x) = \phi(x) \,,
\ee
where $x = (x_0, \vec{x})$, $x_p = (x_0, - \vec{x})$. The notation we have used for
$X_{\bf r}$ emphasises the representation $\bf r$ for the GCP transformations.
}

We note finally that the titles of the following sections 
refer to the residual symmetries of the charged lepton and neutrino mass matrices,
while the titles of the subsections reflect the free complex rotations
contained in the corresponding parametrisation of
$U$, eqs.~(\ref{eq:UZ2eKnu}), (\ref{eq:UZneZ2nu}), (\ref{eq:Z2eZ2nu}), 
(\ref{eq:U0eKnu}) and (\ref{eq:UeZorZZUnu0}).

%%%%%%%%%%%%%%%%%%%%%%%%%%%%%%%%%%%Section 3
%
\section{The Pattern $G_e = Z_2$ and $G_\nu = Z_n$, $n > 2$ or $Z_n \times Z_m$, $n,m \geq 2$}
\label{sec:GeZGnuZZ}
%
%%%%%%%%%%%%%%%%%%%%%%%%%%%%%%%%%%%
%
In this section we derive sum rules for $\cos \delta$ for the cases
given in eq.~(\ref{eq:UZ2eKnu}). 
Recall that the matrix $U_e$
is fixed up to a complex rotation in one plane by the residual 
$G_e = Z_2$ symmetry, while $U_\nu$ 
is completely determined 
(up to multiplication by diagonal phase matrices on the right 
and permutations of columns) 
by the $G_{\nu} = Z_2 \times Z_2$ residual symmetry
in the case of neutrino Majorana mass term,
or by $G_\nu = Z_n$, $n > 2$ or 
$Z_n \times Z_m$, $n,m \geq 2$, 
residual symmetries if the massive neutrinos are Dirac particles.
At the end of this section we will present results 
of a study of the possibility 
of reproducing the observed values 
of the lepton mixing parameters 
$\sin^2\theta_{12}$, $\sin^2\theta_{13}$ and $\sin^2\theta_{23}$  
and of obtaining physically viable predictions for $\cos\delta$
in the cases when the  residual symmetries 
$G_e = Z_2$ and 
 $G_\nu = Z_n$, $n > 2$ or 
$Z_n \times Z_m$, $n,m \geq 2$,
originate from the breaking of the lepton flavour 
symmetries $A_4~(T')$, $S_4$ and $A_5$.

%%%%%%%%%%%%%%%%%%%%%%%
\subsection{The Case with 
$U_{12}(\theta^e_{12}, \delta^e_{12})$ Complex Rotation (Case A1)}
\label{sec:12e}
%%%%%%%%%%%%%%%%%%%%%%%
%
Employing  the parametrisation of the PMNS 
matrix $U$ given in eq.~(\ref{eq:UZ2eKnu})
with $(ij) = (12)$ and the parametrisation of $U^{\circ}$ given as  
%%%%%%%%%%%%%%%%%%%%%%%%%%%%%%%%%
\be
U^\circ(\theta^{\circ}_{12}, \theta^{\circ}_{13}, \theta^{\circ}_{23}, \delta^{\circ}_{12}) = U_{12}(\theta^{\circ}_{12}, \delta^{\circ}_{12}) R_{23}(\theta_{23}^{\circ}) R_{13}(\theta_{13}^{\circ}) \,,
\ee
%%%%%%%%%%%%%%%%%%%%%%%%%%%%%%%%
%
we get for $U$ (see Appendix \ref{app:ParU} for details):
%%%%%%%%%%%%%%%%%%%%%%%%%%%%%%%
\begin{align}
U & = U_{12}(\theta^e_{12}, \delta^e_{12}) U_{12}(\theta^{\circ}_{12}, \delta^{\circ}_{12}) R_{23}(\theta_{23}^{\circ}) R_{13}(\theta_{13}^{\circ}) Q_0 \,.
\label{eq:Uthe12deltae12GeZGnuZZ}
\end{align}
The results derived  in Appendix~\ref{app:ParU} and given 
in eq.~(\ref{eq:trick1}) allow us to 
cast eq.~(\ref{eq:Uthe12deltae12GeZGnuZZ}) in the form:
%%%%%%%%%%%%%%%%%%%%%%%%%
\be
U = R_{12}(\hat \theta_{12}) P_1(\hat \delta_{12}) R_{23}(\theta_{23}^{\circ}) R_{13}(\theta_{13}^{\circ}) Q_0 \,,
\quad
P_1(\hat \delta_{12}) = \diag(e^{i\hat\delta_{12}},1,1)\,,
\label{eq:Uthe12deltae12GeZGnuZZtrick}
\ee
%%%%%%%%%%%%%%%%%%%%%%%%%%%%%%
%
with $\hat \delta_{12} = \alpha - \beta$, where 
$\sin \hat \theta_{12}$, $\alpha$ and $\beta$
are defined as in eqs.~(\ref{eq:trickthhat}) and (\ref{eq:trickalpha}) 
after setting
$i = 1$, $j = 2$, $\theta^a_{12} = \theta^e_{12}$, $\delta^a_{12} = \delta^e_{12}$, 
$\theta^b_{12} = \theta^{\circ}_{12}$ and $\delta^b_{12} = \delta^{\circ}_{12}$.
Using eq.~(\ref{eq:Uthe12deltae12GeZGnuZZtrick}) and the standard 
parametrisation
of the PMNS matrix $U$, we find:
%
%%%%%%%%%%%%%%%%%%%%%%%%%%%%%%%%
\begin{align}
\sin^2 \theta_{13} & = |U_{e3}|^2  = \cos^2 \hat \theta_{12} \sin^2 \theta^{\circ}_{13} + \cos^2 \theta^{\circ}_{13} \sin^2 \hat \theta_{12} \sin^2 \theta^{\circ}_{23} \nonumber \\
& +\dfrac{1}{2} \sin 2 \hat \theta_{12} \sin 2 \theta^{\circ}_{13} \sin \theta^{\circ}_{23} \cos \hat \delta_{12}  \,, 
\label{eq:th13UZ2eKnu12e}\\
\sin^2 \theta_{23} & = \frac{|U_{\mu3}|^2}{1-|U_{e3}|^2} = \dfrac{1}{\cos^2 \theta_{13}} \big[\sin^2 \theta^{\circ}_{13} - \sin^2 \theta_{13} + \cos^2 \theta^{\circ}_{13} \sin^2 \theta^{\circ}_{23} \big] \,, 
\label{eq:th23UZ2eKnu12e}\\
\sin^2 \theta_{12} & = \frac{|U_{e2}|^2}{1-|U_{e3}|^2} = \dfrac{\cos^2 \theta^{\circ}_{23} \sin^2 \hat \theta_{12}}{\cos^2 \theta_{13}}  \,.
\label{eq:th12UZ2eKnu12e}
\end{align}
%%%%%%%%%%%%%%%%%%%%%%%%%%%
%
From eqs.~(\ref{eq:th13UZ2eKnu12e}) and (\ref{eq:th23UZ2eKnu12e}) we get 
the following correlation between the values of 
$\sin^2 \theta_{13}$ and $\sin^2 \theta_{23}$:
%%%%%%%%%%%%%%%%%%%%
\be
\sin^2 \theta_{13} + \cos^2 \theta_{13} \sin^2 \theta_{23} = 
\sin^2 \theta^{\circ}_{13} + \cos^2 \theta^{\circ}_{13} \sin^2 \theta^{\circ}_{23} \,.
\ee
%%%%%%%%%%%%%%%%%%%%
%
Notice that eq.~(\ref{eq:th12UZ2eKnu12e}) implies that
%%%%%%%%%%%%%%%%%%%%%%%%%
\be
\sin^2 \hat \theta_{12} = 
\dfrac{\cos^2 \theta_{13} \sin^2 \theta_{12}}{\cos^2 \theta^{\circ}_{23}} \,.
\ee
%%%%%%%%%%%%%%%%%%%%%%%
%

In order to obtain a sum rule for $\cos \delta$,
we compare the expressions for the absolute value of
the element $U_{\tau 2}$ of the PMNS matrix in the standard 
parametrisation and in the parametrisation defined in 
eq.~(\ref{eq:Uthe12deltae12GeZGnuZZtrick}),
%%%%%%%%%%%%%%%%%%%%%%%%%%%%%%%%%
\begin{align}
|U_{\tau 2}| = 
| \cos \theta_{12} \sin \theta_{23} + \sin \theta_{13} \cos \theta_{23} \sin \theta_{12} e^{i \delta}| = 
|\sin \theta^{\circ}_{23}| \,.
\end{align}
%%%%%%%%%%%%%%%%%%%%%%%%%%%%
%
\noindent From the above equation we get for $\cos \delta$:
%%%%%%%%%%%%%%%%%%%%%%%%%%%%%%%%%%%%%
\begin{align}
\cos \delta = \dfrac{\cos^2 \theta_{13} (\sin^2 \theta^{\circ}_{23} - \cos^2 \theta_{12}) + \cos^2 \theta^{\circ}_{13} \cos^2 \theta^{\circ}_{23} (\cos^2 \theta_{12} - \sin^2 \theta_{12} \sin^2 \theta_{13})}{\sin 2 \theta_{12} \sin \theta_{13} |\cos \theta^{\circ}_{13} \cos \theta^{\circ}_{23}| (\cos^2 \theta_{13} - \cos^2 \theta^{\circ}_{13} \cos^2 \theta^{\circ}_{23})^{\frac{1}{2}}} \,.
\end{align}
%
%%%%%%%%%%%%%%%%%%%%%%%%%%
%
For the considered specific residual symmetries 
$G_e$ and $G_\nu$, the predicted value of 
$\cos\delta$ in the case A1 discussed in this subsection 
depends on the chosen discrete flavour symmetry 
$G_f$ via the values of the angles $\theta^\circ_{13}$ and $\theta^\circ_{23}$.

 The method of derivation of the sum rule for $\cos\delta$ of interest 
employed in the present subsection and consisting, 
in particular, of choosing adequate parametrisations of 
the PMNS matrix $U$ (in terms of the complex 
rotations of $U_e$ and of $U_\nu$) 
and of the matrix $U^\circ$ (determined by the 
symmetries $G_e$, $G_\nu$ and $G_f$), 
which allows to express the PMNS matrix $U$ 
in terms of minimal numbers of angle and phase parameters,
will be used also in all subsequent sections.
The technical details related to the method are 
given in Appendices~\ref{app:ParU} and \ref{app:Gebroken}. 

%
%%%%%%%%%%%%%%%%%%%%%%%%%%
\begin{table}[t!]
\centering
\renewcommand*{\arraystretch}{1.4}
\begin{tabular}{cccc}
\hline
 Mixing & $\theta^{\circ}_{12}$ & $\theta^{\circ}_{23}$ & $\theta^{\circ}_{13}$  \\
\bottomrule
TBM & $\pi/4$ & $-\sin^{-1} (1/\sqrt{3})$ & $\pi/6$ \\ [4pt]
BM & $\sin^{-1} \sqrt{2/3}$ & $-\pi/6$ & $\sin^{-1}(1/\sqrt{3})$ \\  [4pt]
GRA & $\sin^{-1} \sqrt{(7-\sqrt{5})/11}$ & $-\sin^{-1} \sqrt{(5 + \sqrt{5})/20}$ & $\sin^{-1} \sqrt{(7-\sqrt{5})/22}$\\  [4pt]
GRB & $\sin^{-1} \sqrt{2(15-2\sqrt{5})/41}$ & $-\sin^{-1} \sqrt{(3 + \sqrt{5})/16}$ & $\sin^{-1} \sqrt{(15 - 2\sqrt{5})/41}$ \\  [4pt]
HG & $\sin^{-1} \sqrt{2/5}$ & $-\sin^{-1} \sqrt{3/8}$ & $\sin^{-1} \sqrt{1/5}$ \\  [4pt]
\hline
\end{tabular}
\caption{The symmetry forms TBM, BM (LC), GRA,
GRB and HG obtained in terms of the three rotations
$R_{12}(\theta^{\circ}_{12}) R_{23}(\theta^{\circ}_{23}) R_{13}(\theta^{\circ}_{13})$.
}
\label{tab:tho12tho23tho13symm}
\end{table}
%%%%%%%%%%%%%%%%%%%%%%%%%%%%%%%%%%%%%%

We note finally that in the case of $\delta^\circ_{12} = 0$,
the symmetry forms TBM, BM, GRA, GRB and HG
can be obtained from $U^\circ = R_{12}(\theta^{\circ}_{12}) 
R_{23}(\theta_{23}^{\circ}) R_{13}(\theta_{13}^{\circ})$ for 
specific values of the angles given in Table~\ref{tab:tho12tho23tho13symm}.
In this case, the angles $\theta^\circ_{ij}$ 
are related to the angles $\theta^\nu_{ij}$
defined in Section~2.1 of ref.~\cite{Girardi:2015vha} as follows:
%%%%%%%%%%%%%%%%%%%%%%%%%%%%%
\be
\sin^2 \theta^{\circ}_{23} = \cos^2 \theta^{\nu}_{12} \sin^2 \theta^{\nu}_{23} \,,~~~ \sin^2 \theta^{\circ}_{13} = \dfrac{\sin^2 \theta^{\nu}_{23} \sin^2 \theta^{\nu}_{12}}{1- \sin^2 \theta^{\circ}_{23}}\,,~~~
\sin^2 \theta^{\circ}_{12} = \dfrac{\sin^2 \theta^{\nu}_{12}}{1- \sin^2 \theta^{\circ}_{23}} \,.
\label{eq:relparamUZ2eKnu12e}
\ee
%%%%%%%%%%%%%%%%%%%%%%%%%%%%%%%%

%%%%%%%%%%%%%%%%%%%%%%%
\subsection{The Case with $U_{13}(\theta^e_{13}, \delta^e_{13})$ 
Complex Rotation (Case A2)}
\label{sec:13e}
%%%%%%%%%%%%%%%%%%%%%%%
%
Using the parametrisation of the PMNS matrix $U$ given in eq.~(\ref{eq:UZ2eKnu})
with $(ij) = (13)$ and the following parametrisation of $U^{\circ}$, 
%%%%%%%%%%%%%%%%%%%%%%%%%%%%%%%%%%%
\be
U^\circ(\theta^{\circ}_{12}, \theta^{\circ}_{13}, \theta^{\circ}_{23}, \delta^{\circ}_{13} ) = U_{13}(\theta^{\circ}_{13}, \delta^{\circ}_{13}) R_{23}(\theta_{23}^{\circ}) R_{12}(\theta_{12}^{\circ}) \,,
\ee
%%%%%%%%%%%%%%%%%%%%%%%%%%%%%%%
%
we get for $U$ (for details see Appendix \ref{app:ParU}):
%%%%%%%%%%%%%%%%%%%%%%%%%%%%%%%
\begin{align}
U & = U_{13}(\theta^e_{13}, \delta^e_{13}) U_{13}(\theta^{\circ}_{13}, \delta^{\circ}_{13}) R_{23}(\theta_{23}^{\circ}) R_{12}(\theta_{12}^{\circ}) Q_0 \,.
\label{eq:Uthe13deltae13GeZGnuZZ}
\end{align}
The results derived in Appendix \ref{app:ParU} and presented 
in eq.~(\ref{eq:trick1}) allow us to recast 
eq.~(\ref{eq:Uthe13deltae13GeZGnuZZ}) in the following form:
%%%%%%%%%%%%%%%%%%%%%%%%%%%%%
\be
U = R_{13}(\hat \theta_{13}) P_1(\hat \delta_{13}) R_{23}(\theta_{23}^{\circ}) R_{12}(\theta_{12}^{\circ}) Q_0 \,,
\quad
P_1(\hat \delta_{13}) = \diag(e^{i\hat\delta_{13}},1,1)\,.
\label{eq:Uthe13deltae13GeZGnuZZtrick}
\ee
%%%%%%%%%%%%%%%%%%%%%%%%%%%
%
Here $\hat \delta_{13} = \alpha - \beta$, where 
$\sin \hat \theta_{13}$, $\alpha$ and $\beta$
are defined as in 
eqs.~(\ref{eq:trickthhat}) and (\ref{eq:trickalpha}) after setting
$i = 1$, $j = 3$, $\theta^a_{13} = \theta^e_{13}$, $\delta^a_{13} = \delta^e_{13}$, 
$\theta^b_{13} = \theta^{\circ}_{13}$ and $\delta^b_{13} = \delta^{\circ}_{13}$.
Using eq.~(\ref{eq:Uthe13deltae13GeZGnuZZtrick}) and the
standard parametrisation of the PMNS matrix $U$, we find:
%
%%%%%%%%%%%%%%%%%%%%%%%%%%%%%%%%
\begin{align}
\sin^2 \theta_{13} & = |U_{e3}|^2  =  \sin^2 \hat \theta_{13} \cos^2 \theta^{\circ}_{23} \,, 
\label{eq:th13UZ2eKnu13e}\\
\sin^2 \theta_{23} & = \frac{|U_{\mu3}|^2}{1-|U_{e3}|^2} = \dfrac{\sin^2 \theta^{\circ}_{23}}{1 - \sin^2 \theta_{13}}\,, 
\label{eq:th23UZ2eKnu13e}\\
\sin^2 \theta_{12} & = \frac{|U_{e2}|^2}{1-|U_{e3}|^2} = \dfrac{1}{1 - \sin^2 \theta_{13}} \big[ \cos^2 \hat \theta_{13} \sin^2 \theta^{\circ}_{12} + \cos^2 \theta^{\circ}_{12} \sin^2 \hat \theta_{13} \sin^2 \theta^{\circ}_{23} \nonumber \\
& -\dfrac{1}{2} \sin 2 \hat \theta_{13} \sin 2 \theta^{\circ}_{12} \sin \theta^{\circ}_{23} \cos \hat \delta_{13} \big ] \,.
\label{eq:th12UZ2eKnu13e}
\end{align}
%%%%%%%%%%%%%%%%%%%%%%%%%%%
%
Thus, in this scheme, as it follows from eq.~(\ref{eq:th23UZ2eKnu13e}), 
the value of $\sin^2\theta_{23}$ is predicted once the 
symmetry group $G_f$ is fixed.  This prediction, when confronted with 
the measured value of  $\sin^2\theta_{23}$, 
constitutes an important test of the scheme 
considered for any given discrete (lepton flavour) 
symmetry group $G_f$, which contains the residual symmetry groups 
$G_e = Z_2$ and 
$G_\nu = Z_n$, $n > 2$ and/or 
$Z_n \times Z_m$, $n,m \geq 2$
as subgroups.

  As can be easily demonstrated, the case under discussion 
coincides with the one analysed  in Section 2.2 of
ref. \cite{Girardi:2015vha}. The parameters 
$\theta^{\nu}_{23}$ and $\theta^{\nu}_{12}$ in 
\cite{Girardi:2015vha} can be identified with 
$\theta^{\circ}_{23}$ and $\theta^{\circ}_{12}$,
respectively. Therefore the sum rule we obtain coincides with that given 
in eq.~(32) in \cite{Girardi:2015vha}:
%%%%%%%%%%%%%%%%%%%%%%
\begin{align}
\cos \delta = -\dfrac{\cos^2 \theta_{13} (\cos^2 \theta^{\circ}_{12} \cos^2 \theta^{\circ}_{23} - \cos^2 \theta_{12}) + \sin^2 \theta^{\circ}_{23} (\cos^2 \theta_{12} - \sin^2 \theta_{12} \sin^2 \theta_{13})}{\sin 2 \theta_{12} \sin \theta_{13} |\sin \theta^{\circ}_{23}| (\cos^2 \theta_{13} - \sin^2 \theta^{\circ}_{23})^{\frac{1}{2}}} \,.
\end{align}
%%%%%%%%%%%%%%%%%%%%%%%%%%%%%%%%%%%
%
The dependence of $\cos\delta$ on $G_f$ in this case 
is via the values of the angles $\theta^\circ_{12}$ and $\theta^\circ_{23}$.

%%%%%%%%%%%%%%%%%%%%%%%%
\subsection{The Case with $U_{23}(\theta^e_{23}, \delta^e_{23})$ 
Complex Rotation (Case A3)}
\label{sec:23e}
%%%%%%%%%%%%%%%%%%%%%%%%

In the case with $(ij) = (23)$, 
as can be shown, $\cos \delta$ does not satisfy a sum rule, i.e., 
it cannot be expressed in terms of the three neutrino mixing angles
$\theta_{12}$, $\theta_{13}$ and $\theta_{23}$ and the other 
fixed angle parameters of the scheme. 
Indeed, employing the parametrisation of $U^{\circ}$ as
$U^\circ(\theta^{\circ}_{12}, \theta^{\circ}_{13}, \theta^{\circ}_{23}, \delta^{\circ}_{23} ) = 
U_{23}(\theta^{\circ}_{23}, \delta^{\circ}_{23}) R_{13}(\theta_{13}^{\circ}) R_{12}(\theta_{12}^{\circ})$,
we can write the PMNS matrix in the following form:
%%%%%%%%%%%%%%%%%%%%%%%%%%%%%%%
\begin{align}
U & = U_{23}(\theta^e_{23}, \delta^e_{23}) U_{23}(\theta^{\circ}_{23}, \delta^{\circ}_{23}) R_{13}(\theta_{13}^{\circ}) R_{12}(\theta_{12}^{\circ}) Q_0 \,.
\label{eq:Uthe23deltae23GeZGnuZZ}
\end{align}
%%%%%%%%%%%%%%%%%%%%%%%%%%%%%%%%%
%
Using the results derived in Appendix \ref{app:ParU} 
and shown in eq.~(\ref{eq:trick1}), 
we can recast eq.~(\ref{eq:Uthe23deltae23GeZGnuZZ}) as
%%%%%%%%%%%%%%%%%%%%%%%%%%%%%%%
\be
U = R_{23}(\hat \theta_{23}) P_2(\hat \delta_{23}) R_{13}(\theta_{13}^{\circ}) R_{12}(\theta_{12}^{\circ}) Q_0 \,,
\quad
P_2(\hat \delta_{23}) = \diag(1,e^{i\hat\delta_{23}},1)\,,
\label{eq:Uthe23deltae23GeZGnuZZtrick}
\ee
%%%%%%%%%%%%%%%%%%%%%%%%%%%%%%%%
%
with $\hat \delta_{23} = \alpha - \beta$, where 
$\sin \hat \theta_{23}$, $\alpha$ and $\beta$
are defined as in eqs.~(\ref{eq:trickthhat}) and (\ref{eq:trickalpha}) 
after setting
$i = 2$, $j = 3$, $\theta^a_{23} = \theta^e_{23}$, $\delta^a_{23} = \delta^e_{23}$, 
$\theta^b_{23} = \theta^{\circ}_{23}$ and $\delta^b_{23} = \delta^{\circ}_{23}$.
Comparing eq.~(\ref{eq:Uthe23deltae23GeZGnuZZtrick}) and the 
standard parametrisation
of the PMNS matrix, we find that
$\sin^2 \theta_{13} = \sin^2 \theta^{\circ}_{13}$,
$\sin^2 \theta_{23} = \sin^2 \hat \theta_{23}$,
$\sin^2 \theta_{12} = \sin^2 \theta^{\circ}_{12}$
and $\cos \delta = \pm \cos \hat \delta_{23}$.

It follows from the preceding equations, in particular, that 
since, for any given $G_f$ compatible with the considered residual symmetries,
  $\theta^{\circ}_{13}$ and $\theta^{\circ}_{12}$ have fixed values, 
the values of both $\sin^2 \theta_{13}$ and $\sin^2 \theta_{12}$ 
are predicted. The predictions depend on the chosen symmetry $G_f$. 
Due to these predictions the scheme under discussion can be tested 
for any given discrete symmetry candidate $G_f$, 
compatible, in particular, with the considered 
residual symmetries.

 We have also seen that $\delta$ is related only to an unconstrained 
phase parameter of the scheme.
In the case of a flavour symmetry $G_f$ which, in particular,
allows to reproduce correctly the observed values of $\sin^2 \theta_{12}$
and $\sin^2 \theta_{13}$, 
it might be possible to obtain physically viable 
prediction for $\cos\delta$ 
by employing a GCP invariance constraint. 
An example of the effect that GCP invariance has on 
restricting CPV phases is given in Appendix~\ref{app:A5andGCP}.
Investigating the implications of the GCP 
invariance constraint in the charged lepton or the neutrino sector 
in the cases considered by us is, however, 
beyond the scope of the present study.

%%%%%%%%%%%%%%%%%%%%%%%%
\subsection{Results in the Cases of $G_f = A_4~(T^{\prime}),~S_4$ and $A_5$}
\label{sec:ressec2}
%%%%%%%%%%%%%%%%%%%%%%%%
The cases detailed in Sections \ref{sec:12e}~--~\ref{sec:23e} can 
all be obtained from the groups $A_4~(T^{\prime})$, $S_4$ and $A_5$, when 
breaking them to $G_e=Z_2$ and $G_{\nu}=Z_n$ ($n\geq 3$) in the case of Dirac 
neutrinos, or $G_{\nu}=Z_2\times Z_2$ in the case of both Dirac and Majorana neutrinos.\footnote{We only consider $Z_2\times Z_2$  when it is an actual subgroup of $G_f$.}   
We now give an explicit example of 
how these cases can occur in $A_4$.

 In the case of the group $A_4$ (see, e.g., \cite{Altarelli:2005yx}), 
 the structure of the breaking patterns
discussed, e.g.,  
in subsection \ref{sec:12e} can be realised when 
i) the $S$ generator of $A_4$ is preserved in the neutrino sector,
and when, due to an accidental symmetry, the mixing matrix is fixed to be 
tri-bimaximal, $U_{\nu}^{\circ} = U_{\rm TBM}$, 
up to permutations of the columns,
and ii) a $Z_2^{T^2 S T}$ or $Z_2^{T S T^2}$ is preserved
in the charged lepton sector. 
The group element generating the $Z_2$ symmetry 
is diagonalised
by the matrix $U_e^{\circ}$. Therefore the angles 
$\theta_{12}^{\circ}$, $\theta_{13}^{\circ}$
and $\theta_{23}^{\circ}$ are obtained from the product 
$U^{\circ} = (U_e^{\circ})^\dagger U_{\nu}^{\circ}$.
The same structure (the structure discussed in  subsection \ref{sec:13e})
can be obtained in a similar manner from 
the flavour groups $S_4$ and $A_5$ ($A_4$, $S_4$ and $A_5$). 

   We have investigated the possibility of reproducing
 the observed values 
of the lepton mixing parameters 
$\sin^2\theta_{12}$, $\sin^2\theta_{13}$ and $\sin^2\theta_{23}$  
as well as obtaining physically viable predictions for $\cos\delta$ 
in the cases of residual symmetries $G_e = Z_2$ and
$G_\nu = Z_n$, $n > 2$ or $Z_n \times Z_m$, $n,m \geq 2$~\footnote{Note that there are no subgroups of the type
$Z_n \times Z_m$ bigger than $Z_2 \times Z_2$
in the cases of $A_4$, $S_4$ and $A_5$.
 }
(Dirac neutrinos), or $G_{\nu}=Z_2\times Z_2$ (Majorana neutrinos), 
discussed in subsections \ref{sec:12e}, \ref{sec:13e} and  \ref{sec:23e}
denoted further as A1, A2 and A3, assuming that these residual symmetries
originate from the breaking of the flavour symmetries $A_4~(T^{\prime})$, $S_4$ and $A_5$. 
The analysis was performed using the current best fit values 
of the three lepton mixing parameters 
$\sin^2\theta_{12}$, $\sin^2\theta_{13}$ and 
$\sin^2\theta_{23}$.
The results we have obtained for the symmetries  $A_4~(T^{\prime})$, $S_4$ and $A_5$
are summarised below.

We have found that in the cases under discussion, i.e.,
in the cases A1, A2 and A3, 
and  flavour symmetries $G_f = A_4~(T^{\prime})$, $S_4$ and $A_5$,
with the exceptions to be discussed below,
it is impossible either to reproduce at least 
one of the measured values of 
$\sin^2\theta_{12}$, $\sin^2\theta_{13}$ and 
$\sin^2\theta_{23}$ 
even taking into account its 
respective $3\sigma$ uncertainty, 
or to get physically viable
values of $\cos\delta$ satisfying $|\cos\delta| \leq 1$.
In the cases A1 and A2 and the flavour 
groups $A_4$ and $S_4$, for instance, 
the values of $\cos\delta$ are unphysical. 
Using the group  $G_f = A_5$ leads  either to unphysical 
values of $\cos\delta$, or to values of 
$\sin^2\theta_{23}$ which lie outside the corresponding 
current $3\sigma$ allowed interval.
In the case A3 (discussed in 
subsection \ref{sec:23e}),
the symmetry $A_4$, for example,
leads to $(\sin^2\theta_{12},\sin^2\theta_{13}) = (0,0)$ or (1,0).

 As mentioned earlier, there are three 
exceptions in which we can still get phenomenologically 
viable results.  
In the A1 case (A2 case) and $S_4$ flavour symmetry,
one obtains bimaximal mixing corrected 
by a complex rotation in the 
1-2 plane~\footnote{For the case A1 it can been shown that
\be
\diag(-1,1,1) U(\theta^{\circ}_{12},\delta^{\circ}_{12}) R(\theta^{\circ}_{23})
R(\theta^{\circ}_{13}) \diag(1,-1,1) = U_{\rm BM}\,,
\ee
if $\theta^{\circ}_{23} = \sin^{-1}(1/2)$, $\theta^{\circ}_{13} = \sin^{-1}(\sqrt{1/3})$,
$\theta^{\circ}_{12} = \tan^{-1}(\sqrt{3/2} + \sqrt{1/2})$ and $\delta^{\circ}_{12} = 0$.
Therefore, one has 
BM mixing corrected from the left by
a $U(2)$ transformation in the degenerate subspace
in the 1-2 plane. Note that
our results are in agreement with those obtained
in \cite{Li:2014eia}.}
(1-3 plane).
The PMNS angle $\theta_{23}$ is predicted 
to have a value corresponding to 
$\sin^2\theta_{23} = 0.488$ ($\sin^2\theta_{23} = 0.512$).
For the best fit values of 
$\sin^2\theta_{12}$ and $\sin^2\theta_{13}$ we find 
that $\cos\delta = -\,1.29$ ($\cos\delta = +\,1.29$).
However, using the value of $\sin^2\theta_{12}= 0.348$,
which lies in the $3\sigma$ allowed interval, 
one gets the same value of $\sin^2\theta_{23}$ and 
$\cos\delta = - 0.993$ ($\cos\delta = 0.993$),
while in the part of the $3\sigma$ allowed interval of $\sin^2 \theta_{12}$,
$0.348 \leq \sin^2 \theta_{12} \leq 0.359$, we have
$-0.993 \leq \cos\delta \leq -0.915$
($0.993 \geq \cos\delta \geq 0.915$).

Also in the  A1 case (A2 case) but with an $A_5$ flavour symmetry and 
residual symmetry $G_\nu = Z_3$, which is only possible 
if the massive neutrinos are Dirac particles, we get the 
predictions 
 $\sin^2\theta_{23} = 0.553$ ($\sin^2\theta_{23} = 0.447$)
and $\cos\delta = 0.716$ ($\cos\delta = -\,0.716$).
 In the  A1 case (A2 case) with an $A_5$ flavour symmetry and 
residual symmetry $G_\nu = Z_5$, which can be realised for
neutrino Dirac mass term only, for the best fit values of 
$\sin^2\theta_{12}$ and $\sin^2\theta_{13}$
we get the predictions 
 $\sin^2\theta_{23} = 0.630$ ($\sin^2\theta_{23} = 0.370$), 
 which is slightly outside the current $3\sigma$ range)
and $\cos\delta = -\,1.12$ ($\cos\delta = 1.12$). 
However, using the value of  $\sin^2\theta_{12}= 0.321$,
which lies in the $1\sigma$ allowed interval of $\sin^2\theta_{12}$, 
one gets the same value of $\sin^2\theta_{23}$ and 
$\cos\delta = - 0.992$ ($\cos\delta = 0.992$).
In the part of the $3\sigma$ allowed interval of $\sin^2 \theta_{12}$,
$0.321 \leq \sin^2 \theta_{12} \leq 0.359$, one has
$-0.992 \leq \cos\delta \leq -0.633$
($0.992 \geq \cos\delta \geq 0.633$).

%%%%%%%%%%%%%%%%%%%%%%%%%%%%%%%%%%%
%
\section{The Pattern $G_e = Z_n$, $n > 2$ or $Z_n \times Z_m$, $n,m \geq 2$ and $G_{\nu} = Z_2$ }
\label{sec:GeZorZZGnuZ}
%
%%%%%%%%%%%%%%%%%%%%%%%%%%%%%%%%%%%
% 
In this section we derive sum rules for $\cos \delta$ in the case
given in eq.~(\ref{eq:UZneZ2nu}). 
We recall that 
for $G_e = Z_n$, $n > 2$ or 
$Z_n \times Z_m$, $n,m \geq 2$ and 
$G_{\nu} = Z_2$ of interest,  
the matrix $U_e$ is unambiguously determined 
(up to multiplication by diagonal phase matrices on the right and permutations of columns), 
while the matrix $U_\nu$ is determined up to a complex rotation 
in one plane.

%%%%%%%%%%%%%%%%%%%%%%%
\subsection{The Case with $U_{13}(\theta^{\nu}_{13}, \delta^{\nu}_{13})$ 
Complex Rotation (Case B1)}
\label{sec:13nu}
%%%%%%%%%%%%%%%%%%%%%%%
%
% 
Combining the parametrisation of the PMNS matrix $U$ given 
in eq.~(\ref{eq:UZneZ2nu})
with $(ij) = (13)$ and the parametrisation of $U^{\circ}$ 
as
%%%%%%%%%%%%%%%%%%%%%%%%%%%%%%
\be
U^\circ(\theta^{\circ}_{12}, \theta^{\circ}_{13}, \theta^{\circ}_{23}, \delta^{\circ}_{13} ) =  R_{23}(\theta_{23}^{\circ}) R_{12}(\theta_{12}^{\circ}) U_{13}(\theta^{\circ}_{13}, \delta^{\circ}_{13}) \,,
\ee
%%%%%%%%%%%%%%%%%%%%%%%%%%%%%%%%
%
we get for $U$ (the details are given again in
Appendix \ref{app:ParU}):
%%%%%%%%%%%%%%%%%%%%%%%%%%%%%%
\begin{align}
U & = R_{23}(\theta_{23}^{\circ}) R_{12}(\theta_{12}^{\circ}) U_{13}(\theta^{\circ}_{13}, \delta^{\circ}_{13}) U_{13}(\theta^{\nu}_{13}, \delta^{\nu}_{13}) Q_0 \,.
\label{eq:Uthe13deltae12GeZorZZGnuZ}
\end{align}
%%%%%%%%%%%%%%%%%%%%
%
The results derived in Appendix \ref{app:ParU} and reported 
in eq.~(\ref{eq:trick1}) allow us to 
recast eq.~(\ref{eq:Uthe13deltae12GeZorZZGnuZ}) in the form:
%%%%%%%%%%%%%%%%%%%%%%%%%%%%%%%%%%
\be
U = R_{23}(\theta_{23}^{\circ}) R_{12}(\theta_{12}^{\circ}) P_3(\hat \delta_{13}) R_{13}(\hat \theta_{13}) Q_0 \,,
\quad
P_3(\hat \delta_{13}) = \diag(1,1,e^{i\hat\delta_{13}})\,.
\label{eq:Uthe13deltae12GeZorZZGnuZtrick}
\ee
%%%%%%%%%%%%%%%%%%%%%%%%%%%%%%
%
Here $\hat \delta_{13} = -\alpha - \beta$ and we have redefined 
$P_{13}(\alpha, \beta) Q_0$ as $Q_0$,
where  $P_{13}(\alpha,\beta) = \diag (e^{i\alpha},1,e^{i\beta})$
and the expressions for $\sin^2 \hat\theta_{13}$, $\alpha$ and $\beta$
can be obtained from eqs.~(\ref{eq:trickthhat}) and (\ref{eq:trickalpha}), by setting 
$i = 1$, $j = 3$, $\theta^a_{13} = \theta^\circ_{13}$, $\delta^a_{13} = \delta^\circ_{13}$, 
$\theta^b_{13} = \theta^\nu_{13}$ and $\delta^b_{13} = \delta^\nu_{13}$.
Using eq.~(\ref{eq:Uthe13deltae12GeZorZZGnuZtrick}) and 
the standard parametrisation
of the PMNS matrix $U$, we find:
%
%%%%%%%%%%%%%%%%%%%%%%%%%%%%%%%%
\begin{align}
\sin^2 \theta_{13} & = |U_{e3}|^2  = \cos^2 \theta^{\circ}_{12} \sin^2 \hat \theta_{13} \,, 
\label{eq:th13GeZorZZGnuZ13nu}\\
\sin^2 \theta_{23} & = \frac{|U_{\mu3}|^2}{1-|U_{e3}|^2} = \dfrac{1}{\cos^2 \theta_{13}} \big [ \cos^2 \theta^{\circ}_{23} \sin^2 \hat \theta_{13} \sin^2 \theta^{\circ}_{12} 
+ \cos^2 \hat \theta_{13} \sin^2 \theta^{\circ}_{23} \nonumber \\
& - \frac{1}{2} \sin 2 \hat \theta_{13} \sin 2 \theta^{\circ}_{23} \sin \theta^{\circ}_{12} \cos \hat \delta_{13} \big ] \,, 
\label{eq:th23GeZorZZGnuZ13nu}\\
\sin^2 \theta_{12} & = \frac{|U_{e2}|^2}{1-|U_{e3}|^2} = \dfrac{\sin^2 \theta^{\circ}_{12}}{\cos^2 \theta_{13}} \,.
\label{eq:th12GeZorZZGnuZ13nu}
\end{align}
%%%%%%%%%%%%%%%%%%%%%%%%%%%%%%
%
It follows from eq.~(\ref{eq:th12GeZorZZGnuZ13nu})
that in the case under discussion the values of 
$\sin^2 \theta_{12}$ and $\sin^2\theta_{13}$ are correlated.

A sum rule for $\cos \delta$ can be derived by 
comparing the expressions for the absolute value of
the element $U_{\tau 2}$ of the PMNS matrix in the standard 
parametrisation and in the one obtained using 
eq.~(\ref{eq:Uthe13deltae12GeZorZZGnuZtrick}):
%%%%%%%%%%%%%%%%%%%%%%%%%%%%%%%%%
\begin{align}
|U_{\tau 2}| = 
| \cos \theta_{12} \sin \theta_{23} + \sin \theta_{13} \cos \theta_{23} \sin \theta_{12} e^{i \delta}| = 
|\cos \theta^{\circ}_{12} \sin \theta^{\circ}_{23}| \,.
\end{align}
%%%%%%%%%%%%%%%%%%%%%%%%%%%%%%%%%%%%%
%
From this equation we get
%%%%%%%%%%%%%%%%%%%%%%%%%%%%%%%%%%
\begin{align}
\cos \delta = -\dfrac{\cos^2 \theta_{13} (\cos^2 \theta^{\circ}_{12} \cos^2 \theta^{\circ}_{23} - \cos^2 \theta_{23}) + \sin^2 \theta^{\circ}_{12} (\cos^2 \theta_{23} - \sin^2 \theta_{13} \sin^2 \theta_{23})}
{\sin 2 \theta_{23} \sin \theta_{13} |\sin \theta^{\circ}_{12}| (\cos^2 \theta_{13} - \sin^2 \theta^{\circ}_{12})^{\frac{1}{2}}} \,.
\end{align}
%%%%%%%%%%%%%%%%%%%%%%%%%%%%%%%%%%%%%%%
The dependence of the predictions for $\cos\delta$ 
on $G_f$ is in this case via the values of $\theta^\circ_{12}$
and $\theta^\circ_{23}$.

%%%%%%%%%%%%%%%%%%%%%%%
\subsection{The Case with $U_{23}(\theta^{\nu}_{23}, \delta^{\nu}_{23})$ 
Complex Rotation (Case B2)}
\label{sec:23nu}
%%%%%%%%%%%%%%%%%%%%%%%
%
Utilising the parametrisation of the PMNS matrix $U$ 
given in eq.~(\ref{eq:UZneZ2nu})
with $(ij) = (23)$ and the following parametrisation 
of $U^{\circ}$, 
%%%%%%%%%%%%%%%%%%%%%%%%%%
\be
U^\circ(\theta^{\circ}_{12}, \theta^{\circ}_{13}, \theta^{\circ}_{23}, \delta^{\circ}_{23} ) =  
R_{13}(\theta_{13}^{\circ}) R_{12}(\theta_{12}^{\circ}) U_{23}(\theta^{\circ}_{23}, \delta^{\circ}_{23}) \,,
\ee
%%%%%%%%%%%%%%%%%%%%%%%%%%%%%%%%%%
%
we obtain for $U$ 
(Appendix \ref{app:ParU} contains the relevant details):
\begin{align}
U & = R_{13}(\theta_{13}^{\circ}) R_{12}(\theta_{12}^{\circ}) U_{23}(\theta^{\circ}_{23}, \delta^{\circ}_{23}) U_{23}(\theta^{\nu}_{23}, \delta^{\nu}_{23}) Q_0 \,.
\label{eq:Uthe23deltae12GeZorZZGnuZ}
\end{align}
%%%%%%%%%%%%%%%%%%%
%
The results given in  eq.~(\ref{eq:trick1}) in Appendix \ref{app:ParU} 
make it possible to bring eq.~(\ref{eq:Uthe23deltae12GeZorZZGnuZ}) 
to the form:
%%%%%%%%%%%%%%%%%%%%%%%%%%%%%%%%
\be
U = R_{13}(\theta_{13}^{\circ}) R_{12}(\theta_{12}^{\circ}) P_3(\hat \delta_{23}) R_{23}(\hat \theta_{23}) Q_0 \,,
\quad
P_3(\hat \delta_{23}) = \diag(1,1,e^{i\hat\delta_{23}})\,.
\label{eq:Uthe23deltae12GeZorZZGnuZtrick}
\ee
%%%%%%%%%%%%%%%%%%%%%%%%%%
%
Here $\hat \delta_{23} = -\alpha - \beta$ and we have 
redefined $P_{23}(\alpha, \beta) Q_0$ as $Q_0$, 
where $P_{23}(\alpha, \beta) = \diag (1,e^{i\alpha},e^{i\beta})$.
Using eq.~(\ref{eq:Uthe23deltae12GeZorZZGnuZtrick}) 
and the standard parametrisation
of the PMNS matrix $U$, we find:
%
%%%%%%%%%%%%%%%%%%%%%%%%%%%%%%%%
\begin{align}
\sin^2 \theta_{13} & = |U_{e3}|^2  = \cos^2 \theta^{\circ}_{13} \sin^2 \theta^{\circ}_{12} \sin^2 \hat \theta_{23} + \sin^2 \theta^{\circ}_{13} \cos^2 \hat \theta_{23} \nonumber \\
& + \frac{1}{2} \sin 2 \hat \theta_{23} \sin 2 \theta^{\circ}_{13} \sin \theta^{\circ}_{12} \cos \hat \delta_{23} \,, 
\label{eq:th13GeZorZZGnuZ23nu}\\
\sin^2 \theta_{23} & = \frac{|U_{\mu3}|^2}{1-|U_{e3}|^2} = \frac{\cos^2 \theta^{\circ}_{12} \sin^2 \hat \theta_{23}}{\cos^2 \theta_{13}} \,, 
\label{eq:th23GeZorZZGnuZ23nu}\\
\sin^2 \theta_{12} & = \frac{|U_{e2}|^2}{1-|U_{e3}|^2} = \frac{\cos^2 \theta_{13} - \cos^2 \theta^{\circ}_{12} \cos^2 \theta^{\circ}_{13}  }{\cos^2 \theta_{13}} \,.
\label{eq:th12GeZorZZGnuZ23nu}
\end{align}
%%%%%%%%%%%%%%%%%%%%%%%%%%%
%
Equation (\ref{eq:th12GeZorZZGnuZ23nu}) implies that,
as in the case investigated in the preceding subsection,
the values of $\sin^2 \theta_{12}$ and $\sin^2 \theta_{13}$
are correlated.

The sum rule for $\cos \delta$ of interest can be obtained by 
comparing the expressions for the absolute value of
the element $U_{\tau 1}$ of the PMNS matrix in the standard 
parametrisation and in the one obtained using 
eq.~(\ref{eq:Uthe23deltae12GeZorZZGnuZtrick}):
%%%%%%%%%%%%%%%%%%%%%%%%%%%%%%%%%
\begin{align}
|U_{\tau 1}| = | \sin \theta_{12} \sin \theta_{23} - \sin \theta_{13} \cos \theta_{12} \cos \theta_{23} e^{i \delta}| =
| \cos \theta^{\circ}_{12} \sin \theta^{\circ}_{13} | \,.
\end{align}
%%%%%%%%%%%%%%%%%%%%%%%%%%%%
%
From the above equation we get for $\cos \delta$:
%%%%%%%%%%%%%%%%%%%%%%%%%%%%%%%%%%%%%
%
\begin{align}
\cos \delta = 
\dfrac{\cos^2 \theta_{13} (\sin^2 \theta^{\circ}_{12} - \cos^2 \theta_{23}) + \cos^2 \theta^{\circ}_{12} \cos^2 \theta^{\circ}_{13} ( \cos^2 \theta_{23} - \sin^2 \theta_{13} \sin^2 \theta_{23} )}
{ \sin 2 \theta_{23} \sin \theta_{13} | \cos \theta^{\circ}_{12} \cos \theta^{\circ}_{13}| (\cos^2 \theta_{13} - \cos^2 \theta^{\circ}_{12} \cos^2 \theta^{\circ}_{13} )^{\frac{1}{2}}} \,.
\end{align}
%%%%%%%%%%%%%%%%%%%%%%%%%%%%
%
The dependence of $\cos\delta$ on $G_f$ is realised in this case 
through the values of  $\theta^{\circ}_{12}$ and $ \theta^{\circ}_{13}$.

%%%%%%%%%%%%%%%%%%%%%%%%
\subsection{The Case with $U_{12}(\theta^{\nu}_{12}, \delta^{\nu}_{12})$ 
Complex Rotation (Case B3)}
\label{sec:12nu}
%%%%%%%%%%%%%%%%%%%%%%%%
%
In this case, as we show below,  $\cos \delta$ does not satisfy a sum rule,
and thus is, in general, a free parameter.
Indeed, using the parametrisation of $U^{\circ}$ as
$U^\circ(\theta^{\circ}_{12},\theta^{\circ}_{13},\theta^{\circ}_{23},\delta^{\circ}_{12}) 
= R_{23}(\theta_{23}^{\circ}) R_{13}(\theta_{13}^{\circ}) 
U_{12}(\theta^{\circ}_{12}, \delta^{\circ}_{12})$
we get the following expression for $U$:
%%%%%%%%%%%%%%%%%%%%%%%%%%%%%%%
\begin{align}
U & = R_{23}(\theta_{23}^{\circ}) R_{13}(\theta_{13}^{\circ}) 
U_{12}(\theta^{\circ}_{12}, \delta^{\circ}_{12}) 
U_{12}(\theta^{\nu}_{12}, \delta^{\nu}_{12}) Q_0 \,.
\label{eq:Uthe12deltae12GeZorZZGnuZ}
\end{align}
%%%%%%%%%%%%%%%%%%%%%%%%%
%
After recasting  eq.~(\ref{eq:Uthe12deltae12GeZorZZGnuZ}) 
in the form
%%%%%%%%%%%%%%%%%%%%%%%%%%%%%%%%
\be
U = R_{23}(\theta_{23}^{\circ}) R_{13}(\theta_{13}^{\circ}) 
P_2(\hat \delta_{12}) R_{12}(\hat \theta_{12}) Q_0 \,,
\quad
P_2(\hat \delta_{12}) = \diag(1,e^{i\hat\delta_{12}},1)\,,
\label{eq:Uthe12deltae12GeZorZZGnuZtrick}
\ee
%%%%%%%%%%%%%%%%%%%%%%%%%%%%%%%%%%%
%
where $\hat \delta_{12} = -\alpha - \beta$, we find that 
$\sin^2 \theta_{13} = \sin^2 \theta^{\circ}_{13}$,
$\sin^2 \theta_{23} = \sin^2 \theta^{\circ}_{23}$,
$\sin^2 \theta_{12} = \sin^2 \hat \theta_{12}$ and 
$\cos \delta = \pm \cos \hat \delta_{12}$.

It follows from the expressions for the neutrino mixing parameters 
thus derived that, given a discrete  
symmetry $G_f$ which can lead to the 
considered breaking patterns,
the values of $\sin^2 \theta_{13}$ and $\sin^2 \theta_{23}$ 
are predicted. This, in turn, allows to test the 
phenomenological viability of the scheme under discussion 
for any appropriately chosen discrete lepton flavour symmetry $G_f$.  

In what concerns the phase $\delta$, it is expressed in terms 
of an unconstrained phase parameter present in the scheme 
we are considering. 
The comment made at the end of subsection  \ref{sec:23e} 
is valid also in this case. Namely, 
given a non-Abelian discrete flavour symmetry $G_f$ which allows
one to reproduce 
correctly the observed values of 
$\sin^2 \theta_{13}$ and $\sin^2 \theta_{23}$, 
it might be possible to obtain physically viable 
prediction for $\cos\delta$ 
by employing a GCP invariance constraint in 
the charged lepton or the neutrino sector.

%%%%%%%%%%%%%%%%%%%%%%%%
\subsection{Results in the Cases of $G_f = A_4~(T^{\prime}),~S_4$ and $A_5$}
\label{sec:ressec3}
%%%%%%%%%%%%%%%%%%%%%%%%
%
The schemes discussed in Sections~\ref{sec:13nu}~--~\ref{sec:12nu} 
are realised when breaking $G_f= A_4~(T^{\prime})$, $S_4$ and $A_5$, to 
$G_e=Z_n$ ($n\geq 3$) or $Z_2\times Z_2$ and $G_{\nu}=Z_2$, for both Dirac and Majorana  neutrinos.  
As a reminder to the reader, we investigate the case of $Z_2 \times Z_2 $ when 
it is an actual subgroup of $G_f$.  As an explicit example of how this breaking 
can occur, we will consider the case of $G_f=A_4~(T^{\prime})$.
The other cases when $G_f=S_4$ or $A_5$ can be obtained 
from the breaking of $S_4$ 
and $A_5$ to the relevant subgroups as given 
in \cite{Li:2014eia} and \cite{Ding:2011cm}, respectively.

 In the case of the group $A_4$ (see, e.g., \cite{{Altarelli:2005yx}}), 
the structure of the breaking patterns 
discussed, e.g., in subsection 
\ref{sec:13nu} can be obtained by breaking $A_4$ 
i) in the charged lepton sector 
to any of the four $Z_3$ subgroups, 
namely, $Z_3^{T}$, $Z_3^{ST}$, $Z_3^{TS}$, $Z_3^{STS}$, 
and 
ii) to any of the three $Z_2$ subgroups, namely,
$Z_2^{S}$, $Z_2^{T^2ST}$, $Z_2^{TST^2}$, in the neutrino sector.
In this case the matrix $U^{\circ} = U_{\rm TBM}$ gets corrected by 
a complex rotation matrix in the 1-3 plane coming 
from the neutrino sector.

 The results of the study performed by us 
of the phenomenological viability 
of the schemes with residual symmetries 
 $G_e = Z_n$, $n > 2$ or $Z_n \times Z_m$, 
$n,m \geq 2$ and $G_{\nu} = Z_2$, 
discussed in subsections \ref{sec:13nu}, \ref{sec:23nu} and  \ref{sec:12nu},
and denoted further as B1, B2 and B3,  
when the residual symmetries 
result from the breaking of 
the flavour symmetries $A_4~(T^{\prime})$, $S_4$ and $A_5$,
are described below.
We present results only in the cases in which 
we obtain values of $\sin^2\theta_{12}$,  $\sin^2\theta_{13}$ 
and  $\sin^2\theta_{23}$ compatible with their respective measured 
values (including the corresponding $3\sigma$ uncertainties)
and physically acceptable values of $\cos\delta$. 

For $G_f = A_4$, we find that only the case B1 with $G_e = Z_3$ 
is phenomenologically viable. In this case 
we have   $(\sin^2\theta^\circ_{12},\sin^2\theta^\circ_{23}) = 
(1/3,1/2)$, which leads to the predictions 
$\sin^2\theta_{12} = 0.341$ and $\cos\delta = 0.570$.
We find precisely the same results in the case B1 
if $G_f = S_4$ and $G_e = Z_3$.
Phenomenologically viable results are obtained 
for  $G_f = S_4$ and $G_e = Z_3$ in the case B2 
as well. In this case 
$(\sin^2\theta^\circ_{12},\sin^2\theta^\circ_{13}) = (1/6,1/5)$,
implying the predictions  
$\sin^2\theta_{12} = 0.317$ and $\cos\delta = -\,0.269$.
If $G_e = Z_4$ or $Z_2\times Z_2$ results from $G_f = S_4$,
we get in the case B1 
$(\sin^2\theta^\circ_{12},\sin^2\theta^\circ_{23}) = 
(1/4,1/3)$ and correspondingly 
$\sin^2\theta_{12} = 0.256$ (which lies slightly outside the 
current $3\sigma$ allowed range of $\sin^2\theta_{12}$) 
and the unphysical value of $\cos\delta = -\,1.19$.
These two values are obtained for the best 
fit values of $\sin^2\theta_{23}$ and 
$\sin^2\theta_{13}$. However, for 
 $\sin^2\theta_{23} = 0.419$ we find 
the physical value $\cos\delta = -0.990$,
while in the part of the $3\sigma$ allowed interval of 
$\sin^2\theta_{23}$, 
$0.374\leq \sin^2\theta_{23}  \leq 0.419$, we have 
$-0.495 \geq  \cos\delta \geq -0.990$.

If $G_f = A_5$, we find phenomenologically viable 
results  
i) for $G_e = Z_3$, in the case B1,
ii) for $G_e = Z_5$, in the cases B1 and B2, and
iii) for $G_e = Z_2\times Z_2$, in the case B2.
More specifically, if $G_e = Z_3$, we obtain in the case B1
$(\sin^2\theta^\circ_{12},\sin^2\theta^\circ_{23}) = 
(1/3,1/2)$ leading to the predictions 
$\sin^2\theta_{12} = 0.341$ and 
$\cos\delta = 0.570$. 
For $G_e = Z_5$ in the case B1 (case B2) 
we find $(\sin^2\theta^\circ_{12},\sin^2\theta^\circ_{23}) = 
(0.276,1/2)$ 
($(\sin^2\theta^\circ_{12},\sin^2\theta^\circ_{13}) = 
 (0.138,0.160)$), which leads to the predictions 
$\sin^2\theta_{12} = 0.283$ and 
$\cos\delta = 0.655$ ($\sin^2\theta_{12} = 0.259$
and $\cos\delta = -\,0.229$).
Finally, for $G_e = Z_2\times Z_2$ in the case B2 
we have two sets of values for 
$(\sin^2\theta^\circ_{12},\sin^2\theta^\circ_{13})$.
The first one, $(\sin^2\theta^\circ_{12},\sin^2\theta^\circ_{13}) = 
(0.096,0.276)$, together with the best fit values of 
$\sin^2 \theta_{13}$ and $\sin^2 \theta_{23}$,
leads to $\sin^2\theta_{12} = 0.330$ and $\cos\delta = -1.36$.
However,  $\cos\delta$ takes the physical value of 
$\cos \delta = -0.996$ for $\sin^2 \theta_{23} = 0.518$.
In the part of the $3\sigma$ allowed 
interval of values of $\sin^2\theta_{23}$, 
$0.518 \leq \sin^2\theta_{23} \leq 0.641$, 
we have $-0.996 \leq \cos\delta \leq -0.478$.
For the second set of values,
$(\sin^2\theta^\circ_{12},\sin^2\theta^\circ_{13}) = 
(1/4,0.127)$, we get the predictions 
 $\sin^2\theta_{12} = 0.330$ and $\cos\delta = 0.805$.

%%%%%%%%%%%%%%%%%%%%%%%%%%%%%%%%%%%
%
\section{The Pattern $G_e = Z_2$ and 
$G_{\nu} = Z_2$ }
\label{sec:GeZ2GnuZ2}
%
%%%%%%%%%%%%%%%%%%%%%%%%%%%%%%%%%%%
%

In this section we derive sum rules for $\cos \delta$ in the case
given in eq.~(\ref{eq:Z2eZ2nu}).
We recall that when the residual symmetries are 
$G_e = Z_2$ and 
$G_{\nu} = Z_2$, 
each of the matrices $U_e$ and $U_\nu$ 
is determined up to a complex rotation in one plane.

%%%%%%%%%%%%%%%%%%%%%%%
\subsection{The Case with $U_{12}(\theta^e_{12}, \delta^e_{12})$ and $U_{13}(\theta^{\nu}_{13}, \delta^{\nu}_{13})$ Complex Rotations (Case C1)}
\label{sec:12e13nu}
%%%%%%%%%%%%%%%%%%%%%%%
%
Similar to the already considered cases we combine 
the parametrisation of the PMNS matrix $U$ 
given in eq.~(\ref{eq:Z2eZ2nu})
with $(ij) = (12)$ and $(rs) = (13)$, with the parametrisation 
of $U^{\circ}$ given as 
%%%%%%%%%%%%%%%%%%%%%%%
\be
U^\circ(\theta^{\circ}_{12}, \theta^{\circ}_{13}, \theta^{\circ}_{23}, \delta^{\circ}_{12}, \delta^{\circ}_{13} ) = U_{12}(\theta^{\circ}_{12}, \delta^{\circ}_{12}) R_{23}(\theta^{\circ}_{23}) U_{13}(\theta^{\circ}_{13}, \delta^{\circ}_{13}) \,,
\ee
%%%%%%%%%%%%%%%%%%%%%%%%%%%%%%
%
and get the following expression for $U$ 
(as usual, we refer to Appendix \ref{app:ParU} for details):
%%%%%%%%%%%%%%%%%%%%%%%%%%%%%%%%%%%%%%%
\begin{align}
U & = U_{12}(\theta^e_{12}, \delta^e_{12}) U_{12}(\theta^{\circ}_{12}, \delta^{\circ}_{12}) R_{23}(\theta^{\circ}_{23}) U_{13}(\theta^{\circ}_{13}, \delta^{\circ}_{13}) U_{13}(\theta^{\nu}_{13}, \delta^{\nu}_{13}) Q_0 \,.
\label{eq:Uthe12thnu13GeZ2GnuZ2}
\end{align}
%%%%%%%%%%%%%%%%%%%%%%%%%
%
Utilising the results derived in Appendix \ref{app:ParU} 
and reported in eq.~(\ref{eq:trick1}),
we can recast eq.~(\ref{eq:Uthe12thnu13GeZ2GnuZ2}) in the form
%%%%%%%%%%%%%%%%%%%%%%%%%%%%
\be
U = R_{12}(\hat \theta^e_{12}) P_1(\hat \delta) 
R_{23}(\theta^{\circ}_{23}) R_{13}(\hat \theta^{\nu}_{13}) Q_0 \,,
\quad
P_1(\hat \delta) = \diag(e^{i\hat\delta},1,1)\,.
\label{eq:Uthe12thnu13GeZ2GnuZ2trick}
\ee
%%%%%%%%%%%%%%%%%%%%%%%%%%%%%
%
Here $\hat \delta = \alpha^e - \beta^e + \alpha^{\nu} + \beta^{\nu}$ 
and we have redefined the matrix $Q_0$ by absorbing the 
diagonal phase matrix 
$P_{13} (-\beta^\nu,-\alpha^\nu) = \diag (e^{-i\beta^\nu},1,e^{-i\alpha^\nu})$ 
in it.
Using eq.~(\ref{eq:Uthe12thnu13GeZ2GnuZ2trick}) 
and the standard parametrisation
of the PMNS matrix $U$, we find:
%
%%%%%%%%%%%%%%%%%%%%%%%%%%%%%%%%
\begin{align}
\sin^2 \theta_{13} & = |U_{e3}|^2  =  \cos^2 \hat \theta^e_{12} \sin^2 \hat \theta^{\nu}_{13} + \cos^2 \hat \theta^{\nu}_{13} \sin^2 \hat \theta^e_{12} \sin^2 \theta^{\circ}_{23} \nonumber \\
& + \frac{1}{2} \sin 2 \hat \theta^e_{12} \sin 2 \hat \theta^{\nu}_{13} \sin \theta^{\circ}_{23} \cos \hat \delta \,, 
\label{eq:th13Uthe12thnu13GeZ2GnuZ2}\\
\sin^2 \theta_{23} & = \frac{|U_{\mu3}|^2}{1-|U_{e3}|^2} =  \frac{\sin^2 \hat \theta^{\nu}_{13} - \sin^2 \theta_{13} + \cos^2 \hat \theta^{\nu}_{13} \sin^2 \theta^{\circ}_{23}}{1 - \sin^2 \theta_{13}} \,, 
\label{eq:th23Uthe12thnu13GeZ2GnuZ2}\\
\sin^2 \theta_{12} & = \frac{|U_{e2}|^2}{1-|U_{e3}|^2} =  \frac{\sin^2 \hat \theta^e_{12} \cos^2 \theta^{\circ}_{23}}{1 - \sin^2 \theta_{13}}\,.
\label{eq:th12Uthe12thnu13GeZ2GnuZ2}
\end{align}
%%%%%%%%%%%%%%%%%%%%%%%%%%%
%

The sum rule for $\cos \delta$ of interest can be derived 
by comparing the expressions for the absolute value of
the element $U_{\tau 2}$ of the PMNS matrix in the standard 
parametrisation and in the one obtained using 
eq.~(\ref{eq:Uthe12thnu13GeZ2GnuZ2trick}):
%%%%%%%%%%%%%%%%%%%%%%%%%%%%%%%%%
\begin{align}
|U_{\tau 2}| = 
| \cos \theta_{12} \sin \theta_{23} + \sin \theta_{13} \cos \theta_{23} \sin \theta_{12} e^{i \delta}| = 
|\sin \theta^{\circ}_{23}| \,.
\end{align}
%%%%%%%%%%%%%%%%%%%%%%%%%%%%
%
From the above equation we get for $\cos \delta$:
%%%%%%%%%%%%%%%%%%%%%%%%%%%%%%%%%%%%%
%
\begin{align}
\cos \delta = 
\dfrac{\sin^2 \theta^{\circ}_{23} - \cos^2 \theta_{12} \sin^2 \theta_{23} - \cos^2 \theta_{23} \sin^2 \theta_{12} \sin^2 \theta_{13}}
{\sin \theta_{13} \sin 2 \theta_{23} \sin \theta_{12} \cos \theta_{12}} \,.
\label{eq:cosdeltaZ2Z212e13nu}
\end{align}
%%%%%%%%%%%%%%%%%%%%%%%%%%%%%%%%%%%%%%
%
Given the assumed breaking pattern, 
$\cos\delta$ depends on the flavour symmetry 
$G_f$ via the value of $\theta^\circ_{23}$.
Using the best fit values of the standard mixing angles
for the NO neutrino mass spectrum 
and the requirement $|\cos \delta| \leq 1$, we find
that $\sin^2 \theta^{\circ}_{23}$ should lie in the following
interval: $0.236 \leq \sin^2 \theta^{\circ}_{23} \leq 0.377$.
Fixing two of the three angles to their best fit values and varying the
third one in its $3\sigma$ experimentally allowed range 
and considering all the three possible combinations, we
get that $|\cos \delta| \leq 1$ if 
$0.195 \leq \sin^2 \theta^{\circ}_{23} \leq 0.504$.

%%%%%%%%%%%%%%%%%%%%%%%
\subsection{The Case with $U_{13}(\theta^e_{13}, \delta^e_{13})$ and $U_{12}(\theta^{\nu}_{12}, \delta^{\nu}_{12})$ Complex Rotations (Case C2)}
\label{sec:13e12nu}
%%%%%%%%%%%%%%%%%%%%%%%
%
As in the preceding case, we use 
the parametrisation of the PMNS matrix $U$ given in eq.~(\ref{eq:Z2eZ2nu})
but this time with $(ij) = (13)$ and $(rs) = (12)$, and the 
parametrisation of $U^{\circ}$ as
%%%%%%%%%%%%%%%%%%%%%%%%%%%%%%
\be
U^\circ(\theta^{\circ}_{12}, \theta^{\circ}_{13}, \theta^{\circ}_{23}, \delta^{\circ}_{12}, \delta^{\circ}_{13} ) =  U_{13}(\theta^{\circ}_{13}, \delta^{\circ}_{13}) R_{23}(\theta^{\circ}_{23}) U_{12}(\theta^{\circ}_{12}, \delta^{\circ}_{12}) \,,
\ee
%%%%%%%%%%%%%%%%%%%%%%%%%%%%%%%%%%%%%%%%%%%
%
to get for $U$ (again the details can be found in Appendix \ref{app:ParU}):
%%%%%%%%%%%%%%%%%%%%%%%%%%%%%%%%%
\begin{align}
U & = U_{13}(\theta^e_{13}, \delta^e_{13}) U_{13}(\theta^{\circ}_{13}, \delta^{\circ}_{13}) R_{23}(\theta^{\circ}_{23}) U_{12}(\theta^{\circ}_{12}, \delta^{\circ}_{12}) U_{12}(\theta^{\nu}_{12}, \delta^{\nu}_{12}) Q_0 \,.
\label{eq:Uthe13thnu12GeZ2GnuZ2}
\end{align}
%%%%%%%%%%%%%%%%%%%%%%%%%%%%%%%%%%%%%%%%%%%
%
The results derived in Appendix \ref{app:ParU} 
and reported in eq.~(\ref{eq:trick1}) allow us 
to rewrite the expression for $U$ in 
eq.~(\ref{eq:Uthe13thnu12GeZ2GnuZ2}) as follows:
%%%%%%%%%%%%%%%%%%%%%%%%%%%%%%%%%%%%%%%%%%%%%%
\be
U = R_{13}(\hat \theta^e_{13}) P_1(\hat \delta) R_{23}(\theta^{\circ}_{23}) R_{12}(\hat \theta^{\nu}_{12}) Q_0 \,,
\quad
P_1(\hat \delta) = \diag(e^{i\hat\delta},1,1)\,,
\label{eq:Uthe13thnu12GeZ2GnuZ2trick}
\ee
%%%%%%%%%%%%%%%%%%%%%%%%%%%%%%%%%%%%%%
%
where $\hat \delta = \alpha^e - \beta^e + \alpha^{\nu} + \beta^{\nu}$, and also 
in this case we have redefined the matrix $Q_0$
by absorbing the phase matrix
$P_{12} (-\beta^\nu,-\alpha^\nu) = \diag (e^{-i\beta^\nu},e^{-i\alpha^\nu},1)$  
in it.
From eq.~(\ref{eq:Uthe13thnu12GeZ2GnuZ2trick}) 
and the standard parametrisation
of the PMNS matrix $U$ we get:
%
%%%%%%%%%%%%%%%%%%%%%%%%%%%%%%%%
\begin{align}
\sin^2 \theta_{13} & = |U_{e3}|^2  = \cos^2 \theta^{\circ}_{23} \sin^2 \hat \theta^e_{13} \,, 
\label{eq:th13Uthe13thnu12GeZ2GnuZ2}\\
\sin^2 \theta_{23} & = \frac{|U_{\mu3}|^2}{1-|U_{e3}|^2} =  \frac{\sin^2 \theta^{\circ}_{23}}{\cos^2 \theta_{13}} \,, 
\label{eq:th23Uthe13thnu12GeZ2GnuZ2}\\
\sin^2 \theta_{12} & = \frac{|U_{e2}|^2}{1-|U_{e3}|^2} =  \frac{1}{1 - \sin^2 \theta_{13}} \big[ \cos^2 \hat \theta^e_{13} \sin^2 \hat \theta^{\nu}_{12} + \cos^2 \hat \theta^{\nu}_{12} \sin^2 \hat \theta^e_{13} \sin^2 \theta^{\circ}_{23} \nonumber \\
& - \frac{1}{2} \sin 2 \hat \theta^e_{13} \sin 2 \hat \theta^{\nu}_{12} \sin \theta^{\circ}_{23} \cos \hat \delta  \big]\,.
\label{eq:th12Uthe13thnu12GeZ2GnuZ2}
\end{align}
%%%%%%%%%%%%%%%%%%%%%%%%%%%
%
Given the value of $\sin^2 \theta^{\circ}_{23}$,
 eq.~(\ref{eq:th23Uthe13thnu12GeZ2GnuZ2})
implies the existence of a correlation between the values of
$\sin^2 \theta_{23}$ and $\sin^2 \theta_{13}$.

\begin{figure}[h!]
  \begin{center}
 \includegraphics[width=9cm]{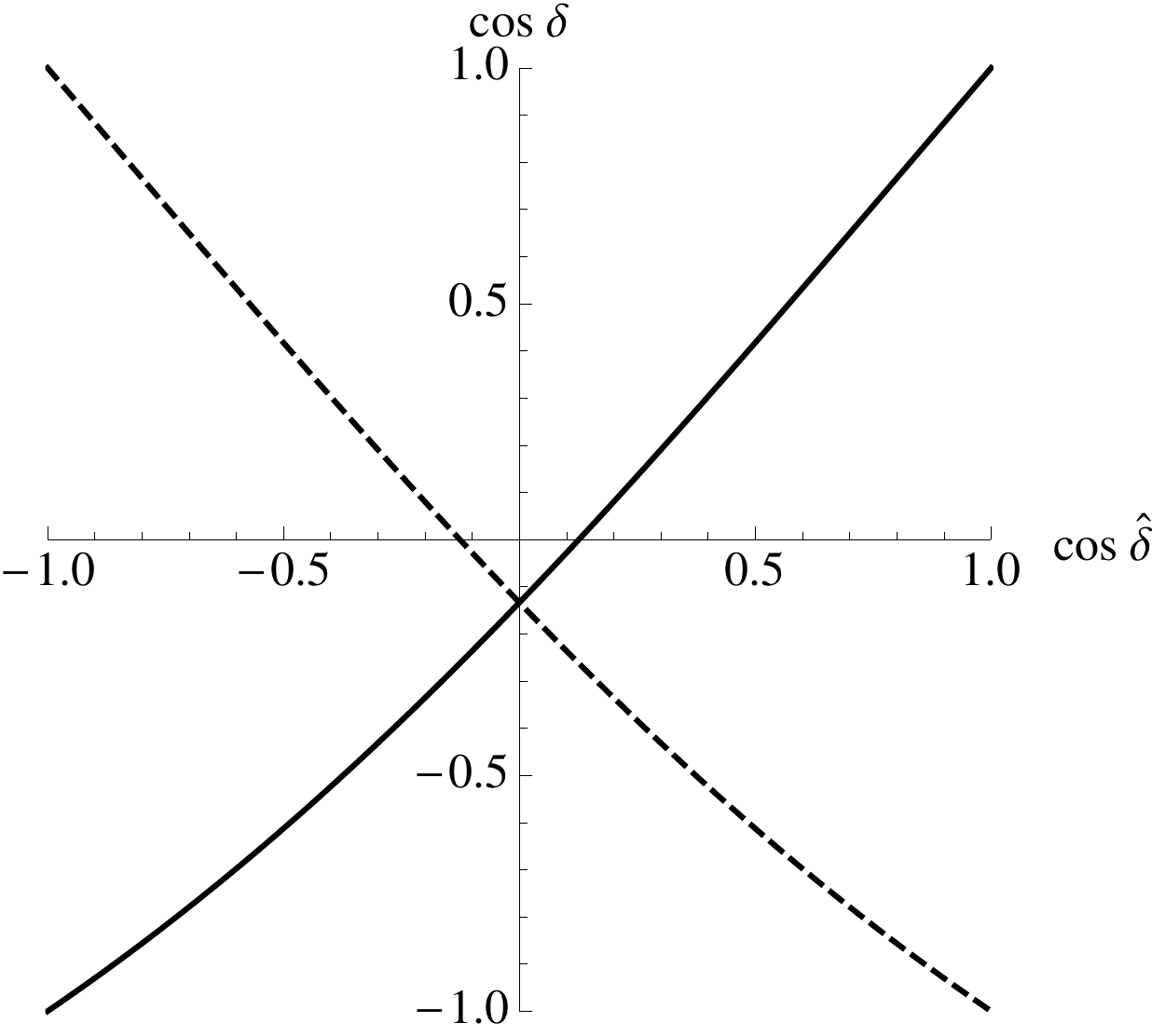}
     \end{center}
\caption{
\label{Fig1}
Dependence of $\cos \delta$ on $\cos \hat\delta$ in the case
of $G_f = S_4$ with $\sin^2 \theta^\circ_{23} = 1/2$.
The mixing parameters $\sin^2 \theta_{12}$ and $\sin^2 \theta_{13}$
have been fixed to their best fit values for the NO neutrino mass spectrum
quoted in eqs.~(\ref{th12values}) and (\ref{th13values}).
The solid (dashed) line is for the case when 
$\sin 2 \hat \theta^e_{13} \sin 2 \hat \theta^{\nu}_{12}$
is positive (negative).}
\end{figure}

Comparing the expressions for the absolute value of
the element $U_{\mu 1}$ of the PMNS matrix in the standard 
parametrisation and in the one obtained using 
eq.~(\ref{eq:Uthe13thnu12GeZ2GnuZ2trick}), 
we have
%%%%%%%%%%%%%%%%%%%%%%%%%%%%%%%%%
\begin{align}
|U_{\mu 1}| = | \sin \theta_{12} \cos \theta_{23} + \sin \theta_{13} \sin \theta_{23} \cos \theta_{12} e^{i \delta}| = 
| \sin \hat \theta^{\nu}_{12} \cos^2 \theta^{\circ}_{23} | \,.
\end{align}
%%%%%%%%%%%%%%%%%%%%%%%%%%%%
%
From the above equations we get for $\cos \delta$:
%%%%%%%%%%%%%%%%%%%%%%%%%%%%%%%%%%%%%
\begin{align}
\cos \delta = 
\dfrac{\cos^2 \theta_{13} (\cos^2 \theta^{\circ}_{23} \sin^2 \hat \theta^{\nu}_{12} - \sin^2 \theta_{12}) + \sin^2 \theta^\circ_{23} (\sin^2 \theta_{12} - \cos^2 \theta_{12} \sin^2 \theta_{13})}
{\sin 2\theta_{12} \sin \theta_{13} |\sin \theta^\circ_{23}| (\cos^2 \theta_{13} - \sin^2 \theta^\circ_{23})^\frac{1}{2}} \,.
\end{align}
%%%%%%%%%%%%%%%%%%%%%%%%%%%%%%%%%%%%
%
In this case $\cos\delta$ is a function of 
the known neutrino mixing angles $\theta_{12}$ 
and  $\theta_{13}$, of the 
angle $\theta^\circ_{23}$ fixed by 
$G_f$ and the assumed symmetry breaking pattern,
as well as of the phase parameter  $\hat \delta$ 
of the scheme. Predictions for $\cos\delta$ 
can only be obtained when  $\hat \delta$ 
is fixed by additional considerations 
of, e.g., GCP invariance,   
symmetries, etc. In view of this 
we show in Fig.~\ref{Fig1} $\cos\delta$ 
as a function of $\cos\hat \delta$ 
for the current best fit values  
of $\sin^2\theta_{12}$ and  $\sin^2\theta_{13}$, 
and for the value 
$\sin^2 \theta^\circ_{23} = 1/2$ corresponding to 
$G_f = S_4$. 
We do not find phenomenologically viable cases for $A_4~(T^{\prime})$ and $A_5$.
Therefore we do not present such a plot for these groups.
%%%%%%%%%%%%%%%%%%%%%%%%%%%%%%%%%
%%%%%%%%%%%%%%%%%%%%%%%%%

%%%%%%%%%%%%%%%%%%%%%%%
\subsection{The Case with $U_{12}(\theta^e_{12}, \delta^e_{12})$ and $U_{23}(\theta^{\nu}_{23}, \delta^{\nu}_{23})$ Complex Rotations (Case C3)}
\label{sec:12e23nu}
%%%%%%%%%%%%%%%%%%%%%%%
%
We get for the PMNS matrix $U$,
%%%%%%%%%%%%%%%%%%%%%%%%%%%%%%%%%
\begin{align}
U & = U_{12}(\theta^e_{12}, \delta^e_{12}) U_{12}(\theta^{\circ}_{12}, \delta^{\circ}_{12}) R_{13}(\theta^{\circ}_{13}) U_{23}(\theta^{\circ}_{23}, \delta^{\circ}_{23}) U_{23}(\theta^{\nu}_{23}, \delta^{\nu}_{23}) Q_0\,,
\label{eq:Uthe12thnu23GeZ2GnuZ2}
\end{align}
%%%%%%
%
utilising the parametrisations of $U$ shown in eq.~(\ref{eq:Z2eZ2nu})
with $(ij) = (12)$ and $(rs) = (23)$ and that of  
$U^{\circ}$ given below (further details can be found 
in Appendix \ref{app:ParU}), 
%%%%%%%%%%%%%%%%%%%%%%%%%%%%%%
\be
U^\circ(\theta^{\circ}_{12}, \theta^{\circ}_{13}, \theta^{\circ}_{23}, \delta^{\circ}_{12}, \delta^{\circ}_{23} ) = U_{12}(\theta^{\circ}_{12}, \delta^{\circ}_{12}) R_{13}(\theta^{\circ}_{13}) U_{23}(\theta^{\circ}_{23}, \delta^{\circ}_{23}) \,.
\ee
%%%%%%%%%%%%%%%%%%%%%%%%%%%%%%
%
With the help of the results derived in Appendix 
\ref{app:ParU} and especially of 
eq.~(\ref{eq:trick1}), the expression in
eq.~(\ref{eq:Uthe12thnu23GeZ2GnuZ2}) for the PMNS matrix $U$ 
can be brought to the form
%%%%%%%%%%%%%%%%%%%%%%%%%%%%%%%%
\be
U = R_{12}(\hat \theta^e_{12}) P_2(\hat \delta) R_{13}(\theta^{\circ}_{13}) R_{23}(\hat \theta^{\nu}_{23}) Q_0 \,,
\quad
P_2(\hat \delta) = \diag(1,e^{i\hat\delta},1)\,,
\label{eq:Uthe12thnu23GeZ2GnuZ2trick}
\ee
%%%%%%%%%%%%%%%%%%%%%%%%%%%%%%%%
%
where $\hat \delta = \beta^e - \alpha^e + \alpha^{\nu} + \beta^{\nu}$
and, as in the preceding cases, we have redefined 
the phase matrix $Q_0$ 
by absorbing the phase matrix 
$P_{23} (-\beta^\nu,-\alpha^\nu) = \diag (1,e^{-i\beta^\nu},e^{-i\alpha^\nu})$ 
in it.
Using eq.~(\ref{eq:Uthe12thnu23GeZ2GnuZ2trick}) 
and the standard parametrisation
of the PMNS matrix $U$, we find:
%%%%%%%%%%%%%%%%%%%%%%%%%%%%%%%%
\begin{align}
\sin^2 \theta_{13} & = |U_{e3}|^2  = \sin^2 \hat \theta^e_{12} \sin^2 \hat \theta^{\nu}_{23} + \cos^2 \hat \theta^e_{12} \cos^2 \hat \theta^{\nu}_{23} \sin^2 \theta^{\circ}_{13} \nonumber \\
& + \frac{1}{2} \sin 2 \hat \theta^e_{12} \sin 2 \hat \theta^{\nu}_{23} \sin \theta^{\circ}_{13} \cos \hat \delta \,, 
\label{eq:th13Uthe12thnu23GeZ2GnuZ2}\\
\sin^2 \theta_{23} & = \frac{|U_{\mu3}|^2}{1-|U_{e3}|^2} = \frac{\sin^2 \hat \theta^{\nu}_{23} - \sin^2 \theta_{13} + \cos^2 \hat \theta^{\nu}_{23} \sin^2 \theta^{\circ}_{13} }{1 - \sin^2 \theta_{13}}  \,, 
\label{eq:th23Uthe12thnu23GeZ2GnuZ2}\\
\sin^2 \theta_{12} & = \frac{|U_{e2}|^2}{1-|U_{e3}|^2} = \frac{\sin^2 \hat \theta^e_{12} - \sin^2 \theta_{13} + \cos^2 \hat \theta^e_{12} \sin^2 \theta^{\circ}_{13} }{1 - \sin^2 \theta_{13}} \,.
\label{eq:th12Uthe12thnu23GeZ2GnuZ2}
\end{align}
%%%%%%%%%%%%%%%%%%%%%%%%%%%
%
The sum rule for  $\cos \delta$ of interest 
can be derived, e.g., by 
comparing the expressions for the absolute value of
the element $U_{\tau 1}$ of the PMNS matrix in the standard 
parametrisation and in the one obtained using 
eq.~(\ref{eq:Uthe12thnu23GeZ2GnuZ2trick}):
%%%%%%%%%%%%%%%%%%%%%%%%%%%%%%%%%
\begin{align}
|U_{\tau 1}| = | \sin \theta_{12} \sin \theta_{23} - \sin \theta_{13} \cos \theta_{12} \cos \theta_{23} e^{i \delta}| =
| \sin \theta^{\circ}_{13} | \,.
\end{align}
%%%%%%%%%%%%%%%%%%%%%%%%%%%%
%
For $\cos \delta$ we get:
%%%%%%%%%%%%%%%%%%%%%%%%%%%%%%%%%%%%%
\begin{align}
\cos \delta = 
\dfrac{\sin^2 \theta_{12} \sin^2 \theta_{23} - \sin^2 \theta^{\circ}_{13} + \cos^2 \theta_{12} \cos^2 \theta_{23} \sin^2 \theta_{13}}
{\sin \theta_{13} \sin 2 \theta_{23} \sin \theta_{12} \cos \theta_{12}} \,.
\end{align}
%%%%%%%%%%%%%%%%%%%%%%%%%%%%%%%
%
In this case, in contrast to that considered 
in the preceding subsection, $\cos\delta$ is 
predicted once the angle $\theta^{\circ}_{13}$, 
i.e., the flavour symmetry $G_f$, is fixed.
Using the best fit values of $\sin^2\theta_{12}$, 
$\sin^2\theta_{13}$ and  $\sin^2\theta_{23}$ 
for the NO neutrino mass spectrum, we find that physical values 
of $\cos\delta$ satisfying $|\cos \delta| \leq 1$
can be obtained only if 
$\sin^2 \theta^{\circ}_{13}$ lies in the following
interval: $0.074 \leq \sin^2 \theta^{\circ}_{13} \leq 0.214$.
Fixing two of the three neutrino mixing 
parameters $\sin^2\theta_{12}$, 
$\sin^2\theta_{13}$ and  $\sin^2\theta_{23}$
 to their best fit values and varying the
third one in its $3\sigma$ experimentally allowed 
range and taking into account all the three possible combinations, we
get that $|\cos \delta| \leq 1$ provided  
$0.056 \leq \sin^2 \theta^{\circ}_{13} \leq 0.267$.

%%%%%%%%%%%%%%%%%%%%%%%
\subsection{The Case with $U_{13}(\theta^e_{13}, \delta^e_{13})$ and $U_{23}(\theta^{\nu}_{23}, \delta^{\nu}_{23})$ Complex Rotations (Case C4)}
\label{sec:13e23nu}
%%%%%%%%%%%%%%%%%%%%%%%
%
The parametrisation of the PMNS matrix $U$, to be used further, 
%%%%%%%%%%%%%%%%%%%%%%%
\begin{align}
U & = U_{13}(\theta^e_{13}, \delta^e_{13}) U_{13}(\theta^{\circ}_{13}, \delta^{\circ}_{13}) R_{12}(\theta^{\circ}_{12}) U_{23}(\theta^{\circ}_{23}, \delta^{\circ}_{23})
U_{23}(\theta^{\nu}_{23}, \delta^{\nu}_{23}) Q_0\,,
\label{eq:Uthe13thnu23GeZ2GnuZ2}
\end{align}
%%%%%%%%%%%%%%%%%%%%%%%
% 
is found in this case from 
the parametrisations of the matrix $U$ given in eq.~(\ref{eq:Z2eZ2nu})
with $(ij) = (13)$ and $(rs) = (23)$ and that of $U^{\circ}$ shown below
(see Appendix \ref{app:ParU} for details), 
%%%%%%%%%%%%%%%%%%%%%%%%%%%%%%
\be
U^\circ(\theta^{\circ}_{12}, \theta^{\circ}_{13}, \theta^{\circ}_{23}, \delta^{\circ}_{13}, \delta^{\circ}_{23} ) = 
U_{13}(\theta^{\circ}_{13}, \delta^{\circ}_{13}) R_{12}(\theta^{\circ}_{12}) U_{23}(\theta^{\circ}_{23}, \delta^{\circ}_{23}) \,.
\ee
%%%%%%%%%%%%%%%%%%%%%%%%%%%%%%%%
%
The results presented in eq.~(\ref{eq:trick1}) of Appendix 
\ref{app:ParU} allow us to 
recast eq.~(\ref{eq:Uthe13thnu23GeZ2GnuZ2}) in the form:
%%%%%%%%%%%%%%%%%%%%%%
\be
U = R_{13}(\hat \theta^e_{13}) P_3(\hat \delta) R_{12}(\theta^{\circ}_{12}) R_{23}(\hat \theta^{\nu}_{23}) Q_0 \,,
\quad
P_3(\hat \delta) = \diag(1,1,e^{i\hat\delta})\,.
\label{eq:Uthe13thnu23GeZ2GnuZ2trick}
\ee
%%%%%%%%%%%%%%%%%%%%%%%%
%
Here $\hat \delta = \beta^e - \alpha^e - \alpha^{\nu} - \beta^{\nu}$ and
we have absorbed the phase matrix
$P_{23} (\alpha^\nu,\beta^\nu) = \diag (1,e^{i\alpha^\nu},e^{i\beta^\nu})$ 
in the matrix $Q_0$.
Using eq.~(\ref{eq:Uthe13thnu23GeZ2GnuZ2trick}) 
and the standard parametrisation
of the PMNS matrix $U$, we find:
%
%%%%%%%%%%%%%%%%%%%%%%%%%%%%%%%%
\begin{align}
\sin^2 \theta_{13} & = |U_{e3}|^2  = \cos^2 \hat \theta^{\nu}_{23} \sin^2 \hat \theta^e_{13} + \cos^2 \hat \theta^e_{13} \sin^2 \hat \theta^{\nu}_{23} \sin^2 \theta^{\circ}_{12} \nonumber \\
& + \frac{1}{2} \sin 2 \hat \theta^e_{13} \sin 2 \hat \theta^{\nu}_{23} \sin \theta^{\circ}_{12} \cos \hat \delta \,, 
\label{eq:th13Uthe13thnu23GeZ2GnuZ2}\\
\sin^2 \theta_{23} & = \frac{|U_{\mu3}|^2}{1-|U_{e3}|^2} = \frac{\cos^2 \theta^{\circ}_{12} \sin^2 \hat \theta^{\nu}_{23}}{1 - \sin^2 \theta_{13}} \,, 
\label{eq:th23Uthe13thnu23GeZ2GnuZ2}\\
\sin^2 \theta_{12} & = \frac{|U_{e2}|^2}{1-|U_{e3}|^2} = \frac{\sin^2 \hat \theta^e_{13} - \sin^2 \theta_{13} + \cos^2 \hat \theta^e_{13} \sin^2 \theta^{\circ}_{12} }{1 - \sin^2 \theta_{13}} \,.
\label{eq:th12Uthe13thnu23GeZ2GnuZ2}
\end{align}
%%%%%%%%%%%%%%%%%%%%%%%%%%%
%
Comparing the expressions for the absolute value of
the element $U_{\mu 1}$ of the PMNS matrix in the standard 
parametrisation and in the one obtained using 
eq.~(\ref{eq:Uthe13thnu23GeZ2GnuZ2trick}), we find
%%%%%%%%%%%%%%%%%%%%%%%%%%%%%%%%%
\begin{align}
|U_{\mu 1}| = | \sin \theta_{12} \cos \theta_{23} + \sin \theta_{13} \sin \theta_{23} \cos \theta_{12} e^{i \delta}| = 
| \sin \theta^{\circ}_{12} | \,.
\end{align}
%%%%%%%%%%%%%%%%%%%%%%%%%%%%
%
From the above equation we get for $\cos \delta$:
%%%%%%%%%%%%%%%%%%%%%%%%%%%%%%%%%%%%%
%
\begin{align}
\cos \delta = 
\dfrac{\sin^2 \theta^{\circ}_{12} - \cos^2 \theta_{23} \sin^2 \theta_{12} - \cos^2 \theta_{12} \sin^2 \theta_{13} \sin^2 \theta_{23}}
{\sin \theta_{13} \sin 2 \theta_{23} \sin \theta_{12} \cos \theta_{12}} \,.
\end{align}
%%%%%%%%%%%%%%%%%%%%%%%%%
%
The predicted value of $\cos\delta$ depends on the 
discrete symmetry $G_f$ through the value of the angle 
$\theta^{\circ}_{12}$.
Using the best fit values 
of the standard mixing angles
for the NO neutrino mass spectrum 
and the requirement $|\cos \delta| \leq 1$, we find
that $\sin^2 \theta^{\circ}_{12}$ should lie in the following
interval: $0.110 \leq \sin^2 \theta^{\circ}_{12} \leq 0.251$.
Fixing two of the three neutrino mixing 
angles to their best fit values and varying the
third one in its $3\sigma$ experimentally allowed range 
and accounting for all the three possible combinations, 
we get that $|\cos \delta| \leq 1$ if 
$0.057 \leq \sin^2 \theta^{\circ}_{12} \leq 0.281$.

%%%%%%%%%%%%%%%%%%%%%%%
\subsection{The Case with $U_{23}(\theta^e_{23}, \delta^e_{23})$ and $U_{13}(\theta^{\nu}_{13}, \delta^{\nu}_{13})$ Complex Rotations (Case C5)}
\label{sec:23e13nu}
%%%%%%%%%%%%%%%%%%%%%%%
%
The parametrisation of the PMNS matrix $U$, 
which is convenient for our further analysis,
%%%%%%%%%%%%%%%%%%%%
\begin{align}
U & = U_{23}(\theta^e_{23}, \delta^e_{23}) U_{23}(\theta^{\circ}_{23}, \delta^{\circ}_{23}) R_{12}(\theta^{\circ}_{12}) U_{13}(\theta^{\circ}_{13}, \delta^{\circ}_{13})
U_{13}(\theta^{\nu}_{13}, \delta^{\nu}_{13}) Q_0\,,
\label{eq:Uthe23thnu13GeZ2GnuZ2}
\end{align}
%%%%%%%%%%%%%%%%%%%%%
%
can be obtained in this case 
utilising the parametrisations of the matrix $U$ given in eq.~(\ref{eq:Z2eZ2nu})
with $(ij) = (23)$ and $(rs) = (13)$ and that of the matrix       
 $U^{\circ}$ given below (for details see Appendix \ref{app:ParU}),
%%%%%%%%%%%%%%%%%%%%%%%%%%%%%%%%% 
\be
U^\circ(\theta^{\circ}_{12}, \theta^{\circ}_{13}, \theta^{\circ}_{23}, \delta^{\circ}_{13}, \delta^{\circ}_{23} ) = 
U_{23}(\theta^{\circ}_{23}, \delta^{\circ}_{23}) R_{12}(\theta^{\circ}_{12}) U_{13}(\theta^{\circ}_{13}, \delta^{\circ}_{13})  \,.
\ee
%%%%%%%%%%%%%%%%%%%%%%%%%%%%%%
%
The expression in eq.~(\ref{eq:Uthe23thnu13GeZ2GnuZ2}) for 
$U$ can further be cast in a ``minimal'' form 
with the help of eq.~(\ref{eq:trick1}) in Appendix \ref{app:ParU}:
%%%%%%%%%%%%%%%%%%%%%%%%%%%%
\be
U = R_{23}(\hat \theta^e_{23}) P_3(\hat \delta) R_{12}(\theta^{\circ}_{12}) R_{13}(\hat \theta^{\nu}_{13}) Q_0 \,,
\quad
P_3(\hat \delta) = \diag(1,1,e^{i\hat\delta})\,,
\label{eq:Uthe23thnu13GeZ2GnuZ2trick}
\ee
%%%%%%%%%%%%%%%%%%%%%%%%%%%%%
%
where $\hat \delta = \beta^e - \alpha^e - \alpha^{\nu} - \beta^{\nu}$
and we have absorbed the matrix
$P_{13} (\alpha^\nu,\beta^\nu) = \diag (e^{i\alpha^\nu},1,e^{i\beta^\nu})$
in the phase matrix $Q_0$.
Using eq.~(\ref{eq:Uthe23thnu13GeZ2GnuZ2trick}) 
and the standard parametrisation
of the PMNS matrix $U$, we find:
%
%%%%%%%%%%%%%%%%%%%%%%%%%%%%%%%%
\begin{align}
\sin^2 \theta_{13} & = |U_{e3}|^2  = \cos^2 \theta^{\circ}_{12} \sin^2 \hat \theta^{\nu}_{13} \,, 
\label{eq:th13Uthe23thnu13GeZ2GnuZ2}\\
\sin^2 \theta_{23} & = \frac{|U_{\mu3}|^2}{1-|U_{e3}|^2} = \frac{1}{1 - \sin^2 \theta_{13}} \big[ \cos^2 \hat \theta^{\nu}_{13} \sin^2 \hat \theta^e_{23} + \cos^2 \hat \theta^e_{23} \sin^2 \hat \theta^{\nu}_{13} \sin^2 \theta^{\circ}_{12} \nonumber \\
& - \frac{1}{2} \sin 2 \hat \theta^e_{23} \sin 2 \hat \theta^{\nu}_{13} \sin \theta^{\circ}_{12} \cos \hat \delta \big ] \,, 
\label{eq:th23Uthe23thnu13GeZ2GnuZ2}\\
\sin^2 \theta_{12} & = \frac{|U_{e2}|^2}{1-|U_{e3}|^2} = \frac{\sin^2 \theta^{\circ}_{12}}{1 - \sin^2 \theta_{13}} \,.
\label{eq:th12Uthe23thnu13GeZ2GnuZ2}
\end{align}
%%%%%%%%%%%%%%%%%%%%%%%%%%%
%
We note that, given $G_f$,
the values of $\sin^2 \theta_{12}$ and  $\sin^2 \theta_{13}$
are correlated. This allows one to perform 
a critical test of the scheme under study once 
the discrete symmetry group $G_f$ has been 
specified.

The sum rule for $\cos \delta$ of interest 
can be derived, e.g., by comparing 
the expressions for the absolute value of
the element $U_{\tau 2}$ of the PMNS matrix in the standard 
parametrisation and in the one obtained using 
eq.~(\ref{eq:Uthe23thnu13GeZ2GnuZ2trick}):
%%%%%%%%%%%%%%%%%%%%%%%%%%%%%%%%%
\begin{align}
|U_{\tau 2}| = 
| \cos \theta_{12} \sin \theta_{23} + \sin \theta_{13} \cos \theta_{23} \sin \theta_{12} e^{i \delta}| = 
|\cos \theta^{\circ}_{12} \sin \hat \theta^e_{23} | \,.
\end{align}
%%%%%%%%%%%%%%%%%%%%%%%%%%%%
%
This leads to
%%%%%%%%%%%%%%%%%%%%%%%%%%%%%%%%%%%%%
\begin{align}
\cos \delta = 
\dfrac{\cos^2 \theta_{13} (\cos^2 \theta^{\circ}_{12} \sin^2 \hat \theta^e_{23} - \sin^2 \theta_{23}) + \sin^2 \theta^\circ_{12} (\sin^2 \theta_{23} - \cos^2 \theta_{23} \sin^2 \theta_{13})}
{\sin 2\theta_{23} \sin \theta_{13} |\sin \theta^\circ_{12}| (\cos^2 \theta_{13} - \sin^2 \theta^\circ_{12})^\frac{1}{2}} \,.
\end{align}
%%%%%%%%%%%%%%%%%%%%%%%%%%%%%%%%%%%%
%

\begin{figure}[h!]
  \begin{center}
 \includegraphics[width=9cm]{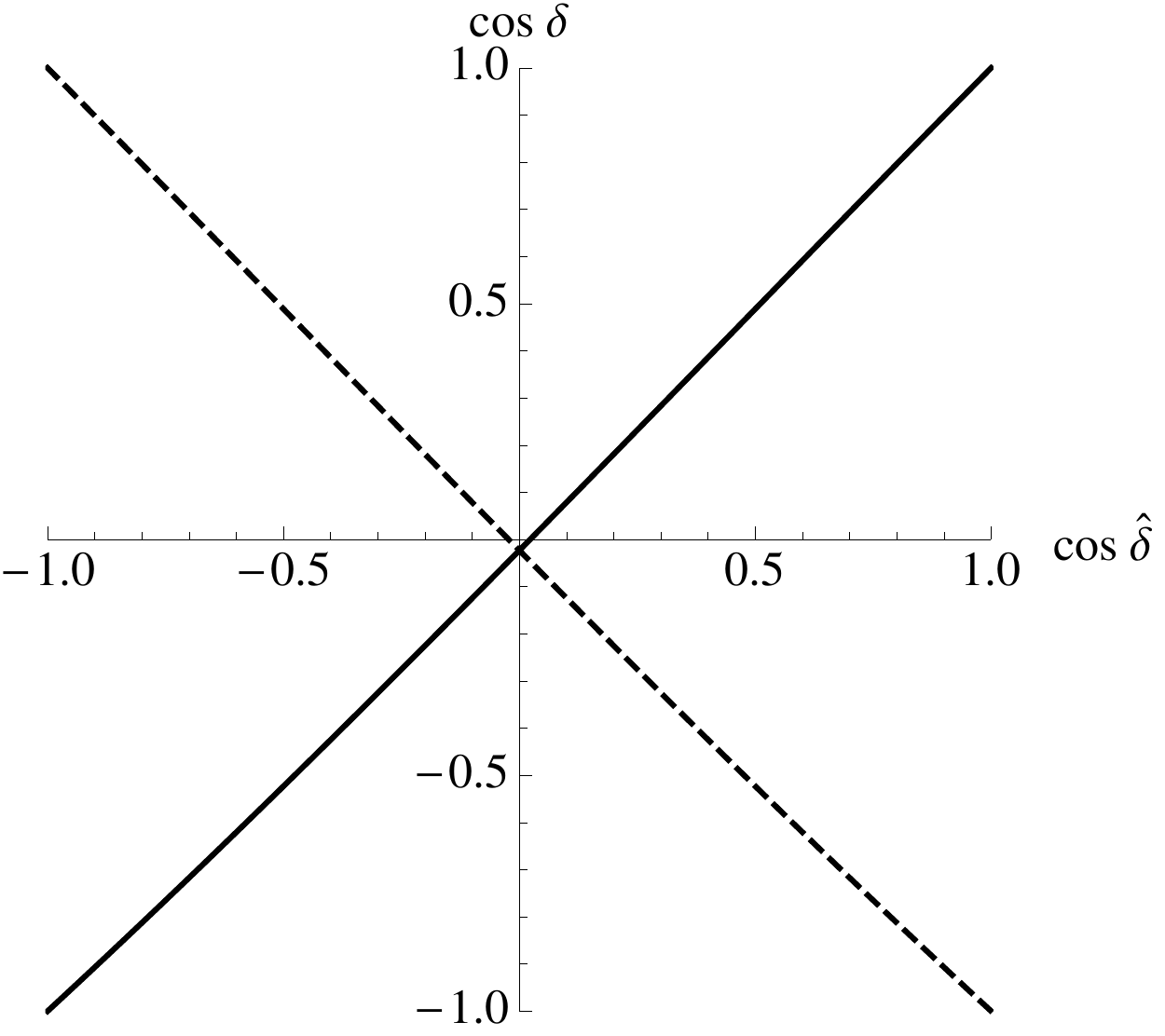}
     \end{center}
\caption{
\label{Fig2}
Dependence of $\cos \delta$ on $\cos \hat\delta$ in the case
of $G_f = S_4$ or $A_5$ with $\sin^2 \theta^\circ_{12} = 1/4$.
The mixing parameters $\sin^2 \theta_{23}$ and $\sin^2 \theta_{13}$
have been fixed to their best fit values for the NO neutrino mass spectrum
quoted in eqs.~(\ref{th23values}) and (\ref{th13values}).
The solid (dashed) line is for the case when 
$\sin 2 \hat \theta^e_{23} \sin 2 \hat \theta^{\nu}_{13}$
is positive (negative).
}
\end{figure}

Similar to the case C2 analysed in subsection 
\ref{sec:13e12nu},  $\cos\delta$ is a function of 
the known neutrino mixing angles 
$\theta_{13}$ and  $\theta_{23}$, of the 
angle $\theta^{\circ}_{12}$ fixed by 
$G_f$ and the assumed symmetry breaking pattern,
as well as 
of the phase parameter  $\hat \delta$ 
of the scheme. Predictions for $\cos\delta$ 
can be obtained if  $\hat \delta$ 
is fixed by additional considerations 
of, e.g., GCP invariance,   
symmetries, etc. In view of this 
we show in Fig.~\ref{Fig2} $\cos\delta$ 
as a function of $\cos\hat \delta$ 
for the current best fit values  
of $\sin^2\theta_{13}$ 
and $\sin^2\theta_{23}$, and for the value 
$\sin^2 \theta^{\circ}_{12} = 1/4$ corresponding to 
$G_f = S_4$ and $A_5$.
We do not find phenomenologically viable cases for
$A_4~(T^{\prime})$. Therefore we do not present such a plot
for these groups.
%%%%%%%%%%%%%%%%%%%%%%%%%%%%%%%%%
%%%%%%%%%%%%%%%%%%%%%%%%%

%%%%%%%%%%%%%%%%%%%%%%%%
\subsection{The Case with $U_{23}(\theta^e_{23}, \delta^e_{23})$ and 
$U_{12}(\theta^{\nu}_{12}, \delta^{\nu}_{12})$ Complex Rotations (Case C6)}
\label{sec:23e12nu}
%%%%%%%%%%%%%%%%%%%%%%%%

 We show below that in this case $\cos\delta$ coincides 
(up to a sign) with the cosine of 
an unconstrained CPV phase parameter of the scheme 
and therefore cannot be determined 
from the values of the neutrino mixing angles and of
the angles determined by the residual symmetries.
Indeed,  using the parametrisation of the matrix $U$ given 
in eq.~(\ref{eq:Z2eZ2nu})
with $(ij) = (23)$ and $(rs) = (12)$ and the parametrisation 
of $U^{\circ}$ as follows (see Appendix \ref{app:ParU} for details), 
%%%%%%%%%%%%%%%%%%%%%%%%%%
\be
U^\circ(\theta^{\circ}_{12},\theta^{\circ}_{13},\theta^{\circ}_{23},\delta^{\circ}_{12},\delta^{\circ}_{23}) = 
U_{23}(\theta^{\circ}_{23}, \delta^{\circ}_{23}) R_{13}(\theta^{\circ}_{13})
U_{12}(\theta^{\circ}_{12}, \delta^{\circ}_{12}) \,,
\ee
%%%%%%%%%%%%%%%%%%%%%%%%%%%%%%
%
we get for $U$:
%%%%%%%%%%%%%%%%%%%%%%%%%
\begin{align}
U & = U_{23}(\theta^e_{23}, \delta^e_{23}) U_{23}(\theta^{\circ}_{23}, \delta^{\circ}_{23}) R_{13}(\theta^{\circ}_{13})
U_{12}(\theta^{\circ}_{12}, \delta^{\circ}_{12}) U_{12}(\theta^{\nu}_{12}, \delta^{\nu}_{12}) Q_0 \,.
\label{eq:Uthe23thnu12GeZ2GnuZ2}
\end{align}
%%%%%%%%%%%%%%%%%%%%%%%%%%%%%%%%%%
%
The results derived in Appendix \ref{app:ParU} in eq.~(\ref{eq:trick1})
make it possible to 
recast eq.~(\ref{eq:Uthe23thnu12GeZ2GnuZ2}) in the form:
%%%%%%%%%%%%%%%%%%%%%%%%%%%%
\be
U = R_{23}( \hat \theta^e_{23}) P_2(\hat \delta) R_{13}(\theta^{\circ}_{13}) 
R_{12}(\hat \theta^{\nu}_{12}) Q_0 \,,
\quad
P_2(\hat \delta) = \diag(1,e^{i\hat\delta},1)\,.
\label{eq:Uthe23thnu12GeZ2GnuZ2trick}
\ee
%%%%%%%%%%%%%%%%%%%%%%%%%%%%%%%%%%%
%
Here $\hat \delta = \alpha^e - \beta^e - \alpha^{\nu} - \beta^{\nu}$ and, 
as in the preceding cases, we have redefined the phase matrix $Q_0$ by 
absorbing the phase
matrix
$P_{12} (\alpha^\nu,\beta^\nu) = \diag (e^{i\alpha^\nu},e^{i\beta^\nu},1)$
in it.
Using eq.~(\ref{eq:Uthe23thnu12GeZ2GnuZ2trick}) 
and the standard parametrisation
of the PMNS matrix $U$, we find:
%%%%%%%%%%%%%%%%%%%%%%%%%%%%%%%%%
\begin{align}
\sin^2 \theta_{13} & = |U_{e3}|^2  = \sin^2 \theta^{\circ}_{13} \,, 
\label{eq:th13Uthe23thnu12GeZ2GnuZ2}\\
\sin^2 \theta_{23} & = \frac{|U_{\mu3}|^2}{1-|U_{e3}|^2} = 
\sin^2 \hat \theta^e_{23} \,, 
\label{eq:th23Uthe23thnu12GeZ2GnuZ2}\\
\sin^2 \theta_{12} & = \frac{|U_{e2}|^2}{1-|U_{e3}|^2} = 
\sin^2 \hat \theta^{\nu}_{12} \,.
\label{eq:th12Uthe23thnu12GeZ2GnuZ2}
\end{align}
%%%%%%%%%%%%%%%%%%%%%%%%%%%%
%
Comparing the absolute value of the element $U_{\tau 1}$
allows us to find that $\cos \delta = \pm \cos \hat \delta$.

 It follows from eq.~(\ref{eq:th13Uthe23thnu12GeZ2GnuZ2}) 
that for a given flavour symmetry $G_f$, 
the value of $\sin^2 \theta_{13}$ is predicted.
This allows to test the phenomenological viability of 
the case under discussion, since the value of  $\sin^2 \theta_{13}$ 
is known experimentally with a relatively high precision.

 A comment, analogous to those made in similar cases 
considered in subsections \ref{sec:23e} and \ref{sec:12nu},
is in order. Namely, for a non-Abelian flavour symmetry $G_f$ 
which allows to reproduce 
correctly the observed values of $\sin^2 \theta_{12}$,
$\sin^2 \theta_{13}$ and $\sin^2 \theta_{23}$, 
it might be possible to obtain physically viable 
prediction for $\cos\delta$ 
by employing GCP invariance in 
the charged lepton or the neutrino sector.

%%%%%%%%%%%%%%%%%%%%%%%%
\subsection{The Case with $U_{12}(\theta^e_{12}, \delta^e_{12})$ and 
$U_{12}(\theta^{\nu}_{12}, \delta^{\nu}_{12})$ Complex Rotations (Case C7)}
\label{sec:12e12nu}
%%%%%%%%%%%%%%%%%%%%%%%%
%
Using the following parametrisation 
of $U^{\circ}$,
%%%%%%%%%%%%%%%%%%%%%%%
\be
U^\circ(\theta^{\circ}_{12}, \tilde\theta^{\circ}_{12}, \theta^{\circ}_{23}, \delta^{\circ}_{12}, \tilde\delta^{\circ}_{12} ) = U_{12}(\theta^{\circ}_{12}, \delta^{\circ}_{12}) R_{23}(\theta^{\circ}_{23}) U_{12}(\tilde\theta^{\circ}_{12}, \tilde\delta^{\circ}_{12}) \,,
\ee
%%%%%%%%%%%%%%%%%%%%%%%%%%%%%%
%
we have for $U$: 
%%%%%%%%%%%%%%%%%%%%%%%%%%%%%%%%%%%%%%%
\begin{align}
U & = U_{12}(\theta^e_{12}, \delta^e_{12}) U_{12}(\theta^{\circ}_{12}, \delta^{\circ}_{12}) R_{23}(\theta^{\circ}_{23}) U_{12}(\tilde\theta^{\circ}_{12}, \tilde\delta^{\circ}_{12}) U_{12}(\theta^{\nu}_{12}, \delta^{\nu}_{12}) Q_0 \,.
\label{eq:Uthe12thnu12GeZ2GnuZ2}
\end{align}
%%%%%%%%%%%%%%%%%%%%%%%%%
%
Utilising the results derived in Appendix \ref{app:ParU} 
and reported in eq.~(\ref{eq:trick1}),
we can recast eq.~(\ref{eq:Uthe12thnu12GeZ2GnuZ2}) in the form:
%%%%%%%%%%%%%%%%%%%%%%%%%%%%
\be
U = R_{12}(\hat \theta^e_{12}) P_1(\hat \delta) 
R_{23}(\theta^{\circ}_{23}) R_{12}(\hat \theta^{\nu}_{12}) Q_0 \,,
\quad
P_1(\hat \delta) = \diag(e^{i\hat\delta},1,1)\,.
\label{eq:Uthe12thnu12GeZ2GnuZ2trick}
\ee
%%%%%%%%%%%%%%%%%%%%%%%%%%%%%
%
Here $\hat \delta = \alpha^e - \beta^e + \alpha^{\nu} + \beta^{\nu}$ 
and we have redefined the matrix $Q_0$ by absorbing the 
diagonal phase matrix 
$P_{12} (-\beta^\nu,-\alpha^\nu) = \diag (e^{-i\beta^\nu},e^{-i\alpha^\nu},1)$ 
in it.
Using eq.~(\ref{eq:Uthe12thnu12GeZ2GnuZ2trick}) 
and the standard parametrisation
of the PMNS matrix $U$, we find:
%
%%%%%%%%%%%%%%%%%%%%%%%%%%%%%%%%
\begin{align}
\sin^2 \theta_{13} & = |U_{e3}|^2  = \sin^2 \theta^\circ_{23} \sin^2 \hat\theta^e_{12}  \,, 
\label{eq:th13Uthe12thnu12GeZ2GnuZ2}\\
\sin^2 \theta_{23} & = \frac{|U_{\mu3}|^2}{1-|U_{e3}|^2} = \frac{\sin^2 \theta^\circ_{23} \cos^2 \hat\theta^e_{12}}{1- \sin^2 \theta_{13}} \,, 
\label{eq:th23Uthe12thnu12GeZ2GnuZ2}\\
\sin^2 \theta_{12} & = \frac{|U_{e2}|^2}{1-|U_{e3}|^2} = \frac{1}{1 - \sin^2 \theta_{13}}
\big[ \cos^2\theta^\circ_{23} \cos^2\hat\theta^\nu_{12} \sin^2\hat\theta^e_{12} 
+ \cos^2\hat\theta^e_{12} \sin^2\hat\theta^\nu_{12} \nonumber \\
& +\frac{1}{2} \sin2\hat\theta^e_{12} \sin2\hat\theta^\nu_{12} \cos\theta^\circ_{23} 
\cos\hat\delta \big] \,.
\label{eq:th12Uthe12thnu12GeZ2GnuZ2}
\end{align}
%%%%%%%%%%%%%%%%%%%%%%%%%%%
%
From eqs.~(\ref{eq:th13Uthe12thnu12GeZ2GnuZ2}) and 
(\ref{eq:th23Uthe12thnu12GeZ2GnuZ2}) we see that the angles
$\theta_{13}$ and $\theta_{23}$ are correlated:
%%%%%%%%%%%%%%%%%%%%%%%%%%%
\be
\sin^2 \theta_{23} = \dfrac{\sin^2 \theta^\circ_{23} - \sin^2 \theta_{13}}
{1 - \sin^2 \theta_{13}}\,.
\ee
%%%%%%%%%%%%%%%%%%%%%%%%%%%

\begin{figure}[h!]
  \begin{center}
 \includegraphics[width=9cm]{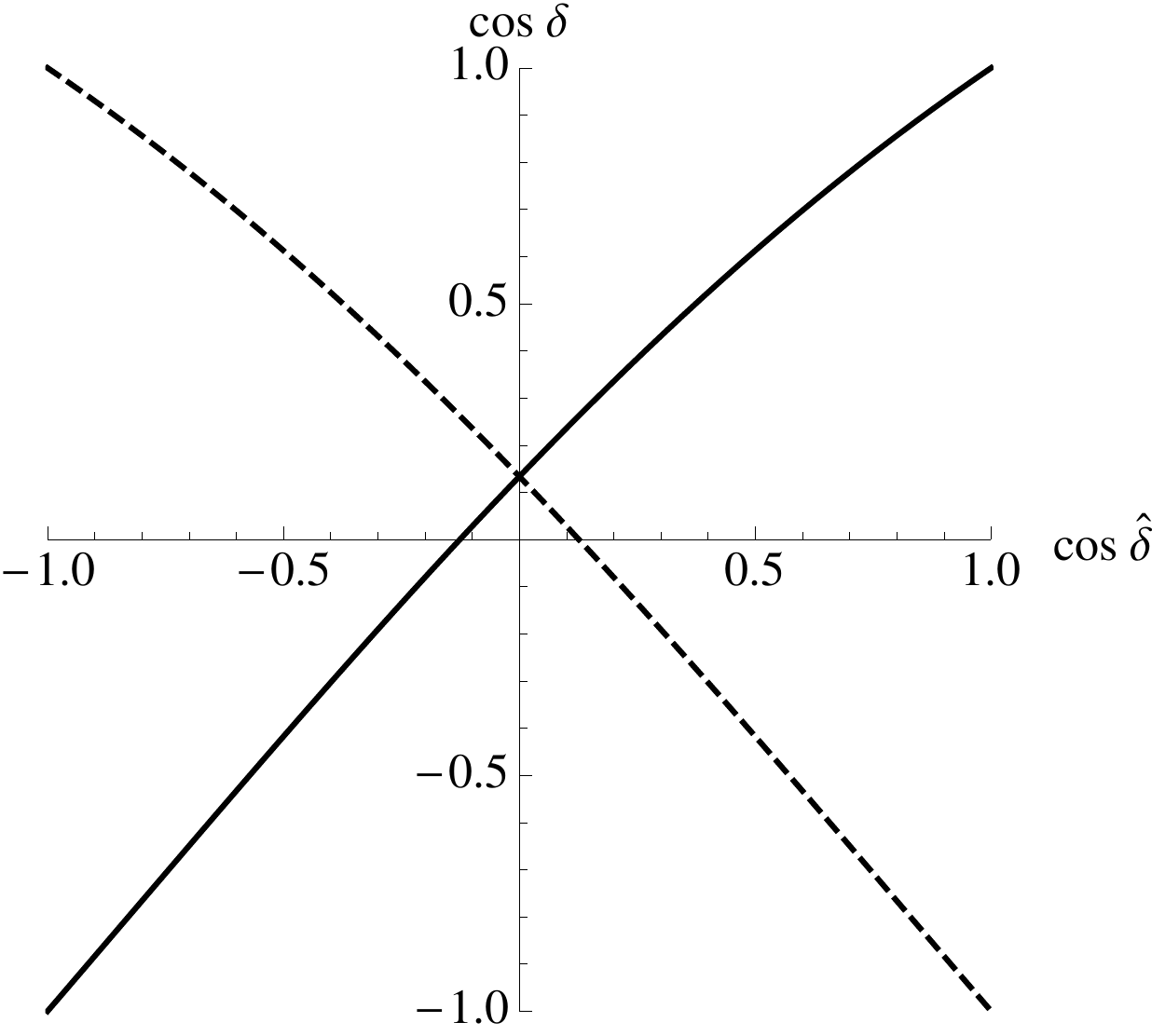}
     \end{center} 
\caption{
\label{Fig3}
Dependence of $\cos \delta$ on $\cos \hat\delta$ in the case
of $G_f = S_4$ with $\sin^2 \theta^\circ_{23} = 1/2$.
The mixing parameters $\sin^2 \theta_{12}$ and $\sin^2 \theta_{13}$
have been fixed to their best fit values for the NO neutrino mass spectrum
quoted in eqs.~(\ref{th12values}) and (\ref{th13values}).
The solid (dashed) line is for the case when 
$\sin 2 \hat \theta^e_{12} \sin 2 \hat \theta^{\nu}_{12}$
is positive (negative).
}
\end{figure}

Comparing the expressions for the absolute value of
the element $U_{\tau 1}$ of the PMNS matrix in the standard 
parametrisation and in the one obtained using 
eq.~(\ref{eq:Uthe12thnu12GeZ2GnuZ2trick}), 
we have
%%%%%%%%%%%%%%%%%%%%%%%%%%%%%%%%%
\begin{align}
|U_{\tau 1}| = | \sin \theta_{12} \sin \theta_{23} - \sin \theta_{13} \cos \theta_{23} \cos \theta_{12} e^{i \delta}| = 
| \sin \hat \theta^{\nu}_{12} \sin \theta^{\circ}_{23} | \,.
\end{align}
%%%%%%%%%%%%%%%%%%%%%%%%%%%%
%
From the above equations we get for $\cos \delta$:
%%%%%%%%%%%%%%%%%%%%%%%%%%%%%%%%%%%%%
\begin{align}
\cos \delta = 
\dfrac{\sin^2 \theta_{13} (\cos^2 \theta_{12} \cos^2 \theta^\circ_{23} - \sin^2 \theta_{12}) + \sin^2 \theta^\circ_{23} (\sin^2 \theta_{12} - \cos^2 \theta_{13} 
\sin^2 \hat \theta^\nu_{12})}
{\sin 2\theta_{12} \sin \theta_{13} |\cos \theta^\circ_{23}| 
(\sin^2 \theta^\circ_{23} - \sin^2 \theta_{13})^\frac{1}{2}} \,.
\end{align}
%%%%%%%%%%%%%%%%%%%%%%%%%%%%%%%%%%%%
%
In this case $\cos\delta$ is a function of 
the known neutrino mixing angles $\theta_{12}$ 
and  $\theta_{13}$, of the 
angle $\theta^\circ_{23}$ fixed by 
$G_f$ and the assumed symmetry breaking pattern,
as well as of the phase parameter  $\hat \delta$ 
of the scheme. Predictions for $\cos\delta$ 
can only be obtained when  $\hat \delta$ 
is fixed by additional considerations 
of, e.g., GCP invariance,   
symmetries, etc. In view of this 
we show in Fig.~\ref{Fig3} $\cos\delta$ 
as a function of $\cos\hat \delta$ 
for the current best fit values  
of $\sin^2\theta_{12}$ and  $\sin^2\theta_{13}$, 
and for the value 
$\sin^2 \theta^\circ_{23} = 1/2$ corresponding to 
$G_f = S_4$. We do not find phenomenologically 
viable cases for $G_f = A_4~(T^{\prime})$ and $A_5$.
%%%%%%%%%%%%%%%%%%%%%%%%%%%%%%%%%
%%%%%%%%%%%%%%%%%%%%%%%%%

%%%%%%%%%%%%%%%%%%%%%%%%
\subsection{The Case with $U_{13}(\theta^e_{13}, \delta^e_{13})$ and 
$U_{13}(\theta^{\nu}_{13}, \delta^{\nu}_{13})$ Complex Rotations (Case C8)}
\label{sec:13e13nu}
%%%%%%%%%%%%%%%%%%%%%%%%
%
Using the following parametrisation 
of $U^{\circ}$,
%%%%%%%%%%%%%%%%%%%%%%%
\be
U^\circ(\theta^{\circ}_{13}, \tilde\theta^{\circ}_{13}, \theta^{\circ}_{23}, \delta^{\circ}_{13}, \tilde\delta^{\circ}_{13} ) = U_{13}(\theta^{\circ}_{13}, \delta^{\circ}_{13}) R_{23}(\theta^{\circ}_{23}) U_{13}(\tilde\theta^{\circ}_{13}, \tilde\delta^{\circ}_{13}) \,,
\ee
%%%%%%%%%%%%%%%%%%%%%%%%%%%%%%
%
we have for $U$: 
%%%%%%%%%%%%%%%%%%%%%%%%%%%%%%%%%%%%%%%
\begin{align}
U & = U_{13}(\theta^e_{13}, \delta^e_{13}) U_{13}(\theta^{\circ}_{13}, \delta^{\circ}_{13}) R_{23}(\theta^{\circ}_{23}) U_{13}(\tilde\theta^{\circ}_{13}, \tilde\delta^{\circ}_{13}) U_{13}(\theta^{\nu}_{13}, \delta^{\nu}_{13}) Q_0 \,.
\label{eq:Uthe13thnu13GeZ2GnuZ2}
\end{align}
%%%%%%%%%%%%%%%%%%%%%%%%%
%
Utilising the results derived in Appendix \ref{app:ParU} 
and reported in eq.~(\ref{eq:trick1}),
we can recast eq.~(\ref{eq:Uthe13thnu13GeZ2GnuZ2}) in the form:
%%%%%%%%%%%%%%%%%%%%%%%%%%%%
\be
U = R_{13}(\hat \theta^e_{13}) P_1(\hat \delta) 
R_{23}(\theta^{\circ}_{23}) R_{13}(\hat \theta^{\nu}_{13}) Q_0 \,,
\quad
P_1(\hat \delta) = \diag(e^{i\hat\delta},1,1)\,.
\label{eq:Uthe13thnu13GeZ2GnuZ2trick}
\ee
%%%%%%%%%%%%%%%%%%%%%%%%%%%%%
%
Here $\hat \delta = \alpha^e - \beta^e + \alpha^{\nu} + \beta^{\nu}$ 
and we have redefined the matrix $Q_0$ by absorbing the 
diagonal phase matrix 
$P_{13} (-\beta^\nu,-\alpha^\nu) = \diag (e^{-i\beta^\nu},1,e^{-i\alpha^\nu})$ 
in it.
Using eq.~(\ref{eq:Uthe13thnu13GeZ2GnuZ2trick}) 
and the standard parametrisation
of the PMNS matrix $U$, we find:
%
%%%%%%%%%%%%%%%%%%%%%%%%%%%%%%%%
\begin{align}
\sin^2 \theta_{13} & = |U_{e3}|^2  = \cos^2\theta^\circ_{23} \cos^2\hat\theta^\nu_{13} \sin^2\hat\theta^e_{13} 
+ \cos^2\hat\theta^e_{13} \sin^2\hat\theta^\nu_{13} \nonumber \\
& +\frac{1}{2} \sin2\hat\theta^e_{13} \sin2\hat\theta^\nu_{13} \cos\theta^\circ_{23} 
\cos\hat\delta \,, 
\label{eq:th13Uthe13thnu13GeZ2GnuZ2}\\
\sin^2 \theta_{23} & = \frac{|U_{\mu3}|^2}{1-|U_{e3}|^2} = \frac{\sin^2 \theta^\circ_{23} \cos^2 \hat\theta^\nu_{13}}{1- \sin^2 \theta_{13}} \,, 
\label{eq:th23Uthe13thnu13GeZ2GnuZ2}\\
\sin^2 \theta_{12} & = \frac{|U_{e2}|^2}{1-|U_{e3}|^2} = \frac{\sin^2 \theta^\circ_{23} \sin^2 \hat\theta^e_{13}}
{1- \sin^2 \theta_{13}} \,.
\label{eq:th12Uthe13thnu13GeZ2GnuZ2}
\end{align}
%%%%%%%%%%%%%%%%%%%%%%%%%%%
%

The sum rule for $\cos \delta$ of interest can be derived 
by comparing the expressions for the absolute value of
the element $U_{\mu 2}$ of the PMNS matrix in the standard 
parametrisation and in the one obtained using 
eq.~(\ref{eq:Uthe13thnu13GeZ2GnuZ2trick}):
%%%%%%%%%%%%%%%%%%%%%%%%%%%%%%%%%
\begin{align}
|U_{\mu 2}| = 
| \cos \theta_{12} \cos \theta_{23} - \sin \theta_{13} \sin \theta_{23} \sin \theta_{12} e^{i \delta}| = 
|\cos \theta^{\circ}_{23}| \,.
\end{align}
%%%%%%%%%%%%%%%%%%%%%%%%%%%%
%
From the above equation we get for $\cos \delta$:
%%%%%%%%%%%%%%%%%%%%%%%%%%%%%%%%%%%%%
%
\begin{align}
\cos \delta = 
\dfrac{\cos^2 \theta_{12} \cos^2 \theta_{23} - \cos^2 \theta^{\circ}_{23} + 
\sin^2 \theta_{12} \sin^2 \theta_{23} \sin^2 \theta_{13}}
{\sin \theta_{13} \sin 2 \theta_{23} \sin \theta_{12} \cos \theta_{12}} \,.
\label{eq:cosdeltaZ2Z212e13nu}
\end{align}
%%%%%%%%%%%%%%%%%%%%%%%%%%%%%%%%%%%%%%
%
Given the assumed breaking pattern, 
$\cos\delta$ depends on the flavour symmetry 
$G_f$ via the value of $\theta^\circ_{23}$.
Using the best fit values of the standard mixing angles 
for the NO neutrino mass spectrum 
and the requirement $|\cos \delta| \leq 1$, we find
that $\sin^2 \theta^{\circ}_{23}$ should lie in the following
interval: $0.537 \leq \sin^2 \theta^{\circ}_{23} \leq 0.677$.
Fixing two of the three angles to their best fit values and varying the
third one in its $3\sigma$ experimentally allowed 
range and considering all the three possible combinations, 
we get that $|\cos \delta| \leq 1$ if 
$0.496 \leq \sin^2 \theta^{\circ}_{23} \leq 0.805$.

%%%%%%%%%%%%%%%%%%%%%%%%
\subsection{The Case with $U_{23}(\theta^e_{23}, \delta^e_{23})$ and 
$U_{23}(\theta^{\nu}_{23}, \delta^{\nu}_{23})$ Complex Rotations (Case C9)}
\label{sec:23e23nu}
%%%%%%%%%%%%%%%%%%%%%%%%
%
Using the following parametrisation 
of $U^{\circ}$,
%%%%%%%%%%%%%%%%%%%%%%%
\be
U^\circ(\theta^{\circ}_{23}, \tilde\theta^{\circ}_{23}, \theta^{\circ}_{12}, \delta^{\circ}_{23}, \tilde\delta^{\circ}_{23} ) = U_{23}(\theta^{\circ}_{23}, \delta^{\circ}_{23}) R_{12}(\theta^{\circ}_{12}) U_{23}(\tilde\theta^{\circ}_{23}, \tilde\delta^{\circ}_{23}) \,,
\ee
%%%%%%%%%%%%%%%%%%%%%%%%%%%%%%
%
we have for $U$: 
%%%%%%%%%%%%%%%%%%%%%%%%%%%%%%%%%%%%%%%
\begin{align}
U & = U_{23}(\theta^e_{23}, \delta^e_{23}) U_{23}(\theta^{\circ}_{23}, \delta^{\circ}_{23}) R_{12}(\theta^{\circ}_{12}) U_{23}(\tilde\theta^{\circ}_{23}, \tilde\delta^{\circ}_{23}) U_{23}(\theta^{\nu}_{23}, \delta^{\nu}_{23}) Q_0 \,.
\label{eq:Uthe23thnu23GeZ2GnuZ2}
\end{align}
%%%%%%%%%%%%%%%%%%%%%%%%%
%
Utilising the results derived in Appendix \ref{app:ParU} 
and reported in eq.~(\ref{eq:trick1}),
we can recast eq.~(\ref{eq:Uthe23thnu23GeZ2GnuZ2}) in the form:
%%%%%%%%%%%%%%%%%%%%%%%%%%%%
\be
U = R_{23}(\hat \theta^e_{23}) P_2(\hat \delta) 
R_{12}(\theta^{\circ}_{12}) R_{23}(\hat \theta^{\nu}_{23}) Q_0 \,,
\quad
P_2(\hat \delta) = \diag(1,e^{i\hat\delta},1)\,.
\label{eq:Uthe23thnu23GeZ2GnuZ2trick}
\ee
%%%%%%%%%%%%%%%%%%%%%%%%%%%%%
%
Here $\hat \delta = \alpha^e - \beta^e + \alpha^{\nu} + \beta^{\nu}$ 
and we have redefined the matrix $Q_0$ by absorbing the 
diagonal phase matrix 
$P_{23} (\alpha^\nu,\beta^\nu) = \diag (1,e^{i\alpha^\nu},e^{i\beta^\nu})$ 
in it.
Using eq.~(\ref{eq:Uthe23thnu23GeZ2GnuZ2trick}) 
and the standard parametrisation
of the PMNS matrix $U$, we find:
%
%%%%%%%%%%%%%%%%%%%%%%%%%%%%%%%%
\begin{align}
\sin^2 \theta_{13} & = |U_{e3}|^2  = \sin^2 \theta^\circ_{12} \sin^2 \hat\theta^\nu_{23} \,, 
\label{eq:th13Uthe23thnu23GeZ2GnuZ2}\\
\sin^2 \theta_{23} & = \frac{|U_{\mu3}|^2}{1-|U_{e3}|^2} = \frac{1}{1 - \sin^2 \theta_{13}}
\big[ \cos^2\theta^\circ_{12} \cos^2\hat\theta^e_{23} \sin^2\hat\theta^\nu_{23} 
+ \cos^2\hat\theta^\nu_{23} \sin^2\hat\theta^e_{23} \nonumber \\
& +\frac{1}{2} \sin2\hat\theta^e_{23} \sin2\hat\theta^\nu_{23} \cos\theta^\circ_{12} 
\cos\hat\delta \big] \,, 
\label{eq:th23Uthe23thnu23GeZ2GnuZ2}\\
\sin^2 \theta_{12} & = \frac{|U_{e2}|^2}{1-|U_{e3}|^2} = \frac{\sin^2 \theta^\circ_{12} \cos^2 \hat\theta^\nu_{23}}{1- \sin^2 \theta_{13}} \,.
\label{eq:th12Uthe23thnu23GeZ2GnuZ2}
\end{align}
%%%%%%%%%%%%%%%%%%%%%%%%%%%
%
From eqs.~(\ref{eq:th13Uthe23thnu23GeZ2GnuZ2}) and 
(\ref{eq:th12Uthe23thnu23GeZ2GnuZ2}) we find that the angles
$\theta_{13}$ and $\theta_{12}$ are correlated:
%%%%%%%%%%%%%%%%%%%%%%%%%%%
\be
\sin^2 \theta_{12} = \frac{\sin^2 \theta^\circ_{12} - \sin^2 \theta_{13}}
{1 - \sin^2 \theta_{13}}\,.
\ee
%%%%%%%%%%%%%%%%%%%%%%%%%%%

\begin{figure}[h!]
  \begin{center}
 \includegraphics[width=9cm]{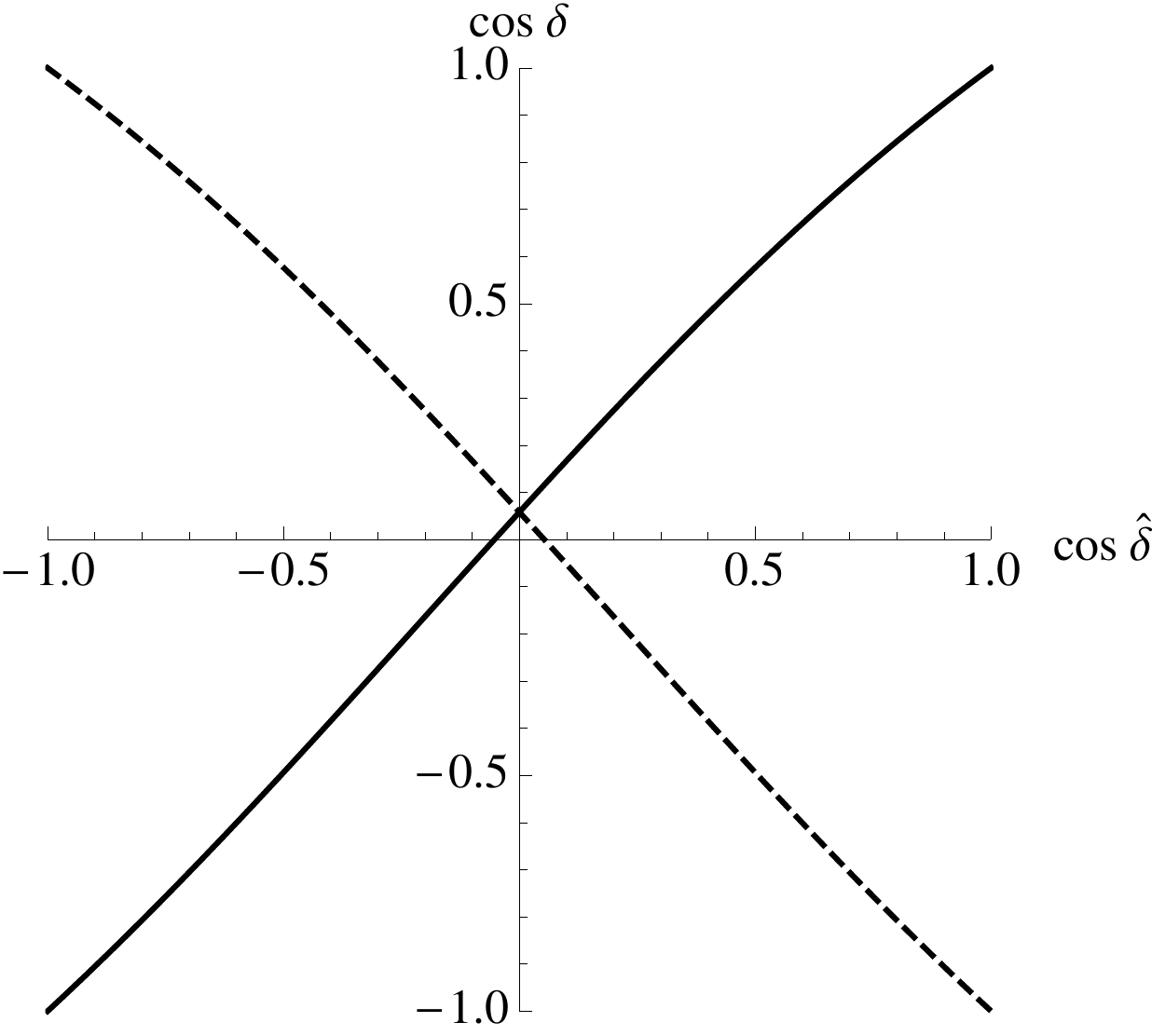}
     \end{center}
\caption{
\label{Fig4}
Dependence of $\cos \delta$ on $\cos \hat\delta$ in the case
of $G_f = A_5$ with $\sin^2 \theta^\circ_{12} = (r + 2)/(4r + 4) \cong 0.345$.
The mixing parameters $\sin^2 \theta_{23}$ and $\sin^2 \theta_{13}$
have been fixed to their best fit values for the NO neutrino mass spectrum
quoted in eqs.~(\ref{th23values}) and (\ref{th13values}).
The solid (dashed) line is for the case when 
$\sin 2 \hat \theta^e_{23} \sin 2 \hat \theta^{\nu}_{23}$
is positive (negative).
}
\end{figure}

Comparing the expressions for the absolute value of
the element $U_{\tau 1}$ of the PMNS matrix in the standard 
parametrisation and in the one obtained using 
eq.~(\ref{eq:Uthe23thnu23GeZ2GnuZ2trick}), 
we have
%%%%%%%%%%%%%%%%%%%%%%%%%%%%%%%%%
\begin{align}
|U_{\tau 1}| = | \sin \theta_{12} \sin \theta_{23} - \sin \theta_{13} \cos \theta_{23} \cos \theta_{12} e^{i \delta}| = 
| \sin \hat \theta^e_{23} \sin \theta^{\circ}_{12} | \,.
\end{align}
%%%%%%%%%%%%%%%%%%%%%%%%%%%%
%
From the above equations we get for $\cos \delta$:
%%%%%%%%%%%%%%%%%%%%%%%%%%%%%%%%%%%%%
\begin{align}
\cos \delta = 
\dfrac{\sin^2 \theta_{13} (\cos^2 \theta_{23} \cos^2 \theta^\circ_{12} - \sin^2 \theta_{23}) + \sin^2 \theta^\circ_{12} (\sin^2 \theta_{23} - \cos^2 \theta_{13} 
\sin^2 \hat \theta^e_{23})}
{\sin 2\theta_{23} \sin \theta_{13} |\cos \theta^\circ_{12}| 
(\sin^2 \theta^\circ_{12} - \sin^2 \theta_{13})^\frac{1}{2}} \,.
\end{align}
%%%%%%%%%%%%%%%%%%%%%%%%%%%%%%%%%%%%
%
In this case $\cos\delta$ is a function of 
the known neutrino mixing angles $\theta_{23}$ 
and  $\theta_{13}$, of the 
angle $\theta^\circ_{12}$ fixed by 
$G_f$ and the assumed symmetry breaking pattern,
as well as of the phase parameter  $\hat \delta$ 
of the scheme. Predictions for $\cos\delta$ 
can only be obtained when  $\hat \delta$ 
is fixed by additional considerations 
of, e.g., GCP invariance,   
symmetries, etc. In view of this 
we show in Fig.~\ref{Fig4} $\cos\delta$ 
as a function of $\cos\hat \delta$ 
for the current best fit values  
of $\sin^2\theta_{23}$ and  $\sin^2\theta_{13}$, 
and for the value 
$\sin^2 \theta^\circ_{12} = (r + 2)/(4r + 4) \cong 0.345$ corresponding to 
$G_f = A_5$.
We do not find phenomenologically 
viable cases for $G_f = A_4~(T^{\prime})$ and $S_4$.
%%%%%%%%%%%%%%%%%%%%%%%%%%%%%%%%%
%%%%%%%%%%%%%%%%%%%%%%%%%

%%%%%%%%%%%%%%%%%%%%%%%%
\subsection{Results in the Cases of $G_f = A_4~(T'),~S_4$ and $A_5$}
\label{sec:ressec4}
%%%%%%%%%%%%%%%%%%%%%%%%
%
The schemes considered in Sections~\ref{sec:12e13nu}~--~\ref{sec:23e23nu} can be applied when considering the breaking $G_f$ to $G_e=Z_2$ and $G_{\nu}=Z_2$, for both Majorana and Dirac neutrinos.  As explicit examples of this, we now consider $G_f=A_4~(T')$, $S_4$ and $A_5$ broken to $G_e=Z_2$ and $G_{\nu}=Z_2$.  As such,
 we have considered all possible combinations of
residual $Z_2$ symmetries for a given flavour symmetry group, namely,
$G_e = Z_2 \mbox{ and } G_{\nu} =  Z_2 \mbox{ for }G_f = A_4~(T^{\prime})$, $S_4$, $A_5$.
For instance, in the cases of the schemes described in subsections 
\ref{sec:12e13nu}~--~\ref{sec:23e13nu}, 
and $G_f = S_4$ broken to $G_e = Z_2^a$ and $G_\nu = Z_2^b$ with
$(a,b) = (T^2U,U)$, $(T^2U,SU)$, $(T^2U,TU)$, $(T^2U,STSU)$, etc. 
(a total of 24 combinations of order two elements), 
the value of the relevant parameter 
contained in the fixed matrix $U^\circ$ yields 
$\sin^2 \theta^\circ_{23} = 1/4$, $\sin^2 \theta^{\circ}_{23} = 1/2$,
$\sin^2 \theta^{\circ}_{13} = 1/4$, $\sin^2 \theta^{\circ}_{12} = 1/4$,
and $\sin^2 \theta^{\circ}_{12} = 1/4$, respectively.
In $A_5$ for the cases C1, C3, C4 and C5 we find the 
sine square of the corresponding fixed angle in the matrix $U^{\circ}$ to be $1/4$,
e.g., for $G_e = Z_2^a$ and $G_\nu = Z_2^b$ with
$(a,b) = (S,S T^2 S T^3 S)$, $(S,S T^3 S T^2 S)$, $(S, T^2 S T^3)$, $(S,T^3 S T^2)$, etc. 
(in total, for 60 combinations of order two elements).

For the symmetry group $A_4$ we find that none of the
combinations of the residual symmetries
$G_e = Z_2$ and $G_{\nu} = Z_2$ provide physical values of $\cos \delta$ 
and phenomenologically viable results for the neutrino mixing angles
simultaneously.

For $G_f = S_4$, using the best fit values of the mixing angles
$\theta_{12}$, $\theta_{13}$ and $\theta_{23}$, we get  
$\cos \delta = -0.806$, $-1.52$ and $0.992$ in the cases C1, C3 and C4,
respectively. Physically acceptable value of $\cos \delta$ in the
case C3 can be obtained for 
$\sin^2 \theta_{23} = 0.562$ allowed at $3\sigma$, for which
$\cos \delta = -0.996$. In the part of the $3\sigma$
allowed range of $\sin^2 \theta_{23}$,
$0.562 \leq \sin^2 \theta_{23} \leq 0.641$,
we have $-0.996 \leq \cos\delta \leq -0.690$.
Further, in the case C2, in which the relevant
parameter $\sin^2 \theta^\circ_{23} = 1/2$, the value of $\cos \delta$
is not fixed, while the atmospheric angle is predicted to have a value 
corresponding to $\sin^2 \theta_{23} = 0.512$.
Similarly, in the case C5 the value of $\cos \delta$ is not fixed,
while $\sin^2 \theta_{12} = 0.256$ (which is slightly outside 
the corresponding $3\sigma$ interval).
In the case C7 we find that $\cos\delta$ is not fixed and 
$\sin^2 \theta_{23} = 0.488$. Finally, for C8 with 
$\sin^2 \theta^\circ_{23} = 1/2$ and $3/4$, using the best fit
values of the neutrino mixing angles for the NO spectrum, 
we have  $\cos\delta = -1.53$ and $2.04$, respectively. 
The physical values of $\cos \delta$
can be obtained, using, e.g., the values of 
$\sin^2 \theta_{23} = 0.380$ and $0.543$, for which
$\cos \delta = -0.995$ and $0.997$, respectively. 
In the parts of the $3\sigma$
allowed range of $\sin^2 \theta_{23}$,
$0.374 \leq \sin^2 \theta_{23} \leq 0.380$ and
$0.543 \leq \sin^2 \theta_{23} \leq 0.641$,
we have $-0.938 \geq \cos\delta \geq -0.995$
and $0.997 \geq \cos\delta \geq 0.045$, respectively.

For the $A_5$ symmetry group the cases 
C1 with $\sin^2 \theta^\circ_{23} = 1/4$,
C3 with $\sin^2 \theta^\circ_{13} = 1/4$  and 
C4 with $\sin^2 \theta^\circ_{12} = 1/4$
lead to the same predictions obtained with $G_f = S_4$,
namely, $\cos \delta = -0.806$, $-1.52$ and $0.992$, respectively.
Moreover, in the case C3 (case C4) the value of 
$\sin^2 \theta^\circ_{13} = 0.096$ 
($\sin^2 \theta^\circ_{12} = 0.096$) is found, which along with 
the best fit values of the mixing angles gives $\cos \delta = 0.688$
($\cos \delta = -1.21$). Using the value of 
$\sin^2 \theta_{23} = 0.487$ allowed at $2\sigma$,
one gets in the case C4 $\cos \delta = -0.997$,
while in the part of the $3\sigma$
allowed range of $\sin^2 \theta_{23}$,
$0.487 \leq \sin^2 \theta_{23} \leq 0.641$,
we have $-0.997 \leq \cos\delta \leq -0.376$.
Note also, if $\sin^2 \theta_{23}$ is fixed to its best fit value,
one can obtain the physical value of $\cos\delta = -0.999$
using $\sin^2 \theta_{12} = 0.277$. 
For the part of the $3\sigma$
allowed range of $\sin^2 \theta_{12}$,
$0.259 \leq \sin^2 \theta_{12} \leq 0.277$,
one gets $-0.871 \geq \cos\delta \geq -0.999$.
The cases C5 and C8 are the same as for the $S_4$ symmetry group. 
Finally, in the case C9
the value of $\cos\delta$ is not fixed, while
using the best fit value of the reactor angle,
we get $\sin^2 \theta_{12} = 0.330$.

%%%%%%%%%%%%%%%%%%%%%%%%
\section{Summary of 
the Results of Sections 3, 4 and 5}
\label{sec:sumcom}
%%%%%%%%%%%%%%%%%%%%%%%%
%

The sum rules derived in Sections 3, 4 and 5  
are summarised in Tables~\ref{tab:summarysumrules} and \ref{tab:summarysumrulesB}.
The formulae for $\sin^2 \theta_{12}$, $\sin^2 \theta_{13}$ and 
$\sin^2 \theta_{23}$, which lead to predictions for 
the indicated neutrino mixing parameters once 
the discrete flavour symmetry $G_f$ is fixed, 
are given in Tables~\ref{tab:summarysin2th23_12} and \ref{tab:summarysin2th23_12B}.
In the cases
in Tables~\ref{tab:summarysin2th23_12} and \ref{tab:summarysin2th23_12B} in which 
$\cos \delta$ is unconstrained, a relatively precise 
measurement of $\sin^2 \theta_{12}$, $\sin^2 \theta_{13}$ or $\sin^2 \theta_{23}$ can
provide a critical test of the corresponding schemes 
due to constraints satisfied by the indicated neutrino mixing parameters.

A general comment on the results 
derived in Sections 3, 4 and 5 is in order.
Since we do not have any information on the mass matrices, 
we have the freedom to
permute the columns of the matrices $U_e$ and $U_{\nu}$, 
or equivalently, the columns and the rows
of the PMNS matrix $U$. 
The results in Tables~\ref{tab:summarysumrules} and \ref{tab:summarysumrulesB} 
cover all the possibilities
because, as we demonstrate below, 
the permutations bring one of the considered cases 
into another considered case.
For example, consider the case of
$U = U_{13}(\theta^e_{13}, \delta^e_{13}) U^{\circ} U_{23}(\theta^{\nu}_{23}, \delta^{\nu}_{23}) Q_0$.
The permutation of the second and the third rows of $U$ is given by
$\pi_{23} U = \pi_{23} U_{13}(\theta^e_{13}, \delta^e_{13}) \pi_{23} \pi_{23} U^{\circ} U_{23}(\theta^{\nu}_{23}, \delta^{\nu}_{23}) Q_0$,
where we have defined
%%%%%%%%%%%%%%%%%%%%%%%%%%%%%%%
\begin{align}
\pi_{23} = \begin{pmatrix}
1  & 0 & 0 \\[2pt]
0  & 0 & 1 \\[2pt]
0 &  1 & 0 \\[2pt]
\end{pmatrix} \;.
\end{align}
%%%%%%%%%%%%%%%%%%%%%%%%%%%%
%
Since the combination $\pi_{23} U_{13}(\theta^e_{13}, \delta^e_{13}) \pi_{23}$ 
gives a unitary matrix
$U_{12}(\theta^e_{13}, \delta^e_{13})$, 
the result after the redefinition,
$\theta^e_{13} \rightarrow \theta^e_{12}$, $\delta^e_{13} \rightarrow \delta^e_{12}$
and $\pi_{23} U^{\circ} \rightarrow U^{\circ}$, 
yields 
$$U = U_{12}(\theta^e_{12}, \delta^e_{12}) U^{\circ} 
U_{23}(\theta^{\nu}_{23}, \delta^{\nu}_{23}) Q_0\,,$$
which represents another case present in 
Table~\ref{tab:summarysumrulesB}.
It is worth noting that the freedom in redefining the matrix $U^{\circ}$
follows from the fact that $U^{\circ}$ is a general $3 \times 3$ unitary
matrix and hence can be parametrised as described in Section~\ref{sec:prelim}
and in Appendix~\ref{app:ParU}.
All the other permutations should be treated 
in the same way and lead to similar results.\\

\newpage
\begin{landscape}
\pagestyle{empty}
%%%%%%%%%%%%%%%%%%%%%%%%%%
\begin{table}[h]
\renewcommand*{\arraystretch}{1.2}
\vspace{-0.8cm}
\hspace{0.5cm}
\begin{tabular}{lll}
\hline
& \\ [-10pt]
Case & Parametrisation of $U$ & Sum rule for $\cos \delta$ \\ [8pt]
\bottomrule
& \\ [-10pt]
A1 & $U_{12}(\theta^e_{12}, \delta^e_{12}) U_{12}(\theta^{\circ}_{12}, \delta^{\circ}_{12}) R_{23}(\theta_{23}^{\circ}) R_{13}(\theta_{13}^{\circ}) Q_0 $ & 
$\phantom{-}\dfrac{\cos^2 \theta_{13} (\sin^2 \theta^{\circ}_{23} - \cos^2 \theta_{12}) + \cos^2 \theta^{\circ}_{13} \cos^2 \theta^{\circ}_{23} (\cos^2 \theta_{12} - \sin^2 \theta_{12} \sin^2 \theta_{13})}{\sin 2 \theta_{12} \sin \theta_{13} |\cos \theta^{\circ}_{13} \cos \theta^{\circ}_{23}| (\cos^2 \theta_{13} - \cos^2 \theta^{\circ}_{13} \cos^2 \theta^{\circ}_{23})^{\frac{1}{2}}}$ \\ [15pt]
\hline
& \\ [-10pt]
A2 & $U_{13}(\theta^e_{13}, \delta^e_{13}) U_{13}(\theta^{\circ}_{13}, \delta^{\circ}_{13}) R_{23}(\theta_{23}^{\circ}) R_{12}(\theta_{12}^{\circ}) Q_0 $ & 
$-\dfrac{\cos^2 \theta_{13} (\cos^2 \theta^{\circ}_{12} \cos^2 \theta^{\circ}_{23} - \cos^2 \theta_{12}) + \sin^2 \theta^{\circ}_{23} (\cos^2 \theta_{12} - \sin^2 \theta_{12} \sin^2 \theta_{13})}{\sin 2 \theta_{12} \sin \theta_{13} |\sin \theta^{\circ}_{23}| (\cos^2 \theta_{13} - \sin^2 \theta^{\circ}_{23})^{\frac{1}{2}}}$ \\ [15pt]
\hline
& \\ [-10pt]
A3 & $U_{23}(\theta^e_{23}, \delta^e_{23}) U_{23}(\theta^{\circ}_{23}, \delta^{\circ}_{23}) R_{13}(\theta_{13}^{\circ}) R_{12}(\theta_{12}^{\circ}) Q_0$ & 
$\pm \cos \hat \delta_{23}$ \\ [15pt]
\hline
& \\ [-10pt]
B1 & $R_{23}(\theta_{23}^{\circ}) R_{12}(\theta_{12}^{\circ}) U_{13}(\theta^{\circ}_{13}, \delta^{\circ}_{13}) U_{13}(\theta^{\nu}_{13}, \delta^{\nu}_{13}) Q_0$ & 
$-\dfrac{\cos^2 \theta_{13} (\cos^2 \theta^{\circ}_{12} \cos^2 \theta^{\circ}_{23} - \cos^2 \theta_{23}) + \sin^2 \theta^{\circ}_{12} (\cos^2 \theta_{23} - \sin^2 \theta_{13} \sin^2 \theta_{23})}
{\sin 2 \theta_{23} \sin \theta_{13} |\sin \theta^{\circ}_{12}| (\cos^2 \theta_{13} - \sin^2 \theta^{\circ}_{12})^{\frac{1}{2}}}$ \\ [15pt]
\hline
& \\ [-10pt]
B2 & $R_{13}(\theta_{13}^{\circ}) R_{12}(\theta_{12}^{\circ}) U_{23}(\theta^{\circ}_{23}, \delta^{\circ}_{23}) U_{23}(\theta^{\nu}_{23}, \delta^{\nu}_{23}) Q_0$ & 
$\phantom{-}\dfrac{\cos^2 \theta_{13} (\sin^2 \theta^{\circ}_{12} - \cos^2 \theta_{23}) + \cos^2 \theta^{\circ}_{12} \cos^2 \theta^{\circ}_{13} ( \cos^2 \theta_{23} - \sin^2 \theta_{13} \sin^2 \theta_{23} )}
{ \sin 2 \theta_{23} \sin \theta_{13} | \cos \theta^{\circ}_{12} \cos \theta^{\circ}_{13}| (\cos^2 \theta_{13} - \cos^2 \theta^{\circ}_{12} \cos^2 \theta^{\circ}_{13} )^{\frac{1}{2}}}$ \\ [15pt]
 \hline
 & \\ [-10pt]
B3 & $R_{23}(\theta_{23}^{\circ}) R_{13}(\theta_{13}^{\circ}) 
U_{12}(\theta^{\circ}_{12}, \delta^{\circ}_{12}) 
U_{12}(\theta^{\nu}_{12}, \delta^{\nu}_{12}) Q_0$ & 
$\pm \cos \hat \delta_{12}$ \\ [15pt]
\hline
\end{tabular}
\caption{Summary of the sum rules for $\cos\delta$.
The cases A1, A2 and A3 correspond to $G_e = Z_2$ and $G_{\nu} = Z_n$, $n > 2$ or $Z_n \times Z_m$, $n,m \geq 2$,
while B1, B2 and B3 correspond to $G_e = Z_n$, $n > 2$ or $Z_n \times Z_m$, $n,m \geq 2$ and $G_{\nu} = Z_2$.
See text for further details.
}
\label{tab:summarysumrules}
\end{table}
%%%%%%%%%%%%%%%%%%%%%%%%%% 
\end{landscape}

\newpage
\begin{landscape}
\pagestyle{empty}
%%%%%%%%%%%%%%%%%%%%%%%%%%
\begin{table}[h]
\renewcommand*{\arraystretch}{1.2}
\vspace{-0.8cm}
\hspace{-0.5cm}
\begin{tabular}{lll}
\hline
& \\ [-10pt]
Case & Parametrisation of $U$ & Sum rule for $\cos \delta$ \\ [8pt]
\bottomrule
& \\ [-10pt]
 C1 & $U_{12}(\theta^e_{12}, \delta^e_{12}) U_{12}(\theta^{\circ}_{12}, \delta^{\circ}_{12}) R_{23}(\theta^{\circ}_{23}) U_{13}(\theta^{\circ}_{13}, \delta^{\circ}_{13}) U_{13}(\theta^{\nu}_{13}, \delta^{\nu}_{13}) Q_0$ & 
$\dfrac{\sin^2 \theta^{\circ}_{23} - \cos^2 \theta_{12} \sin^2 \theta_{23} - \cos^2 \theta_{23} \sin^2 \theta_{12} \sin^2 \theta_{13}}
{\sin \theta_{13} \sin 2 \theta_{23} \sin \theta_{12} \cos \theta_{12}}$
\\ [15pt]
\hline
 & \\ [-10pt]
C2 & $U_{13}(\theta^e_{13}, \delta^e_{13}) U_{13}(\theta^{\circ}_{13}, \delta^{\circ}_{13}) R_{23}(\theta^{\circ}_{23}) U_{12}(\theta^{\circ}_{12}, \delta^{\circ}_{12}) U_{12}(\theta^{\nu}_{12}, \delta^{\nu}_{12}) Q_0$ & 
$\dfrac{\cos^2 \theta_{13} (\cos^2 \theta^{\circ}_{23} \sin^2 \hat \theta^{\nu}_{12} - \sin^2 \theta_{12}) + \sin^2 \theta^\circ_{23} (\sin^2 \theta_{12} - \cos^2 \theta_{12} \sin^2 \theta_{13})}
{\sin 2\theta_{12} \sin \theta_{13} |\sin \theta^\circ_{23}| (\cos^2 \theta_{13} - \sin^2 \theta^\circ_{23})^\frac{1}{2}}$
\\ [15pt]
\hline
 & \\ [-10pt]
C3 & $U_{12}(\theta^e_{12}, \delta^e_{12}) U_{12}(\theta^{\circ}_{12}, \delta^{\circ}_{12}) R_{13}(\theta^{\circ}_{13}) U_{23}(\theta^{\circ}_{23}, \delta^{\circ}_{23}) U_{23}(\theta^{\nu}_{23}, \delta^{\nu}_{23}) Q_0$ & 
$\dfrac{\sin^2 \theta_{12} \sin^2 \theta_{23} - \sin^2 \theta^{\circ}_{13} + \cos^2 \theta_{12} \cos^2 \theta_{23} \sin^2 \theta_{13}}
{\sin \theta_{13} \sin 2 \theta_{23} \sin \theta_{12} \cos \theta_{12}}$
\\ [15pt]
\hline
 & \\ [-10pt]
C4 & $U_{13}(\theta^e_{13}, \delta^e_{13}) U_{13}(\theta^{\circ}_{13}, \delta^{\circ}_{13}) R_{12}(\theta^{\circ}_{12}) U_{23}(\theta^{\circ}_{23}, \delta^{\circ}_{23}) U_{23}(\theta^{\nu}_{23}, \delta^{\nu}_{23}) Q_0$ & 
$\dfrac{\sin^2 \theta^{\circ}_{12} - \cos^2 \theta_{23} \sin^2 \theta_{12} - \cos^2 \theta_{12} \sin^2 \theta_{13} \sin^2 \theta_{23}}
{\sin \theta_{13} \sin 2 \theta_{23} \sin \theta_{12} \cos \theta_{12}}$
\\ [15pt]
\hline
 & \\ [-10pt]
C5 & $U_{23}(\theta^e_{23}, \delta^e_{23}) U_{23}(\theta^{\circ}_{23}, \delta^{\circ}_{23}) R_{12}(\theta^{\circ}_{12}) U_{13}(\theta^{\circ}_{13}, \delta^{\circ}_{13}) U_{13}(\theta^{\nu}_{13}, \delta^{\nu}_{13}) Q_0$ & 
$\dfrac{\cos^2 \theta_{13} (\cos^2 \theta^{\circ}_{12} \sin^2 \hat \theta^e_{23} - \sin^2 \theta_{23}) + \sin^2 \theta^\circ_{12} (\sin^2 \theta_{23} - \cos^2 \theta_{23} \sin^2 \theta_{13})}
{\sin 2\theta_{23} \sin \theta_{13} |\sin \theta^\circ_{12}| (\cos^2 \theta_{13} - \sin^2 \theta^\circ_{12})^\frac{1}{2}}$
\\ [15pt]
\hline
 & \\ [-10pt]
C6 & $U_{23}(\theta^e_{23}, \delta^e_{23}) U_{23}(\theta^{\circ}_{23}, \delta^{\circ}_{23}) R_{13}(\theta^{\circ}_{13})
U_{12}(\theta^{\circ}_{12}, \delta^{\circ}_{12}) U_{12}(\theta^{\nu}_{12}, \delta^{\nu}_{12}) Q_0$ & $\pm \cos \hat \delta$
\\ [15pt]
\hline
 & \\ [-10pt]
C7 & $U_{12}(\theta^e_{12}, \delta^e_{12}) U_{12}(\theta^{\circ}_{12}, \delta^{\circ}_{12}) R_{23}(\theta^{\circ}_{23}) U_{12}(\tilde\theta^{\circ}_{12}, \tilde\delta^{\circ}_{12}) U_{12}(\theta^{\nu}_{12}, \delta^{\nu}_{12}) Q_0$ & $\dfrac{\sin^2 \theta_{13} (\cos^2 \theta_{12} \cos^2 \theta^\circ_{23} - \sin^2 \theta_{12}) + \sin^2 \theta^\circ_{23} (\sin^2 \theta_{12} - \cos^2 \theta_{13} 
\sin^2 \hat \theta^\nu_{12})}
{\sin 2\theta_{12} \sin \theta_{13} |\cos \theta^\circ_{23}| 
(\sin^2 \theta^\circ_{23} - \sin^2 \theta_{13})^\frac{1}{2}}$
\\ [15pt]
\hline
 & \\ [-10pt]
C8 & $U_{13}(\theta^e_{13}, \delta^e_{13}) U_{13}(\theta^{\circ}_{13}, \delta^{\circ}_{13}) R_{23}(\theta^{\circ}_{23}) U_{13}(\tilde\theta^{\circ}_{13}, \tilde\delta^{\circ}_{13}) U_{13}(\theta^{\nu}_{13}, \delta^{\nu}_{13}) Q_0$ & $\dfrac{\cos^2 \theta_{12} \cos^2 \theta_{23} - \cos^2 \theta^{\circ}_{23} + 
\sin^2 \theta_{12} \sin^2 \theta_{23} \sin^2 \theta_{13}}
{\sin \theta_{13} \sin 2 \theta_{23} \sin \theta_{12} \cos \theta_{12}} $
\\ [15pt]
\hline
 & \\ [-10pt]
C9 & $U_{23}(\theta^e_{23}, \delta^e_{23}) U_{23}(\theta^{\circ}_{23}, \delta^{\circ}_{23}) R_{12}(\theta^{\circ}_{12}) U_{23}(\tilde\theta^{\circ}_{23}, \tilde\delta^{\circ}_{23}) U_{23}(\theta^{\nu}_{23}, \delta^{\nu}_{23}) Q_0$ & $\dfrac{\sin^2 \theta_{13} (\cos^2 \theta_{23} \cos^2 \theta^\circ_{12} - \sin^2 \theta_{23}) + \sin^2 \theta^\circ_{12} (\sin^2 \theta_{23} - \cos^2 \theta_{13} 
\sin^2 \hat \theta^e_{23})}
{\sin 2\theta_{23} \sin \theta_{13} |\cos \theta^\circ_{12}| 
(\sin^2 \theta^\circ_{12} - \sin^2 \theta_{13})^\frac{1}{2}}$
\\ [15pt]
\hline
\end{tabular}
\caption{Summary of the sum rules for $\cos\delta$.
The cases C1~--~C9 correspond to 
$G_e = Z_2$ and $G_{\nu} = Z_2$. See text for further details.
}
\label{tab:summarysumrulesB}
\end{table}
%%%%%%%%%%%%%%%%%%%%%%%%%%
\end{landscape}

\begin{landscape}
\pagestyle{empty}
%%%%%%%%%%%%%%%%%%%%%%%%%%
\begin{table}[h]
\centering
\renewcommand*{\arraystretch}{1.2}
\vspace{-0.8cm}
\hspace{-0.5cm}
\begin{tabular}{lll}
\hline
& \\ [-10pt]
Case & Parametrisation of $U$ & Sum rule for $\sin^2 \theta_{12}$ and/or $\sin^2 \theta_{13}$ and/or $\sin^2 \theta_{23}$ \\ [8pt]
\bottomrule
& \\ [-10pt]
A1 & $U_{12}(\theta^e_{12}, \delta^e_{12}) U_{12}(\theta^{\circ}_{12}, \delta^{\circ}_{12}) R_{23}(\theta_{23}^{\circ}) R_{13}(\theta_{13}^{\circ}) Q_0$ & 
$\sin^2 \theta_{23} = \dfrac{\sin^2 \theta^{\circ}_{13} - \sin^2 \theta_{13} + \cos^2 \theta^{\circ}_{13} \sin^2 \theta^{\circ}_{23}}{1 - \sin^2 \theta_{13}}$  \\ [15pt]
\hline
& \\ [-10pt]
A2 & $U_{13}(\theta^e_{13}, \delta^e_{13}) U_{13}(\theta^{\circ}_{13}, \delta^{\circ}_{13}) R_{23}(\theta_{23}^{\circ}) R_{12}(\theta_{12}^{\circ}) Q_0$ & 
$\sin^2 \theta_{23} = \dfrac{\sin^2 \theta^{\circ}_{23}}{1 - \sin^2 \theta_{13}}$  \\ [15pt]
\hline
& \\ [-10pt]
A3 & $U_{23}(\theta^e_{23}, \delta^e_{23}) U_{23}(\theta^{\circ}_{23}, \delta^{\circ}_{23}) R_{13}(\theta_{13}^{\circ}) R_{12}(\theta_{12}^{\circ}) Q_0$ & 
$\sin^2 \theta_{13} = \sin^2 \theta^{\circ}_{13}$, $\sin^2 \theta_{12} = \sin^2 \theta^{\circ}_{12}$ \\ [15pt]
\hline
& \\ [-10pt]
B1 & $R_{23}(\theta_{23}^{\circ}) R_{12}(\theta_{12}^{\circ}) U_{13}(\theta^{\circ}_{13}, \delta^{\circ}_{13}) U_{13}(\theta^{\nu}_{13}, \delta^{\nu}_{13}) Q_0$ &
$\sin^2 \theta_{12} = \dfrac{\sin^2 \theta^{\circ}_{12}}{1 - \sin^2 \theta_{13}}$ \\ [15pt]
\hline
& \\ [-10pt]
B2 & $R_{13}(\theta_{13}^{\circ}) R_{12}(\theta_{12}^{\circ}) U_{23}(\theta^{\circ}_{23}, \delta^{\circ}_{23}) U_{23}(\theta^{\nu}_{23}, \delta^{\nu}_{23}) Q_0$ & 
$\sin^2 \theta_{12} = \dfrac{\cos^2 \theta_{13} - \cos^2 \theta^{\circ}_{12} \cos^2 \theta^{\circ}_{13}  }{1 - \sin^2 \theta_{13}}$  \\ [15pt]
\hline
& \\ [-10pt]
B3 & $R_{23}(\theta_{23}^{\circ}) R_{13}(\theta_{13}^{\circ}) 
U_{12}(\theta^{\circ}_{12}, \delta^{\circ}_{12}) 
U_{12}(\theta^{\nu}_{12}, \delta^{\nu}_{12}) Q_0$ & 
$\sin^2 \theta_{13} = \sin^2 \theta^{\circ}_{13}$, $\sin^2 \theta_{23} = \sin^2 \theta^{\circ}_{23}$ \\ [15pt]
\hline
\end{tabular}
\caption{Summary of the formulae for $\sin^2 \theta_{12}$ and/or $\sin^2 \theta_{13}$ and/or $\sin^2 \theta_{23}$. 
The cases A1, A2 and A3 correspond to $G_e = Z_2$ and $G_{\nu} = Z_n$, $n > 2$ or $Z_n \times Z_m$, $n,m \geq 2$,
while B1, B2 and B3 correspond to $G_e = Z_n$, $n > 2$ or $Z_n \times Z_m$, $n,m \geq 2$ and $G_{\nu} = Z_2$.
See text for further details.
}
\label{tab:summarysin2th23_12}
\end{table}
%%%%%%%%%%%%%%%%%%%%%%%%%%
\end{landscape}

\begin{landscape}
\pagestyle{empty}
%%%%%%%%%%%%%%%%%%%%%%%%%%
\begin{table}[h]
\centering
\renewcommand*{\arraystretch}{1.2}
\vspace{-0.8cm}
\hspace{-0.5cm}
\begin{tabular}{lll}
\hline
& \\ [-10pt]
Case & Parametrisation of $U$ & Sum rule for $\sin^2 \theta_{12}$ and/or $\sin^2 \theta_{13}$ and/or $\sin^2 \theta_{23}$ \\ [8pt]
\bottomrule
 & \\ [-10pt]
C1 & $U_{12}(\theta^e_{12}, \delta^e_{12}) U_{12}(\theta^{\circ}_{12}, \delta^{\circ}_{12}) R_{23}(\theta^{\circ}_{23}) U_{13}(\theta^{\circ}_{13}, \delta^{\circ}_{13}) U_{13}(\theta^{\nu}_{13}, \delta^{\nu}_{13}) Q_0$ & 
not fixed \\ [15pt]
 \hline
 & \\ [-10pt]
C2 & $U_{13}(\theta^e_{13}, \delta^e_{13}) U_{13}(\theta^{\circ}_{13}, \delta^{\circ}_{13}) R_{23}(\theta^{\circ}_{23}) U_{12}(\theta^{\circ}_{12}, \delta^{\circ}_{12}) U_{12}(\theta^{\nu}_{12}, \delta^{\nu}_{12}) Q_0$ & 
$\sin^2 \theta_{23} = \dfrac{\sin^2 \theta^{\circ}_{23}}{1 - \sin^2 \theta_{13}}$  \\ [15pt]
\hline
 & \\ [-10pt]
C3 & $U_{12}(\theta^e_{12}, \delta^e_{12}) U_{12}(\theta^{\circ}_{12}, \delta^{\circ}_{12}) R_{13}(\theta^{\circ}_{13}) U_{23}(\theta^{\circ}_{23}, \delta^{\circ}_{23}) U_{23}(\theta^{\nu}_{23}, \delta^{\nu}_{23}) Q_0$ & 
not fixed \\ [15pt]
\hline
 & \\ [-10pt]
C4 & $U_{13}(\theta^e_{13}, \delta^e_{13}) U_{13}(\theta^{\circ}_{13}, \delta^{\circ}_{13}) R_{12}(\theta^{\circ}_{12}) U_{23}(\theta^{\circ}_{23}, \delta^{\circ}_{23})U_{23}(\theta^{\nu}_{23}, \delta^{\nu}_{23}) Q_0$ & 
not fixed \\ [15pt]
\hline
 & \\ [-10pt]
C5 & $U_{23}(\theta^e_{23}, \delta^e_{23}) U_{23}(\theta^{\circ}_{23}, \delta^{\circ}_{23}) R_{12}(\theta^{\circ}_{12}) U_{13}(\theta^{\circ}_{13}, \delta^{\circ}_{13}) U_{13}(\theta^{\nu}_{13}, \delta^{\nu}_{13}) Q_0$ & 
$\sin^2 \theta_{12} = \dfrac{\sin^2 \theta^{\circ}_{12}}{1 - \sin^2 \theta_{13}}$ \\ [15pt]
\hline
 & \\ [-10pt]
C6 & $U_{23}(\theta^e_{23}, \delta^e_{23}) U_{23}(\theta^{\circ}_{23}, \delta^{\circ}_{23}) R_{13}(\theta^{\circ}_{13})
U_{12}(\theta^{\circ}_{12}, \delta^{\circ}_{12}) U_{12}(\theta^{\nu}_{12}, \delta^{\nu}_{12}) Q_0$ & 
$\sin^2 \theta_{13} = \sin^2 \theta^{\circ}_{13}$ \\ [15pt]
\hline
 & \\ [-10pt]
C7 & $U_{12}(\theta^e_{12}, \delta^e_{12}) U_{12}(\theta^{\circ}_{12}, \delta^{\circ}_{12}) R_{23}(\theta^{\circ}_{23}) U_{12}(\tilde\theta^{\circ}_{12}, \tilde\delta^{\circ}_{12}) U_{12}(\theta^{\nu}_{12}, \delta^{\nu}_{12}) Q_0$ & 
$\sin^2 \theta_{23} = \dfrac{\sin^2 \theta^\circ_{23} - \sin^2 \theta_{13}}
{1 - \sin^2 \theta_{13}}$ \\ [15pt]
\hline
 & \\ [-10pt]
C8 & $U_{13}(\theta^e_{13}, \delta^e_{13}) U_{13}(\theta^{\circ}_{13}, \delta^{\circ}_{13}) R_{23}(\theta^{\circ}_{23}) U_{13}(\tilde\theta^{\circ}_{13}, \tilde\delta^{\circ}_{13}) U_{13}(\theta^{\nu}_{13}, \delta^{\nu}_{13}) Q_0$ & 
not fixed \\ [15pt]
\hline
 & \\ [-10pt]
C9 & $U_{23}(\theta^e_{23}, \delta^e_{23}) U_{23}(\theta^{\circ}_{23}, \delta^{\circ}_{23}) R_{12}(\theta^{\circ}_{12}) U_{23}(\tilde\theta^{\circ}_{23}, \tilde\delta^{\circ}_{23}) U_{23}(\theta^{\nu}_{23}, \delta^{\nu}_{23}) Q_0$ & 
$\sin^2 \theta_{12} = \dfrac{\sin^2 \theta^\circ_{12} - \sin^2 \theta_{13}}
{1 - \sin^2 \theta_{13}}$ \\ [15pt]
\hline
\end{tabular}
\caption{Summary of the formulae for $\sin^2 \theta_{12}$ and/or $\sin^2 \theta_{13}$ and/or $\sin^2 \theta_{23}$. 
The cases C1~--~C9 correspond to 
$G_e = Z_2$ and $G_{\nu} = Z_2$. See text for further details.
}
\label{tab:summarysin2th23_12B}
\end{table}
%%%%%%%%%%%%%%%%%%%%%%%%%%
\end{landscape}

%%%%%%%%%%%%%%%%%%%%%%
%
\section{The Case of Fully Broken $G_e$}
\label{sec:Gebroken}
%
%%%%%%%%%%%%%%%%%%%%%%%%%%%%%
%
If the discrete flavour symmetry $G_f$ is fully broken 
in the charged lepton sector
the matrix $U_e$ is unconstrained and includes, in general, 
three rotation angle and three CPV phase parameters.
It is impossible to derive predictions for the 
mixing angles and CPV phases in the PMNS matrix 
in this case. Therefore, we will consider 
in this section forms of $U_e$ 
corresponding to one of the rotation 
angle parameters being equal to zero. 
Some of these forms of  $U_e$ correspond to a class of models 
of neutrino mass generation (see, e.g., \cite{Gehrlein:2014wda,Meroni:2012ty,
Marzocca:2011dh,Antusch:2012fb,Chen:2009gf,Girardi:2013sza})
and lead, in particular, to sum rules 
for $\cos\delta$.

We give in Appendix~\ref{app:Gebroken}  the most general 
parametrisations of $U$ under the
assumption that in the case 
of fully broken $G_e$ one rotation angle 
in the matrix $U_e$ vanishes.
The second case in Table~\ref{tab:parUeGebroken}
with $\theta^{\circ}_{13} = 0$ have been analysed in  
\cite{Petcov:2014laa,Girardi:2014faa,Girardi:2015vha}, while 
the third case with
$U_{12}(\theta^e_{12}, \delta^e_{12}) U_{13}(\theta^e_{13}, \delta^e_{13})$ 
has been investigated in \cite{Girardi:2015vha}. 

%%%%%%%%%%%%%%%%%%%%%%
%
\subsection{The Scheme with 
$U_{23}(\theta^e_{23},\delta^e_{23}) U_{12}(\theta^e_{12},\delta^e_{12})$ (Case D1)}
\label{sec:23e12e}
\begin{figure}[t!]
  \begin{center}
 \includegraphics[width=14cm]{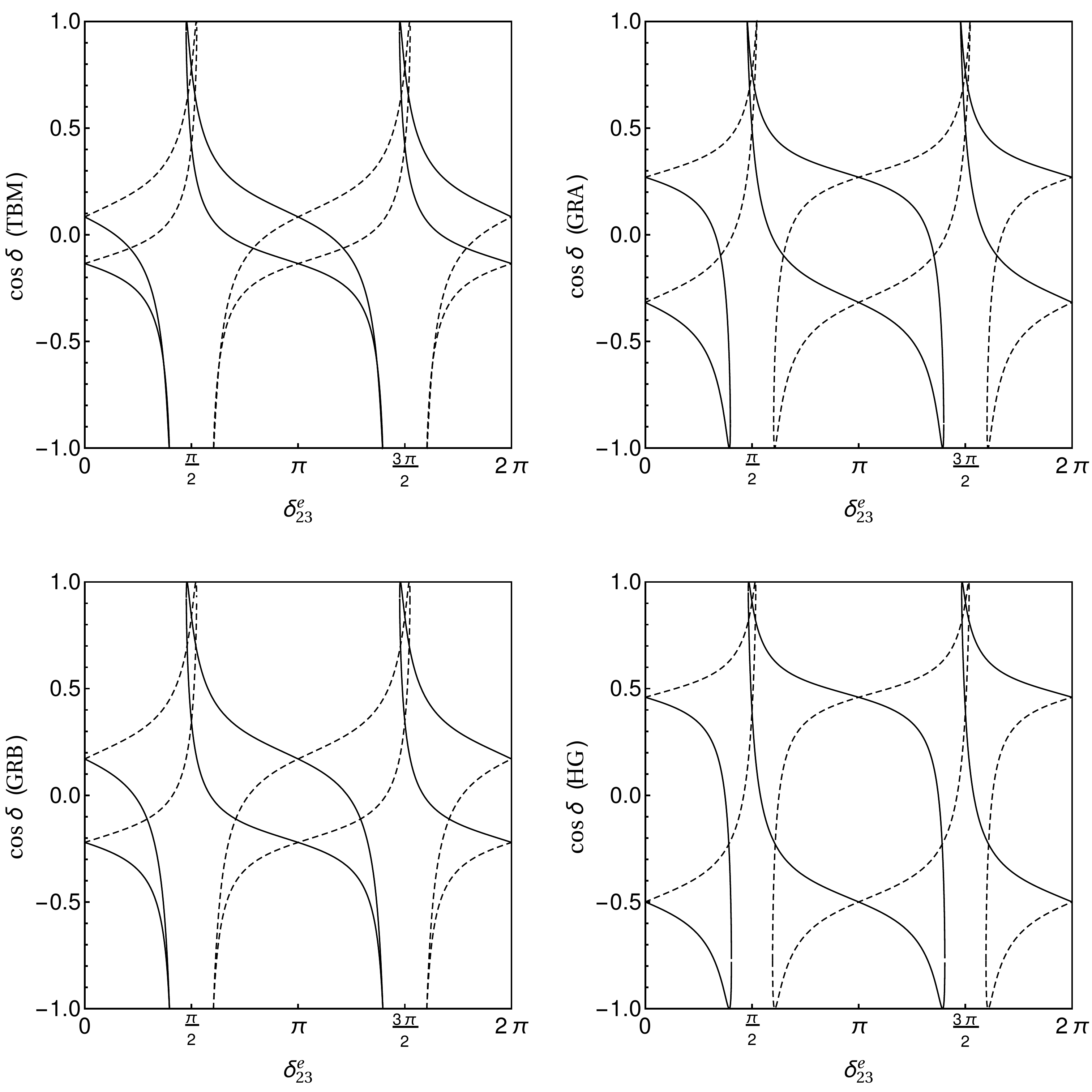}
     \end{center}
\caption{
\label{Fig5}
Dependence of $\cos \delta$ on $\delta^e_{23}$ in the cases
of the TBM, GRA, GRB and HG symmetry forms.
The mixing parameters $\sin^2 \theta_{12}$, $\sin^2 \theta_{23}$ and $\sin^2 \theta_{13}$
have been fixed to their best fit values for the NO neutrino mass spectrum
quoted in eqs.~(\ref{th12values})~--~(\ref{th13values}).
The angle $\hat\theta_{12}$ is assumed to belong to the first quadrant. 
The solid lines correspond to $\hat\delta = \cos^{-1} (\cos\hat\delta)$,
where $\cos\hat\delta$ is the solution of eq.~(\ref{eq:e23e12th13}),
while the dashed lines correspond to $\hat\delta =2\pi - \cos^{-1} (\cos\hat\delta)$.
See text for further details.
}
\end{figure}

%%%%%%%%%%%%%%%%%%%%%%
%
We consider the following parametrisation of the PMNS matrix 
(see Appendix~\ref{app:Gebroken}, first case in Table~\ref{tab:parUeGebroken}):
%%%%%%%%%%%%%%%%%%%%%%%%%%%%%%%%
\be
U = U_{23}(\theta^e_{23}, \delta^e_{23}) R_{12}(\hat \theta_{12}) 
P_1(\hat \delta) R_{23}(\theta^{\circ}_{23}) R_{13}(\theta^{\circ}_{13}) Q_0 \,,
\quad
P_1(\hat \delta) = \diag (e^{i\hat\delta},1,1).
\ee
%%%%%%%%%%%%%%%%%%%%%%%%%%%%%%%%%%%%%%
%
We find that:
%%%%%%%%%%%%%%%%%%%%%%%%%%%%%%%%
\begin{align}
\sin^2 \theta_{13} & = |U_{e3}|^2  = \sin^2 \theta_{13}(\hat \theta_{12}, \hat \delta,\theta^{\circ}_{13}, \theta^{\circ}_{23}) \,, \label{eq:e23e12th13}\\
\sin^2 \theta_{23} & = \frac{|U_{\mu3}|^2}{1-|U_{e3}|^2} = \sin^2 \theta_{23}(\hat \theta_{12}, \hat \delta, \theta^e_{23}, \delta^e_{23}, \theta^{\circ}_{13}, \theta^{\circ}_{23})  \,, \label{eq:e23e12th23} \\
\sin^2 \theta_{12} & = \frac{|U_{e2}|^2}{1-|U_{e3}|^2} =  \dfrac{\cos^2 \theta^{\circ}_{23} \sin^2 \hat \theta_{12}}{\cos^2 \theta_{13}} \label{eq:e23e12th12} \,.
\end{align}
%%%%%%%%%%%%%%%%%%%%%%%%%%%
As it can be seen from the previous equations and the 
absolute value of the element $U_{\mu2}$, 
\be
|U_{\mu 2}| = |\cos \theta^e_{23} \cos \hat \theta_{12} \cos \theta^{\circ}_{23} - e^{-i \delta^e_{23}} \sin \theta^e_{23} \sin \theta^{\circ}_{23}| \,,
\ee
a sum rule for $\cos \delta$ might be derived in the case of fixed $\delta^e_{23}$.
In the general case of free $\delta^e_{23}$ we find that $\cos \delta$ is a function
of $\delta^e_{23}$. Since in this case the analytical expression of $\cos \delta$
in terms of $\delta^e_{23}$ is rather complicated,
we do not present this result here.
Note that imposing either $\theta^{\circ}_{23} = 0$ or $\theta^{\circ}_{13} = 0$
is not enough to fix the value of $\cos \delta$.
As eqs.~(\ref{eq:e23e12th13}) and (\ref{eq:e23e12th23})
suggest, in the case of fixed $\delta^e_{23}$
 there exist
multiple solutions for the value of $\cos \delta$ 
for any given value of $\delta^e_{23}$.
This is demonstrated in Fig.~\ref{Fig5}, in which we
plot $\cos\delta$ versus $\delta^e_{23}$, assuming that the angles
$\theta^{\circ}_{13}$ and $\theta^{\circ}_{23}$ have the values
corresponding to the TBM, GRA, GRB and HG symmetry forms
given in Table~\ref{tab:tho12tho23tho13symm}.
The figure is obtained for $\hat\theta_{12}$ belonging to
the first quadrant. The solid lines correspond to 
$\hat\delta = \cos^{-1} (\cos\hat\delta)$,
where $\cos\hat\delta$ is the solution of eq.~(\ref{eq:e23e12th13}),
while the dashed lines correspond to $\hat\delta =2\pi - \cos^{-1} (\cos\hat\delta)$.
Multiple lines reflect the fact that eq.~(\ref{eq:e23e12th23}) 
for $\theta^e_{23}$ has several solutions.
We note that Fig.~\ref{Fig5} remains the same for $\hat\theta_{12}$ 
belonging to the third quadrant, while for $\hat\theta_{12}$ lying
in the second or fourth quadrant the solid and dashed lines interchange.
For the BM (LC) symmetry form $\cos\hat\delta$ has an unphysical value,
which indicates that the considered scheme with the BM (LC) form
of the matrix diagonalising the neutrino mass matrix
does not provide a good description of the current data on
the neutrino mixing angles \cite{Marzocca:2013cr}.\footnote{Note that the scheme under discussion corresponds to
inverse ordering of the charged lepton corrections, i.e.,
$U_e^\dagger = U_{23}(\theta^e_{23},\delta^e_{23}) U_{12}(\theta^e_{12},\delta^e_{12})$
(see \cite{Marzocca:2013cr}).}
%%%%%%%%%%%%%%%%%%%%%%%%%%%%%%%%%
%%%%%%%%%%%%%%%%%%%%%%%%%
Thus, we do not present such a plot in this case.
If $\delta^e_{23}$ turns out to be fixed (by GCP invariance, 
symmetries, etc.), then, as can be seen from Fig.~\ref{Fig5}, $\cos\delta$
is predicted to take a value from a discrete set. 
For instance, when $\delta^e_{23} = 0$ or $\pi$, we have
\begin{align}
& \cos\delta = \{-0.135, 0.083\} ~~ \text{for TBM;} \\
& \cos\delta = \{-0.317, 0.269\} ~~ \text{for GRA;} \\
& \cos\delta = \{-0.221, 0.170\} ~~ \text{for GRB;} \\
& \cos\delta = \{-0.500, 0.459\} ~~ \text{for HG.}
\end{align}
In the case of $\delta^e_{23} = \pi/2$ or $3\pi/2$, we find
\begin{align}
& \cos\delta = \{0.418, 0.779\} ~~ \text{for TBM;} \\
& \cos\delta = \{0.498, 0.761\} ~~ \text{for GRA;} \\
& \cos\delta = \{0.346, 0.837\} ~~ \text{for GRB;} \\
& \cos\delta = \{0.394, 0.906\} ~~ \text{for HG.}
\end{align}

%%%%%%%%%%%%%%%%%%%%%%
%
\subsection{The Scheme with $U_{13}(\theta^e_{13},\delta^e_{13}) U_{12}(\theta^e_{12},\delta^e_{12})$ (Case D2)}
%%%%%%%%%%%%%%%%%%%%%%%
%
We consider the following parametrisation of the PMNS 
matrix (see Appendix~\ref{app:Gebroken},
first case in Table~\ref{tab:parUeGebroken}):
%%%%%%%%%%%%%%%%%%%%%%%%%%%%%%%%%%%%%%%
\be
U = U_{13}(\theta^e_{13}, \delta^e_{13}) R_{12}(\hat \theta_{12}) P_1(\hat \delta) R_{23}(\theta^{\circ}_{23}) R_{13}(\theta^{\circ}_{13}) Q_0 \,, \quad P_1(\hat \delta) = \diag(e^{i \hat \delta},1,1)\,.
\label{sec72U}
\ee
%%%%%%%%%%%%%%%%%%%%%%%%%%%%%%%%%%
%
 A sum rule for $\cos\delta$ is obtained 
in the cases of either 
$\theta^{\circ}_{23} = k\pi$, $k=0,1,2$, or
$\theta^{\circ}_{13} = q\pi/2$, $q=0,1,2,3,4$.
For the general form of $U$ we find for the 
absolute value of the element $U_{\mu2}$:
%%%%%%%%%%%%%%%%%%%%%%%%
\be
|U_{\mu 2}| = |\cos \hat \theta_{12} \cos \theta^{\circ}_{23}|\,,
\label{sec72Umu2}
\ee
%%%%%%%%%%%%%%%%%%%%%%%%%%%%%
%
which in each of the two limits indicated above 
is fixed because $|\cos\hat \theta_{12}|$ 
can be expressed in 
terms of the PMNS neutrino mixing angles. 
This can be seen from the following relation, 
which is obtained using the expressions for
$|U_{\mu 3}|^2$ in the standard parametrisation 
of the PMNS matrix $U$ and in the parametrisation 
given in eq.~(\ref{sec72U}):
%%%%%%%%%%%%%%%%%%%%%%%%%%%%%%%%
\be
\cos^2 \theta_{13} \sin^2 \theta_{23} = 
|-e^{i \hat \delta} \sin \hat \theta_{12} \sin \theta^{\circ}_{13} 
+ \cos \hat \theta_{12} \cos \theta^{\circ}_{13} \sin \theta^{\circ}_{23}|^2 \,.
\ee
%%%%%%%%%%%%%%%%%%%%%%%%%%%%%%%%
%
Equating the expression for $|U_{\mu 2}|$ given in eq.~(\ref{sec72Umu2})
with the one in the standard parametrisation, we find
\begin{equation}
\cos \delta  = \dfrac{\cos^2 \theta_{23} \cos^2 \theta_{12} + \sin^2 \theta_{12} \sin^2 \theta_{13} \sin^2 \theta_{23}  - \cos^2\hat\theta_{12} \cos^2\theta^\circ_{23}}
{\sin 2 \theta_{23} \sin \theta_{12} \cos \theta_{12} \sin \theta_{13}}\,.
\end{equation}

%%%%%%%%%%%%%%%%%%%%%%
%
\subsection{The Scheme with $U_{12}(\theta^e_{12},\delta^e_{12}) 
U_{23}(\theta^e_{23},\delta^e_{23})$ (Case D3)}
%
%%%%%%%%%%%%%%%%%%%%%%%%%%%%%%%%%%%%%%%
%
We consider the following parametrisation of the PMNS matrix 
(see Appendix~\ref{app:Gebroken}, second case in Table~\ref{tab:parUeGebroken}):
%%%%%%%%%%%%%%%%%%%%%%%%%%%%%%
\be
U = U_{12}(\theta^e_{12}, \delta^e_{12}) R_{23}(\hat \theta_{23}) 
P_2(\hat \delta) R_{13}(\theta_{13}^{\circ}) R_{12}(\theta_{12}^{\circ}) Q_0 \,,
\quad P_2(\hat \delta) = \diag(1, e^{i \hat \delta},1)\,.
\label{sec73U}
\ee
%%%%%%%%%%%%%%%%%%%%%%%%%%%
%
A sum rule for $\cos\delta$ can be derived 
in the cases of either $\theta^{\circ}_{13} = k\pi$, 
$k=0,1,2$, or 
$\theta^{\circ}_{12} =q\pi/2$, $q=0,1,2,3,4$.
Indeed, the relation $\cos^2 \theta_{13} \cos^2 \theta_{23} = 
\cos^2 \hat \theta_{23} \cos^2 \theta^{\circ}_{13}$ 
(which can be obtained from the  expressions for the 
element $U_{\tau 3}$ of the PMNS matrix $U$ 
in the standard parametrisation and in the one given in 
eq.~(\ref{sec73U})), 
allows us to express 
$\cos^2 \hat \theta_{23}$ in terms of the known product  
 $\cos^2 \theta_{13} \cos^2 \theta_{23}$ and the 
parameter $\cos^2 \theta^{\circ}_{13}$ which, in principle, is fixed 
by the symmetries $G_f$ and $G_\nu$. 
We have also
%%%%%%%%%%%%%%%%%%%%%%%%%%%%
\be
|U_{\tau 2}| = |e^{i \hat \delta} \cos \theta^{\circ}_{12} \sin \hat \theta_{23} 
+ \cos \hat \theta_{23} \sin \theta^{\circ}_{12} \sin \theta^{\circ}_{13}|\,.
\label{sec73Utau2}
\ee
%%%%%%%%%%%%%%%%%%%%%%%%%%%%%
%
In the limits
of either $\theta^{\circ}_{13} = k\pi$, $k=0,1,2$, or 
$\theta^{\circ}_{12} =q\pi/2$, $q=0,1,2,3,4$, $|U_{\tau 2}|$ does not depend on $\hat \delta$
and is also fixed. This makes it possible to 
derive  a sum rule for $\cos \delta$.
In the general case, $\cos\delta$ is a function of $\hat\delta$:
%%%%%%%%%%%%%%%%%%%%%%%%%%%%
\begin{align}
\cos \delta &  = \dfrac{2}{\sin2\theta_{12} \sin2\theta_{23} \sin\theta_{13} \cos^2\theta^\circ_{13}}
\bigg[ \cos^2\theta^\circ_{12} \left( \cos^2\theta^\circ_{13} - \cos^2\theta_{13} \cos^2\theta_{23} \right) \nonumber \\
& - \cos^2\theta_{12} \sin^2\theta_{23} \cos^2 \theta^\circ_{13}
+ \cos^2\theta_{23} \left( \cos^2\theta_{13} \sin^2\theta^\circ_{12} \sin^2 \theta^\circ_{13} - \sin^2\theta_{12} \sin^2\theta_{13} \cos^2 \theta^\circ_{13} \right) \nonumber \\
& + \kappa \cos\hat\delta \cos\theta_{13} \cos\theta_{23} \sin2\theta^\circ_{12} \sin\theta^\circ_{13}
\left( \cos^2\theta^\circ_{13} - \cos^2\theta_{13} \cos^2\theta_{23} \right)^\frac{1}{2} \bigg]\,,
\end{align}
%%%%%%%%%%%%%%%%%%%%%%%%%%%%%%%
where $\kappa = 1$ if $\hat\theta_{23}$ belongs to the first or third quadrant,
and $\kappa = -1$ otherwise.
For $\theta^{\circ}_{13} = 0$
the sum rule reduces to the one derived 
in \cite{Petcov:2014laa} and discussed in detail in 
\cite{Petcov:2014laa,Girardi:2014faa,Girardi:2015vha}.

%%%%%%%%%%%%%%%%%%%%%%
%
\subsection{The Scheme with $U_{13}(\theta^e_{13},\delta^e_{13}) U_{23}(\theta^e_{23},\delta^e_{23})$ (Case D4)}
%
%%%%%%%%%%%%%%%%%%%%%%%
%
We consider the following parametrisation of the PMNS matrix 
(see Appendix~\ref{app:Gebroken}, second case in Table~\ref{tab:parUeGebroken}):
%%%%%%%%%%%%%%%%%%%%%%%%%%%%%%%%%%%%%
\be
U = U_{13}(\theta^e_{13}, \delta^e_{13}) R_{23}(\hat \theta_{23}) P_2(\hat \delta) R_{13}(\theta_{13}^{\circ}) R_{12}(\theta_{12}^{\circ}) Q_0 \,, \quad
P_2(\hat \delta) = \diag(1, e^{i \hat \delta},1)\,.
\ee
%%%%%%%%%%%%%%%%%%%%%%%%%%%%%%%%
%
In this case a sum rule for $\cos\delta$ exists 
provided  either $\theta^{\circ}_{13} = k\pi$, $k=0,1,2$,
or $\theta^{\circ}_{12} = q\pi/2$, $q=0,1,2,3,4$.
This follows from the relation 
$|U_{\mu 3}|^2 = 
\cos^2 \theta_{13} \sin^2 \theta_{23} = 
\cos^2 \theta^{\circ}_{13} \sin^2 \hat \theta_{23}$
and the expression for $|U_{\mu 2}|$:
%
%%%%%%%%%%%%%%%%%%%%%%%%%%%%%%%%%%
\be
|U_{\mu 2}| = |e^{i \hat \delta} \cos \theta^{\circ}_{12} \cos \hat \theta_{23} - 
\sin \hat \theta_{23} \sin \theta^{\circ}_{12} \sin \theta^{\circ}_{13}|\,.
\label{sec74Umu2}
\ee
%%%%%%%%%%%%%%%%%%%%%%%%%%%%%%%%
%
The sum rule of interest for $\cos \delta$ reads
%%%%%%%%%%%%%%%%%%%%%%%%%%%%
\begin{align}
\cos \delta &  = - \dfrac{2}{\sin2\theta_{12} \sin2\theta_{23} \sin\theta_{13} \cos^2\theta^\circ_{13}}
\bigg[ \cos^2\theta^\circ_{12} \left( \cos^2\theta^\circ_{13} - \cos^2\theta_{13} \sin^2\theta_{23} \right) \nonumber \\
& - \cos^2\theta_{12} \cos^2\theta_{23} \cos^2 \theta^\circ_{13}
+ \sin^2\theta_{23} \left( \cos^2\theta_{13} \sin^2\theta^\circ_{12} \sin^2 \theta^\circ_{13} - \sin^2\theta_{12} \sin^2\theta_{13} \cos^2 \theta^\circ_{13} \right) \nonumber \\
& - \kappa \cos\hat\delta \cos\theta_{13} \sin\theta_{23} \sin2\theta^\circ_{12} \sin\theta^\circ_{13}
\left( \cos^2\theta^\circ_{13} - \cos^2\theta_{13} \sin^2\theta_{23} \right)^\frac{1}{2} \bigg]\,,
\label{cosdsec74}
\end{align}
%%%%%%%%%%%%%%%%%%%%%%%%%%%%%%%
where $\kappa = 1$ if $\hat\theta_{23}$ belongs to the first or third quadrant,
and $\kappa = -1$ otherwise. As in the previous case, $\cos\delta$ is a function
of $\hat\delta$.
For $\theta^{\circ}_{13} = 0$ the sum rule in eq.~(\ref{cosdsec74})  
reduces to the one derived in \cite{Girardi:2015vha}.

%%%%%%%%%%%%%%%%%%%%%%
%
\subsection{The Scheme with $U_{23}(\theta^e_{23},\delta^e_{23}) U_{13}(\theta^e_{13},\delta^e_{13})$ (Case D5)}
\label{sec:23e13e}
\begin{figure}[t!]
  \begin{center}
 \includegraphics[width=14cm]{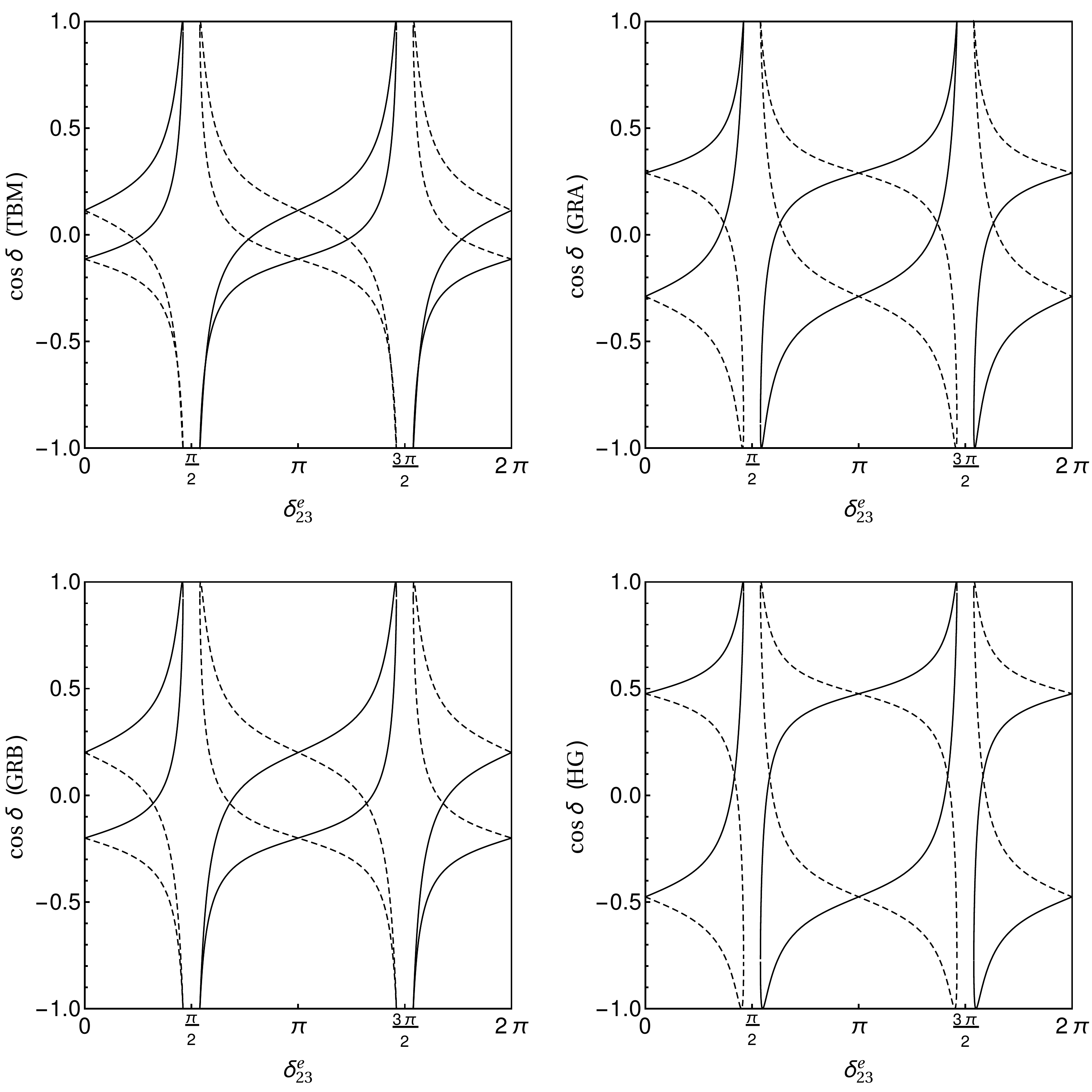}
     \end{center}
\caption{
\label{Fig6}
Dependence of $\cos \delta$ on $\delta^e_{23}$ in the cases
of the TBM, GRA, GRB and HG symmetry forms.
The mixing parameters $\sin^2 \theta_{12}$, $\sin^2 \theta_{23}$ and $\sin^2 \theta_{13}$
have been fixed to their best fit values for the NO neutrino mass spectrum
quoted in eqs.~(\ref{th12values})~--~(\ref{th13values}).
The angle $\hat\theta_{13}$ is assumed to belong to the first quadrant. 
The solid lines correspond to $\hat\delta = \cos^{-1} (\cos\hat\delta)$,
where $\cos\hat\delta$ is the solution of eq.~(\ref{eq:e23e13th12}),
while the dashed lines correspond to $\hat\delta =2\pi - \cos^{-1} (\cos\hat\delta)$.
See text for further details.
}
\end{figure}

%
%%%%%%%%%%%%%%%%%%%%%%%%%%%%%%%%%%%%%%%%
%
In this case we consider the following parametrisation of the PMNS matrix 
(see Appendix~\ref{app:Gebroken}, third case in Table~\ref{tab:parUeGebroken}):
%%%%%%%%%%%%%%%%%%%%%%%%%%
\be
U = U_{23}(\theta^e_{23}, \delta^e_{23}) R_{13}(\hat \theta_{13}) P_1(\hat \delta) R_{23}(\theta_{23}^{\circ}) R_{12}(\theta_{12}^{\circ}) Q_0 \,, \quad 
P_1(\hat \delta) = \diag(e^{i \hat \delta},1,1)\,.
\ee
%%%%%%%%%%%%%%%%%%%%%%%%%%%
%
We find that:
%%%%%%%%%%%%%%%%%%%%%%%%%%%%%%%%
\begin{align}
\sin^2 \theta_{13} & = |U_{e3}|^2  =  \cos^2 \theta^{\circ}_{23} \sin^2 \hat \theta_{13} \,, \label{eq:e23e13th13} \\
\sin^2 \theta_{23} & = \frac{|U_{\mu3}|^2}{1-|U_{e3}|^2} = \sin^2 \theta_{23}(\hat \theta_{13}, \theta^e_{23}, \delta^e_{23}, \theta^{\circ}_{23}) \,, \label{eq:e23e13th23}\\
\sin^2 \theta_{12} & = \frac{|U_{e2}|^2}{1-|U_{e3}|^2} =  \sin^2 \theta_{12}(\hat \theta_{13}, \hat \delta, \theta^{\circ}_{12}, \theta^{\circ}_{23}) \,. \label{eq:e23e13th12}
\end{align} 
%%%%%%%%%%%%%%%%%%%%%%%%%%%
Since, as can be shown, $|U_{\mu 2}|$ is a function of the parameters 
$\theta^e_{23}$, $\delta^e_{23}$, $\hat \delta$, $\hat \theta_{13}$, $\theta^{\circ}_{12}$ and $\theta^{\circ}_{23}$, 
and $\hat \theta_{13}$, and $\cos \hat \delta$
can be extracted from 
eqs.~(\ref{eq:e23e13th13}) and (\ref{eq:e23e13th12}), respectively,
it might be possible to find a sum
rule for $\cos \delta$ in the case of fixed $\delta^e_{23}$.
Since in this case the analytical expression of $\cos \delta$
in terms of $\delta^e_{23}$ is rather complicated,
we do not present it here.
Note that imposing either $\theta^{\circ}_{12} = 0$ or $\theta^{\circ}_{23} = 0$
is not enough to fix the value of $\cos \delta$.
Even in the case of fixed $\delta^e_{23}$ it follows from eqs.~(\ref{eq:e23e13th23}) and (\ref{eq:e23e13th12})
that for any given value of $\delta^e_{23}$,
$\cos \delta$ can take several values.
This can be understood, e.g., from eq.~(\ref{eq:e23e13th12}) which allows
to fix $\cos \hat \delta$, but not $\sin \hat \delta$. This ambiguity,
in particular,
leads to multiple solutions for $\cos \delta$.
In Fig.~\ref{Fig6} we show these solutions in the cases of the
TBM, GRA, GRB and HG symmetry forms.
We remind that for these forms 
$\theta^\circ_{23} = -\pi/4$ and 
$\theta^\circ_{12} = \sin^{-1} (1/\sqrt{3})$ (TBM),
$\theta^\circ_{12} = \sin^{-1} (1/\sqrt{2+r})$ (GRA),
$r = (1+\sqrt{5})/2$ being the golden ratio,
$\theta^\circ_{12} = \sin^{-1} (\sqrt{3-r}/2)$ (GRB), and
$\theta^\circ_{12} = \pi/6$ (HG).
We assume $\hat\theta_{13}$ to lie in
the first quadrant. The solid lines correspond to 
$\hat\delta = \cos^{-1} (\cos\hat\delta)$,
where $\cos\hat\delta$ is the solution of eq.~(\ref{eq:e23e13th12}),
while the dashed lines correspond to $\hat\delta =2\pi - \cos^{-1} (\cos\hat\delta)$.
Multiple lines reflect the fact that eq.~(\ref{eq:e23e13th23}) 
for $\theta^e_{23}$ has several solutions.
We note that Fig.~\ref{Fig6} does not change in the case of $\hat\theta_{13}$ 
belonging to the third quadrant, while for $\hat\theta_{13}$ lying
in the second or fourth quadrant the solid and dashed lines interchange.
For $\delta^e_{23} = 0$ or $\pi$, we find
\begin{align}
& \cos\delta = \{-0.114, 0.114\} ~~ \text{for TBM;} \\
& \cos\delta = \{-0.289, 0.289\} ~~ \text{for GRA;} \\
& \cos\delta = \{-0.200, 0.200\} ~~ \text{for GRB;} \\
& \cos\delta = \{-0.476, 0.476\} ~~ \text{for HG.}
\end{align}
It is worth noting that in the scheme under consideration 
the values of $\delta^e_{23}$ in a vicinity
of $\pi/2$ ($3\pi/2$) do not provide physical values of $\cos\delta$ 
(see Fig.~\ref{Fig6}).
%%%%%%%%%%%%%%%%%%%%%%%%%%%%%%%%%
%%%%%%%%%%%%%%%%%%%%%%%%% 

%%%%%%%%%%%%%%%%%%%%%%
%
\subsection{The Scheme with $U_{12}(\theta^e_{12},\delta^e_{12}) U_{13}(\theta^e_{13},\delta^e_{13})$ (Case D6)}
%
%%%%%%%%%%%%%%%%%%%%%%%%%%%%%%%
%
It is convenient to consider the following  
parametrisation of the PMNS matrix $U$ 
(see Appendix~\ref{app:Gebroken}, 
third case in Table~\ref{tab:parUeGebroken}):
%%%%%%%%%%%%%%%%%%%%%%%%%%%%
\be
U = U_{12}(\theta^e_{12}, \delta^e_{12}) R_{13}(\hat \theta_{13}) P_1(\hat \delta) R_{23}(\theta_{23}^{\circ}) R_{12}(\theta_{12}^{\circ}) Q_0 \,, \quad
P_1(\hat \delta) = \diag(e^{i \hat \delta},1,1)\,.
\label{eq:parUe12Ue13}
\ee
%%%%%%%%%%%%%%%%%%%%%%%%%%%%%%%%%
%
We find that a sum rule for $\cos\delta$ can be derived 
if either $\theta^{\circ}_{12} =q\pi/2$, $q=0,1,2,3,4$,
or $\theta^{\circ}_{23} = k\pi$, $k=0,1,2$.
Indeed, the relation 
$|U_{\tau 3}|^2 =  
\cos^2 \theta_{13} \cos^2 \theta_{23} = 
\cos^2 \hat \theta_{13} \cos^2 \theta^{\circ}_{23}$,
allows us to determine $\cos^2\hat \theta_{13}$ in terms of
the known quantity $\cos^2 \theta_{13} \cos^2 \theta_{23}$ 
and the parameter $\cos^2 \theta^{\circ}_{23}$, which is fixed 
once $G_f$ and $G_\nu$ are fixed.   
Further, we have
%%%%%%%%%%%%%%%%%%%%%%%%%%
\be
|U_{\tau 2}| = |e^{i \hat \delta} \sin \theta^{\circ}_{12} \sin \hat \theta_{13} 
+ \cos \hat \theta_{13} \cos \theta^{\circ}_{12} \sin \theta^{\circ}_{23}| \,,
\label{sec76Utau2}
\ee
%%%%%%%%%%%%%%%%%%%%%%%%%%%
%
where the only unconstrained parameter is the phase $\hat \delta$.
In the cases indicated above with either 
 $\theta^{\circ}_{12} =q\pi/2$, $q=0,1,2,3,4$,
or $\theta^{\circ}_{23} = k\pi$, $k=0,1,2$, 
the absolute value of
the element $U_{\tau 2}$ does not depend on 
 $\hat \delta$, which in turn allows   
 a sum rule for $\cos \delta$
to be derived. 
In general, $\cos\delta$ is a function of $\hat\delta$:
%%%%%%%%%%%%%%%%%%%%%%%%%%%%
\begin{align}
\cos \delta &  = \dfrac{2}{\sin2\theta_{12} \sin2\theta_{23} \sin\theta_{13} \cos^2\theta^\circ_{23}}
\bigg[ \sin^2\theta^\circ_{12} \left( \cos^2\theta^\circ_{23} - \cos^2\theta_{13} \cos^2\theta_{23} \right) \nonumber \\
& - \cos^2\theta_{12} \sin^2\theta_{23} \cos^2 \theta^\circ_{23}
+ \cos^2\theta_{23} \left( \cos^2\theta_{13} \cos^2\theta^\circ_{12} \sin^2 \theta^\circ_{23} - \sin^2\theta_{12} \sin^2\theta_{13} \cos^2 \theta^\circ_{23} \right) \nonumber \\
& + \kappa \cos\hat\delta \cos\theta_{13} \cos\theta_{23} \sin2\theta^\circ_{12} \sin\theta^\circ_{23}
\left( \cos^2\theta^\circ_{23} - \cos^2\theta_{13} \cos^2\theta_{23} \right)^\frac{1}{2} \bigg]\,,
\end{align}
%%%%%%%%%%%%%%%%%%%%%%%%%%%%%%%
where $\kappa = 1$ if $\hat\theta_{13}$ belongs to the first or third quadrant,
and $\kappa = -1$ otherwise.
In this case the sum rule for $\cos \delta$ has been derived first
in \cite{Girardi:2015vha} assuming $\theta^{\circ}_{13} = 0$,
but as we can see this result holds also for any fixed value
of $\theta^{\circ}_{13}$, since the parametrisation given in
eq.~(\ref{eq:parUe12Ue13}) and the corresponding one 
in \cite{Girardi:2015vha} are the same after a redefinition 
of the parameters.

The sum rules derived in Section~\ref{sec:Gebroken} 
are summarised in Table~\ref{tab:summarysumrulesGebroken}.

%%%%%%%%%%%%%%%%%%%%%%%%%%%%%%%%%%%%%%
\begin{landscape}
\pagestyle{empty}
%%%%%%%%%%%%%%%%%%%%%%%%%%
\begin{table}[h]
\renewcommand*{\arraystretch}{1.2}
\hspace{-0.7cm}
\begin{tabular}{lll}
\hline
& \\ [-10pt]
Case & Parametrisation of $U$ & Sum rule for $\cos \delta$ \\ [8pt]
\bottomrule
& \\ [-10pt]
D2 & $U_{13}(\theta^e_{13}, \delta^e_{13}) R_{12}(\hat \theta_{12}) P_1(\hat \delta) R_{23}(\theta^{\circ}_{23}) R_{13}(\theta^{\circ}_{13}) Q_0$ & 
$\dfrac{\cos^2 \theta_{23} \cos^2 \theta_{12} + \sin^2 \theta_{12} \sin^2 \theta_{13} \sin^2 \theta_{23}  - \cos^2\hat\theta_{12} \cos^2\theta^\circ_{23}}
{\sin 2 \theta_{23} \sin \theta_{12} \cos \theta_{12} \sin \theta_{13}}$
 \\ [10pt]
\hline
& \\ [-10pt]
D3 & $U_{12}(\theta^e_{12}, \delta^e_{12}) R_{23}(\hat \theta_{23}) P_2(\hat \delta) R_{13}(\theta_{13}^{\circ}) R_{12}(\theta_{12}^{\circ}) Q_0$ & 
$\dfrac{2}{\sin2\theta_{12} \sin2\theta_{23} \sin\theta_{13} \cos^2\theta^\circ_{13}}
\bigg[ \cos^2\theta^\circ_{12} \left( \cos^2\theta^\circ_{13} - \cos^2\theta_{13} \cos^2\theta_{23} \right)$ \\ [10pt]
& &  $- \cos^2\theta_{12} \sin^2\theta_{23} \cos^2 \theta^\circ_{13}
+ \cos^2\theta_{23} \left( \cos^2\theta_{13} \sin^2\theta^\circ_{12} \sin^2 \theta^\circ_{13} - \sin^2\theta_{12} \sin^2\theta_{13} \cos^2 \theta^\circ_{13} \right)$ \\ [10pt]
& & $+ \kappa \cos\hat\delta \cos\theta_{13} \cos\theta_{23} \sin2\theta^\circ_{12} \sin\theta^\circ_{13}
\left( \cos^2\theta^\circ_{13} - \cos^2\theta_{13} \cos^2\theta_{23} \right)^\frac{1}{2} \bigg]$
 \\ [10pt]
\hline
& \\ [-10pt]
D4 & $U_{13}(\theta^e_{13}, \delta^e_{13}) R_{23}(\hat \theta_{23}) P_2(\hat \delta) R_{13}(\theta_{13}^{\circ}) R_{12}(\theta_{12}^{\circ}) Q_0$ & 
$- \dfrac{2}{\sin2\theta_{12} \sin2\theta_{23} \sin\theta_{13} \cos^2\theta^\circ_{13}}
\bigg[ \cos^2\theta^\circ_{12} \left( \cos^2\theta^\circ_{13} - \cos^2\theta_{13} \sin^2\theta_{23} \right)$ \\ [10pt]
& & $- \cos^2\theta_{12} \cos^2\theta_{23} \cos^2 \theta^\circ_{13}
+ \sin^2\theta_{23} \left( \cos^2\theta_{13} \sin^2\theta^\circ_{12} \sin^2 \theta^\circ_{13} - \sin^2\theta_{12} \sin^2\theta_{13} \cos^2 \theta^\circ_{13} \right)$ \\ [10pt]
& & $- \kappa \cos\hat\delta \cos\theta_{13} \sin\theta_{23} \sin2\theta^\circ_{12} \sin\theta^\circ_{13}
\left( \cos^2\theta^\circ_{13} - \cos^2\theta_{13} \sin^2\theta_{23} \right)^\frac{1}{2} \bigg]$
 \\ [10pt]
\hline
& \\ [-10pt]
D6 & $U_{12}(\theta^e_{12}, \delta^e_{12}) R_{13}(\hat \theta_{13}) P_1(\hat \delta) R_{23}(\theta_{23}^{\circ}) R_{12}(\theta_{12}^{\circ}) Q_0$ & 
$\dfrac{2}{\sin2\theta_{12} \sin2\theta_{23} \sin\theta_{13} \cos^2\theta^\circ_{23}}
\bigg[ \sin^2\theta^\circ_{12} \left( \cos^2\theta^\circ_{23} - \cos^2\theta_{13} \cos^2\theta_{23} \right)$ \\ [10pt]
& & $- \cos^2\theta_{12} \sin^2\theta_{23} \cos^2 \theta^\circ_{23}
+ \cos^2\theta_{23} \left( \cos^2\theta_{13} \cos^2\theta^\circ_{12} \sin^2 \theta^\circ_{23} - \sin^2\theta_{12} \sin^2\theta_{13} \cos^2 \theta^\circ_{23} \right)$ \\ [10pt]
& & $+ \kappa \cos\hat\delta \cos\theta_{13} \cos\theta_{23} \sin2\theta^\circ_{12} \sin\theta^\circ_{23}
\left( \cos^2\theta^\circ_{23} - \cos^2\theta_{13} \cos^2\theta_{23} \right)^\frac{1}{2} \bigg]$
 \\ [10pt]
\hline
\end{tabular}
\caption{Summary of the sum rules for $\cos\delta$ 
in the case of fully broken $G_e$
under the assumption that the matrix $U_e$ consists of two complex rotation
matrices. 
The parameter $\kappa = 1$ if the corresponding hat angle 
belongs to the first or third quadrant,
and $\kappa = -1$ otherwise.
The cases D3 and D4 have been analysed for $\theta^{\circ}_{13} = 0$
in \cite{Petcov:2014laa,Girardi:2015vha}. 
In the case D6 
the sum rule for $\cos \delta$ has been derived first
in \cite{Girardi:2015vha} assuming $\theta^{\circ}_{13} = 0$,
but this result holds also for any fixed value
of $\theta^{\circ}_{13}$. See text for further details.
}
\label{tab:summarysumrulesGebroken}
\end{table}
%%%%%%%%%%%%%%%%%%%%%%%%%%
\end{landscape}

%%%%%%%%%%%%%%%%%%%%%%%%%%%%%%%%%%
%
\section{The Case of Fully Broken $G_{\nu}$}
\label{sec:Gnubroken}
%
%%%%%%%%%%%%%%%%%%%%%%%%%%%%%
%
When the discrete flavour symmetry $G_f$ is fully broken 
in the neutrino sector, 
the matrix $U_\nu$ is unconstrained and includes, in general, 
three complex rotations and three phases, i.e.,  
three angle and six CPV phase parameters.
It is impossible to derive predictions for the 
mixing angles and CPV phases in the PMNS matrix 
in this case. Therefore we will consider 
in this section forms of $U_\nu$ 
corresponding to one of the rotation 
angle parameters being equal to zero. 
Some of these forms of  $U_\nu$ correspond to a class of models 
of neutrino mass generation or phenomenological studies
(see, e.g., \cite{Shimizu:2014ria}) 
and lead, in particular, to sum rules 
for $\cos\delta$.
Since in this case $G_f$ is fully broken in the neutrino sector,
the $Z_2 \times Z_2$ symmetry of the Majorana mass
term does arise accidentally. Therefore the matrix 
$U_{\nu}$ is not constrained by the symmetry 
group $G_f$.
We give in Table~\ref{tab:parUeGebroken} in Appendix~\ref{app:Gebroken}
the most general parametrisations of $U$ under the
assumption that for fully broken $G_\nu$
one rotation angle vanishes in the matrix $U_{\nu}$.

%%%%%%%%%%%%%%%%%%%%%%
%
\subsection{The Scheme with $U_{12}(\theta^\nu_{12},\delta^\nu_{12}) U_{13}(\theta^\nu_{13},\delta^\nu_{13})$ (Case E1)}
%
%%%%%%%%%%%%%%%%%%%%%%%%%%%%%
%
It proves convenient to
consider the following parametrisation of the 
PMNS matrix $U$ in this case 
(see Appendix~\ref{app:Gebroken}, fourth case in Table~\ref{tab:parUeGebroken}):
%%%%%%%%%%%%%%%%%%%%%%%%%%%%%
\be
U = R_{23}(\theta^{\circ}_{23}) R_{13}(\theta^{\circ}_{13}) P_1(\hat \delta) R_{12}(\hat \theta_{12}) U_{13}(\theta^\nu_{13}, \delta^\nu_{13}) Q_0 \,, \quad
P_1(\hat \delta) = \diag(e^{i \hat \delta},1,1)\,.
\ee
%%%%%%%%%%%%%%%%%%%%%%%%%%%%%%%%%%%
%
Consider first the case of $\theta^{\circ}_{13} = 0$.
In this case the phase 
$\hat \delta$ is unphysical.
Comparing this parametrisation of $U$ with the standard parametrisation, we find:
%
%%%%%%%%%%%%%%%%%%%%%%%%%%%%%%%%
\begin{align}
\sin^2 \theta_{13} & = |U_{e3}|^2  = \sin^2 \theta^{\nu}_{13} \cos^2 \hat \theta_{12} \,, 
\label{eq:th13Uthnu12thnu13GeZ3Gnu0}\\
\sin^2 \theta_{23} & = \frac{|U_{\mu3}|^2}{1-|U_{e3}|^2} =  \frac{1}{\cos^2 \theta_{13}} \big[
\sin^2 \theta^{\circ}_{23} \cos^2 \theta^\nu_{13} + \cos^2 \theta^\circ_{23} \sin^2 \theta^\nu_{13} \sin^2 \hat \theta_{12} \nonumber \\
& - \frac{1}{2} \sin 2 \theta^\circ_{23} \sin 2 \theta^{\nu}_{13} \sin \hat \theta_{12} \cos \delta^\nu_{13}  \big] \,, 
\label{eq:th23Uthnu12thnu13GeZ3Gnu0}\\
\sin^2 \theta_{12} & = \frac{|U_{e2}|^2}{1-|U_{e3}|^2} =  \frac{\sin^2 \hat \theta_{12}}{\cos^2 \theta_{13}} \,.
\label{eq:th12Uthnu12thnu13GeZ3Gnu0}
\end{align}
%%%%%%%%%%%%%%%%%%%%%%%%%%%
%
From the ratio 
%%%%%%%%%%%%%%%%%%%%%%%%%%%%
\be
\left|\frac{U_{\tau2}}{U_{\mu2}}\right|^2 = \tan^2 \theta^\circ_{23}\,,
\ee
%%%%%%%%%%%%%%%%%%%%%%%%%%%%%%
%
we get the following sum rule for $\cos \delta$:
%%%%%%%%%%%%%%%%%%%%%%%%%%%%%%%%%%%%
\be
\cos\delta =  - \frac{\tan\theta_{12}}{\sin2\theta_{23}\sin\theta_{13}}
\left [\cos2\theta^\circ_{23} \sin^2 \theta_{13} + 
\left (\sin^2\theta_{23} - \sin^2\theta^\circ_{23} \right )
 \left (\cot^2 \theta_{12} - \sin^2\theta_{13}\right )\right ]\,.
\label{eq:cosdeltaUthnu12thnu13GeZ3Gnu0}
\ee
%%%%%%%%%%%%%%%%%%%%%%%%%%%%
%
Substituting the best fit values of the neutrino mixing angles 
for the NO neutrino mass spectrum and the value of $\theta^\circ_{23} = -\pi/4$, 
which corresponds to the TBM, BM, GRA, GRB and HG symmetry forms,
we obtain $\cos \delta = 0.616$. We note that in the considered scheme
the predictions for $\cos \delta$ are all the same for the symmetry forms 
mentioned above, since these forms are characterised by 
different values of the angle $\theta^\circ_{12}$, which
has been absorbed by the free parameter $\hat \theta_{12}$. 
This ``degeneracy" can be lifted in specific models
in which the value of $\theta^{\nu}_{12}$ is fixed.
Using the best fit values 
and the requirement $|\cos \delta| \leq 1$, we find
that the allowed values of $\sin^2 \theta^{\circ}_{23}$ belong to
the following interval: $0.338 \leq \sin^2 \theta^{\circ}_{23} \leq 0.538$.

In order to give the general result for $\cos \delta$ in the case of 
$\theta^{\circ}_{13} \neq 0$,
we use the expression for $\sin^2 \theta_{12}$ for non-zero
$\theta^{\circ}_{13}$:
%%%%%%%%%%%%%%%%%%%%%%%%%%%%%%
\begin{align}
\sin^2 \theta_{12} & = \frac{|U_{e2}|^2}{1-|U_{e3}|^2} =  \frac{\cos^2 \theta^{\circ}_{13} \sin^2 \hat \theta_{12}}{\cos^2 \theta_{13}} \,.
\end{align}
%%%%%%%%%%%%%%%%%%%%%%%%%%%
%
Employing this relation in the expression for $|U_{\tau2}|^2$,
we get
%%%%%%%%%%%%%%%%%%%%%%%%%%%%
\begin{align}
\cos \delta &  = - \dfrac{2}{\sin2\theta_{12} \sin2\theta_{23} \sin\theta_{13} \cos^2\theta^\circ_{13}}
\bigg[ \cos^2\theta^\circ_{23} \left( \cos^2\theta^\circ_{13} - \sin^2\theta_{12} \cos^2\theta_{13} \right) \nonumber \\
& - \cos^2\theta_{12} \cos^2\theta_{23} \cos^2 \theta^\circ_{13}
+ \sin^2\theta_{12} \left( \cos^2\theta_{13} \sin^2\theta^\circ_{13} \sin^2 \theta^\circ_{23} - \sin^2\theta_{13} \sin^2\theta_{23} \cos^2 \theta^\circ_{13} \right) \nonumber \\
& - \kappa \cos\hat\delta \sin\theta_{12} \cos\theta_{13} \sin\theta^\circ_{13} \sin2\theta^\circ_{23}
\left( \cos^2\theta^\circ_{13} - \sin^2\theta_{12} \cos^2\theta_{13} \right)^\frac{1}{2} \bigg]\,,
\label{eq:cosdeltaSect81}
\end{align}
%%%%%%%%%%%%%%%%%%%%%%%%%%%%%%%
where $\kappa = 1$ if $\hat\theta_{12}$ belongs to the first or third quadrant,
and $\kappa = -1$ otherwise.

Similar to the cases C2, C5, C7 and C9 analysed in subsections 
\ref{sec:13e12nu}, \ref{sec:23e13nu}, \ref{sec:12e12nu} and \ref{sec:23e23nu}, 
$\cos\delta$ is a function of 
the known neutrino mixing angles $\theta_{12}$, 
$\theta_{13}$ and  $\theta_{23}$, of the 
angles $\theta^\circ_{13}$ and $\theta^\circ_{23}$  fixed by 
$G_f$ and the assumed symmetry breaking pattern,
as well as of the phase parameter  $\hat \delta$ 
of the scheme. Predictions for $\cos\delta$ 
can be obtained if  $\hat \delta$ 
is fixed by additional considerations 
of, e.g., GCP invariance,   
symmetries, etc. 

For $\theta^\circ_{13} = k \pi$, $k = 0, 1, 2$,
and/or $\theta^\circ_{23} = k' \pi/2$, $k' = 0, 1, 2, 3, 4$,
$\cos\delta$ does not depend on $\hat\delta$ and $\kappa$.
In the first case the expression in eq.~(\ref{eq:cosdeltaSect81})
reduces to the sum rule given in eq.~(\ref{eq:cosdeltaUthnu12thnu13GeZ3Gnu0}).

%%%%%%%%%%%%%%%%%%%%%%
%
\subsection{The Scheme with $U_{12}(\theta^\nu_{12},\delta^\nu_{12}) U_{23}(\theta^\nu_{23},\delta^\nu_{23})$ (Case E2)}
%
%%%%%%%%%%%%%%%%%%%%%%%%%%%%%
%
In this case it is convenient to use 
another possible parametrisation of the PMNS matrix, 
the fourth case in Table~\ref{tab:parUeGebroken}
given in Appendix~\ref{app:Gebroken}. Namely,
%%%%%%%%%%%%%%%%%%%%%%%%%%%%%%%%%%%
\be
U = R_{23}(\theta^{\circ}_{23}) R_{13}(\theta^{\circ}_{13}) P_1(\hat \delta) R_{12}(\hat \theta_{12}) U_{23}(\theta^\nu_{23}, \delta^\nu_{23}) Q_0 \,, \quad
P_1(\hat \delta) = \diag(e^{i \hat \delta},1,1)\,.
\label{eq:USect.8.2}
\ee
%%%%%%%%%%%%%%%%%%%%%%%%%%%%%%%%%%%
%
Consider first the possibility of $\theta^\circ_{13} = 0$.
Under this assumption we find:
%%%%%%%%%%%%%%%%%%%%%%%%%%%%%%%%
\begin{align}
\sin^2 \theta_{13} & = |U_{e3}|^2  = \sin^2 \theta^{\nu}_{23} \sin^2 \hat \theta_{12} \,, 
\label{eq:th13Uthnu12thnu23GeZ3Gnu0}\\
\sin^2 \theta_{23} & = \frac{|U_{\mu3}|^2}{1-|U_{e3}|^2} =  \frac{1}{\cos^2 \theta_{13}} \big[
\sin^2 \theta^{\circ}_{23} \cos^2 \theta^\nu_{23} + \cos^2 \theta^\circ_{23} \sin^2 \theta^\nu_{23} \cos^2 \hat \theta_{12} \nonumber \\
& + \frac{1}{2} \sin 2 \theta^\circ_{23} \sin 2 \theta^{\nu}_{23} \cos \hat \theta_{12} \cos \delta^\nu_{23}  \big] \,, 
\label{eq:th23Uthnu12thnu23GeZ3Gnu0}\\
\sin^2 \theta_{12} & = \frac{|U_{e2}|^2}{1-|U_{e3}|^2} =  \frac{\cos^2 \theta^\nu_{23} \sin^2 \hat \theta_{12}}{\cos^2 \theta_{13}} \,.
\label{eq:th12Uthnu12thnu23GeZ3Gnu0}
\end{align}
%%%%%%%%%%%%%%%%%%%%%%%%%%%
%
The sum rule of interest for $\cos\delta$ 
can be derived in this case using the ratio
%%%%%%%%%%%%%%%%%%%%%%%%%%%%%%%%%%%%%%%%%%%
\be
\left|\frac{U_{\tau1}}{U_{\mu1}}\right|^2 = \tan^2 \theta^\circ_{23}\,.
\ee
%%%%%%%%%%%%%%%%%%%%%%%%%%%%%%%%%%%%%%%%%%
%
We get
%%%%%%%%%%%%%%%%%%%%%%%%%%
\be
\cos\delta =  \frac{\cot\theta_{12}}{\sin2\theta_{23}\sin\theta_{13}}
\left [\cos2\theta^\circ_{23} \sin^2 \theta_{13} + 
\left (\sin^2\theta_{23} - \sin^2\theta^\circ_{23} \right )
 \left (\tan^2 \theta_{12} - \sin^2\theta_{13}\right )\right ]\,.
\label{eq:cosdeltaUthnu12thnu23GeZ3Gnu0}
\ee
%%%%%%%%%%%%%%%%%%%%%%%%%%%
%
This sum rule can be formally obtained 
from the r.h.s. of eq.~(\ref{eq:cosdeltaUthnu12thnu13GeZ3Gnu0}) by
interchanging $\tan \theta_{12}$ and $\cot \theta_{12}$
and by multiplying it by $(-1)$.
Substituting the best fit values of the neutrino mixing angles 
for the NO neutrino mass spectrum and the value of $\theta^\circ_{23} = -\pi/4$,
we get $\cos \delta = - 0.262$. Using the best fit values 
and the requirement $|\cos \delta| \leq 1$, we find
that the allowed values of $\sin^2 \theta^{\circ}_{23}$ belong to
the following interval: $0.227 \leq \sin^2 \theta^{\circ}_{23} \leq 0.659$.

In order to find a general 
result for $\cos \delta$
for arbitrary fixed $\theta^{\circ}_{13} \neq 0$, we 
use the following relation: 
%%%%%%%%%%%%%%%%%%%%%%%%%%%%%%
\begin{equation}
\cos^2\theta_{12} \cos^2\theta_{13} = \cos^2\hat\theta_{12} \cos^2\theta^\circ_{13} \,,
\end{equation}
%%%%%%%%%%%%%%%%%%%%%%%%%%%%%%%%%%
%
which follows from the expressions for $|U_{e1}|^2$
in the standard parametrisation and in the parametrisation
given in eq.~(\ref{eq:USect.8.2}).
With the help of %the above 
this relation, using $|U_{\mu1}|$, we get
%%%%%%%%%%%%%%%%%%%%%%%%%%%%
\begin{align}
\cos \delta &  = \dfrac{2}{\sin2\theta_{12} \sin2\theta_{23} \sin\theta_{13} \cos^2\theta^\circ_{13}}
\bigg[ \cos^2\theta^\circ_{23} \left( \cos^2\theta^\circ_{13} - \cos^2\theta_{12} \cos^2\theta_{13} \right) \nonumber \\
& - \sin^2\theta_{12} \cos^2\theta_{23} \cos^2 \theta^\circ_{13}
+ \cos^2\theta_{12} \left( \cos^2\theta_{13} \sin^2\theta^\circ_{13} \sin^2 \theta^\circ_{23} - \sin^2\theta_{13} \sin^2\theta_{23} \cos^2 \theta^\circ_{13} \right) \nonumber \\
& + \kappa \cos\hat\delta \cos\theta_{12} \cos\theta_{13} \sin\theta^\circ_{13} \sin2\theta^\circ_{23}
\left( \cos^2\theta^\circ_{13} - \cos^2\theta_{12} \cos^2\theta_{13} \right)^\frac{1}{2} \bigg]\,,
\label{eq:cosdeltaSect82}
\end{align}
%%%%%%%%%%%%%%%%%%%%%%%%%%%%%%%
where $\kappa = 1$ if $\hat\theta_{12}$ belongs to the first or third quadrant,
and $\kappa = -1$ otherwise.
Also in this case $\cos\delta$ is a function of 
the unconstrained phase parameter  $\hat \delta$ 
of the scheme. Predictions for $\cos\delta$ 
can be obtained if $\hat \delta$ 
is fixed by additional considerations 
(e.g., GCP invariance,   
symmetries, etc.).

As like in the case E1, for 
$\theta^\circ_{13} = k \pi$, $k = 0, 1, 2$,
and/or $\theta^\circ_{23} = k' \pi/2$, $k' = 0, 1, 2, 3, 4$,
$\cos\delta$ does not depend on $\hat\delta$ and $\kappa$.
For $\theta^\circ_{13} = 0, \pi, 2\pi$, the sum rule in eq.~(\ref{eq:cosdeltaSect82})
coincides with the sum rule given in eq.~(\ref{eq:cosdeltaUthnu12thnu23GeZ3Gnu0}).

%%%%%%%%%%%%%%%%%%%%%%
%
\subsection{The Scheme with $U_{23}(\theta^\nu_{23},\delta^\nu_{23}) U_{12}(\theta^\nu_{12},\delta^\nu_{12})$ (Case E3)}
\label{sec:23nu12nu}
%
%%%%%%%%%%%%%%%%%%%%%%%%%%%%%%%
%
The convenient parametrisation for $U$ to use in this case 
is that of the fifth case in Table~\ref{tab:parUeGebroken}
given in Appendix~\ref{app:Gebroken}:
%%%%%%%%%%%%%%%%%%%%%%%%%%%%%%%%%%%%
\beq
U = R_{13}(\theta^{\circ}_{13}) R_{12}(\theta^{\circ}_{12}) P_2(\hat \delta) R_{23}(\hat \theta_{23}) U_{12}(\theta^\nu_{12}, \delta^\nu_{12}) Q_0 \,, \quad
P_2(\hat \delta) = \diag(1,e^{i \hat \delta},1)\,.
\eeq
%%%%%%%%%%%%%%%%%%%%%%%%%%%%%
%
We find that:
%%%%%%%%%%%%%%%%%%%%%%%%%%%%%%%%
\begin{align}
\sin^2 \theta_{13} & = |U_{e3}|^2  =  \sin^2 \theta_{13} (\hat \theta_{23}, \hat\delta, \theta^\circ_{12}, \theta^\circ_{13}) \,, \label{eq:thnu23thnu12:th13} \\
\sin^2 \theta_{23} & = \frac{|U_{\mu3}|^2}{1-|U_{e3}|^2} =  \frac{\cos^2 \theta^\circ_{12} \sin^2 \hat \theta_{23}}{\cos^2 \theta_{13}} \,, \label{eq:thnu23thnu12:th23} \\
\sin^2 \theta_{12} & = \frac{|U_{e2}|^2}{1-|U_{e3}|^2} = \sin^2 \theta_{12}(\hat \theta_{23}, \hat\delta, \theta^{\nu}_{12}, \delta^\nu_{12}, \theta^\circ_{12}, \theta^\circ_{13}) \,. \label{eq:thnu23thnu12:th12}
\end{align}
%%%%%%%%%%%%%%%%%%%%%%%%%%%
%
However, a sum rule for $\cos \delta$ 
cannot be obtained because $\cos \delta$ 
turns out to depend, in particular, on 
$\delta^\nu_{12}$ which is
an unconstrained
phase parameter of the scheme 
considered, which can be seen
from the expression for $|U_{\mu 1}|$: 
%%%%%%%%%%%%%%%%%%%%%%%%%%%%%%%%%%%%
\be
|U_{\mu 1}| = |\cos \theta^{\nu}_{12} \sin \theta^{\circ}_{12} + e^{i(\hat \delta + \delta^{\nu}_{12})} \cos \hat \theta_{23} \cos \theta^{\circ}_{12} \sin \theta^{\nu}_{12} |\,.
\ee
%%%%%%%%%%%%%%%%%%%%%%%%%%%%%%%%
%
The situation here is analogous to the cases analysed
in subsections \ref{sec:23e12e} and \ref{sec:23e13e}. Namely,
considering a certain residual symmetry group $G_e$, 
from eq.~(\ref{eq:thnu23thnu12:th23}) we find that
$\sin^2\hat\theta_{23}$ is fixed. Then, $\cos\hat\delta$ is fixed
(up to a sign) by eq.~(\ref{eq:thnu23thnu12:th13}). Hence, $\theta^\nu_{12}$
can be expressed in terms of $\delta^\nu_{12}$ by virtue of
eq.~(\ref{eq:thnu23thnu12:th12}).
Thus, numerical predictions for $\cos \delta$ can be obtained
if $\delta^{\nu}_{12}$ is fixed.

%%%%%%%%%%%%%%%%%%%%%%
%
\subsection{The Scheme with $U_{23}(\theta^\nu_{23},\delta^\nu_{23}) U_{13}(\theta^\nu_{13},\delta^\nu_{13})$ (Case E4)}
%
%%%%%%%%%%%%%%%%%%%%%%%%%%%%%%%%%%%%%%%%%%
%
Employing  the parametrisation for $U$ given in 
Appendix~\ref{app:Gebroken}, namely the fifth case 
in Table~\ref{tab:parUeGebroken},
%%%%%%%%%%%%%%%%%%%%%%%%%%%%%%%%
\beq
U = R_{13}(\theta^{\circ}_{13}) R_{12}(\theta^{\circ}_{12}) P_2(\hat \delta) R_{23}(\hat \theta_{23}) U_{13}(\theta^\nu_{13}, \delta^\nu_{13}) Q_0 \,, \quad
P_2(\hat \delta) = \diag(1,e^{i \hat \delta},1)\,,
\eeq
%%%%%%%%%%%%%%%%%%%%%%%%%%%%%%
%
we find that $\cos \delta$ is a function of $\hat \theta_{23}$, $\theta^\circ_{12}$ 
and the PMNS mixing angles. Therefore,
$\cos \delta$ can be determined only in those cases when 
$\hat \theta_{23}$ is fixed. Using the result
%%%%%%%%%%%%%%%%%%%%%%%%%%%%%
\begin{align}
\sin^2 \theta_{12} & = \frac{|U_{e2}|^2}{1-|U_{e3}|^2} =  \frac{1}{\cos^2 \theta_{13}} \bigg[
\cos^2 \hat \theta_{23} \cos^2 \theta^{\circ}_{13} \sin^2 \theta^{\circ}_{12} + \sin^2 \hat \theta_{23} \sin^2 \theta^{\circ}_{13} \nonumber \\
& - \dfrac{1}{2} \cos \hat \delta \sin 2 \hat \theta_{23} \sin 2 \theta^{\circ}_{13} \sin \theta^{\circ}_{12} \bigg]\,,
\end{align}
%%%%%%%%%%%%%%%%%%%%%%%%%%%%%%%%%%
%
we find these cases to be, for example: 
i) $\theta^{\circ}_{12} = 0$, $\pi$, leading to the relation
 $\sin^2 \theta_{12} \cos^2 \theta_{13} 
= \sin^2 \hat \theta_{23} \sin^2 \theta^{\circ}_{13}$,
ii) $\theta^{\circ}_{13} = 0$, $\pi$, implying 
$\sin^2 \theta_{12} \cos^2 \theta_{13} 
= \cos^2 \hat \theta_{23} \sin^2 \theta^{\circ}_{12}$,
iii) $\theta^{\circ}_{13} = \pi/2$, $3\pi/2$, 
giving $\sin^2 \theta_{12} \cos^2 \theta_{13} = \sin^2 \hat \theta_{23}$.
For this reason we give $\cos \delta$ 
as a function of the angle  $\hat \theta_{23}$.
Namely, the sum rule of interest, which is obtained using 
$|U_{\mu 2}| = |\cos \hat \theta_{23} \cos \theta^{\circ}_{12}|$,
reads
%%%%%%%%%%%%%%%%%%%%%%%%%%%%%%%%%%%
\begin{align}
& \cos \delta  = \dfrac{\cos^2 \theta_{12} \cos^2 \theta_{23} 
+ \sin^2 \theta_{12} \sin^2 \theta_{13} \sin^2 \theta_{23} 
- \cos^2 \hat \theta_{23} \cos^2 \theta^{\circ}_{12}}
{\sin 2 \theta_{23} \sin \theta_{12} \cos \theta_{12} \sin \theta_{13}}\,.
\end{align}
%%%%%%%%%%%%%%%%%%%%%%%%%%%%%%%%
%
The dependence of $\cos \delta$ on $G_f$ is realised via the values of 
the angles $\theta^{\circ}_{12}$ and $\theta^\circ_{13}$.

%%%%%%%%%%%%%%%%%%%%%%%%%%%%%%%%%%%
%
\subsection{The Scheme with $U_{13}(\theta^\nu_{13},\delta^\nu_{13}) U_{12}(\theta^\nu_{12},\delta^\nu_{12})$ (Case E5)}
%
%%%%%%%%%%%%%%%%%%%%%%%%%%%%%%%%%%%%
%
The parametrisation for the PMNS matrix $U$ employed by us 
in this subsection is the sixth case 
in Table~\ref{tab:parUeGebroken} given in Appendix~\ref{app:Gebroken}:
%%%%%%%%%%%%%%%%%%%%%%%%%%%%%%%%%%
\beq
U = R_{23}(\theta^{\circ}_{23}) R_{12}(\theta^{\circ}_{12}) P_1(\hat \delta) R_{13}(\hat \theta_{13}) U_{12}(\theta^\nu_{12}, \delta^\nu_{12}) Q_0 \,, \quad
P_1(\hat \delta) = \diag(e^{i \hat \delta},1,1)\,.
\eeq
%%%%%%%%%%%%%%%%%%%%%%%%
%
We find that:
%%%%%%%%%%%%%%%%%%%%%%%%%%%%%%%%
\begin{align}
\sin^2 \theta_{13} & = |U_{e3}|^2  = \cos^2 \theta^{\circ}_{12} \sin^2 \hat \theta_{13}  \,, \label{eq:thnu13thnu12:th13} \\
\sin^2 \theta_{23} & = \frac{|U_{\mu3}|^2}{1-|U_{e3}|^2} =  \sin^2 \theta_{23}(\hat \theta_{13}, \hat\delta, \theta^\circ_{12}, \theta^\circ_{23})\,, 
\label{eq:thnu13thnu12:th23}\\
\sin^2 \theta_{12} & = \frac{|U_{e2}|^2}{1-|U_{e3}|^2} = \sin^2 \theta_{12}(\hat \theta_{13}, \hat\delta, \theta^{\nu}_{12}, \delta^\nu_{12}, \theta^\circ_{12})  \,.
\label{eq:thnu13thnu12:th12}
\end{align}
%%%%%%%%%%%%%%%%%%%%%%%%%%%
However, a sum rule for $\cos \delta$ 
cannot be obtained because $\cos \delta$ 
turns out to depend, in particular, on 
$\delta^\nu_{12}$ which is
an unconstrained 
phase parameter of the scheme 
considered. This can be seen, e.g., from the expression 
for $|U_{\mu 1}|$:
\be
|U_{\mu 1}| = |\cos \theta^{\nu}_{12} (e^{i \hat \delta}  \sin \theta^{\circ}_{12} \cos \theta^{\circ}_{23} \cos \hat \theta_{13} 
+ \sin \hat \theta_{13} \sin \theta^{\circ}_{23}) + e^{i \delta^{\nu}_{12}} \cos \theta^{\circ}_{12} \cos \theta^{\circ}_{23} \sin \theta^{\nu}_{12}|.
\ee
Similarly to the case analysed
in subsection \ref{sec:23nu12nu},
for a certain residual symmetry group $G_e$, 
from eq.~(\ref{eq:thnu13thnu12:th13}) we find that
$\sin^2\hat\theta_{13}$ is fixed. Then, $\cos\hat\delta$ is fixed
(up to a sign) by eq.~(\ref{eq:thnu13thnu12:th23}), and so the angle $\theta^\nu_{12}$
can be expressed in terms of $\delta^\nu_{12}$ by virtue of
eq.~(\ref{eq:thnu13thnu12:th12}).
Therefore, numerical predictions for $\cos \delta$ can be obtained
if $\delta^{\nu}_{12}$ is fixed.

%%%%%%%%%%%%%%%%%%%%%%
%
\subsection{The Scheme with $U_{13}(\theta^\nu_{13},\delta^\nu_{13}) U_{23}(\theta^\nu_{23},\delta^\nu_{23})$ (Case E6)}
%
%%%%%%%%%%%%%%%%%%%%%%%%%%%%%%%%%
%
The parametrisation of the PMNS matrix $U$ utilised by us 
in the present subsection is that of the sixth case in Table~\ref{tab:parUeGebroken}
given in Appendix~\ref{app:Gebroken}:
%%%%%%%%%%%%%%%%%%%%%%%%%%%%%%
\beq
U = R_{23}(\theta^{\circ}_{23}) R_{12}(\theta^{\circ}_{12}) P_1(\hat \delta) R_{13}(\hat \theta_{13}) U_{23}(\theta^\nu_{23}, \delta^\nu_{23}) Q_0 \,, \quad
P_1(\hat \delta) = \diag(e^{i \hat \delta},1,1)\,.
\eeq
%%%%%%%%%%%%%%%%%%%%%%%%%%%%%%%
%
A sum rule and predictions for $\cos\delta$ can be derived 
in the cases of either $\theta^{\circ}_{23} =q\pi/2$, $q = 0,1,2,3,4$,
or $\theta^{\circ}_{12} = k\pi$, $k=0,1,2$.
Indeed, using the relation 
%%%%%%%%%%%%%%%%%%%%%%%%%%%%%%%%%%%%%%%
\be
|U_{e1}|^2 = \cos^2 \theta_{12} \cos^2 \theta_{13} = 
\cos^2 \hat \theta_{13} \cos^2 \theta^{\circ}_{12} \,,
\label{sec86Ue1}
\ee
%%%%%%%%%%%%%%%%%%%%%%%%%%%%%%%%%%%
%
we can express $\cos^2 \hat \theta_{13}$
in terms of the product 
of PMNS neutrino mixing parameters 
$\cos^2 \theta_{12}$ $\cos^2 \theta_{13}$
and, the fixed by $G_f$ parameter, 
$\cos^2 \theta^{\circ}_{12}$.
The sum rule of interest for $\cos\delta$ 
can be derived, e.g., from the expression 
for the absolute value of the element $U_{\mu 1}$:
%%%%%%%%%%%%%%%%%%%%%%%%%%%%%
\be
|U_{\mu 1}|
= |e^{-i \hat \delta} \cos \hat \theta_{13} \cos \theta^{\circ}_{23} 
\sin \theta^{\circ}_{12} + \sin \hat \theta_{13} \sin \theta^{\circ}_{23}|\,,
\label{sec86Umu1}
\ee
%%%%%%%%%%%%%%%%%%%%%%%%%%%%
%
since in any of the two limits indicated above,
$\theta^{\circ}_{23} =q\pi/2$, $q = 0,1,2,3,4$,
or $\theta^{\circ}_{12} = k\pi$, $k=0,1,2$,
$|U_{\mu 1}|$  does not depend on $\hat \delta$. In fact, it 
is given only in terms of the known PMNS neutrino 
mixing parameters and an angle (either 
$\theta^{\circ}_{23}$ or $\theta^{\circ}_{12}$)
which is fixed by the symmetry $G_f$.
In the general case, $\cos\delta$ is a function of $\hat\delta$.
Using eqs.~(\ref{sec86Ue1}) and (\ref{sec86Umu1}), we get
%%%%%%%%%%%%%%%%%%%%%%%%%%%%
\begin{align}
\cos \delta &  = \dfrac{2}{\sin2\theta_{12} \sin2\theta_{23} \sin\theta_{13} \cos^2\theta^\circ_{12}}
\bigg[ \sin^2\theta^\circ_{23} \left( \cos^2\theta^\circ_{12} - \cos^2\theta_{12} \cos^2\theta_{13} \right) \nonumber \\
& - \sin^2\theta_{12} \cos^2\theta_{23} \cos^2 \theta^\circ_{12}
+ \cos^2\theta_{12} \left( \cos^2\theta_{13} \sin^2\theta^\circ_{12} \cos^2 \theta^\circ_{23} - \sin^2\theta_{13} \sin^2\theta_{23} \cos^2 \theta^\circ_{12} \right) \nonumber \\
& + \kappa \cos\hat\delta \cos\theta_{12} \cos\theta_{13} \sin\theta^\circ_{12} \sin2\theta^\circ_{23}
\left( \cos^2\theta^\circ_{12} - \cos^2\theta_{12} \cos^2\theta_{13} \right)^\frac{1}{2} \bigg]\,,
\end{align}
%%%%%%%%%%%%%%%%%%%%%%%%%%%%%%%
%
where $\kappa = 1$ if $\hat\theta_{13}$ lies in the first or third quadrant,
and $\kappa = -1$ otherwise.
For $\theta^\circ_{12} = k \pi$, $k = 0, 1, 2$,
and/or $\theta^\circ_{23} = k' \pi/2$, $k' = 0, 1, 2, 3, 4$,
$\cos\delta$ does not depend on $\hat\delta$ and $\kappa$.

The sum rules derived in Section~\ref{sec:Gnubroken}
are summarised in Table~\ref{tab:summarysumrulesGnubroken}.

%%%%%%%%%%%%%%%%%%%%%%%%%%
\begin{landscape}
\pagestyle{empty}
%%%%%%%%%%%%%%%%%%%%%%%%%%
\begin{table}[h]
\renewcommand*{\arraystretch}{1.2}
\hspace{-0.7cm}
\begin{tabular}{lll}
\hline
& \\ [-10pt]
Case & Parametrisation of $U$ & Sum rule for $\cos \delta$ \\ [8pt]
\bottomrule
& \\ [-10pt]
E1 & $R_{23}(\theta^{\circ}_{23}) R_{13}(\theta^{\circ}_{13}) P_1(\hat \delta) R_{12}(\hat \theta_{12}) U_{13}(\theta^\nu_{13}, \delta^\nu_{13}) Q_0$ & 
$- \dfrac{2}{\sin2\theta_{12} \sin2\theta_{23} \sin\theta_{13} \cos^2\theta^\circ_{13}}
\bigg[ \cos^2\theta^\circ_{23} \left( \cos^2\theta^\circ_{13} - \sin^2\theta_{12} \cos^2\theta_{13} \right)$ \\ [10pt]
& & $- \cos^2\theta_{12} \cos^2\theta_{23} \cos^2 \theta^\circ_{13}
+ \sin^2\theta_{12} \left( \cos^2\theta_{13} \sin^2\theta^\circ_{13} \sin^2 \theta^\circ_{23} - \sin^2\theta_{13} \sin^2\theta_{23} \cos^2 \theta^\circ_{13} \right)$ \\ [10pt]
& & $- \kappa \cos\hat\delta \sin\theta_{12} \cos\theta_{13} \sin\theta^\circ_{13} \sin2\theta^\circ_{23}
\left( \cos^2\theta^\circ_{13} - \sin^2\theta_{12} \cos^2\theta_{13} \right)^\frac{1}{2} \bigg]$
\\ [10pt]
\hline
& \\ [-10pt]
E2 & $R_{23}(\theta^{\circ}_{23}) R_{13}(\theta^{\circ}_{13}) P_1(\hat \delta) R_{12}(\hat \theta_{12}) U_{23}(\theta^\nu_{23}, \delta^\nu_{23}) Q_0$ & 
$\dfrac{2}{\sin2\theta_{12} \sin2\theta_{23} \sin\theta_{13} \cos^2\theta^\circ_{13}}
\bigg[ \cos^2\theta^\circ_{23} \left( \cos^2\theta^\circ_{13} - \cos^2\theta_{12} \cos^2\theta_{13} \right)$ \\ [10pt]
& & $- \sin^2\theta_{12} \cos^2\theta_{23} \cos^2 \theta^\circ_{13}
+ \cos^2\theta_{12} \left( \cos^2\theta_{13} \sin^2\theta^\circ_{13} \sin^2 \theta^\circ_{23} - \sin^2\theta_{13} \sin^2\theta_{23} \cos^2 \theta^\circ_{13} \right)$ \\ [10pt]
& & $+ \kappa \cos\hat\delta \cos\theta_{12} \cos\theta_{13} \sin\theta^\circ_{13} \sin2\theta^\circ_{23}
\left( \cos^2\theta^\circ_{13} - \cos^2\theta_{12} \cos^2\theta_{13} \right)^\frac{1}{2} \bigg]$
\\ [10pt]
\hline
& \\ [-10pt]
E4 & $R_{13}(\theta^{\circ}_{13}) R_{12}(\theta^{\circ}_{12}) P_2(\hat \delta) R_{23}(\hat \theta_{23}) U_{13}(\theta^\nu_{13}, \delta^\nu_{13}) Q_0$ & 
$\dfrac{\cos^2 \theta_{12} \cos^2 \theta_{23} 
+ \sin^2 \theta_{12} \sin^2 \theta_{13} \sin^2 \theta_{23} 
- \cos^2 \hat \theta_{23} \cos^2 \theta^{\circ}_{12}}
{\sin 2 \theta_{23} \sin \theta_{12} \cos \theta_{12} \sin \theta_{13}}$
\\ [10pt]
\hline
& \\ [-10pt]
E6 & $R_{23}(\theta^{\circ}_{23}) R_{12}(\theta^{\circ}_{12}) P_1(\hat \delta) R_{13}(\hat \theta_{13}) U_{23}(\theta^\nu_{23}, \delta^\nu_{23}) Q_0$ & 
$\dfrac{2}{\sin2\theta_{12} \sin2\theta_{23} \sin\theta_{13} \cos^2\theta^\circ_{12}}
\bigg[ \sin^2\theta^\circ_{23} \left( \cos^2\theta^\circ_{12} - \cos^2\theta_{12} \cos^2\theta_{13} \right)$ \\ [10pt]
& & $- \sin^2\theta_{12} \cos^2\theta_{23} \cos^2 \theta^\circ_{12}
+ \cos^2\theta_{12} \left( \cos^2\theta_{13} \sin^2\theta^\circ_{12} \cos^2 \theta^\circ_{23} - \sin^2\theta_{13} \sin^2\theta_{23} \cos^2 \theta^\circ_{12} \right)$ \\ [10pt]
& & $+ \kappa \cos\hat\delta \cos\theta_{12} \cos\theta_{13} \sin\theta^\circ_{12} \sin2\theta^\circ_{23}
\left( \cos^2\theta^\circ_{12} - \cos^2\theta_{12} \cos^2\theta_{13} \right)^\frac{1}{2} \bigg]$
\\ [10pt]
\hline
\end{tabular}
\caption{Summary of the sum rules for $\cos\delta$ in the case of fully
broken $G_{\nu}$ under the assumption that the matrix $U_{\nu}$ consists of two complex rotation
matrices. 
The parameter $\kappa = 1$ if the corresponding hat angle 
belongs to the first or third quadrant,
and $\kappa = -1$ otherwise.
See text for further details.
}
\label{tab:summarysumrulesGnubroken}
\end{table}
%%%%%%%%%%%%%%%%%%%%%%%%%%
\end{landscape}

%%%%%%%%%%%%%%%%%%%%%%
%
\section{Summary of the Predictions for $G_f = A_4~(T^{\prime}),~S_4$ and $A_5$}
\label{sec:predictions}
%
%%%%%%%%%%%%%%%%%%%%%%%%%%%%%

In this section we summarise the numerical results obtained in the cases
of the discrete flavour symmetry groups $A_4~(T^{\prime})$, $S_4$ and $A_5$, which
have been already discussed in subsections~\ref{sec:ressec2}, 
\ref{sec:ressec3} and \ref{sec:ressec4}.
In Tables~\ref{tab:A4s}\,--\,\ref{tab:A5s} we give the values 
of the fixed angles,
obtained from the diagonalisation of the corresponding group elements which lead
to physical values of $\cos\delta$ and phenomenologically viable results 
for the ``standard'' mixing angles 
$\theta_{12}$, $\theta_{13}$ and $\theta_{23}$. 
In the cases when the standard mixing angles are not fixed by the schemes in 
Tables~\ref{tab:A4s}\,--\,\ref{tab:A5s}, we use their best fit 
values for the NO spectrum 
quoted in eqs.~(\ref{th12values})~--~(\ref{th13values}). 
For the cases in the tables marked with an 
asterisk, physical values of $\cos \delta$, i.e., $|\cos \delta| \leq 1$, 
cannot be obtained employing the
best fit values of the neutrino mixing angles 
$\theta_{12}$, $\theta_{13}$ and $\theta_{23}$, 
but they can be achieved for values
of the relevant mixing parameters allowed at $3\sigma$.  
Note that unphysical values of $\cos \delta$, $|\cos \delta| > 1$,
occur when the relations between the parameters of
the scheme and the standard parametrisation mixing angles
cannot be fulfilled for given values of $\sin^2 \theta_{12}$,
$\sin^2 \theta_{13}$ and $\sin^2 \theta_{23}$. 
Indeed the parameter space of 
$\sin^2 \theta_{12}$, $\sin^2 \theta_{13}$ and $\sin^2 \theta_{23}$
is reduced by these constraints coming from the schemes.

%%%%%%%%%%%%%%%%%%%%%%%%%%
\begin{table}[h!]
\centering
\renewcommand*{\arraystretch}{1.2}
\hspace{-0.2cm}
\begin{tabular}{ccc}
\hline
& &  \\ [-10pt] 
$(G_e,G_{\nu}) = (Z_3,Z_2)$ & $\cos \delta$ & $\sin^2 \theta_{12}$ \\ [6pt] 
\bottomrule
 & &  \\ [-10pt] 
B1 $(\sin^2 \theta^{\circ}_{12},\sin^2 \theta^{\circ}_{23}) = (1/3,1/2)$ & $0.570$ & $0.341$  \\ [6pt] 
\hline 
 \end{tabular}
\caption{The phenomenologically viable case for the symmetry group $A_4$.
The values of $\cos \delta$ and $\sin^2 \theta_{12}$ predicted by the
scheme B1, which refers to the corresponding parametrisation in 
Tables~\ref{tab:summarysumrules} and \ref{tab:summarysin2th23_12}, 
have been obtained using the best fit values of 
 $\sin^2 \theta_{13}$ and $\sin^2 \theta_{23}$ 
for the NO spectrum quoted in eqs.~(\ref{th12values})~--~(\ref{th13values}).
}
\label{tab:A4s}
\end{table}
%%%%%%%%%%%%%%%%%%%%%%%%%%
%

For the symmetry group $A_4$ we find that the residual symmetries
\begin{itemize}
\item $(G_e,G_{\nu}) = (Z_2,Z_2)$ in the cases C1~--~C9;
\item $(G_e,G_{\nu}) = (Z_3,Z_2)$ in the cases B2 and B3; 
\item $(G_e,G_{\nu}) = (Z_2 \times Z_2,Z_2)$ in the cases B1, B2 and B3;
\item $(G_e,G_{\nu}) = (Z_2,Z_3)$ or $(Z_2,Z_2 \times Z_2)$ 
in the cases A1, A2 and A3 
\end{itemize}
do not provide phenomenologically viable results for
$\cos \delta$ and/or the standard mixing angles.
It is worth noticing that the predicted value of $\sin^2 \theta_{12} = 0.341$
in Table~\ref{tab:A4s} is within the $2\sigma$ allowed range. 
Varying  $\sin^2 \theta_{13}$, 
which enters into the expression for  $\sin^2 \theta_{12}$, 
within its respective $3\sigma$ allowed range for the NO neutrino mass
spectrum, we find $0.339 \leq \sin^2 \theta_{12} \leq 0.343$.
%%%%%%%%%%%%%%%%%%%%%%%%%%
\begin{table}[h]
\centering
\renewcommand*{\arraystretch}{1.2}
\begin{tabular}{ccc}
\hline 
 & &  \\ [-10pt] 
 $(G_e,G_{\nu}) = (Z_2,Z_2)$ & $\cos \delta$ & $\sin^2 \theta_{ij}$\\ [6pt] 
\bottomrule
 & &  \\ [-10pt] 
C1 $\sin^2 \theta^{\circ}_{23} = 1/4$ & $-0.806$ & not fixed\\   
C2 $\sin^2 \theta^{\circ}_{23} = 1/2$ & not fixed &$\sin^2 \theta_{23} = 0.512$ \\    
C3 $\sin^2 \theta^{\circ}_{13} = 1/4$ & 
$-1^*$ & not fixed\\   
C4 $\sin^2 \theta^{\circ}_{12} = 1/4$ & $0.992$ & not fixed \\
C5 $\sin^2 \theta^{\circ}_{12} = 1/4$ & not fixed & $\sin^2 \theta_{12} = 0.256$ \\     
C7 $\sin^2 \theta^{\circ}_{23} = 1/2$ & not fixed & $\sin^2 \theta_{23} = 0.488$ \\
C8 $\sin^2 \theta^{\circ}_{23} = \{1/2,3/4\}$ & $\{-1^*,1^*\}$ & not fixed \\ [6pt]
\hline 
& &  \\ [-10pt] 
$(G_e,G_{\nu}) = (Z_3,Z_2)$ & $\cos \delta$ & $\sin^2 \theta_{12}$ \\ [6pt] 
\bottomrule
 & &  \\ [-10pt] 
\begin{tabular}{c} 
 B1 $(\sin^2 \theta^{\circ}_{12},\sin^2 \theta^{\circ}_{23}) = (1/3,1/2)$ \\
 \end{tabular} 
 & $0.570$ & $0.341$ \\   
\begin{tabular}{c} 
 B2 $(\sin^2 \theta^{\circ}_{12},\sin^2 \theta^{\circ}_{13}) = (1/6,1/5)$ \\ 
 \end{tabular} 
 & $-0.269$ & $0.317$ \\ [6pt]  
\hline 
%%%%%%%%%%%%%%%%%%%%%%%%%%%%%%%%%%%
& &  \\ [-10pt] 
$(G_e,G_{\nu}) = (Z_4,Z_2), ~(Z_2 \times Z_2,Z_2)$ & $\cos \delta$ & $\sin^2 \theta_{12}$ \\ [6pt] 
\bottomrule
 & &  \\ [-10pt] 
\begin{tabular}{c} 
 B1 $(\sin^2 \theta^{\circ}_{12},\sin^2 \theta^{\circ}_{23}) = (1/4,1/3)$ \\ 
 \end{tabular} 
 & \begin{tabular}{c} 
 $-1^*$ \\ 
\end{tabular} 
& \begin{tabular}{c} 
 $0.256$ \\ 
 \end{tabular} 
\\ [6pt]  
\hline 
%%%%%%%%%%%%%%%%%%%%%%%%%%%%%%%%%%%
& &  \\ [-10pt] 
$(G_e,G_{\nu}) = (Z_2,Z_4),~(Z_2,Z_2 \times Z_2)$ & $\cos \delta$ & $\sin^2 \theta_{23}$ \\ [6pt] 
\bottomrule
 & &  \\ [-10pt] 
\begin{tabular}{c} A1 $(\sin^2 \theta^{\circ}_{13},\sin^2 \theta^{\circ}_{23}) = (1/3,1/4)$ \\ 
 \end{tabular} 
 & \begin{tabular}{c}
 $-1^*$ \\ 
 \end{tabular} 
 & \begin{tabular}{c} $0.488$ \\ 
 \end{tabular} 
\\   
\begin{tabular}{c} A2 $(\sin^2 \theta^{\circ}_{12},\sin^2 \theta^{\circ}_{23}) = (1/2,1/2)$\\ 
 \end{tabular} 
 & \begin{tabular}{c} 
 $\phantom{-}1^*$ \\ 
 \end{tabular} 
 & \begin{tabular}{c} $0.512$ \\ 
 \end{tabular} 
\\ [6pt]  
\hline 
%%%%%%%%%%%%%%%%%%%%%%%%%%%%%%%%%%%
 \end{tabular}
\caption{The phenomenologically viable cases for the symmetry group $S_4$.
The values of $\cos \delta$ and $\sin^2 \theta_{12}$ or $\sin^2 \theta_{23}$ predicted by the schemes A1, A2, etc., which refer to the 
corresponding parametrisations in 
Tables~\ref{tab:summarysumrules}~--~\ref{tab:summarysin2th23_12B}, 
have been obtained using the best fit values for the NO spectrum of the 
other two (not fixed) neutrino mixing parameters 
($\sin^2 \theta_{13}$ and $\sin^2 \theta_{23}$, or 
$\sin^2 \theta_{12}$ and $\sin^2 \theta_{13}$)
quoted in eqs.~(\ref{th12values})~--~(\ref{th13values}). 
In the cases marked with an asterisk, physical values of $\cos \delta$  
cannot be obtained employing the best fit values of the mixing angles, 
but are possible for values of the relevant neutrino 
mixing parameters lying in their respective $3\sigma$ allowed 
intervals.
See text for further details.
}
\label{tab:S4s}
\end{table}
%%%%%%%%%%%%%%%%%%%%%%%%%%

For the symmetry group $S_4$ we find that the residual symmetries
\begin{itemize} 
\item $(G_e,G_{\nu}) = (Z_2,Z_2)$ in the cases C6 and C9; 
\item $(G_e,G_{\nu}) = (Z_3,Z_2)$ in the case B3; 
\item $(G_e,G_{\nu}) = (Z_4,Z_2)$ or $(Z_2 \times Z_2,Z_2)$ in the cases B2 and B3;
\item $(G_e,G_{\nu}) = (Z_2,Z_3)$ in the cases A1, A2 and A3; 
\item $(G_e,G_{\nu}) = (Z_2,Z_4)$ or $(Z_2,Z_2 \times Z_2)$ in the case A3 
\end{itemize}
do not provide phenomenologically 
viable results for $\cos \delta$ and/or for the standard mixing angles.

The cases in Table~\ref{tab:S4s} marked with an asterisk are discussed below.
Firstly, using the best fit values of $\sin^2 \theta_{12}$ 
and $\sin^2 \theta_{13}$
we get a physical value of $\cos \delta$ in the case C3 for the minimal
value of $\sin^2 \theta_{23} = 0.562$, for which $\cos \delta = -0.996$.
For C8 with 
$\sin^2 \theta^\circ_{23} = 1/2$ and $3/4$, using the best fit
values of the neutrino mixing angles for the NO spectrum, 
we have  $\cos\delta = -1.53$ and $2.04$, respectively. 
The physical values of $\cos \delta$
can be obtained, using, e.g., the values of 
$\sin^2 \theta_{23} = 0.380$ and $0.543$, for which
$\cos \delta = -0.995$ and $0.997$, respectively. 
In the parts of the $3\sigma$
allowed range of $\sin^2 \theta_{23}$,
$0.374 \leq \sin^2 \theta_{23} \leq 0.380$ and
$0.543 \leq \sin^2 \theta_{23} \leq 0.641$,
we have $-0.938 \geq \cos\delta \geq -0.995$
and $0.997 \geq \cos\delta \geq 0.045$, respectively.
Secondly, in the case B1 we obtain $\cos\delta = -0.990$ 
employing the best fit value
of $\sin^2 \theta_{13}$ and the maximal value of $\sin^2 \theta_{23} = 0.419$.
%
%
%
%%%%%%%%%%%%%%%%%%%%%%%%%%
\begin{table}[h!]
\centering
\renewcommand*{\arraystretch}{1.2}
\begin{tabular}{ccc}
\hline 
 & &  \\ [-10pt] 
 $(G_e,G_{\nu}) = (Z_2,Z_2)$ & $\cos \delta$ & $\sin^2 \theta_{ij}$\\ [6pt]
 \bottomrule 
 & &  \\ [-10pt] 
C1 $\sin^2 \theta^{\circ}_{23} = 1/4$ & $-0.806$ & not fixed\\      
C3 $\sin^2 \theta^{\circ}_{13} = 0.0955$, $1/4$ & $0.688$, $-1^*$ & not fixed\\   
C4 $\sin^2 \theta^{\circ}_{12} = 0.0955$, $1/4$ & $-1^*$, 
$0.992$ & not fixed \\
C5 $\sin^2 \theta^{\circ}_{12} = 1/4$ & not fixed &$\sin^2 \theta_{12} = 0.256$ \\    
C8 $\sin^2 \theta^{\circ}_{23} = 3/4$ & $1^*$ & not fixed \\
C9 $\sin^2 \theta^{\circ}_{12} = 0.3455$ & not fixed & $\sin^2 \theta_{12} = 0.330$ \\  [6pt]
\hline 
& &  \\ [-10pt] 
$(G_e,G_{\nu}) = (Z_3,Z_2)$ & $\cos \delta$ & $\sin^2 \theta_{12}$ \\ [6pt]
\bottomrule 
 & &  \\ [-10pt] 
\begin{tabular}{c} 
 B1 $(\sin^2 \theta^{\circ}_{12},\sin^2 \theta^{\circ}_{23}) = (1/3,1/2)$ \\ 
 \end{tabular} 
 & $0.570$ & $0.341$ 
\\ [6pt]  
\hline 
& &  \\ [-10pt] 
$(G_e,G_{\nu}) = (Z_5,Z_2)$ & $\cos \delta$ & $\sin^2 \theta_{12}$ \\ [6pt]
\bottomrule 
 & &  \\ [-10pt] 
\begin{tabular}{c} 
 B1 $(\sin^2 \theta^{\circ}_{12},\sin^2 \theta^{\circ}_{23}) = (0.2764,1/2)$ \\ 
 \end{tabular} 
 & \begin{tabular}{c} 
 $0.655$ \\ 
 \end{tabular} 
 & \begin{tabular}{c} 
 $0.283$ \\ 

 \end{tabular} 
 \\   
\begin{tabular}{c} 
 B2 $(\sin^2 \theta^{\circ}_{12},\sin^2 \theta^{\circ}_{13}) = (0.1382,0.1604)$ \\ 
 \end{tabular} 
 & \begin{tabular}{c} 
 $-0.229$ \\ 
 \end{tabular} 
 & \begin{tabular}{c} 
 $0.259$ \\ 
 \end{tabular} 
\\ [6pt]  
\hline 
 & &  \\ [-10pt]  
$(G_e,G_{\nu}) = (Z_2 \times Z_2,Z_2)$ & $\cos \delta$ & $\sin^2 \theta_{12}$ \\ [6pt]
\bottomrule 
 & &  \\ [-10pt]   
\begin{tabular}{c} 
 B2 $(\sin^2 \theta^{\circ}_{12},\sin^2 \theta^{\circ}_{13}) = (0.0955,0.2764)$ \\
\phantom{\hspace{3.5cm}} $(1/4,0.1273)$ \\ 
 \end{tabular} 
 & \begin{tabular}{c} 
  $-1^*$ \\ 
 $0.805$ \\ 
 \end{tabular} 
 & \begin{tabular}{c} 
 $0.330$\\
 $0.330$ \\ 
 \end{tabular} 
\\ [6pt]  
\hline 
   & &  \\ [-10pt] 
$(G_e,G_{\nu}) = (Z_2,Z_3)$ & $\cos \delta$ & $\sin^2 \theta_{23}$ \\ [6pt]
\bottomrule 
 & &  \\ [-10pt] 
\begin{tabular}{c} A1 $(\sin^2 \theta^{\circ}_{13},\sin^2 \theta^{\circ}_{23}) = (0.2259,0.4363)$\\ 
 \end{tabular} 
 & \begin{tabular}{c} $0.716$ \\ 
 \end{tabular} 
 & \begin{tabular}{c} $0.553$ \\ 
 \end{tabular} 
\\   
\begin{tabular}{c} A2 $(\sin^2 \theta^{\circ}_{12},\sin^2 \theta^{\circ}_{23}) = (0.2259,0.4363)$ \\ 
 \end{tabular} 
 & \begin{tabular}{c} $-0.716$ \\ 
 \end{tabular} 
 & \begin{tabular}{c} $0.447$ \\ 
 \end{tabular} 
\\ [6pt]  
   \hline 
& &  \\ [-10pt] 
$(G_e,G_{\nu}) = (Z_2,Z_5)$ & $\cos \delta$ & $\sin^2 \theta_{23}$ \\ [6pt] 
\bottomrule
 & &  \\ [-10pt] 
\begin{tabular}{c} A1 $(\sin^2 \theta^{\circ}_{13},\sin^2 \theta^{\circ}_{23}) = (0.4331,0.3618)$ \\ 
 \end{tabular} 
 & \begin{tabular}{c} $-1^*$ 
 \\ 
 \end{tabular} 
 & \begin{tabular}{c} $0.630$ \\ 
 \end{tabular} 
\\   
\begin{tabular}{c} A2 $(\sin^2 \theta^{\circ}_{12},\sin^2 \theta^{\circ}_{23}) = (0.4331,0.3618)$ \\ 
 \end{tabular} 
 & \begin{tabular}{c} $\phantom{-}1^*$ 
 \\ 
 \end{tabular} 
 & \begin{tabular}{c} $0.370$ \\ 
 \end{tabular} 
\\ [6pt]  
\hline 
 \end{tabular}
\caption{The phenomenologically viable cases for the symmetry group $A_5$.
The values of $\cos \delta$ and $\sin^2 \theta_{12}$ or $\sin^2 \theta_{23}$ predicted by the
schemes A1, A2, etc., which refer to the corresponding parametrisations in 
Tables~\ref{tab:summarysumrules}~--~\ref{tab:summarysin2th23_12B}, 
have been obtained using the best fit values of the other standard mixing angles
for the NO spectrum quoted in eqs.~(\ref{th12values})~--~(\ref{th13values}).
In the cases marked with an asterisk, the predicted values of 
$\cos \delta$, obtained for the best fit values of the neutrino 
mixing angles $\theta_{12}$,  $\theta_{13}$ and  $\theta_{23}$,
are unphysical; physical values of $\cos \delta$ can be 
obtained for values of the neutrino mixing parameters 
$\sin^2 \theta_{12}$, 
$\sin^2 \theta_{13}$ and $\sin^2 \theta_{23}$ 
lying in their respective $3\sigma$ allowed intervals.
See text for further details.
}
\label{tab:A5s}
\end{table}
%%%%%%%%%%%%%%%%%%%%%%%%%%
%
Finally, utilising the best fit value of $\sin^2 \theta_{13}$, we 
get physical values
of $\cos \delta$ in the cases A1 and A2 for the minimal value of
$\sin^2 \theta_{12} = 0.348$, for which $\cos \delta = -0.993$ and $0.993$,
respectively. Note that for the cases in which $\sin^2 \theta_{23}$ is fixed,
the predicted values are within the corresponding $2\sigma$ range,
while in the cases in which $\sin^2 \theta_{12}$ is fixed,
the values of $\sin^2 \theta_{12} = 0.341$ and $0.317$
are within $2\sigma$ and $1\sigma$, respectively.
The value of $\sin^2 \theta_{12} = 0.256$ lies slightly outside the current
$3\sigma$ allowed range.

For the symmetry group $A_5$ we find that the residual symmetries 
\begin{itemize}
\item $(G_e,G_{\nu}) = (Z_2,Z_2)$ in the cases C2, C6 and C7; 
\item $(G_e,G_{\nu}) = (Z_3,Z_2)$ in the cases B2 and B3;
\item $(G_e,G_{\nu}) = (Z_5,Z_2)$ in the case B3; 
\item $(G_e,G_{\nu}) = (Z_2 \times Z_2,Z_2)$ in the cases B1 and B3; 
\item $(G_e,G_{\nu}) = (Z_2,Z_3)$ or $(Z_2,Z_5)$ in the case A3; 
\item $(G_e,G_{\nu}) = (Z_2,Z_2 \times Z_2)$ in the cases A1, A2 and A3 
\end{itemize}
do not provide phenomenologically 
viable results for $\cos \delta$ and/or for 
the standard mixing angles $\theta_{12}$, 
$\theta_{13}$ and $\theta_{23}$. 

We will describe next the cases in Table~\ref{tab:A5s} 
marked with an asterisk,
apart from those which have also been found for $G_f = S_4$ 
and discussed earlier. 
Using the best fit values of $\sin^2 \theta_{12}$ and $\sin^2 \theta_{13}$
we get a physical value of $\cos \delta$ in the case C4
for the minimal value of $\sin^2 \theta_{23} = 0.487$, 
for which $\cos \delta = -0.997$.
Instead using the best fit values of $\sin^2 \theta_{13}$ and $\sin^2 \theta_{23}$
one gets the physical values of $\cos \delta = -1$ for the maximal
value of $\sin^2 \theta_{12} = 0.277$.
Employing the best fit value of $\sin^2 \theta_{13}$ we find a physical value
of $\cos \delta$ in the case B2 with residual symmetries
$(G_e,G_{\nu}) = (Z_2 \times Z_2,Z_2)$ for the minimal value of
$\sin^2 \theta_{23} = 0.518$, for which $\cos \delta = -0.996$.
Similarly for the cases A1 and A2 with residual symmetries
$(G_e,G_{\nu}) = (Z_2,Z_5)$, the values of $\cos \delta = -0.992$ and $0.992$
are obtained using the minimal value of $\sin^2 \theta_{12} = 0.321$.

 The values of $\sin^2 \theta^{\circ}_{ij}$ in Table~\ref{tab:A5s}
used to compute $\cos \delta$ and $\sin^2 \theta_{ij}$ are the following ones:
$1/(4 r^2) \cong 0.0955$, $(3 - r)/4 \cong 0.3455$, 
$1/(2 + r) \cong 0.2764$, $1/(4 + 2 r) \cong 0.1382$,
$1/(3 + 2 r) \cong 0.1604$, $1/(3 + 3 r) \cong 0.1273$, $2/(4 r^2 - r) \cong 0.2259$,
$r/(6r - 6) \cong 0.4363$, $(6 r - 4)/(10 r - 3) \cong 0.4331$, 
$(1 - r)/(8 - 6 r) \cong 0.3618$.

\cleardoublepage

%%%%%%%%%%%%%%%%%%%%%%%%%%%%%%%%%
\section{Conclusions}
\label{sec:summary}
%%%%%%%%%%%%%%%%%%%%%%%%%%%%%%%%%

 In the present article we have employed the 
discrete symmetry approach to understanding 
the observed pattern of 3-neutrino mixing 
and, within this approach, have derived 
sum rules and predictions for the Dirac phase 
$\delta$ present in the PMNS neutrino mixing matrix $U$. 
The approach is based on the assumption 
of the existence at some energy scale of a (lepton) flavour  
symmetry corresponding to a non-Abelian discrete group $G_f$. 
The flavour symmetry group $G_f$ can be broken, in general, 
to different ``residual symmetry'' subgroups $G_e$ and $G_{\nu}$ 
of the charged lepton and neutrino mass terms,
respectively. Given $G_f$, typically there are more than one 
(but still a finite number of) possible 
residual symmetries $G_e$ and $G_{\nu}$.
The residual symmetries can constrain the forms of the 
$3\times 3$ unitary matrices $U_{e}$ and $U_{\nu}$, 
which diagonalise the charged lepton and 
neutrino mass matrices, 
and the product of which represents the 
PMNS neutrino mixing matrix $U$,
$U = U_e^{\dagger}\, U_{\nu}$.
Thus, by constraining the form of the matrices 
$U_{e}$ and $U_{\nu}$, the residual symmetries 
constrain also the form of the PMNS matrix $U$. 
This can lead, in particular, 
to a correlation between 
the values of the PMNS neutrino mixing angles $\theta_{12}$, 
$\theta_{13}$ and $\theta_{23}$, which have been 
determined experimentally with a rather good precision, 
and the value of the cosine of the Dirac CP violation 
phase $\delta$ present in $U$, 
i.e., to a ``sum rule'' for $\cos\delta$.  
The sum rule for $\cos\delta$ thus obtained depends 
on residual symmetries $G_e$ and $G_{\nu}$ 
and in some cases can involve, in addition to 
$\theta_{12}$, $\theta_{13}$ and $\theta_{23}$, parameters 
which cannot be constrained even when $G_f$ is fixed.
For a given fixed $G_f$, 
unambiguous predictions for the value of $\cos\delta$ can 
be derived in the cases when, apart from the parameters 
determined by $G_f$ (and  $G_e$ and $G_{\nu}$),
only $\theta_{12}$, $\theta_{13}$ and $\theta_{23}$ enter into the 
expression for the respective sum rule.

 In the present article we have 
derived sum rules for  $\cos\delta$ considering the following
discrete residual symmetries:
i) $G_e = Z_2$ and $G_{\nu} = Z_n$, $n > 2$ or $Z_n \times Z_m$, $n,m \geq 2$ 
(Section~\ref{sec:GeZGnuZZ});
ii)~$G_e = Z_n$, $n > 2$ or $Z_n \times Z_m$, $n,m \geq 2$ and 
$G_{\nu} = Z_2$ (Section~\ref{sec:GeZorZZGnuZ});
iii) $G_e = Z_2$ and $G_{\nu} = Z_2$ (Section~\ref{sec:GeZ2GnuZ2});
iv) $G_e$ is fully broken and $G_{\nu} = Z_n$, $n > 2$ or 
$Z_n \times Z_m$, $n,m \geq 2$ (Section~\ref{sec:Gebroken}); and 
v) $G_e = Z_n$, $n > 2$ or $Z_n \times Z_m$, 
$n,m \geq 2$ and $G_{\nu}$ is fully broken (Section~\ref{sec:Gnubroken}). 
The sum rules are summarised in Tables~\ref{tab:summarysumrules}, \ref{tab:summarysumrulesB},
\ref{tab:summarysumrulesGebroken} and \ref{tab:summarysumrulesGnubroken}.
For given $G_e$ and $G_\nu$, the sum rules for $\cos\delta$ 
we have derived are exact, within the approach employed, and 
are valid, in particular, for any $G_f$ containing $G_e$ and $G_\nu$ as subgroups. 
We have identified the cases when the value of $\cos\delta$ 
cannot be determined, or cannot be uniquely determined,  
from the sum rule without making additional assumptions 
on unconstrained parameters (cases A3 in Section~\ref{sec:GeZGnuZZ} and B3 in Section~\ref{sec:GeZorZZGnuZ} 
(see also Table~\ref{tab:summarysumrules}); cases C2, C5, C6, C7 and C9 in Section~\ref{sec:GeZ2GnuZ2}
(see also Table~\ref{tab:summarysumrulesB});
the cases discussed in Sections~\ref{sec:Gebroken} and \ref{sec:Gnubroken}). 
In the majority  of the phenomenologically viable 
cases we have considered
the value of $\cos\delta$ can be unambiguously predicted 
once the flavour symmetry $G_f$ is fixed. 
In certain cases of fixed $G_f$, $G_e$ and $G_{\nu}$, 
correlations between the values of some of the measured 
neutrino mixing parameters 
$\sin^2\theta_{12}$, $\sin^2\theta_{13}$ and $\sin^2\theta_{23}$, 
are predicted, and/or the values of some of these parameters,
typically of $\sin^2\theta_{12}$ or $\sin^2\theta_{23}$, 
are fixed. These correlations and ``predictions'' are summarised 
in Tables~\ref{tab:summarysin2th23_12}
and \ref{tab:summarysin2th23_12B}. We have found that a relatively large 
number of these cases are not phenomenologically viable, i.e., 
they lead to 
results which are not compatible with the 
existing data on neutrino mixing. 
We have derived predictions for $\cos\delta$   
for the flavour symmetry groups $G_f = S_4$, $A_4$, $T^\prime$ and $A_5$ 
using the best fit values of $\sin^2\theta_{12}$, 
$\sin^2\theta_{13}$ and $\sin^2\theta_{23}$,   
when  $\cos\delta$ is unambiguously determined by the 
corresponding sum rule. We have presented the predictions 
for $\cos\delta$ 
only in the phenomenologically viable cases, 
i.e., when the measured values of the 3-neutrino mixing parameters 
$\sin^2\theta_{12}$, $\sin^2\theta_{13}$ and $\sin^2\theta_{23}$, 
taking into account their respective $3\sigma$ uncertainties, are 
successfully reproduced.
These predictions, together with the 
predictions for the value of one of the mixing parameters 
$\sin^2\theta_{12}$ and $\sin^2\theta_{23}$, 
in the cases when it is fixed by the symmetries,  
are summarised in Tables~\ref{tab:A4s}~--~\ref{tab:A5s}.   

  The results derived in the present study show, 
in particular, that  
with the accumulation of more precise data on the 
PMNS neutrino mixing parameters $\sin^2\theta_{12}$, 
$\sin^2\theta_{13}$ and $\sin^2\theta_{23}$, 
and with the measurement of the Dirac phase $\delta$ 
present in the neutrino mixing matrix $U$,
it will be possible to critically test 
the predictions of the current
phenomenologically viable theories,
models and schemes of neutrino mixing based on different 
non-Abelian discrete (lepton) flavour symmetries $G_f$   
and sets of their non-trivial subgroups of 
residual symmetries $G_e$ and $G_\nu$,  
operative respectively in the charged lepton 
and neutrino sectors,                
and thus critically test the discrete symmetry approach 
to understanding the 
observed pattern of neutrino mixing.

%%%%%%%%%%%%%%%%%%%%%%%%%%%%%%%%
 \section*{Acknowledgements}
A.S.~thanks L.~Everett for useful discussions.
This work was supported in part by the European Union FP7
ITN INVISIBLES (Marie Curie Actions, PITN-GA-2011-289442-INVISI\-BLES),
by the INFN program on Theoretical Astroparticle Physics (TASP),
by the research grant  2012CPPYP7 ({\sl  Theoretical Astroparticle Physics})
under the program  PRIN 2012 funded by the Italian Ministry 
of Education, University and Research (MIUR), 
by the World Premier International Research Center
Initiative (WPI Initiative, MEXT), Japan (S.T.P.), 
and by a grant from the Simons Foundation (S.T.P.).
Part of this work was done at the Aspen Center for Physics, 
which is supported by National Science Foundation grant PHY-1066293.

\appendix

%%%%%%%%%%%%%%%%%%%%%%%%%%%%%%%%%
\section{Appendix: The Discrete Groups $A_4,~T^{\prime},~S_4$ and $A_5$}
\label{app:A4}
%%%%%%%%%%%%%%%%%%%%%%%%%%%%%%%%%

$A_4$ is the symmetry group of even permutations of four objects (see, e.g., \cite{Ishimori:2010au}).  It is isomorphic to the tetrahedral symmetry group, i.e., the group of rotational symmetries of a regular tetrahedron.  As such it can be
defined in terms of two generators $S$ and $T$, satisfying 
$S^2 = T^3 = (ST)^3 = 1$.  In this work, we choose to work in the Altarelli-Feruglio basis\cite{Altarelli:2005yx} for the 3-dimensional representation of the $S$ and $T$ generators,  see Table~\ref{tab:groupsconv}.

%%%%%%%%%%%%%%%%%%%%%%%%%%
\begin{table}[h!]
\centering
\renewcommand{\tabcolsep}{4.4pt}
\begin{tabular}{llll}
\hline
&  \\ [-6pt]
\multicolumn{1}{l}{Group} & \multicolumn{3}{l}{3-dimensional representation of the generators} \\ [4pt]
\bottomrule
&  \\ [-6pt]
$A_4$ & 
$S = \dfrac{1}{3} \begin{pmatrix}
-1  & 2  & 2 \\
2  & -1 &  2 \\
2 & 2 & -1 \\
\end{pmatrix}$
&
$T = 
\begin{pmatrix}
1 & 0 & 0 \\
0 & \omega^2 & 0 \\
0 & 0 & \omega \\
\end{pmatrix}
$
 \\
 &  \\ [-6pt]
 $T^{\prime}$ & 
$S  = \dfrac{1}{3}
\begin{pmatrix}
 - 1& 2 \omega & 2 \omega^2 \\
 2 \omega^2 & -1 & 2 \omega \\
 2 \omega & 2 \omega^2 & -1 \\
\end{pmatrix}$
&
$T =
 \begin{pmatrix}
 1 & 0 & 0 \\
 0 & \omega  & 0 \\
 0 & 0 & \omega ^2 \\
\end{pmatrix}$
&
\\
&  \\ [-6pt]
$S_4$ & 
$S  = 
\begin{pmatrix}
-1  & 0  & 0 \\
0  & 1 &  0 \\
0 & 0 & -1 \\
\end{pmatrix}$
& 
$T =
\dfrac{1}{2} \begin{pmatrix}
i & -\sqrt{2} i & -i \\
\sqrt{2} & 0 & \sqrt{2} \\
i & \sqrt{2} i & -i \\
\end{pmatrix}$
&
$U  = 
\begin{pmatrix}
0  & 0  & i \\
0  & -1 & 0 \\
-i & 0 & 0 \\
\end{pmatrix}$
\\
&  \\ [-6pt]
$A_5$ & 
$S  = 
\dfrac{1}{\sqrt{5}}
\begin{pmatrix}
 1& -\sqrt{2} & -\sqrt{2} \\
 -\sqrt{2} & -r & 1/r \\ 
 -\sqrt{2} & 1/r & -r \\
\end{pmatrix}$
& 
$T =
 \begin{pmatrix}
 1 & 0 & 0 \\
 0 & \rho  & 0 \\
 0 & 0 & \rho^4 \\
\end{pmatrix}$\\
&  \\ [-6pt]
\hline
  \end{tabular}
%%%%%%%%%%%%%%%%%%%%%%%%%%%
\caption{3-dimensional representation of the generators of
$A_4$, $T^{\prime}$, $S_4$ and $A_5$.
We have defined $\omega = e^{ 2 \pi i/3}$, $r = (1 + \sqrt{5})/2$
and $\rho = e^{2 \pi i/5}$.
}
\label{tab:groupsconv}
\end{table}

The group $T^{\prime}$ is the double covering group of $A_4$ (see, e.g., \cite{Ishimori:2010au}),
which can be defined in terms of two generators $S$ and $T$
through the algebraic relations: $R^2 = T^3 = (ST)^3 = 1$, $RT= TR$, where $R = S^2$.
We use the basis for the 3-dimensional representation of the generators $S$ and $T$
from \cite{Feruglio:2007uu}, summarised in Table~\ref{tab:groupsconv}.
Since we restrict ourselves to the triplet representation for the LH
charged lepton and neutrino fields, there is no way to distinguish $T^{\prime}$
from $A_4$ \cite{Feruglio:2007uu}.\footnote{It is worth noting that
$A_4$ is not a subgroup of $T^{\prime}$.}
Note that matrices representing $S$ and $T$ in Table~\ref{tab:groupsconv} for $A_4$, are related
with those for $T^{\prime}$ by the following redefinition
$S \rightarrow T ST^2$, $T \rightarrow T^2$, where $S$ and $T$
before (after) the arrows
are the matrices presented in Table~\ref{tab:groupsconv}
for $A_4$ ($T^{\prime}$).

$S_4$ is the group of permutations of four objects, i.e., the rotational symmetry group of a cube (see, e.g., \cite{Ishimori:2010au}).  It can be defined in terms of three generators 
$S$, $T$ and $U$, satisfying \cite{Li:2014eia}: $S^2 = T^3 = U^2 = (ST)^3 = (SU)^2 = (TU)^2 = (STU)^4 = 1$.
 We employ for the 3-dimensional representation of the $S$, $T$ and $U$ generators the
basis given in \cite{Li:2014eia} and summarised in Table~\ref{tab:groupsconv}. As it was also shown in 
\cite{Li:2014eia}, this basis is equivalent to the basis widely used in the literature \cite{Altarelli:2009gn}.

$A_5$ is the group of even permutations of five objects (see, e.g., \cite{Ishimori:2010au}), 
i.e., the rotational symmetry group of an icosahedron,
which
can be defined in terms of two generators $S$ and $T$,
satisfying $S^2 = T^5 = (ST)^3 = 1$.
We employ the basis defined in  \cite{Ding:2011cm}, which for the 3-dimensional
representation of the generators $S$ and $T$ is summarised in Table~\ref{tab:groupsconv}.  

We conclude this appendix by noting that a list of the Abelian subgroups of $A_4$, $T^{\prime}$, $S_4$ and $A_5$
can be found in \cite{Ding:2013bpa}, \cite{Girardi:2013sza}, \cite{Li:2014eia} and \cite{Ding:2011cm}, respectively.

%%%%%%%%%%%%%%%%%%%%%%%%%%%%%%%%%
\section{Appendix: Parametrisations of a $3 \times 3$ Unitary Matrix}
\label{app:ParU}
%%%%%%%%%%%%%%%%%%%%%%%%%%%%%%%%%

Parametrisations of a $3 \times 3$ unitary matrix $W$ 
(see, e.g., \cite{Fritzsch:1997st,Rasin:1997pn,Rodejohann:2011uz}) can be obtained,
e.g., from one of the six permutations of a product of three complex rotations
and diagonal phase matrices, e.g., as follows:
\be
W = \Psi_1 \Psi_2 \Psi_3 \overline W = \Psi_1 \Psi_2 \Psi_3 \, U_{ij} \, U_{kl} \, U_{rs} \,,
\label{eq:parUPMNS}
\ee
where we have assumed $ij \neq kl \neq rs$. It is worth noticing that sometimes
it is convenient to use the parametrisations of $\overline{W}$ of the following form:
\be
\overline{W} = U_{ij} \, U_{kl} \, \tilde U_{ij}\,.
\ee
As shown in \cite{Fritzsch:1997st}, the number of distinctive parametrisations of a CKM-like 
matrix is nine. 
We have defined the phase matrices $\Psi_i$ in eq.~(\ref{eq:Psimatricesfree}) and 
the complex rotation matrix in the $i$-$j$ plane $U_{ij} \equiv U_{ij}(\theta_{ij},\delta_{ij})$
in eq.~(\ref{eq:U12comrot}). The latter can be always parametrised
as a product of diagonal phase matrices and the rotation matrix 
$R_{ij} \equiv R_{ij}(\theta_{ij}) = U_{ij}(\theta_{ij},0)$, i.e.,
\be
U_{ij} = P_i(\delta)^* \, R_{ij} \, P_i(\delta) = P_j(-\delta)^* \, R_{ij} \, P_j(-\delta)\,,
\ee
where $P_i(\delta)$ are diagonal matrices 
defined as follows:
\be
P_1(\delta) = \diag(e^{i \delta},1,1)\,, \quad
P_2(\delta) = \diag(1,e^{i \delta},1)\,, \quad
P_3(\delta) = \diag(1,1,e^{i \delta})\,. 
\ee
Defining $P_{ij}(\alpha,\beta)$ as 
a product $P_{ij}(\alpha,\beta) \equiv P_i(\alpha) P_j(\beta)$,
the following relation holds:
\be
U_{ij}(\theta_{ij}, \delta_{ij}) \, P_{ij}(\alpha,\beta) = P_{ij}(\alpha,\beta) U_{ij}  (\theta_{ij}, \delta_{ij}^{\prime}) \,,
\label{eq:UijPij}
\ee
with $\delta_{ij}^{\prime} = \delta_{ij} + \alpha - \beta$.

Starting from the general parametrisation of $W$
in eq.~(\ref{eq:parUPMNS}) and the relation
in eq.~(\ref{eq:UijPij}), we find 
convenient parametrisations for $\overline W$.
They are summarised in Table~\ref{tab:parV}.
The parametrisations of the matrix 
$U^\circ(\theta^{\circ}_{12}, \theta^{\circ}_{13}, \theta^{\circ}_{23}, \{ \delta^{\circ}_{kl} \})$
defined in Section~\ref{sec:prelim} have been obtained from 
Table~\ref{tab:parV} after a redefinition of the phases 
$\{ \delta^{\circ}_{kl} \}$. For example, in the first case when
$U^\circ(\theta^{\circ}_{12}, \theta^{\circ}_{13}, \theta^{\circ}_{23}, \{ \delta^{\circ}_{kl} \})$ 
is represented by the product
$U_{12}(\theta^{\circ}_{12},\delta^{\circ}_{12}) U_{23}(\theta^{\circ}_{23},\delta^{\circ}_{23}) U_{13}(\theta^{\circ}_{13},\delta^{\circ}_{13})$
%
%
%%%%%%%%%%%%%%%%%%%%%%%%%%
\begin{table}[h!]
\centering
\renewcommand*{\arraystretch}{1.4}
\begin{tabular}{lll}
\hline
&  \\ [-14pt]
Case & Initial form of $\overline W$ \phantom{12} & Final parametrisation of $\overline W$ \\ [4pt]
\bottomrule
&  \\ [-14pt]
A1 & $U_{12} U_{23} U_{13}$ &  $P^*_{12} (\delta_{13}, \delta_{23}) U_{12}(\theta_{12}, \delta_{12} - \delta_{13} + \delta_{23}) R_{23} R_{13} P_{12} (\delta_{13}, \delta_{23})$ \\  [4pt]
A2 & $U_{13} U_{23} U_{12}$ & $P^*_{13}(\delta_{12}, -\delta_{23}) U_{13}(\theta_{13}, \delta_{13} - \delta_{12} - \delta_{23}) R_{23} R_{12} P_{13}(\delta_{12}, -\delta_{23})$ \\ [4pt]
A3 & $U_{23} U_{13} U_{12}$ & $P_{23}(\delta_{12}, \delta_{13}) U_{23}(\theta_{23}, \delta_{23} + \delta_{12} - \delta_{13}) R_{13} R_{12} P^*_{23}(\delta_{12}, \delta_{13})$ \\ [4pt]
B1 & $U_{23} U_{12} U_{13}$ & $P_{13}^*(\delta_{12},-\delta_{23}) R_{23} R_{12} U_{13}(\theta_{13}, \delta_{13} - \delta_{12} - \delta_{23}) P_{13}(\delta_{12},-\delta_{23})$ \\ [4pt]
B2 & $U_{13} U_{12} U_{23}$ & $P_{23}(\delta_{12},\delta_{13}) R_{13} R_{12} U_{23}(\theta_{23}, \delta_{23} + \delta_{12} - \delta_{13}) P_{23}^*(\delta_{12},\delta_{13})$ \\ [4pt]
B3 & $U_{23} U_{13} U_{12}$ & $P_{12}^*(\delta_{13},\delta_{23}) R_{23} R_{13} U_{12}(\theta_{12}, \delta_{12} - \delta_{13} + \delta_{23}) P_{12}(\delta_{13},\delta_{23})$ \\ [4pt]
C1 & $U_{12} U_{23} U_{13}$ & $P_{3}(\delta_{23}) U_{12}(\theta_{12}, \delta_{12}) R_{23} U_{13}(\theta_{13}, \delta_{13} - \delta_{23}) P_{3}^*(\delta_{23})$ \\ [4pt]
C2 & $U_{13} U_{23} U_{12}$ & $P_{3}(\delta_{23}) U_{13}(\theta_{13}, \delta_{13} - \delta_{23}) R_{23} U_{12}(\theta_{12}, \delta_{12}) P_{3}^*(\delta_{23})$ \\ [4pt]
C3 & $U_{12} U_{13} U_{23}$ & $P_{3}(\delta_{13}) U_{12}(\theta_{12}, \delta_{12}) R_{13} U_{23}(\theta_{23}, \delta_{23} - \delta_{13}) P_{3}^*(\delta_{13})$ \\ [4pt]
C4 & $U_{13} U_{12} U_{23}$ & $P_{2}(\delta_{12}) U_{13}(\theta_{13}, \delta_{13}) R_{12} U_{23}(\theta_{23}, \delta_{23} + \delta_{12}) P_{2}^*(\delta_{12})$ \\ [4pt]
C5 & $U_{23} U_{12} U_{13}$ & $P_{2}(\delta_{12}) U_{23}(\theta_{23}, \delta_{23} + \delta_{12}) R_{12} U_{13}(\theta_{13}, \delta_{13}) P_{2}^*(\delta_{12})$ \\ [6pt]
C6 & $U_{23} U_{13} U_{12}$ & $P_{3}(\delta_{13}) U_{23}(\theta_{23}, \delta_{23} - \delta_{13}) R_{13} U_{12}(\theta_{12}, \delta_{12}) P_{3}^*(\delta_{13})$ \\ [4pt]
C7 & $U_{12} U_{23} \tilde{U}_{12}$ & $P_3(\delta_{23}) U_{12}(\theta_{12}, \delta_{12}) R_{23} U_{12}(\tilde \theta_{12}, \tilde \delta_{12}) P_3^*(\delta_{23})$ \\ [4pt]
C8 & $U_{13} U_{23} \tilde{U}_{13}$ & $P^*_2(\delta_{23}) U_{13}(\theta_{13}, \delta_{13}) R_{23} U_{13}(\tilde \theta_{13}, \tilde \delta_{13}) P_2(\delta_{23})$ \\ [4pt]
C9 & $U_{23} U_{12} \tilde{U}_{23}$ & $P^*_1(\delta_{12}) U_{23}(\theta_{23}, \delta_{23}) R_{12} U_{23}(\tilde \theta_{23}, \tilde \delta_{23}) P_1(\delta_{12})$ \\ [4pt]
\hline
  \end{tabular}
%%%%%%%%%%%%%%%%%%%%%%%%%%%
\caption{Equivalent parametrisations of $\overline W$ obtained using
the result in eq.~(\ref{eq:UijPij}), which allows us to find the convenient 
form of the matrix $U^\circ(\theta^{\circ}_{12}, \theta^{\circ}_{13}, \theta^{\circ}_{23}, \{ \delta^{\circ}_{kl} \})$
defined in Section~\ref{sec:prelim}.}
\label{tab:parV}
\end{table}
%%%%%%%%%%%%%%%%%%%%%%%%%%
the following redefinition is used: $\delta^{\circ}_{12} - \delta^{\circ}_{13} + \delta^{\circ}_{23} \, \rightarrow \, \delta^{\circ}_{12}$.

The product of two complex rotations in the $i$-$j$ plane
can always be written as
\begin{align}
\label{eq:trick1}
U_{ij}(\theta^a_{ij}, \delta^a_{ij}) U_{ij}(\theta^b_{ij}, \delta^b_{ij}) & = P_{ij}(\beta, -\alpha) R_{ij}(\hat \theta_{ij}) P_i(\alpha - \beta) = P_j(-\alpha - \beta) R_{ij}(\hat \theta_{ij}) P_{ij}(\alpha, \beta) \\
& =  P_{ij}(\alpha, -\beta) R_{ij}(\hat \theta_{ij}) P_j(\beta - \alpha) = P_i(\alpha + \beta) R_{ij}(\hat \theta_{ij}) P_{ij}(-\beta, -\alpha) \nonumber \,,
\end{align}
where we have defined the angle $\hat \theta_{ij}$ as
\begin{align}
\label{eq:trickthhat}
\sin \hat \theta_{ij} & = | s^a_{ij} c^b_{ij} e^{-i \delta^a_{ij}} + c^a_{ij} s^b_{ij} e^{-i \delta^b_{ij}} | \,, 
\end{align}
and the phases $\alpha$, $\beta$ as
\begin{align}
\label{eq:trickalpha}
\alpha & = {\rm{arg}} \big[ c^a_{ij} c^b_{ij} - s^a_{ij} s^b_{ij} e^{i(\delta^b_{ij} - \delta^a_{ij})} \big] \,,\quad
\beta  = {\rm{arg}} \big [ s^a_{ij} c^b_{ij} e^{-i \delta^a_{ij}} + c^a_{ij} s^b_{ij} e^{-i \delta^b_{ij}} \big] \,,
\end{align}
with $s^{a(b)}_{ij} = \sin \theta^{a(b)}_{ij}$ and $c^{a(b)}_{ij} = \cos \theta^{a(b)}_{ij}$.

%%%%%%%%%%%%%%%%%%%%%%%%%%%%%%%%%
\section{Appendix: The Case of Fully Broken $G_e$ or $G_{\nu}$}
\label{app:Gebroken}
%%%%%%%%%%%%%%%%%%%%%%%%%%%%%%%%%
In the case when the group $G_e$ is fully broken and 
$G_{\nu} = Z_n$, $n > 2$ or $Z_n \times Z_m$, $n,m \geq 2$,
there are cases in which one can express
$\cos \delta$ as a function of $\theta_{12}$, $\theta_{13}$, $\theta_{23}$
and $\theta^{\circ}_{12}$, $\theta^{\circ}_{13}$, $\theta^{\circ}_{23}$.
In the cases 
\begin{align*}
\mbox{i) } U_e^\dagger & = U_{23(13)}(\theta^e_{23(13)},\delta^e_{23(13)}) U_{12}(\theta^e_{12},\delta^e_{12}), \\
\mbox{ii) } U_e^\dagger & = U_{12(13)}(\theta^e_{12(13)}, \delta^e_{12(13)}) U_{23}(\theta^e_{23}, \delta^e_{23}), \\
\mbox{iii) } U_e^\dagger & = U_{23(12)}(\theta^e_{23(12)}, \delta^e_{23(12)}) U_{13}(\theta^e_{13}, \delta^e_{13}),
\end{align*}
we choose for convenience, respectively:
\begin{align*}
\mbox{i) } U^\circ(\theta^{\circ}_{12}, \theta^{\circ}_{13}, \theta^{\circ}_{23}, \delta^{\circ}_{12}) & = 
U_{12}(\theta^{\circ}_{12}, \delta^{\circ}_{12}) R_{23}(\theta^{\circ}_{23}) R_{13}(\theta^{\circ}_{13}) \,, \\
\mbox{ii) } U^\circ(\theta^{\circ}_{12}, \theta^{\circ}_{13}, \theta^{\circ}_{23}, \delta^{\circ}_{23} ) & = 
U_{23}(\theta^{\circ}_{23}, \delta^{\circ}_{23}) R_{13}(\theta_{13}^{\circ}) R_{12}(\theta_{12}^{\circ}) \,, \\
\mbox{iii) } U^\circ(\theta^{\circ}_{12}, \theta^{\circ}_{13}, \theta^{\circ}_{23}, \delta^{\circ}_{13} ) & = 
U_{13}(\theta^{\circ}_{13}, \delta^{\circ}_{13}) R_{23}(\theta_{23}^{\circ}) R_{12}(\theta_{12}^{\circ}) \,.
\end{align*}
The possible parametrisations of $U$ presented
in Table~\ref{tab:parUeGebroken} can be obtained
from i), ii) and iii) using eqs.~(\ref{eq:trick1})~--~(\ref{eq:trickalpha}).
The angles $\theta^e_{ij}$, $\hat \theta_{ij}$ and the phases
$\delta^e_{ij}$, $\hat \delta$ are free parameters.
It can be seen from Table~\ref{tab:parUeGebroken} that if one of the fixed angles turns out
to be zero, the number of free parameters reduces from four to three. 
The same situation happens if one of the two free phases is fixed.
Thus, in some of these cases a sum rule for $\cos \delta$ can be derived.

%%%%%%%%%%%%%%%%%%%%%%%%%%
\begin{table}[h!]
\centering
\renewcommand*{\arraystretch}{1.4}
\renewcommand{\tabcolsep}{5.6pt}
\begin{tabular}{ll}
\hline
&  \\ [-14pt]
$U(\theta^e_{12},\theta^e_{13},\theta^e_{23}, \{ \delta^e_{kl} \})$ & Parametrisation of $U$ for fully broken $G_e$ \\ [4pt]
\bottomrule
&  \\ [-14pt]
$U_{23(13)}(\theta^e_{23(13)},\delta^e_{23(13)}) U_{12}(\theta^e_{12},\delta^e_{12})$ &  $U_{23(13)}(\theta^e_{23(13)}, \delta^e_{23(13)}) R_{12}(\hat \theta_{12}) P_1(\hat \delta) R_{23}(\theta^{\circ}_{23}) R_{13}(\theta^{\circ}_{13}) Q_0$ \\  [4pt]
$U_{12(13)}(\theta^e_{12(13)}, \delta^e_{12(13)}) U_{23}(\theta^e_{23}, \delta^e_{23})$ &  $U_{12(13)}(\theta^e_{12(13)}, \delta^e_{12(13)}) R_{23}(\hat \theta_{23}) P_2(\hat \delta) R_{13}(\theta_{13}^{\circ}) R_{12}(\theta_{12}^{\circ}) Q_0$ \\  [4pt]
$U_{23(12)}(\theta^e_{23(12)}, \delta^e_{23(12)}) U_{13}(\theta^e_{13}, \delta^e_{13})$ &  $U_{23(12)}(\theta^e_{23(12)}, \delta^e_{23(12)}) R_{13}(\hat \theta_{13}) P_1(\hat \delta) R_{23}(\theta_{23}^{\circ}) R_{12}(\theta_{12}^{\circ}) Q_0$ \\  [6pt]
\hline
&  \\ [-14pt]
$U(\theta^\nu_{12},\theta^\nu_{13},\theta^\nu_{23}, \{ \delta^\nu_{kl} \})$ & Parametrisation of $U$ for fully broken $G_{\nu}$ \\ [4pt]
\bottomrule
&  \\ [-14pt]
$U_{12}(\theta^\nu_{12},\delta^\nu_{12}) U_{13(23)}(\theta^\nu_{13(23)},\delta^\nu_{13(23)})$ &  $R_{23}(\theta^{\circ}_{23}) R_{13}(\theta^{\circ}_{13}) P_1(\hat \delta) R_{12}(\hat \theta_{12}) U_{13(23)}(\theta^\nu_{13(23)}, \delta^\nu_{13(23)}) Q_0$ \\  [4pt]
$U_{23}(\theta^\nu_{23},\delta^\nu_{23}) U_{12(13)}(\theta^\nu_{12(13)},\delta^\nu_{12(13)})$ &  $R_{13}(\theta^{\circ}_{13}) R_{12}(\theta^{\circ}_{12}) P_2(\hat \delta) R_{23}(\hat \theta_{23}) U_{12(13)}(\theta^\nu_{12(13)}, \delta^\nu_{12(13)}) Q_0$ \\  [4pt]
$U_{13}(\theta^\nu_{13},\delta^\nu_{13}) U_{12(23)}(\theta^\nu_{12(23)},\delta^\nu_{12(23)})$ &  $R_{23}(\theta^{\circ}_{23}) R_{12}(\theta^{\circ}_{12}) P_1(\hat \delta) R_{13}(\hat \theta_{13}) U_{12(23)}(\theta^\nu_{12(23)}, \delta^\nu_{12(23)}) Q_0$ \\  [6pt]
\hline
  \end{tabular}
%%%%%%%%%%%%%%%%%%%%%%%%%%%
\caption{Upper (lower) part. Parametrisations of $U$ in the case of 
fully broken $G_e$ ($G_{\nu}$) and $G_{\nu} = Z_n$, $n > 2$ or $Z_n \times Z_m$, $n,m \geq 2$
($G_e = Z_n$, $n > 2$ or $Z_n \times Z_m$, $n,m \geq 2$)
when $U_e$ ($U_{\nu}$) has particular forms.}
\label{tab:parUeGebroken}
\end{table}
%%%%%%%%%%%%%%%%%%%%%%%%%%

In the case when the group 
$G_e = Z_n$, $n > 2$ or $Z_n \times Z_m$, $n,m \geq 2$ 
and $G_{\nu}$ is fully broken, we consider the following 
forms of the matrix $U_{\nu}$,
\begin{align*}
\mbox{iv) } U_{\nu} & = U_{12}(\theta^\nu_{12},\delta^\nu_{12}) U_{13(23)}(\theta^\nu_{13(23)},\delta^\nu_{13(23)}) Q_0, \\
\mbox{v) } U_{\nu} & = U_{23}(\theta^\nu_{23},\delta^\nu_{23}) U_{12(13)}(\theta^\nu_{12(13)},\delta^\nu_{12(13)}) Q_0, \\
\mbox{vi) } U_{\nu} & = U_{13}(\theta^\nu_{13},\delta^\nu_{13}) U_{12(23)}(\theta^\nu_{12(23)},\delta^\nu_{12(23)}) Q_0 ,
\end{align*}
for which we choose for convenience, respectively:
\begin{align*}
\mbox{iv) } U^\circ(\theta^{\circ}_{12}, \theta^{\circ}_{13}, \theta^{\circ}_{23}, \delta^{\circ}_{12}) & = 
R_{23}(\theta_{23}^{\circ}) R_{13}(\theta_{13}^{\circ}) U_{12}(\theta^{\circ}_{12}, \delta^{\circ}_{12}) \,, \\
\mbox{v) } U^\circ(\theta^{\circ}_{12}, \theta^{\circ}_{13}, \theta^{\circ}_{23}, \delta^{\circ}_{23} ) & = 
R_{13}(\theta_{13}^{\circ}) R_{12}(\theta_{12}^{\circ}) U_{23}(\theta^{\circ}_{23}, \delta^{\circ}_{23}) \,, \\
\mbox{vi) } U^\circ(\theta^{\circ}_{12}, \theta^{\circ}_{13}, \theta^{\circ}_{23}, \delta^{\circ}_{13} ) & = 
R_{23}(\theta_{23}^{\circ}) R_{12}(\theta_{12}^{\circ}) U_{13}(\theta^{\circ}_{13}, \delta^{\circ}_{13}) \,.
\end{align*}
The parametrisations of $U$ in the cases iv), v) and vi)
presented in Table~\ref{tab:parUeGebroken}
have been obtained with eqs.~(\ref{eq:trick1})~--~(\ref{eq:trickalpha}).
The angles $\theta^\nu_{ij}$, $\hat \theta_{ij}$ and the phases
$\delta^\nu_{ij}$, $\hat \delta$ are free parameters.
It can be seen from Table~\ref{tab:parUeGebroken} that if one of the fixed angles turns out
to be zero, the number of free parameters reduces from four to three. 
The same situation happens if one of the two free phases is fixed.
Thus, in some of these cases a sum rule for $\cos \delta$ can be derived.

%%%%%%%%%%%%%%%%%%%%%%%%%%%%%%%%%
\section{Appendix: Results for $G_f = A_5$ and Generalised CP}
\label{app:A5andGCP}
%%%%%%%%%%%%%%%%%%%%%%%%%%%%%%%%%

Models with $A_5$ and GCP symmetry have been recently developed 
by several authors \cite{Ballett:2015wia,DiIura:2015kfa,Li:2015jxa}.
We show that our results for the symmetry group $A_5$
under the same assumptions of \cite{Ballett:2015wia} and 
the same breaking patterns reduce to the one derived in \cite{Ballett:2015wia}.
The results in eqs.~(10), (11), (12) and (14) in \cite{Ballett:2015wia}
lead to the following phenomenologically viable cases:
\begin{align*}
\mbox{i) } U & = \diag (1,i,-i) R_{23} (\theta^{\circ}_{23}) R_{12}(\theta^{\circ}_{12}) \diag (1,-i,i) R_{13} (\theta^{\nu}_{13})\,, \mbox{ for $G_e = Z_3$, $G_{\nu} = Z_2$,} \\
\mbox{ii) } U & = \diag (1,i,-i) R_{23} (\theta^{\circ}_{23}) R_{12}(\theta^{\circ}_{12}) \diag (1,-i,i) R_{13} (\theta^{\nu}_{13}) \,,   \mbox{ for $G_e = Z_5$, $G_{\nu} = Z_2$,} \\
\mbox{iii) } U & = \diag (1,1,-1) R_{23} (\theta^{\circ}_{23}) R_{12}(\theta^{\circ}_{12}) \diag (1,1,-1) R_{13} (\theta^{\nu}_{13}) \,, \mbox{ for $G_e = Z_5$, $G_{\nu} = Z_2$,} \\
\mbox{iv) } U & = R_{13}(\theta^{\circ}_{13}) R_{12}(\theta^{\circ}_{12}) R_{23} (\theta^{\circ}_{23}) \diag (1,1,-1) R_{23} (\theta^{\nu}_{23}) \,, \mbox{ for $G_e = Z_2 \times Z_2$, $G_{\nu} = Z_2$,}
\end{align*}
where we have in i) $\theta^{\circ}_{12} = \sin^{-1}(1/\sqrt{3})$ and $\theta^{\circ}_{23} = -\pi/4$,
ii) $\theta^{\circ}_{12} = \sin^{-1}(1/\sqrt{2 + r})$ and $\theta^{\circ}_{23} = -\pi/4$,
iii) $\theta^{\circ}_{12} = \sin^{-1}(1/\sqrt{2 + r})$ and $\theta^{\circ}_{23} = -\pi/4$,
iv) $\theta^{\circ}_{12} = \sin^{-1} (1/(2 r))$, $\theta^{\circ}_{13} = \sin^{-1}(1/\sqrt{2 + r})$ 
and $\theta^{\circ}_{23} = \sin^{-1}(r/\sqrt{2 + r})$.

Using $(\sin^2 \theta^{\circ}_{12}, \sin^2 \theta^{\circ}_{13}, 
\sin^2 \theta^{\circ}_{23}) = (1/3,0,1/2)$ in the case i), 
the results in eqs.~(\ref{eq:th13GeZorZZGnuZ13nu})~--~(\ref{eq:th12GeZorZZGnuZ13nu}), 
after defining $\hat \theta_{13} = \theta^{\nu}_{13} = \theta$
and setting $\hat \delta_{13} = \delta^{\nu}_{13} = \pi/2$, reduce to
\begin{equation*}
\sin^2 \theta_{13} = \dfrac{2}{3} \sin^2 \theta\,, \quad
\sin^2 \theta_{12} = \dfrac{1}{3 - 2 \sin^2 \theta}\,, \quad
\sin^2 \theta_{23} = \dfrac{1}{2} \quad \mbox{and} \quad 
\cos \delta = 0\,. 
\end{equation*}
Denoting $\hat \theta_{13} = \theta^{\nu}_{13} = \theta$
and setting $\hat \delta_{13} = \delta^{\nu}_{13} = \pi/2$
in the case ii), the results in 
eqs.~(\ref{eq:th13GeZorZZGnuZ13nu})~--~(\ref{eq:th12GeZorZZGnuZ13nu}) reduce to
\begin{equation*}
\sin^2 \theta_{13} = \dfrac{\sin^2 \theta}{1 + (1 - r)^2}\,, \quad
\sin^2 \theta_{12} = \dfrac{1}{1 + r^2 \cos^2 \theta}\,, \quad
\sin^2 \theta_{23} = \dfrac{1}{2} \quad \mbox{and} \quad \cos \delta = 0\,. 
\end{equation*}
The difference between the case iii) and the case ii) 
consists only in the phase $\hat \delta_{13}$ which now is equal to 
$\pi$, $\hat \delta_{13} = \delta^{\nu}_{13} = \pi$.
Therefore while $\sin^2 \theta_{13}$ and $\sin^2 \theta_{12}$ remain unchanged,
we find 
\begin{equation*}
\sin^2 \theta_{23} = \dfrac{1}{2} \dfrac{(\sin \theta- \sqrt{1+ r^2} \cos \theta)^2}{1 + r^2 \cos^2 \theta} \quad
\mbox{and}  \quad |\cos \delta| = 1\,.
\end{equation*}
Finally, in the case iv) from 
eqs.~(\ref{eq:th13GeZorZZGnuZ23nu})~--~(\ref{eq:th12GeZorZZGnuZ23nu}),
defining $\hat \theta_{23} = \theta^{\circ}_{23} - \theta^{\nu}_{23} 
= \theta^{\circ}_{23} - \theta$ and  $\hat \delta_{23} = 0$, we get: 
\begin{align*}
 \sin^2 \theta_{13} & = \dfrac{1 + (1 - r) f(\theta)}{4} \,, \quad \sin^2 \theta_{23}  = \dfrac{1 + r(\cos^2 \theta - \sin 2 \theta)}{3 - (1 - r) f(\theta)}\,, \\ 
\sin^2 \theta_{12} & = \dfrac{1 + (1 - r) (\cos^2 \theta + \sin 2 \theta)}{3 - (1 - r) f(\theta)}\quad 
\mbox{and}  \quad |\cos \delta| = 1\,,
\end{align*}
where $f(\theta) = (\sin^2 \theta - \sin 2 \theta)$.
Therefore the general results derived in Sections~\ref{sec:13nu} and \ref{sec:23nu} 
with the choices as in i), ii), iii) and iv) 
and the additional restriction of the parameters 
due to the presence of GCP allow one
to find the formulae derived in \cite{Ballett:2015wia}.

%%%%%%%%%%%%%%%%%%%%%%
%
\section{Appendix: General Statement}
\label{sec:genstatement}
%
%%%%%%%%%%%%%%%%%%%%%%%%%%%%%

In this appendix we prove the general statement that {\it $Z_2$ symmetries preserved 
in the neutrino and charged lepton sectors can lead to phenomenologically viable
predictions, only if their generators do not belong to the same $Z_2 \times Z_2$ 
subgroup of the original flavour symmetry group}.
We compute the form of $U^{\circ}$
in a model independent way. Given a $Z_2 \times Z_2$
symmetry with elements $Z_2 \times Z_2 = \{ 1, g_1, g_2 , g_3 \}$
and a unitary matrix $V$ such that $V^{\dagger} g_1 V = \diag(1,-1,-1)$,
$V^{\dagger} g_2 V = \diag(-1,1,-1)$, $V^{\dagger} g_3 V = \diag(-1,-1,1)$,
we consider first the case of $G_e = Z_2 = \{ 1, g_i \} $ and $G_{\nu} = Z_2 = \{ 1, g_j \}$
with $i,j = 1,2,3$ for all the cases
C1~--~C9 in Table~\ref{tab:summarysumrulesB}.
In the case C1 (C2) we find that the matrix $U^{\circ}$ reads
\begin{align}
U^{\circ} = \pi_{23} \equiv
\begin{pmatrix}
1  & 0 & 0 \\[2pt]
0  & 0 & 1 \\[2pt]
0 & 1 & 0 \\[2pt]
\end{pmatrix} \mbox{ for $i = j$} \;, \quad
U^{\circ} = 
\begin{pmatrix}
1  & 0 & 0 \\[2pt]
0  & 1 & 0 \\[2pt]
0 & 0 & 1 \\[2pt]
\end{pmatrix} \mbox{ for $i \neq j$} \;,
\label{eq:proof1}
\end{align}
defined up to permutations of the 1st and 3rd (1st and 2nd) columns and
the 1st and 2nd (1st and 3rd) rows. These permutations are not relevant 
because they correspond to a redefinition of the free 
parameters in the transformations $U_{12}(\theta^e_{12}, \delta^e_{12})$,
$U_{13}(\theta^{\nu}_{13},\delta^{\nu}_{13})$ and phase
matrices contributing to the Majorana phases
or removed with a redefinition of the charged lepton fields.
In the case C3 (C6) we find that the matrix $U^{\circ}$ reads
\begin{align}
U^{\circ} = \pi_{13} \equiv
\begin{pmatrix}
0  & 0 & 1 \\[2pt]
0  & 1 & 0 \\[2pt]
1 & 0 & 0 \\[2pt]
\end{pmatrix} \mbox{ for $i = j$}
\;, \quad
U^{\circ} = 
\begin{pmatrix}
1  & 0 & 0 \\[2pt]
0  & 1 & 0 \\[2pt]
0 &  0 & 1 \\[2pt]
\end{pmatrix} \mbox{ for $i \neq j$} \;,
\label{eq:proof2}
\end{align}
defined up to permutations of the 2nd and 3rd (1st and 2nd) columns and
the 1st and 2nd (2nd and 3rd) rows.
For the case C4 (C5)
we find that the matrix $U^{\circ}$ reads
\begin{align}
U^{\circ} = \pi_{12} \equiv
\begin{pmatrix}
0  & 1 & 0 \\[2pt]
1  & 0 & 0 \\[2pt]
0 & 0 & 1 \\[2pt]
\end{pmatrix} \mbox{ for $i = j$}
\;, \quad
U^{\circ} = 
\begin{pmatrix}
1  & 0 & 0 \\[2pt]
0  & 1 & 0 \\[2pt]
0 & 0 & 1 \\[2pt]
\end{pmatrix} \mbox{ for $i \neq j$} \;,
\label{eq:proof3}
\end{align}
defined up to permutations of the 2nd and 3rd (1st and 3rd) columns and
the 1st and 3rd (2nd and 3rd) rows. 
The freedom in permuting the columns
and rows as we described above does not have physical implications
because it represents the freedom to perform a fixed $U(2)$
transformation in the degenerate subspace of the generator of
the corresponding $Z_2$ symmetry. 
For the other cases we find similar results.
Namely, 
\begin{align}
& \mbox{$U^{\circ} = \diag(1,1,1)$ for $i = j$ and $U^{\circ} = \pi_{23(13)}$ for $i \neq j$ for case C7,}\\
& \mbox{$U^{\circ} = \diag(1,1,1)$ for $i = j$ and $U^{\circ} = \pi_{23(12)}$ for $i \neq j$ for case C8,}\\
& \mbox{$U^{\circ} = \diag(1,1,1)$ for $i = j$ and $U^{\circ} = \pi_{13(12)}$ for $i \neq j$ for case C9.}
\label{eq:AppElast}
\end{align}
The cases in eqs.~(\ref{eq:proof1})~--~(\ref{eq:AppElast})
do not lead to phenomenologically viable results because 
some of the elements of the resulting PMNS
mixing matrix equal zero.
The cases when a)~$G_e = Z_2 \times Z_2 = \{ 1, g_1, g_2 , g_3 \}$ and $G_{\nu} = Z_2 = \{ 1, g_j \}$,
b) $G_{\nu} = Z_2 \times Z_2 = \{ 1, g_1, g_2 , g_3 \}$ and  $G_e = Z_2 = \{ 1, g_i \} $,
c) $G_e = Z_2 \times Z_2 = \{ 1, g_1, g_2 , g_3 \}$ and $G_{\nu} = Z_2 \times Z_2 = \{ 1, g_1, g_2 , g_3 \}$
are not phenomenologically viable as well. This can be seen trivially setting one or two of the 
free rotation angles, $\theta^e_{ij}$, $\theta^{\nu}_{kl}$, to zero.

%%%%%%%%%%%%%%%%%%%%%%%%%%%%%%%%%%%%%%%%%%%%%


\begin{thebibliography}{99}

\bibitem{PDG2014}
K.~Nakamura and S.~T.~Petcov,
% {\it ``Neutrino Masses, Mixing and Oscillations''};
in K.~A.~Olive {\it et al.} (Particle Data Group),
Chin.\ Phys.\ C {\bf 38} (2014) 090001.
%%%CITATION = CHPHD,C38,090001;%%

  
  %\cite{Ishimori:2010au}
\bibitem{Ishimori:2010au}
H.~Ishimori, T.~Kobayashi, H.~Ohki, Y.~Shimizu, H.~Okada and M.~Tanimoto,
  %``Non-Abelian Discrete Symmetries in Particle Physics,''
  Prog.\ Theor.\ Phys.\ Suppl.\  {\bf 183} (2010) 1.
%  [arXiv:1003.3552 [hep-th]].
  %%CITATION = ARXIV:1003.3552;%%
  
  
  %\cite{Altarelli:2010gt}
\bibitem{Altarelli:2010gt}
 G.~Altarelli and F.~Feruglio,
  %``Discrete Flavor Symmetries and Models of Neutrino Mixing,''
  Rev.\ Mod.\ Phys.\  {\bf 82} (2010) 2701.
 % [arXiv:1002.0211 [hep-ph]].
  %%CITATION = ARXIV:1002.0211;%%
  
  
 %\cite{King:2013eh}
\bibitem{King:2013eh}
S.~F.~King and C.~Luhn,
  %``Neutrino Mass and Mixing with Discrete Symmetry,''
  Rept.\ Prog.\ Phys.\  {\bf 76} (2013) 056201.
% [arXiv:1301.1340 [hep-ph]].


%\cite{King:2014nza}
\bibitem{King:2014nza}
S.~F.~King, A.~Merle, S.~Morisi, Y.~Shimizu and M.~Tanimoto,
  %``Neutrino Mass and Mixing: from Theory to Experiment,''
  New J.\ Phys.\  {\bf 16} (2014) 045018.
 % [arXiv:1402.4271 [hep-ph]].
  %%CITATION = ARXIV:1402.4271;%% 
  
  
  %\cite{King:2013vna}
\bibitem{King:2013vna}
 S.~F.~King, T.~Neder and A.~J.~Stuart,
  %``Lepton mixing predictions from $\Delta(6n^2)$ family Symmetry,''
  Phys.\ Lett.\ B {\bf 726} (2013) 312.
%  [arXiv:1305.3200 [hep-ph]].
  %%CITATION = ARXIV:1305.3200;%%
  

    %\cite{Hagedorn:2014wha}
\bibitem{Hagedorn:2014wha}
  C.~Hagedorn, A.~Meroni and E.~Molinaro,
  %``Lepton mixing from ?(3$n^2$) and ?(6$n^2$) and CP,''
  Nucl.\ Phys.\ B {\bf 891} (2015) 499.
 % [arXiv:1408.7118 [hep-ph]].
 
 
%\cite{Li:2015jxa}
\bibitem{Li:2015jxa}
  C.~C.~Li and G.~J.~Ding,
  %``Lepton Mixing in $A_5$ Family Symmetry and Generalized CP,''
  JHEP {\bf 1505} (2015) 100.
  %[arXiv:1503.03711 [hep-ph]].
  %%CITATION = ARXIV:1503.03711;%%
 
 
%\cite{DiIura:2015kfa}
\bibitem{DiIura:2015kfa}
  A.~Di Iura, C.~Hagedorn and D.~Meloni,
  %``Lepton mixing from the interplay of the alternating group A$_{5}$ and CP,''
  JHEP {\bf 1508} (2015) 037.
  %[arXiv:1503.04140 [hep-ph]].
  %%CITATION = ARXIV:1503.04140;%%
  %9 citations counted in INSPIRE as of 02 Oct 2015
  
  
%\cite{Ballett:2015wia}
\bibitem{Ballett:2015wia}
  P.~Ballett, S.~Pascoli and J.~Turner,
  %``Mixing angle and phase correlations from A5 with generalised CP and their prospects for discovery,''
  arXiv:1503.07543 [hep-ph].
  
 
%\cite{Petcov:2014laa}
\bibitem{Petcov:2014laa}
S.~T.~Petcov,
  %``Predicting the values of the leptonic CP violation phases in theories with discrete flavour symmetries,''
  Nucl.\ Phys.\ B {\bf 892} (2015) 400.
 % [arXiv:1405.6006 [hep-ph]].
  %%CITATION = ARXIV:1405.6006;%%
  
  
%\cite{Marzocca:2013cr}
\bibitem{Marzocca:2013cr}
D.~Marzocca, S.~T.~Petcov, A.~Romanino and M.~C.~Sevilla,
  %``Nonzero |U_e3| from Charged Lepton Corrections
  % and the Atmospheric Neutrino Mixing Angle,''
  JHEP {\bf 1305} (2013) 073.
 % [arXiv:1302.0423 [hep-ph]].
%%CITATION = ARXIV:1302.0423;%%  


   %\cite{Girardi:2014faa}
\bibitem{Girardi:2014faa}
  I.~Girardi, S.~T.~Petcov and A.~V.~Titov,
  %``Determining the Dirac CP Violation Phase in the Neutrino Mixing Matrix from Sum Rules,''
  Nucl.\ Phys.\ B {\bf 894} (2015) 733;
% [arXiv:1410.8056 [hep-ph]].
%%\cite{Girardi:2015zva}
%\bibitem{Girardi:2015zva}
  I.~Girardi, S.~T.~Petcov and A.~V.~Titov,
  %``Predictions for the Dirac CP Violation Phase in the Neutrino Mixing Matrix,''
  Int.\ J.\ Mod.\ Phys.\ A {\bf 30} (2015) 13,  1530035.
%  [arXiv:1504.02402 [hep-ph]].
  %%CITATION = ARXIV:1504.02402;%%
  
    
%\cite{Girardi:2015vha}
\bibitem{Girardi:2015vha}
  I.~Girardi, S.~T.~Petcov and A.~V.~Titov,
  %``Predictions for the Leptonic Dirac CP Violation Phase: a Systematic Phenomenological Analysis,''
  Eur.\ Phys.\ J.\ C {\bf 75} (2015) 7,  345.
 % [arXiv:1504.00658 [hep-ph]].
  %%CITATION = ARXIV:1504.00658;%%

  
  %\cite{Turner:2015uta}
\bibitem{Turner:2015uta}
  J.~Turner,
  %``Predictions for Leptonic Mixing Angle Correlations and Non-trivial Dirac CP Violation from $A_5$ with Generalised CP Symmetry,''
  arXiv:1507.06224 [hep-ph].
  %%CITATION = ARXIV:1507.06224;%%
  
  
  %\cite{Bilenky:1980cx}
\bibitem{Bilenky:1980cx}
  S.~M.~Bilenky, J.~Hosek and S.~T.~Petcov,
%``On Oscillations of Neutrinos with Dirac and Majorana Masses,''
Phys.\ Lett.\ B {\bf 94} (1980) 495.
%%CITATION = PHLTA,B94,495;%%


%\cite{Girardi:2013sza}
\bibitem{Girardi:2013sza}
I.~Girardi, A.~Meroni, S.~T.~Petcov and M.~Spinrath,
  %``Generalised geometrical CP violation in a T' lepton flavour model,''
  JHEP {\bf 1402} (2014) 050.
 % [arXiv:1312.1966 [hep-ph]].
 
 
\bibitem{TBM} 
%\cite{Harrison:2002er}
%\bibitem{Harrison:2002er} 
  P.~F.~Harrison, D.~H.~Perkins and W.~G.~Scott,
  %``Tri-bimaximal mixing and the neutrino oscillation data,''
  Phys.\ Lett.\ B {\bf 530}, 167 (2002);
 % [hep-ph/0202074];
  %%CITATION = HEP-PH/0202074;%%
%\cite{Harrison:2002kp}
%\bibitem{Harrison:2002kp} 
  P.~F.~Harrison and W.~G.~Scott,
  %``Symmetries and generalizations of tri - bimaximal neutrino mixing,''
  Phys.\ Lett.\ B {\bf 535}, 163 (2002);
%  [hep-ph/0203209];
  %%CITATION = HEP-PH/0203209;%%
%\cite{Xing:2002sw}
%\bibitem{Xing:2002sw} 
  Z.~z.~Xing,
  %``Nearly tri bimaximal neutrino mixing and CP violation,''
  Phys.\ Lett.\ B {\bf 533}, 85 (2002);
 % [hep-ph/0204049];
  %%CITATION = HEP-PH/0204049;%%
%\cite{He:2003rm}
%\bibitem{He:2003rm} 
  X.~G.~He and A.~Zee,
  %``Some simple mixing and mass matrices for neutrinos,''
  Phys.\ Lett.\ B {\bf 560}, 87 (2003);
 % [hep-ph/0301092];
  %%CITATION = HEP-PH/0301092;%%
see also:
  L.~Wolfenstein,
  Phys.\ Rev.\ D {\bf 18} (1978) 958.
  %%CITATION = PHRVA,D18,958;%%
  
  
    \bibitem{BM}
%% \bibitem{bima}
F.~Vissani, hep-ph/9708483;
%%[arXiv:{hep-ph/9708483}];
% %%CITATION = HEP-PH 9708483;%%
V.~D.~Barger, S.~Pakvasa, T.~J.~Weiler and K.~Whisnant,
Phys.\ Lett.\ B {\bf 437} (1998) 107;
%% [arXiv:{hep-ph/9806387}];
%%CITATION = HEP-PH 9806387;%%
A.~J.~Baltz, A.~S.~Goldhaber and M.~Goldhaber,
Phys.\ Rev.\ Lett.\  {\bf 81} (1998) 5730.
%% [arXiv:{hep-ph/9806540}].


  \bibitem{SPPD82}
%%\cite{Petcov:1982ya}
%%\bibitem{Petcov:1982ya}
  S.~T.~Petcov,
  %``On Pseudodirac Neutrinos, Neutrino Oscillations and Neutrinoless Double beta Decay,''
  Phys.\ Lett.\ B {\bf 110} (1982) 245.
  %%CITATION = PHLTA,B110,245;%%
    

%\cite{Datta:2003qg}
\bibitem{Datta:2003qg}
A.~Datta, F.~S.~Ling and P.~Ramond,
  %``Correlated hierarchy, Dirac masses and large mixing angles,''
  Nucl.\ Phys.\ B {\bf 671} (2003) 383.
 % [hep-ph/0306002].
  %%CITATION = HEP-PH/0306002;%%

%%\cite{Everett:2008et}
\bibitem{Everett:2008et}
  L.~L.~Everett and A.~J.~Stuart,
  %``Icosahedral (A(5)) Family Symmetry and the
  % Golden Ratio Prediction for Solar Neutrino Mixing,''
  Phys.\ Rev.\ D {\bf 79} (2009) 085005.
 % [arXiv:0812.1057 [hep-ph]].
  %%CITATION = ARXIV:0812.1057;%%
  
\bibitem{GRAM}
%% %\cite{Kajiyama:2007gx}
%%% \bibitem{Kajiyama:2007gx}
  Y.~Kajiyama, M.~Raidal and A.~Strumia,
  %``The Golden ratio prediction for the solar neutrino mixing,''
  Phys.\ Rev.\ D {\bf 76} (2007) 117301.
%%  [arXiv:0705.4559 [hep-ph]].
 %%CITATION = ARXIV:0705.4559;%%


\bibitem{GRBM}
%%%\cite{Rodejohann:2008ir}
%%% \bibitem{Rodejohann:2008ir}
  W.~Rodejohann,
  %``Unified Parametrization for Quark and Lepton Mixing Angles,''
  Phys.\ Lett.\ B {\bf 671} (2009) 267;
  % [arXiv:0810.5239 [hep-ph]].
  %%CITATION = ARXIV:0810.5239;%%
%%% \cite{Adulpravitchai:2009bg}
%%% \bibitem{Adulpravitchai:2009bg}
  A.~Adulpravitchai, A.~Blum and W.~Rodejohann,
  %``Golden Ratio Prediction for Solar Neutrino Mixing,''
  New J.\ Phys.\  {\bf 11} (2009) 063026.
  % [arXiv:0903.0531 [hep-ph]].
  %%CITATION = ARXIV:0903.0531;%%

\bibitem{HGM}
%%% \cite{Albright:2010ap}
%%% \bibitem{Albright:2010ap}
  C.~H.~Albright, A.~Dueck and W.~Rodejohann,
  %``Possible Alternatives to Tri-bimaximal Mixing,''
  Eur.\ Phys.\ J.\ C {\bf 70} (2010) 1099;
  %[arXiv:1004.2798 [hep-ph]].
  %%CITATION = ARXIV:1004.2798;%%
%%\cite{Kim:2010zub}
%%\bibitem{Kim:2010zub}
  J.~E.~Kim and M.~S.~Seo,
  %``Quark and lepton mixing angles with a dodeca-symmetry,''
  JHEP {\bf 1102} (2011) 097.
  % [arXiv:1005.4684 [hep-ph]].
  %%CITATION = ARXIV:1005.4684;%%


%%\cite{Frampton:2004ud}
\bibitem{Frampton:2004ud}
  P.~H.~Frampton, S.~T.~Petcov and W.~Rodejohann,
%``On deviations from bimaximal neutrino mixing,''
Nucl.\ Phys.\ B {\bf 687} (2004) 31.
%  %[hep-ph/0401206].
%%CITATION = HEP-PH/0401206;%%


%\cite{Molinaro:2008rg}
\bibitem{Molinaro:2008rg}
  E.~Molinaro and S.~T.~Petcov,
  %``The Interplay Between the 'Low' and 'High' Energy CP-Violation 
  % in Leptogenesis,''
  Eur.\ Phys.\ J.\ C {\bf 61} (2009) 93.
 % [arXiv:0803.4120 [hep-ph]].
  %%CITATION = ARXIV:0803.4120;%%


%\cite{Capozzi:2013csa}
\bibitem{Capozzi:2013csa}
  F.~Capozzi, G.~L.~Fogli, E.~Lisi, A.~Marrone, D.~Montanino and A.~Palazzo,
  %``Status of three-neutrino oscillation parameters, circa 2013,''
  Phys.\ Rev.\ D {\bf 89} (2014) 093018.
  %[arXiv:1312.2878 [hep-ph]].

  
%\cite{Gonzalez-Garcia:2014bfa}
\bibitem{Gonzalez-Garcia:2014bfa}
  M.~C.~Gonzalez-Garcia, M.~Maltoni and T.~Schwetz,
  %``Updated fit to three neutrino mixing: status of leptonic CP violation,''
  JHEP {\bf 1411} (2014) 052.
  %[arXiv:1409.5439 [hep-ph]].


  %\cite{Feruglio:2007uu}
\bibitem{Feruglio:2007uu}
F.~Feruglio, C.~Hagedorn, Y.~Lin and L.~Merlo,
  %``Tri-bimaximal Neutrino Mixing and Quark Masses from a Discrete Flavour Symmetry,''
  Nucl.\ Phys.\ B {\bf 775} (2007) 120
   [Nucl.\ Phys.\ B {\bf 836} (2010) 127].
%  [hep-ph/0702194].
  %%CITATION = HEP-PH/0702194;%%

  
  %\cite{Altarelli:2009gn}
\bibitem{Altarelli:2009gn}
G.~Altarelli, F.~Feruglio and L.~Merlo,
  %``Revisiting Bimaximal Neutrino Mixing in a Model with S(4) Discrete Symmetry,''
  JHEP {\bf 0905} (2009) 020.
%  [arXiv:0903.1940 [hep-ph]].
  %%CITATION = ARXIV:0903.1940;%%

  
  %\cite{Gehrlein:2014wda}
\bibitem{Gehrlein:2014wda}
J.~Gehrlein, J.~P.~Oppermann, D.~Sch\"{a}fer and M.~Spinrath,
  %``An SU(5) $\times$ A$_5$ golden ratio flavour model,''
  Nucl.\ Phys.\ B {\bf 890} (2014) 539.
 % [arXiv:1410.2057 [hep-ph]].

  
%\cite{Meroni:2012ty}
\bibitem{Meroni:2012ty}
A.~Meroni, S.~T.~Petcov and M.~Spinrath,
  %``A SUSY SU(5)xT' Unified Model of Flavour with large $\theta_{13}$,''
  Phys.\ Rev.\ D {\bf 86} (2012) 113003.
%[arXiv:1205.5241 [hep-ph]].


%\cite{Marzocca:2011dh}
\bibitem{Marzocca:2011dh}
D.~Marzocca, S.~T.~Petcov, A.~Romanino and M.~Spinrath,
  %``Sizeable $\theta_{13}$ from the Charged Lepton Sector in SU(5), (Tri-)Bimaximal Neutrino Mixing and Dirac CP Violation,''
  JHEP {\bf 1111} (2011) 009.
  %[arXiv:1108.0614 [hep-ph]].


%\cite{Antusch:2012fb}
\bibitem{Antusch:2012fb}
S.~Antusch, C.~Gross, V.~Maurer and C.~Sluka,
  %``\theta^PMNS_13 = \theta_C / \sqrt2 from GUTs,''
  Nucl.\ Phys.\ B {\bf 866} (2013) 255.
 % [arXiv:1205.1051 [hep-ph]].


 %\cite{Chen:2009gf}
\bibitem{Chen:2009gf}
M.~C.~Chen and K.~T.~Mahanthappa,
  %``Group Theoretical Origin of CP Violation,''
  Phys.\ Lett.\ B {\bf 681} (2009) 444;
  M.~C.~Chen, J.~Huang, K.~T.~Mahanthappa and A.~M.~Wijangco,
  %``Large $\theta_{13}$ in a SUSY SU(5) x T' Model,''
  JHEP {\bf 1310} (2013) 112.
  % [arXiv:1307.7711].
%%CITATION = ARXIV:1307.7711;%%


%\cite{Marzocca:2014tga}
\bibitem{Marzocca:2014tga}
  D.~Marzocca and A.~Romanino,
  %``Stable fermion mass matrices and the charged lepton contribution to neutrino mixing,''
  JHEP {\bf 1411} (2014) 159.
%  [arXiv:1409.3760 [hep-ph]].
  %%CITATION = ARXIV:1409.3760;%%
    
  
\bibitem{PKSP3nu88}
%\cite{Krastev:1988yu}
%\bibitem{Krastev:1988yu}
P.~I.~Krastev and S.~T.~Petcov,
  %``Resonance Amplification and t Violation Effects in Three Neutrino Oscillations in the Earth,''
  Phys.\ Lett.\ B {\bf 205} (1988) 84. 

  
  %\cite{Branco:1986gr}
\bibitem{Branco:1986gr}
  G.~C.~Branco, L.~Lavoura and M.~N.~Rebelo,
  %``Majorana Neutrinos and {CP} Violation in the Leptonic Sector,''
  Phys.\ Lett.\ B {\bf 180} (1986) 264.
  %%CITATION = PHLTA,B180,264;%%
  
  
  %\cite{Feruglio:2012cw}
\bibitem{Feruglio:2012cw}
  F.~Feruglio, C.~Hagedorn and R.~Ziegler,
  %``Lepton Mixing Parameters from Discrete and CP Symmetries,''
  JHEP {\bf 1307} (2013) 027.
%  [arXiv:1211.5560 [hep-ph]].
  %%CITATION = ARXIV:1211.5560;%%
  

  %\cite{Chen:2014wxa}
\bibitem{Chen:2014wxa}
  P.~Chen, C.~C.~Li and G.~J.~Ding,
  %``Lepton Flavor Mixing and CP Symmetry,''
  Phys.\ Rev.\ D {\bf 91} (2015) 3,  033003.
%  [arXiv:1412.8352 [hep-ph]].
  %%CITATION = ARXIV:1412.8352;%%
  
  
  %\cite{Everett:2015oka}
\bibitem{Everett:2015oka}
  L.~L.~Everett, T.~Garon and A.~J.~Stuart,
  %``A Bottom-Up Approach to Lepton Flavor and CP Symmetries,''
  JHEP {\bf 1504} (2015) 069.
%  [arXiv:1501.04336 [hep-ph]].
  %%CITATION = ARXIV:1501.04336;%%
  
  
  %\cite{Holthausen:2012dk}
\bibitem{Holthausen:2012dk}
  M.~Holthausen, M.~Lindner and M.~A.~Schmidt,
  %``CP and Discrete Flavour Symmetries,''
  JHEP {\bf 1304} (2013) 122.
%  [arXiv:1211.6953 [hep-ph]].
  %%CITATION = ARXIV:1211.6953;%%
  
  
  %\cite{Chen:2014tpa}
\bibitem{Chen:2014tpa}
  M.~C.~Chen, M.~Fallbacher, K.~T.~Mahanthappa, M.~Ratz and A.~Trautner,
  %``CP Violation from Finite Groups,''
  Nucl.\ Phys.\ B {\bf 883} (2014) 267.
%  [arXiv:1402.0507 [hep-ph]].
  %%CITATION = ARXIV:1402.0507;%%
  
     
  %\cite{Altarelli:2005yx}
\bibitem{Altarelli:2005yx}
G.~Altarelli and F.~Feruglio,
  %``Tri-bimaximal neutrino mixing, A(4) and the modular symmetry,''
  Nucl.\ Phys.\ B {\bf 741} (2006) 215.
  %[hep-ph/0512103].
  %%CITATION = HEP-PH/0512103;%%
  

%\cite{Li:2014eia}
\bibitem{Li:2014eia}
  C.~C.~Li and G.~J.~Ding,
  %``Deviation from bimaximal mixing and leptonic CP phases in S$_{4}$ family symmetry and generalized CP,''
  JHEP {\bf 1508} (2015) 017.
  %[arXiv:1408.0785 [hep-ph]].
  %%CITATION = ARXIV:1408.0785;%%
  
  
  %\cite{Ding:2011cm}
\bibitem{Ding:2011cm}
G.~J.~Ding, L.~L.~Everett and A.~J.~Stuart,
  %``Golden Ratio Neutrino Mixing and $A_5$ Flavor Symmetry,''
  Nucl.\ Phys.\ B {\bf 857} (2012) 219.
%  [arXiv:1110.1688 [hep-ph]].
  %%CITATION = ARXIV:1110.1688;%%
  

  %\cite{Shimizu:2014ria}
\bibitem{Shimizu:2014ria}
 Y.~Shimizu, M.~Tanimoto and K.~Yamamoto,
  %``Predicting CP violation in Deviation from Tri-bimaximal mixing of Neutrinos,''
  Mod.\ Phys.\ Lett.\ A {\bf 30} (2015) 1550002.
%  [arXiv:1405.1521 [hep-ph]].
  %%CITATION = ARXIV:1405.1521;%%


  %\cite{Ding:2013bpa}
\bibitem{Ding:2013bpa}
G.~J.~Ding, S.~F.~King and A.~J.~Stuart,
  %``Generalised CP and $A_4$ Family Symmetry,''
  JHEP {\bf 1312} (2013) 006.
%  [arXiv:1307.4212 [hep-ph]].
  %%CITATION = ARXIV:1307.4212;%%
 

%\cite{Fritzsch:1997st}
\bibitem{Fritzsch:1997st}
  H.~Fritzsch and Z.~z.~Xing,
  %``On the parametrization of flavor mixing in the standard model,''
  Phys.\ Rev.\ D {\bf 57} (1998) 594.
%  [hep-ph/9708366].
  %%CITATION = HEP-PH/9708366;%%
  
  
  %\cite{Rasin:1997pn}
\bibitem{Rasin:1997pn}
  A.~Rasin,
  %``Diagonalization of quark mass matrices and the Cabibbo-Kobayashi-Maskawa matrix,''
  hep-ph/9708216.
  %%CITATION = HEP-PH/9708216;%%
  
  
  %\cite{Rodejohann:2011uz}
\bibitem{Rodejohann:2011uz}
  W.~Rodejohann, H.~Zhang and S.~Zhou,
  %``Systematic search for successful lepton mixing patterns with nonzero $\theta_{13}$,''
  Nucl.\ Phys.\ B {\bf 855} (2012) 592.
%  [arXiv:1107.3970 [hep-ph]].
  %%CITATION = ARXIV:1107.3970;%%
  
    


\end{thebibliography}
\end{document}